Research Article

# On Laplace Equation Solution in Orthogonal Similar Oblate Spheroidal Coordinates

**Pavel Strunz**

Nuclear Physics Institute of the CAS, 25068, Řež, Czech Republic
E-mail: strunz@ujf.cas.cz

**Abstract:** Orthogonal coordinate systems enable expressing the boundary conditions of differential equations in accord with the physical boundaries of the problem. It can significantly simplify calculations. The orthogonal similar oblate spheroidal (SOS) coordinate system can be particularly useful for a physical processes description inside or in the vicinity of the bodies or particles with the geometry of an oblate spheroid. The interior solution of the Laplace equation in the SOS coordinates was recently found; however, the exterior solution was missing. The exterior solution of the azimuthally symmetric Laplace equation in the SOS coordinates is derived. In the steps leading to this solution, important formulas of the SOS algebra are found. Various forms of the Laplace operator in the SOS coordinates in azimuthally symmetric case are shown. General transformation between two different SOS coordinate systems is derived. It is determined that the SOS harmonics are physically the same as the solid harmonics. Further, a formula expressing any generalized Legendre polynomial as a finite sum of monomials is found. The reported relations have potential application in geophysics, astrophysics, electrostatics, electromagnetism, fluid flow and solid state physics (e.g., ferroic inclusions).

## 1. Introduction

Curvilinear coordinate systems are valuable tools particularly for differential equations solutions, especially in the potential theory. They enable expressing the boundary conditions of differential equations in a way greatly simplifying the calculations when the coordinate surfaces fit the physical boundaries of the problem [1]. A range of field problems that can be handled effectively depends on the number of well developed coordinate systems. Amongst them, the orthogonal curvilinear coordinate systems [2] are the most useful.

Recently, the earlier suggested [3] similar oblate spheroidal (SOS) orthogonal coordinate system was finalized [4, 5]. The similar oblate spheroidal coordinates are distinct from all the well known standard orthogonal coordinate systems [2, 6], the confocal oblate spheroidal (COS) system not excluding. The SOS system can be a powerful tool for a description of field or physical processes inside or in the vicinity of the bodies or particles with an oblate spheroidal geometry. Such objects range (but are not limited to) from ferroic nanoparticles through planets up to galaxies.

In electrostatics and solid state physics, the SOS coordinates could find application in description of electric field potential in ferroelectric materials (e.g., ferroelectric nanocomposites) containing dielectric inclusions (or vice versa-ferroelectric inclusions in dielectric matrix) [7], particularly the inclusions of spheroidal shape [8].

Other areas where the novel coordinates could facilitate solution of physical problems are fluid flow and electromagnetic wave scattering in an environment exhibiting special geometry. The use of specific orthogonal coordinate systems is of a large advantage in such cases [9–11]. It was also shown that using non-Cartesian coordinates (particularly spheroidal) can bring advantages for solution of Stokes flow [12]. Solution of differential equations (continuity, momentum, motion), including Laplace and Poisson equations, connected to such problems can be obviously achieved more simply within a proper coordinate system.

In the field of atmospheric physics, the SOS coordinates could be of help for better modeling of geopotential surfaces, allowing for a better description of the spatial variation of the apparent gravity [3]. In astrophysics, similar oblate spheroids are frequently used for modeling of iso-density levels inside galaxies [13, 14].

Although the derived basic relations for the SOS system are analytical, they cannot be expressed in a closed form. They employ convergent infinite power series with generalized binomial coefficients, which can be-nevertheless-handled with the use of combinatorial identities. This fact makes from calculations in the SOS system an interesting mathematical topic.

The analytical coordinate transformation from the SOS coordinates to the Cartesian system and the metric scale factors were already determined [4, 5], as well as the formulas necessary for the transformation of a vector field between the SOS system and the Cartesian coordinates [5].

Recently, the Laplace equation solution for the interior space was derived [15] in the SOS coordinates. Nevertheless, the found harmonic functions cannot be used or easily extended to the exterior space (i.e., for the space ranging up to infinity). Therefore, a separate solution valid in the exterior has to be searched for.

This article deals with solution of Laplace differential equation in the SOS coordinates in the exterior space for azimuthally symmetric case. A full solution of the Laplace equation would represent an important step in the development of the SOS coordinate system and its applicability in physics and other fields. The complete determination of harmonic functions would also help to find solutions of more complex differential equations in SOS coordinates.

The organization of the text is following. First, a summary of the SOS coordinate system is provided together with the already found relations relevant for the present derivation. Then, preparatory considerations which enable achieving the aim of the article are carried out. Finally, the exterior solution is found, and the complete solution of the Laplace equation in the azimuthally symmetric case in the SOS coordinates is reported.

As some derivations within the article are very lengthy when carried out in a full detail, a large part of them is moved to Supplements to this article.

# 2. Summary of the previous results

## 2.1 *SOS coordinates*

The similar oblate spheroidal coordinates, introduced by White et al. [3], are qualitatively different from the standard orthogonal coordinate systems [2, 6] thanks to the fact that the second coordinate surfaces family is not of the second degree or of the fourth degree but it is formed by general power functions $z \sim x^{1+\mu}$ (i.e., with a real-number exponent $1+\mu$) rotated around the $z$-axis [3, 4]. These are orthogonal to the similar oblate spheroids representing the first set of the coordinate surfaces (For terminological clarity, a spheroid means in this text an ellipsoid of revolution or rotational ellipsoid. An oblate spheroid is a quadric surface obtained by rotating an ellipse about the shorter principal axis).

For the SOS coordinate system $(R, \nu, \lambda)$ [4], the basic coordinate surfaces of the $R$ coordinate are similar oblate spheroids given in 3D Cartesian coordinates $(x_{3D}, y_{3D}, z_{3D})$ by the formula

$$x_{3D}^2 + y_{3D}^2 + (1+\mu) z_{3D}^2 = R^2 \tag{1}$$

The $R$ coordinate value is equal to the equatorial radius of the particular spheroid from the family. The parameter $\mu$ characterizes the oblateness of the whole family of the similar oblate spheroids (the larger the parameter $\mu$, the flatter the spheroid). The parameter $\mu > 0$ for oblate spheroids. The minor and the major semi-axes of each member of the spheroid family have the ratio $(1+\mu)^{1/2}$. As a limit (when $\mu = 0$), a sphere (and spherical coordinate surfaces) is obtained by (1). A special reference spheroid is introduced with the equatorial radius $R_0$, usually coinciding with the reference surface of the object for which the SOS coordinate system is to be applied.

The second set of the coordinate surfaces, orthogonal to the similar oblate spheroids defined above, are power functions in 3D of the form [3, 4]

$$z_{3D} = \frac{1}{\sqrt{1+\mu}} \frac{1}{R_0^{\mu}} \frac{\sin \nu}{\cos^{1+\mu} \nu} \left( \sqrt{x_{3D}^2 + y_{3D}^2} \right)^{1+\mu} \tag{2}$$

The labeling, i.e., the coordinate $\nu$ corresponding to these surfaces, is equivalent to the so called parametric latitude [4]. The coordinate $\nu$ is also equivalent to the parameter used for the standard parametric equation of the ellipse, representing the meridional section of the reference spheroid mentioned above, having a special equatorial radius (major semi-axis) equal to $R_0$, i.e.,

$$\sqrt{x_{3D}^2 + y_{3D}^2} = R_0 \cos \nu \quad \text{and} \quad z_{3D} = \frac{R_0}{\sqrt{1+\mu}} \sin \nu \tag{3}$$

Finally, the third set of the coordinate surfaces, orthogonal to the previous two, are semi-infinite planes containing the rotation axis. The associated coordinate is the longitude angle $\lambda$, which is the same as its equivalent coordinate in the spherical coordinate system.

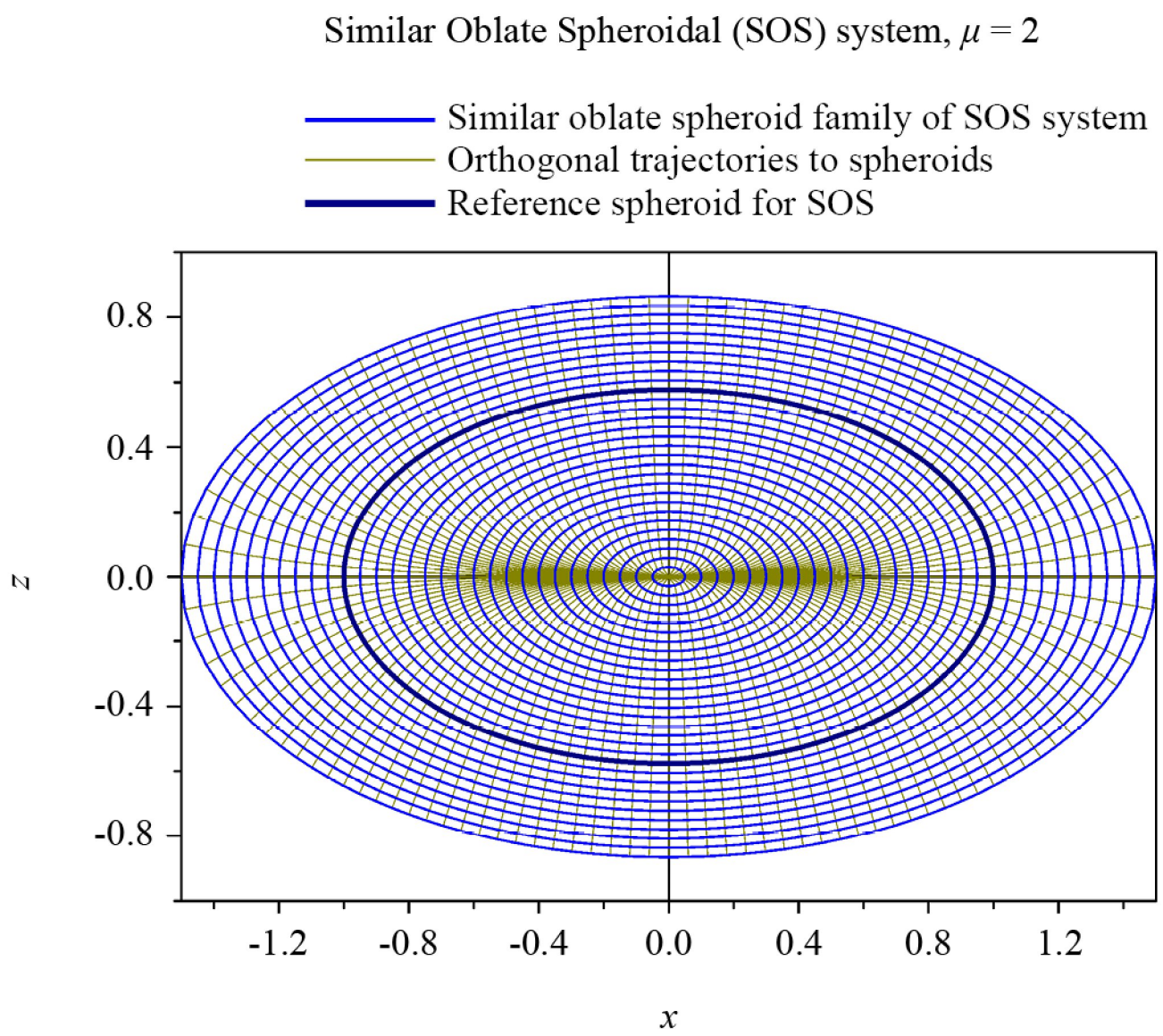

**Figure 1.** The cross section ($x$-$z$) of the SOS coordinate system along the meridian. The sets of the orthogonal SOS coordinate-system lines for the coordinates $R$ and $\nu$ are displayed. The section of the reference spheroid with the equatorial radius $R_0 = 1$ is shown as well

Figure 1 displays the $x$-$z$ cross section of the SOS coordinate system along the meridional plane for the oblateness parameter $\mu = 2$. The full description of the analytical solution of the SOS coordinates can be found in [4]. A shortened summary is reported in [5]. A limited description (to the extent needed for the tasks of this article) is given also further in this section.

## 2.2 *Basic relations of the SOS coordinate system*

A key role in the derivation of the SOS coordinate algebra (e.g., the coordinates transformation between the SOS and the Cartesian system [4]) plays the dimensionless real-number parameter $W$ defined as

$$W = \left(\frac{R}{R_0}\right)^{\mu} \frac{\sin \nu}{\cos^{1+\mu} \nu} \tag{4}$$

The constant-$W$ surfaces are straight half-lines starting at the origin and rotated around the system axis, i.e., cones [4]. In what follows, the calculation is restricted to the parametric latitude $\nu \in \left[0, \frac{\pi}{2}\right]$, i.e., to the first quadrant in Figure 1. With this limitation, the parameter $W$, see (4), is always non-negative which simplifies further derivations. Due to the symmetry (reflection with respect to the equator), expressions and solutions relevant for the SOS system in the complementary range $\nu \in \left[-\frac{\pi}{2}, 0\right)$ can be easily obtained.

The SOS coordinates can be transformed to the Cartesian coordinates using analytical expressions including infinite power series with generalized binomial coefficients [4, 5]. Relations enabling such an approach were first reported by Pólya and Szegö [16]. Due to the convergence limits of the involved power series, the expressions have to be derived separately in two regions: the so called "small-$\nu$ region" and separately in the "large-$\nu$ region" [4]. The boundary between the two regions is defined by the parameter $W$ (see (5)) value fulfilling the following relation:

$$W_b(R, \nu) = \sqrt{\frac{\mu^{\mu}}{(1+\mu)^{1+\mu}}} = \text{constant} \tag{5}$$

This boundary surface is a straight half-line starting at the origin and rotated around the symmetry axis (and forming thus the surface of a cone) [4, 5]. Although $W_b$ is a constant for a fixed $\mu$, the coordinates $R$ and $\nu$ vary along this half-line.

The metric scale factors-separately in the small-$\nu$ region and in the large-$\nu$ region-for $R$ coordinate,

$$h_R = \sqrt{\sum_{k=0}^{\infty} \binom{-\mu k}{k} (W^2)^k} \text{ and } h_R = \frac{1}{\sqrt{1+\mu}} \sqrt{\sum_{k=0}^{\infty} \binom{\frac{\mu}{1+\mu}k}{k} \left(W^{-\frac{2}{1+\mu}}\right)^k} \tag{6}$$

as well as for the $\nu$ coordinate,

$$h_\nu = \frac{R}{\sqrt{1+\mu}} \frac{\partial W}{\partial \nu} \sqrt{\sum_{k=0}^{\infty} \binom{-(\mu+2)-\mu k}{k} (W^2)^k} \text{ and}$$
$$h_\nu = \frac{R}{1+\mu} W^{-\frac{2+\mu}{1+\mu}} \frac{\partial W}{\partial \nu} \sqrt{\sum_{k=0}^{\infty} \binom{-\frac{2+\mu}{1+\mu} + \frac{\mu}{1+\mu}k}{k} \left(W^{-\frac{2}{1+\mu}}\right)^k} \tag{7}$$

were determined [4]. The left side relations show the formulae that hold for the small-$\nu$ region while the right side relations show the formulae that hold for the large-$\nu$ region.

The functions for the generalized cosine ($f_C$) and for the generalized sine ($f_S$) in the frame of the SOS coordinate system were found as well [5]:

$$f_C = \sqrt{\sum_{k=0}^{\infty} \binom{-1-\mu k}{k} (W^2)^k} \quad \text{and} \quad f_C = \frac{W^{-\frac{1}{1+\mu}}}{\sqrt{1+\mu}} \sqrt{\sum_{k=0}^{\infty} \binom{-\frac{1}{1+\mu} + \frac{\mu}{1+\mu}k}{k} \left(W^{-\frac{2}{1+\mu}}\right)^k} \tag{8}$$

$$f_S = W\sqrt{1+\mu}\sqrt{\sum_{k=0}^{\infty} \binom{-(1+\mu)-\mu k}{k} (W^2)^k} \quad \text{and} \quad f_S = \sqrt{\sum_{k=0}^{\infty} \binom{-1+\frac{\mu}{1+\mu}k}{k} \left(W^{-\frac{2}{1+\mu}}\right)^k} \tag{9}$$

The left sides show the formulae that hold for the small-$\nu$ region while the right sides show the formulae that hold for the large-$\nu$ region. Note that $f_S$, $f_C$ depend solely on $W$ (not separately on $R$ and $\nu$).

For the solution of the Laplace equation, it is of advantage to define a new function $s(W)$ as

$$s \equiv \frac{f_S}{h_R} \tag{10}$$

Using (9) and (6) and the known binomial identities [5], the power-series expressions of $s$ are as follows:

$$s = W\sqrt{1+\mu}\sqrt{\sum_{k=0}^{\infty} \frac{-(1+\mu)}{-(1+\mu)-\mu k} \binom{-(1+\mu)-\mu k}{k} (W^2)^k} \tag{11}$$

in the small-$\nu$ region, and

$$s = \sqrt{1+\mu}\sqrt{\sum_{k=0}^{\infty} \frac{-1}{-1+\frac{\mu}{1+\mu}k} \binom{-1+\frac{\mu}{1+\mu}k}{k} \left(W^{-\frac{2}{1+\mu}}\right)^k} \tag{12}$$

in the large-$\nu$ region [5].

The functions listed in this sub-section contain infinite power series with generalized binomial coefficients. It is possible to deal with relations involving them with a help of known combinatorial identities [5, 16–20].

### 2.3 *Laplacian in the SOS coordinates*

The Laplace operator of a general scalar function $V(R, \nu)$, for example a scalar potential, in the SOS coordinates in the azimuthally symmetric case with the spheroid-coordinate surfaces of the $R$ coordinate described by the oblateness parameter $\mu$ is

$$\Delta V = \frac{1}{\mathfrak{I}}\left[\frac{\partial}{\partial R}\left(\frac{\mathfrak{I}}{h_R^2}\frac{\partial V(R,\nu)}{\partial R}\right) + \frac{\partial}{\partial \nu}\left(\frac{\mathfrak{I}}{h_\nu^2}\frac{\partial V(R,\nu)}{\partial \nu}\right)\right] \tag{13}$$

In the case when $V$ can be separated to an $R$-dependent part and a $W$-dependent part, i.e., $V = r(R)F(W)$, the Laplacian in the SOS coordinates in the azimuthally symmetric case can be rewritten with the help of product and chain rules for derivatives to the form

$$\Delta V = \frac{1}{h_R^2}\frac{d^2r(R)}{dR^2}F(W) + \frac{1}{R}\frac{dr(R)}{dR}\left[(\mu+3) - \frac{1}{h_R^2}\right]F(W) + \frac{1}{R^2}r(R)\left[\frac{(\mu-2)\mu}{h_R^2} + (\mu+3)\mu + \frac{h_R^2(1+\mu)^2}{f_C^2}\right]$$
$$W\frac{dF(W)}{dW} + 2\frac{1}{R}\frac{dr(R)}{dR}\frac{\mu W}{h_R^2}\frac{dF(W)}{dW} + \frac{1}{R^2}r(R)\left(\frac{\mu^2}{h_R^2} + \frac{h_R^2(1+\mu)^2}{f_C^2 f_S^2}\right)W^2\frac{d^2F(W)}{dW^2} \tag{14}$$

(deduce it from (B20) in the Supplement B of the interior-solution article [15] and from the steps leading to it). The Laplacian (14) is a sum of terms which are always a product of the $W$-dependent expression and the $R$-dependent expression.

The Laplacian can be also expressed in $s$-terms:

$$\Delta V = \frac{1}{1+\mu}\left((1+\mu)+\mu s^2\right)F(s)\frac{d^2r(R)}{dR^2} + \left[(\mu+2) - \frac{\mu}{1+\mu}s^2\right]F(s)\frac{1}{R}\frac{dr(R)}{dR} + 2\frac{\mu}{1+\mu}\left[(1+\mu)-s^2\right]$$
$$s\frac{dF(s)}{ds}\frac{1}{R}\frac{dr(R)}{dR} + \frac{3\mu s^2 - \left(3\mu^2+5\mu+2\right)}{1+\mu}s\frac{dF(s)}{ds}\frac{r(R)}{R^2} + \frac{\left[(1+\mu)-s^2\right]\left[(1+\mu)^2-\mu s^2\right]}{1+\mu}\frac{d^2F(s)}{ds^2}\frac{r(R)}{R^2} \tag{15}$$

This relation is derived in Appendix A.

When using the relations (14) or (15) for the Laplacian, the Laplace equation cannot be separated by a standard separation procedure (see e.g., [2]). Nevertheless, a special separation procedure described in [15] can be successfully employed.

For completeness, the Laplacian in the SOS coordinates is derived for the axially symmetric case for the non-separable function $V_N$ in Supplement 1 as well (see (S1.23) in the Supplement). The steps leading to the result are similar to the ones described in Appendix A for the separable function $V$. The result for the non-separable function $V_N$ is

$$\Delta V_N = \left[\frac{\mu s^2+(1+\mu)}{1+\mu}\frac{\partial^2 V_N}{\partial R^2} + (1+\mu)^2\frac{(1+\mu)-s^2}{\mu s^2+(1+\mu)}\frac{1}{R^2}\frac{\partial^2 V_N}{\partial s^2}\right]$$
$$+\left[\frac{(1+\mu)(\mu+2)-\mu s^2}{1+\mu}\frac{1}{R}\frac{\partial V_N}{\partial R} - 2(1+\mu)^4\frac{s}{\left[\mu s^2+(1+\mu)\right]^2}\frac{1}{R^2}\frac{\partial V_N}{\partial s}\right] \tag{16}$$

This represents the Laplacian expressed in $R$ and $s$ terms. Going still further, one can rewrite the derivatives in (16) from the form “with respect to $s$” to the form “with respect to $v$” (see Supplement 1, relation (S1.39)):

$$\Delta V_N = \frac{\mu s^2+(1+\mu)}{1+\mu}\frac{\partial^2 V_N}{\partial R^2} + \frac{(1+\mu)(\mu+2)-\mu s^2}{1+\mu}\frac{1}{R}\frac{\partial V_N}{\partial R} + \frac{(1+\mu)^2}{s^2}\frac{\mu s^2+(1+\mu)}{(1+\mu)-\ s^2}\frac{\sin^2\nu\ \cos^2\nu}{\left(1+\mu\ \sin^2\nu\right)^2}\frac{1}{R^2}\frac{\partial^2 V_N}{\partial \nu^2}$$

$$+\frac{(1+\mu)^2}{[(1+\mu)-\ s^2]}\frac{\sin\nu\ \cos\nu}{\left(1+\mu\ \sin^2\nu\right)}\left((1+\mu)-\frac{(1+\mu)+\mu s^2}{s^2}\frac{\left[2+3\mu+\mu^2\sin^2\nu\right]\sin^2\nu}{\left(1+\mu\ \sin^2\nu\ \right)^2}\right)\frac{1}{R^2}\frac{\partial V_N}{\partial \nu} \tag{17}$$

## 2.4 *The solution of Laplace equation in the interior space*

The Laplace equation in the orthogonal similar oblate spheroidal coordinates in the azimuthally symmetric case can be written as

$$\Delta V = \frac{\partial}{\partial R}\left(\frac{\Im}{h_R^2}\frac{\partial V(R,\,\nu)}{\partial R}\right)+\frac{\partial}{\partial \nu}\left(\frac{\Im}{h_\nu^2}\frac{\partial V(R,\,\nu)}{\partial \nu}\right)=0 \tag{18}$$

where the Laplacian is to be expressed in the form suitable for the special variable separation, i.e., (14) or (15).

The radial part of the separated Laplace equation has the solution (see [15], Equation (33))

$$r(R)\sim R^{K_d} \tag{19}$$

where the separation constant $K_d$ of the Laplace equation is in principle not restricted to a limited range for this, radial, part of the Laplace equation solution. Therefore, it can be used both for the interior as well as for the exterior solution solid harmonics.

The angular part of the Laplace equation expressed in $s$ terms in the case of variables separation is (see [15], Equation (48))

$$\left[(1+\mu)-s^2\right]\left[(1+\mu)^2-\mu s^2\right]\frac{d^2F(s)}{ds^2}\ +\ s\left[-(3\mu+2)(1+\mu)+2\mu(1+\mu)K_d+\mu\,(3-2K_d)\,s^2\right]\frac{dF(s)}{ds}$$

$$+\ K_d\left[(K_d-2)\,\mu s^2+(1+\mu)K_d+(1+\mu)^2\right]F(s)=0 \tag{20}$$

In [15], this equation was solved for positive values of the separation constant $K_d$ (i.e., for the interior space) with a help of newly defined generalized Legendre functions. It was shown that a solution of (20) certainly exists when the separation constant $K_d$ is a non-negative integer $n$. Then, the function $F(s)$ is proportional to the generalized Legendre function of the first kind, $P_n^{SI}(s)$, or of the second kind, $Q_n^{SI}(s)$.

In [15], Equation (102), the complete interior solution of the Laplace equation

$$V(R,\,\nu)=V(R,\,s)=\sum_{n=0}^{\infty}a_nR^nP_n^{SI}(s)+\sum_{n=0}^{\infty}b_nR^nQ_n^{SI}(s) \tag{21}$$

is presented. $a_n$ and $b_n$ are arbitrary real coefficients, which can be determined for the particular boundary conditions. The relation (21) thus represents the interior harmonic functions in the SOS coordinates. Several low-index generalized

Legendre functions can be seen in Tables 1 and 2 of [15]. For a better understanding, the first five generalized Legendre functions of the first kind (i.e., the generalized Legendre polynomials) are listed here as well:

$$P_0^{SI}(s) = 1, \quad P_1^{SI}(s) = \frac{1}{1+\mu}s, \quad P_2^{SI}(s) = \frac{1}{(1+\mu)^2}\frac{1}{2}\left[(\mu+3)s^2 - (1+\mu)^2\right],$$

$$P_3^{SI}(s) = \frac{1}{(1+\mu)^3}\frac{1}{2}\left[(3\mu+5)s^3 - 3(1+\mu)^2 s\right], \tag{22}$$

$$P_4^{SI}(s) = \frac{1}{(1+\mu)^4}\cdot\frac{1}{2^3}\cdot\left[\left(3\mu^2+30\mu+35\right)s^4 - 6(\mu+5)(1+\mu)^2 s^2 + 3(1+\mu)^4\right]$$

The higher degree polynomials can be retrieved by the Bonnet-like recursion formula [15]

$$P_{n+1}^{SI}(s) = \frac{2n+1}{n+1}\frac{1}{1+\mu}sP_n^{SI}(s) - \frac{n}{n+1}\left(1 - \frac{\mu}{(1+\mu)^2}s^2\right)P_{n-1}^{SI}(s) \tag{23}$$

The same formula holds also for the generalized Legendre functions of the second kind $Q_n^{SI}(s)$.

However, the solution of (20) for negative integer values of the separation constant $K_d = -1-n$, $n \geq 0$, i.e., in fact for the exterior space, was not yet found, and it is the topic of the following sections.

# 3. Preparatory considerations for the exterior solution

## 3.1 *Transformation between two different SOS systems*

It will be of advantage to find-in addition to the relation for the position vector magnitude $v_r$ reported in Equation (53) of [15]-also a relation for the angle $\chi$ between the position vector of a point $(R, \nu)$ and the equatorial plane. The angle $\chi$ is equivalent to the latitude, and $(v_r, \chi, \lambda)$ are thus in fact spherical coordinates (although not the polar ones). The sine of the angle $\chi$ satisfies clearly the relation

$$\sin\chi = \frac{z_{3D}}{v_r} \tag{24}$$

where, according to Equations (49) and (53) of the article [15],

$$z_{3D} = \frac{1}{1+\mu}Rs \quad \text{and} \quad v_r = R\sqrt{1 - \frac{\mu}{(1+\mu)^2}s^2} \tag{25}$$

Combination of (24) and (25) results in

$$\sin\chi = \frac{sR}{(1+\mu)v_r} \quad \Rightarrow \quad v_r\sin\chi = \frac{1}{1+\mu}Rs \tag{26}$$

Then also (using (25) and (26))

$$\sin\chi = \frac{\frac{1}{1+\mu}Rs}{R\sqrt{1-\frac{\mu}{(1+\mu)^2}s^2}} = \frac{s}{\sqrt{(1+\mu)^2-\mu s^2}} \tag{27}$$

These relations between the SOS and the spherical coordinates also follow from the transformation between two different SOS systems, part of which is reported in [4]. As such transformation relations are significant, the complete formulae derivation of such general transformation between two different SOS systems is shown in the following paragraphs.

Assume two SOS systems: The first SOS system, ($R$, $\nu$, $\lambda$), has the parameter of oblateness $\mu$. The second SOS coordinate system, ($R_V$, $\nu_V$, $\lambda_V$), has the parameter of oblateness $\mu_V \neq \mu$. Both systems are assumed to have the same equatorial radius $R_0$ of their reference spheroids. According to Equation (123) in [4] for the transformation of radial coordinate we have that

$$R_V = R\sqrt{\frac{1+\mu_V}{1+\mu} - \frac{\mu_V-\mu}{1+\mu}\sum_{M=0}^{\infty}\frac{-1}{-1-\mu M}\binom{-1-\mu M}{M}\left(W^2\right)^M} \tag{28}$$

in the small-$\nu$ region. As (see [4], Equations (10) and (47), or [5], Equations (6), (7) and (43))

$$\begin{aligned} x_{3D}^2+y_{3D}^2 &= \left(\cos\lambda\; R\sum_{k=0}^{\infty}\binom{-\frac{1}{2}-\mu k}{k}\frac{-\frac{1}{2}}{-\frac{1}{2}-\mu k}\left(W^2\right)^k\right)^2 \\ &\quad + \left(\sin\lambda\; R\sum_{k=0}^{\infty}\binom{-\frac{1}{2}-\mu k}{k}\frac{-\frac{1}{2}}{-\frac{1}{2}-\mu k}\left(W^2\right)^k\right)^2 \\ &= R^2\left(\sum_{k=0}^{\infty}\binom{-\frac{1}{2}-\mu k}{k}\frac{-\frac{1}{2}}{-\frac{1}{2}-\mu k}\left(W^2\right)^k\right)^2 = R^2\left(p^{-\frac{1}{2}}\right)^2 = R^2p^{-1} \\ &= R^2\sum_{M=0}^{\infty}\frac{-1}{-1-\mu M}\binom{-1-\mu M}{M}\left(W^2\right)^M \end{aligned} \tag{29}$$

and as (see [15], Equation (52), and consider that $s = f_S/h_R$) the distance from the axis is also

$$x_{3D}^2+y_{3D}^2 = R^2\left(1-\frac{1}{1+\mu}s^2\right) \tag{30}$$

the following identity holds

$$\sum_{M=0}^{\infty} \frac{-1}{-1-\mu M} \binom{-1-\mu M}{M} \left(W^2\right)^M = 1 - \frac{1}{1+\mu} s^2 \tag{31}$$

in the small-$\nu$ region. Then, (28) becomes

$$R_V = R\sqrt{\frac{1+\mu_V}{1+\mu} - \frac{\mu_V - \mu}{1+\mu}\left(1 - \frac{1}{1+\mu}s^2\right)} = R\sqrt{1 - \frac{\mu - \mu_V}{(1+\mu)^2}s^2} \tag{32}$$

Further (see [15], Equation (49), and considering that $s = f_S/h_\mathrm{R}$)

$$z_{3D} = \frac{1}{1+\mu} Rs \tag{33}$$

in the first SOS system ($R$, $\nu$, $\lambda$), whereas it is

$$z_{3D} = \frac{1}{1+\mu_V} R_V s_V \tag{34}$$

in the second SOS coordinate system ($R_V$, $\nu_V$, $\lambda_V$). Equating (33) and (34) gives

$$\frac{1}{1+\mu_V} R_V s_V = \frac{1}{1+\mu} Rs \tag{35}$$

With the help of (32), we obtain

$$s_V = \frac{1+\mu_V}{1+\mu} \frac{R}{R_V} s = \frac{1+\mu_V}{1+\mu} \frac{1}{\sqrt{1 - \frac{\mu - \mu_V}{(1+\mu)^2}s^2}} s = \frac{1+\mu_V}{\sqrt{(1+\mu)^2 - (\mu - \mu_V)s^2}} s \tag{36}$$

The Equations (32) and (36) are sumarized as follows,

$$R_V = \frac{R}{1+\mu}\sqrt{(1+\mu)^2 - (\mu - \mu_V)s^2} \qquad \text{and} \qquad s_V = \frac{1+\mu_V}{\sqrt{(1+\mu)^2 - (\mu - \mu_V)s^2}} s \tag{37}$$

and they represent the transformation between any two SOS coordinate systems having the same equatorial radius $R_0$ of their reference spheroids. The same relation can be derived also in the large-$\nu$ region.

In the special case when the parameter of oblateness $\mu_V$ of the second SOS system is equal to zero, i.e., it is equivalent to the spherical coordinate system, the transformation (37) reads

$$v_r = \left. R_V \right|_{\mu_V=0} = R\sqrt{1 - \frac{\mu}{(1+\mu)^2}s^2} \qquad \text{and} \qquad \sin\chi = \left. s_V \right|_{\mu_V=0} = \frac{s}{\sqrt{(1+\mu)^2 - \mu s^2}} \tag{38}$$

These relations are equivalent to the ones in (25) and (27) derived previously for this special case $\mu_V = 0$. From (27) and (38), the following relations can be easily found:

$$s^2 = \frac{(1+\mu)^2}{\frac{1}{\sin^2\chi}+\mu} = \frac{(1+\mu)^2\sin^2\chi}{1+\mu\sin^2\chi} \quad \Rightarrow \quad s = \frac{(1+\mu)\sin\chi}{\sqrt{1+\mu\sin^2\chi}} \quad \text{and} \quad \frac{1}{(1+\mu)^2-\mu s^2} = \frac{1+\mu\sin^2\chi}{(1+\mu)^2} \tag{39}$$

$$R = \frac{(1+\mu)v_r}{\sqrt{(1+\mu)^2-\mu s^2}} = v_r\sqrt{1+\mu\sin^2\chi} \tag{40}$$

Then, the following relation holds between the coordinates of the SOS and of the spherical systems:

$$R^n s^m = \left(v_r\sqrt{1+\mu\sin^2\chi}\right)^n \left(\frac{(1+\mu)\sin\chi}{\sqrt{1+\mu\sin^2\chi}}\right)^m = v_r^n\left(\sqrt{1+\mu\sin^2\chi}\right)^{n-m}(1+\mu)^m\sin^m\chi \tag{41}$$

The Laplace equation solution of the first kind in the interior space, $R^n P_n^{SI}(s)$, is in fact a series of terms involving $R^n s^m$, see (21) and (22). As we know (see Table 1 of [15]) that $n-m$ is always an even number, we can rewrite (41) in the form

$$R^n s^m = (1+\mu)^m v_r^n\left(1+\mu\sin^2\chi\right)^{\frac{n-m}{2}}\sin^m\chi = (1+\mu)^m v_r^n\left[\sum_{k=0}^{\frac{(n-m)}{2}}\begin{pmatrix}\frac{n-m}{2}\\ k\end{pmatrix}\left(\mu\sin^2\chi\right)^k\right]$$
$$\sin^m\chi = (1+\mu)^m v_r^n\sum_{k=0}^{\frac{(n-m)}{2}}\begin{pmatrix}\frac{n-m}{2}\\ k\end{pmatrix}(\sin\chi)^{2k+m} \tag{42}$$

where the binomial expansion was used.

## 3.2 *Equivalence between the spherical solid harmonics and the SOS harmonics*

To proceed with the Laplace equation solution for the exterior space, it is worth to use the already found [15] interior solution (21) involving the generalized Legendre polynomials $P_n^{SI}(s)$. We can write them in the following form:

$$P_n^{SI}(s) = \sum_{N=0}^{\left[\frac{n}{2}\right]} s^{n-2N}A_{n,\,n-2N} \tag{43}$$

Here, $[n/2]$ denotes the floor function, i.e., the integer part of $n/2$. Note, that all $A_{n,\,k}$ coefficients are zero for $n-k$ being odd (such $A$'s are, moreover, not included in the sum in the above formula). This corresponds to the shape of the already determined interior solutions $n = 0$ to 6, see Table 1 of [15]. Further note, that a general formula for the coefficients $A_{n,\,n-2N}$ is not yet known and, presently, a Bonet-like recursion described by (23) is to be used to find a particular shape of a generalized Legendre function of degree $n$ from the lower-degree functions. The formula (43) is, nevertheless, similar to the expresion for the standard Legendre polynomials $P_n(x)$ derived from the Rodriguez formula, i.e.,

$$P_n(x) = \sum_{L=0}^{\left[\frac{n}{2}\right]} x^{n-2L} \frac{1}{2^n} (-1)^L \binom{n}{L} \binom{2n-2L}{n} = \sum_{L=0}^{\left[\frac{n}{2}\right]} x^{n-2L} a_{n,\, n-2L} \tag{44}$$

(according to [17], Equation 3.133). The formula for the coefficients $a_{n,\, n-2L}$ is thus known for this case:

$$a_{n,\, n-2L} \equiv \frac{1}{2^n} (-1)^L \binom{n}{L} \binom{2n-2L}{n} \tag{45}$$

The interior harmonic function in the spherical coordinates is then

$$v_r^n P_n(x) = v_r^n \sum_{L=0}^{\left[\frac{n}{2}\right]} x^{n-2L} a_{n,\, n-2L} \tag{46}$$

The individual interior solution of $n$-th degree of the Laplace equation in the SOS coordinates is, on the other hand, proportional to

$$R^n P_n^{SI}(s) = R^n \sum_{N=0}^{[n/2]} s^{n-2N} A_{n,\, n-2N} = \sum_{N=0}^{[n/2]} R^n s^{n-2N} A_{n,\, n-2N} \tag{47}$$

We now insert, for $R^n s^{n-2N}$, the relation according to (42) and get

$$\begin{aligned} R^n P_n^{SI}(s) &= \sum_{N=0}^{[n/2]} A_{n,\, n-2N} (1+\mu)^{n-2N} v_r^n \sum_{k=0}^{N} \binom{N}{k} \mu^k (\sin\chi)^{2k+n-2N} \\ &= v_r^n \sum_{N=0}^{[n/2]} A_{n,\, n-2N} (1+\mu)^{n-2N} \sum_{k=0}^{N} \binom{N}{k} \mu^k (\sin\chi)^{2k+n-2N} \end{aligned} \tag{48}$$

Now, substitute $k$ according to the relation $L = N - k$:

$$R^n P_n^{SI}(s) = v_r^n \sum_{N=0}^{[n/2]} A_{n,\, n-2N} (1+\mu)^{n-2N} \sum_{L=N}^{0} \binom{N}{N-L} \mu^{N-L} (\sin\chi)^{n-2L} \tag{49}$$

Further, change the direction of summation in the inner sum and take into account that $\binom{N}{N-L} = \binom{N}{L}$. Then

$$R^n P_n^{SI}(s) = v_r^n \sum_{N=0}^{[n/2]} \sum_{L=0}^{N} A_{n,\, n-2N} (1+\mu)^{n-2N} \binom{N}{L} \mu^{N-L} (\sin\chi)^{n-2L} \tag{50}$$

As the binomial coefficient would be zero for $L > N$ (calling negative number in binomial coefficient results in zero), we can expand counting in the inner sum up from $N$ to $[n/2]$ without a change of the result:

$$R^n P_n^{SI}(s) = v_r^n \sum_{N=0}^{[n/2]} \sum_{L=0}^{[n/2]} A_{n,\, n-2N} (1+\mu)^{n-2N} \binom{N}{L} \mu^{N-L} (\sin\chi)^{n-2L} \tag{51}$$

Then, we can exchange the order of the two sums,

$$R^n P_n^{SI}(s) = v_r^n \sum_{L=0}^{[n/2]} \sum_{N=0}^{[n/2]} A_{n,\, n-2N} (1+\mu)^{n-2N} \binom{N}{L} \mu^{N-L} (\sin\chi)^{n-2L} \tag{52}$$

and put the sine power (depending on $L$, not $N$) out of the inner sum:

$$R^n P_n^{SI}(s) = v_r^n \sum_{L=0}^{[n/2]} (\sin\chi)^{n-2L} \sum_{N=0}^{[n/2]} A_{n,\, n-2N} \mu^{N-L} (1+\mu)^{n-2N} \binom{N}{L} \tag{53}$$

Equipped with the above derived relations (46) and (53), we can declare a significant statement which is to be confirmed later as output of the following analysis. Particularly, that the form of the individual Laplace equation interior solution of the first kind in the SOS coordinates is physically the same as the form of the individual Laplace equation solution of the first kind in the spherical coordinates. If this assumption is valid, then

$$R^n P_n^{SI}(s) = v_r^n P_n(\sin\chi), \quad n \geq 0 \tag{54}$$

i.e., that the solid harmonics of the first kind (the right side of (54)) and the SOS harmonics (the left side of (54)) are, each individually, equivalent for the interior space. That means that the form of the individual Laplace equation solutions of the first kind in the SOS coordinates, and in the spherical coordinates are physically the same (although expressed in different coordinates). They are not a linear combination of the others, but they are equal to each other one by one (Note, that such rule is not the case when the solid harmonics are compared for the solutions in the spherical coordinates and in the well known confocal oblate spheroidal coordinates).

The right-hand side of (54) is given by (46), whereas the left-hand side by (53). By comparison, we can see that the inner sum in (53) would be thus equal to the coefficients of the standard Legendre polynomials $a_{n,\, n-2L}$, see (45). Then, we get a relation between the coefficients of the generalized Legendre polynomials and the Legendre polynomials in the form

$$\sum_{N=0}^{[n/2]} A_{n,\, n-2N} \mu^{N-L} (1+\mu)^{n-2N} \binom{N}{L} = \frac{1}{2^n} (-1)^L \binom{n}{L} \binom{2n-2L}{n} = a_{n,\, n-2L} \tag{55}$$

As the binomial coefficient on the left side is non-zero only when $N \geq L$, we restrict the sum range:

$$\sum_{N=L}^{[n/2]} A_{n,\, n-2N} \mu^{N-L} (1+\mu)^{n-2N} \binom{N}{L} = \frac{1}{2^n} (-1)^L \binom{n}{L} \binom{2n-2L}{n} = a_{n,\, n-2L} \tag{56}$$

Note, that the right-hand side is independent of $\mu$. Then, if our assumption holds, the terms containing $\mu$ need to cancel mutually on the left side. Using (56), it can be checked with the already found coefficients $A_{n,\, n-2N}$ for $n = 0$ to 6 whether we can obtain the coefficients of the standard Legendre polynomials $a_{n,\, n-2L}$ from the coefficients of the generalized Legendre polynomials $A_{n,\, n-2N}$. In the positive case, our assumption that the form of the individual Laplace equation solution of the first kind in SOS is physically the same as the form of the individual Laplace equation solution of the first kind in spherical coordinates would have a strong support.

The detailed check of the validity of (56) for $n$ up to 6 is given in Supplement 2 with a positive output: its validity is confirmed. It indicates (although it is not proven for all cases) that the relation (53) could hold generally for any $n \geq 0$, and the solid harmonics are thus generally the same as the SOS harmonics of the first kind. Note, that such an equivalence is not the case for the confocal oblate spheroidal (COS) coordinates harmonics, which differ physically from the solid harmonics in the spherical coordinates: the COS harmonics can be obtained only as a linear combination of several spherical harmonics [21].

Note also that, although the solid harmonics are solutions of the Laplace equation in the spherical as well as in the SOS coordinates, the spherical harmonics (i.e., the solutions of the angular part of the separated Laplace equation in spherical coordinates) are not the solutions of the angular part of the separated Laplace equation in the SOS coordinates (20). The generalized Legendre functions have to be used instead.

We will use the assumption (54) in what follows for the determination of the exterior solution of the Laplace equation in the SOS coordinates. If the assumption holds, the final check with the model function assembled in this way and inserted into the equation has to lead to the fulfillment of the equation regardless of the value of $n$.

### 3.3 *Expressing generalized legendre polynomials as sum of monomials*

For testing of a model solution of the Laplace equation, it would be of large advantage to have generalized Legendre polynomials expressed as a sum of monomials with known coefficients. As a continuation of the analysis from the previous sub-section, the following formula for the coefficients of the generalized Legendre polynomials is derived (see Supplement 3, relation (S3.66)):

$$A_{n,\, n-2N} = \frac{1}{(1+\mu)^{n-2N}} \frac{1}{2^n} (-1)^N \binom{n}{N} \sum_{j=0}^{\left[\frac{n}{2}\right]-N} \binom{n-N}{j} \binom{2(n-N-j)}{n} \mu^j \tag{57}$$

The formula is obtained as an estimated generalization of several low-index coefficients. If this formula is later successfully used in the Laplace equation solution, it would confirm its general validity.

An arbitrary generalized Legendre polynomial of degree $n$ can be, therefore, written as a sum of monomials without a need of any reccurence relation, similarly as the Rodriguez formula is used for the expression of the standard Legendre polynomials as a sum of monomials. Such a formula is (using (43) and (57)):

$$P_n^{SI}(s) = \frac{1}{2^n} \sum_{N=0}^{\left[\frac{n}{2}\right]} s^{n-2N} \frac{(-1)^N}{(1+\mu)^{n-2N}} \binom{n}{N} \sum_{j=0}^{\left[\frac{n}{2}\right]-N} \binom{n-N}{j} \binom{2(n-N-j)}{n} \mu^j \tag{58}$$

For $\mu = 0$, the relation (58) reduces to (44).

## 4. The exterior solution

### 4.1 *Model for the exterior solution*

In the previous section, it was strongly indicated that the solid harmonics and the SOS harmonics of the first kind in the interior are physically the same (see (54)). The present model for the exterior solution supposes that the same scheme is valid for the exterior solution, i.e.,

$$R^{-1-n}F_n^{SE}(s) = v_r^{-1-n}P_n(\sin\chi), \qquad n \geq 0 \tag{59}$$

where $n \geq 0$ clearly means negative exponents of $R$ (and thus non-diverging functions on the left side for $R$ approaching infinity), and $F_n^{SE}(s)$ is the foreseen exterior solution of the angular part of the Laplace Equation (20). The right side of (59) represents the classical solution of the Laplace equation in the exterior space in the spherical coordinates (the non-polar ones), whereas we use the still unknown function $F_n^{SE}$ on the left side as a prospective solution of the angular part of the Laplace equation in the SOS coordinates, which is to be found. This function is not necessarily a polynomial. This approach also means that we expect separability of the exterior space solutions, similarly as was previously expected for the interior solution.

By multiplying (59) by $v_r^{2n+1}$, we get

$$v_r^{2n+1}R^{-1-n}F_n^{SE}(s) = v_r^n P_n(x) \tag{60}$$

We know (see (54)) that the right side of (60) is equal to $R^n P_n^{SI}(s)$. Thus

$$v_r^{2n+1}R^{-1-n}F_n^{SE}(s) = R^n P_n^{SI}(s) \tag{61}$$

Then

$$F_n^{SE}(s) = \frac{R^n}{v_r^{2n+1}R^{-1-n}}P_n^{SI}(s) = \frac{R^{2n+1}}{v_r^{2n+1}}P_n^{SI}(s) \tag{62}$$

When (25) is inserted into (62), we obtain for the angular part of the exterior solution in the SOS case

$$F_n^{SE}(s) = \frac{R^{2n+1}}{\left(\dfrac{R}{1+\mu}\sqrt{(1+\mu)^2-\mu s^2}\right)^{2n+1}}P_n^{SI}(s) = \frac{(1+\mu)^{2n+1}}{\left(\sqrt{(1+\mu)^2-\mu s^2}\right)^{2n+1}}P_n^{SI}(s) \tag{63}$$

It can be seen that the exterior solution model is based on the generalized Legendre polynomial but itself it is not a polynomial in $s$ (except the case $\mu = 0$). It is a more general function.

### 4.2 *Proof for the exterior solution*

Further, we have to test if our assumptions are correct, i.e., whether (63) is in reality the solution of the angular part of the Laplace Equation (20) where the separation constant $K_d$ is negative, particularly $K_d = -1-n$, $n \geq 0$. It would also confirm that all the assumptions leading to this solution model were correct. The proof is to be done generally, i.e., for any non-negative $n$, leading thus to the exterior space solution, see (59), with the negative exponents of the radial part (19) in such a case.

The proof strategy is simple: insertion of (63) into the angular part of the Laplace Equation (20) with plugged-in the polynomial $P_n^{SI}(s)$ according to (58). Nevertheless, the detailed performance of the proof is rather lengthy, and it is thus given in Supplement 4. It is finally proved in the Supplement that the model function (63) is an exterior space solution of the angular part of the Laplace equation in the SOS coordinates for any non-negative $n$. Therefore, according to (59),

$$R^{-1-n}\frac{(1+\mu)^{2n+1}}{\left(\sqrt{(1+\mu)^2-\mu s^2}\right)^{2n+1}}P_n^{SI}(s)=v_r^{-1-n}P_n(\sin\chi) \tag{64}$$

Then, also the assumption that the physical form of the SOS solution of the Laplace equation is the same as the solution in the spherical coordinates is confirmed.

## 5. The complete solution of the azimuthally symmetric Laplace equation and simple example of its use

The full Laplace equation solution of the first kind (both in the exterior as well as in the interior space) is thus

$$\begin{aligned} V(R,\,\nu)=V(R,\,s)&=\sum_{n=0}^{\infty}\left[a_{nI}R^nP_n^{SI}(s)+a_{nE}R^{-1-n}\frac{(1+\mu)^{2n+1}}{\left(\sqrt{(1+\mu)^2-\mu s^2}\right)^{2n+1}}P_n^{SI}(s)\right]\\ &=\sum_{n=0}^{\infty}\left[a_{nI}R^n+a_{nE}R^{-n-1}\frac{(1+\mu)^{2n+1}}{\left(\sqrt{(1+\mu)^2-\mu s^2}\right)^{2n+1}}\right]P_n^{SI}(s) \end{aligned} \tag{65}$$

For the exterior space, all the $a_{nI}$ coefficients are zero (except perhaps $a_{0I}$, depending on the level to which the function $V$ is set in infinity), while-for the interior space-all $a_{nE}$ are equal to zero. In this relation, the generalized Legendre functions of the first kind are given by (58).

To demonstrate a simple application of the found solution, consider an example from the field of electrostatics. We intend to find the electrostatic potential outside an oblate spheroid with the equatorial radius $b$ whose surface is held at a potential given in the SOS coordinates (the reference surface equatorial radius $R_0=b$) by

$$V(R_0,\,\nu)=V(b,\,\nu)=V_0+C\frac{1}{\sqrt{1-\dfrac{\mu}{1+\mu}\sin^2\nu}} \tag{66}$$

where $V_0$ and $C$ are constants and $\nu$ is the parametric latitude. The relation (66) is the boundary condition of the problem.

The full Laplace equation solution of the first kind (65) applied in the exterior is to be used to find the exterior potential. As the solution cannot diverge in the infinity, all coefficients $a_{nI}$ except $a_{0I}$ have to be zero:

$$V(R,\,\nu)=V(R,\,s)=a_{0I}+\sum_{n=0}^{\infty}\left[a_{nE}R^{-1-n}\frac{(1+\mu)^{2n+1}}{\left(\sqrt{(1+\mu)^2-\mu s^2}\right)^{2n+1}}P_n^{SI}(s)\right] \tag{67}$$

It was previously found (see [15], Equation (63)) that the function $s$ on the reference surface is

$$s|_{R=R_0}=\sqrt{1+\mu}\sin\nu \tag{68}$$

Considering this relation and the form of the generalized Legendre polynomials (see (22) or (58)), it can be concluded that the solution derived from (67) can have no term in the numerator contain any power of $s$. Then, the only non-zero term in the sum in (67) can be the one containing the zero-degree generalized Legendre polynomial, i.e. the one with $n=0$. The solution (67) in such a case simplifies to

$$V(R,\,\nu)=V(R,\,s)=a_{0I}+a_{0E}\frac{1}{R\sqrt{1-\dfrac{\mu}{(1+\mu)^2}s^2}} \tag{69}$$

This solution, used at the boundary and compared with the boundary condition (66), provides the values for the parameters $a_{0I}$ and $a_{0E}$:

$$a_{0I}=V_0,\;\; a_{0E}=CR_0=Cb \tag{70}$$

Therefore, the particular solution is finally

$$V(R,\,\nu)=V(R,\,s)=V_0+C\frac{b}{R}\frac{1}{\sqrt{1-\dfrac{\mu}{(1+\mu)^2}s^2}} \tag{71}$$

Note, that the solution in this case conains only one term plus a constant. Other boundary conditions would lead to a more complex solutions derived from the general one given by (65), possibly also with infinite number of terms.

## 6. Conclusions

The exterior solution of the azimuthally symmetric Laplace equation in the SOS coordinates was found. The approach leading to this result was of different nature than the one used previously for the interior solution determination [15] and proved the exterior solutions of the first kind for any degree of generalized Legendre polynomial.

In the steps leading to the solution, important formulae were derived which hold for the SOS algebra. Various forms of the Laplace operator in the SOS coordinates in the azimuthally symmetric case were shown. General transformation between two different SOS coordinate systems was derived (see (37)). Its special case-the transformation between the SOS coordinate system and the spherical coordinate system (see (38))-was used to prove that the SOS harmonics are physically the same as the solid harmonics in the system of spherical coordinates, only expressed in a different form. Further, the formula expressing any generalized Legendre polynomial of degree $n$ as a finite sum of monomials (i.e., without a need of a recurrence relation) is derived (see (58) with coefficients given by (57)).

The final proof of the validity of the solution justifies the assumptions made during the derivation. All the abovementioned results are novel and were not-to the author's knowledge-yet published. Finally, the general solution of the azimuthally symmetric Laplace equation in the SOS coordinates, both in the interior as well as in the exterior space, is presented.

Note however, that while the proof of the equivalence of solid harmonics (see (59) and (64)) was given for the exterior space harmonics generally, the equivalence for the interior solid harmonics (see (54)) was proven only up to the degree $n = 6$. To carry it out generally also in the interior, the angular part of the Laplace Equation (20) needs to be tested with generalized Legendre functions $P_n^{SI}(s)$ expressed as a sum of monomials (58) in a similar way as done in Supplement 4 for the exterior space. Such a general proof is, nevertheless, left for future derivations. Also, the second kind exterior solution of the Laplace equation is not derived in the present text. It can be hypothesized that it is of the same form as (65) which is valid with the generalized Legendre functions of the first kind, only with $P_n^{SI}(s)$ substituted by $Q_n^{SI}(s)$, i.e. the second kind generalized Legendre functions. Nevertheless, the functions $Q_n^{SI}(s)$ still need to be derived generally, similarly as done for $P_n^{SI}(s)$ in (58), and tested in the Laplace equation.

The found relations for the SOS coordinate system can have application in several fields of physics where fields inside or around objects of oblate spheroidal shape are investigated, including geophysics, astrophysics, electromagnetism, fluid flow, electrostatics and solid state physics (e.g., ferroic inclusions).

## Data availability

This article deals with the derivation of theoretical relations within the similar oblate spheroidal coordinate system. There are thus no data sets to be disclosed.

## Acknowledgments

The author acknowledges support from the long-term conceptual development project RVO 61389005 of the Nuclear Physics Institute of the Czech Academy of Sciences, and from the OP JAK project of MEYS Nr. CZ.02.01.01/00/22_008/0004591. The author thanks the teacher Ivana Malá form "ZŠ Strossmayerovo náměstí" who attracted his interest to mathematics.

## Conflict of interest

The author declares no conflict of interests.

## Appendix A
### *The Laplacian in s-terms*

When we want to express the Laplacian in the SOS coordinates in terms of $R$ and $s$, we need several relations, first of all derivatives. With the help of the Supplement C of the interior-solution paper ([15], Equations (C25) and (C28)), we know the derivatives of the function $F(W)$ with respect to $W$ expressed in terms of $F$ derivatives with respect to $s$:

$$\frac{dF(W)}{dW} = \frac{dF(s)}{ds} \frac{\left[(1+\mu)-s^2\right]^{\frac{\mu+3}{2}}}{(1+\mu)^{\frac{\mu}{2}}\left[(1+\mu)+\mu s^2\right]} \tag{A1}$$

and

$$\frac{d^2F(W)}{dW^{\,2}} = \frac{d^2F(s)}{ds^2} \frac{\left[(1+\mu)-s^2\right]^{(\mu+3)}}{(1+\mu)^{\mu}\left[(1+\mu)+\mu s^2\right]^2} - \frac{dF(s)}{ds} s \frac{(1+\mu)^{(1-\mu)}\left[(1+\mu)-s^2\right]^{(\mu+2)}\left[3(1+\mu)+\mu s^2\right]}{\left[(1+\mu)+\mu s^2\right]^3} \tag{A2}$$

Also (see Equation (C21) of the Supplement C of [15]), $W$ expressed in $s$ terms is

$$W = \sqrt{\frac{\frac{1}{1+\mu}\quad s^2}{\left(1 \;-\; \frac{1}{1+\mu}\; s^2\right)^{1+\mu}}} = \sqrt{\frac{(1+\mu)^{\mu} s^2}{\left[(1+\mu) \;-\; s^2\right]^{1+\mu}}} \tag{A3}$$

Further, $\frac{1}{h_R^2}$ can be expressed according to Equation (A19) from the Appendix A of [15] with the result

$$\frac{1}{h_R^2} = \left(\frac{\mu}{1+\mu} s^2 + 1\right) \tag{A4}$$

Since further (see [15])

$$f_C^2 = 1 - f_S^2, \qquad f_S^2 = \frac{1+\mu}{\mu}\left(1-h_R^2\right), \qquad f_C^2 = \frac{(1+\mu)h_R^2 - 1}{\mu} \tag{A5}$$

then (14) can be written as

$$\Delta V = \left(\frac{\mu}{1+\mu}s^2+1\right)\frac{d^2r(R)}{dR^2}F(s)+\left[\frac{1}{R}\frac{dr(R)}{dR}(\mu+3)F(s)-\frac{1}{R}\frac{dr(R)}{dR}\left(\frac{\mu}{1+\mu}s^2+1\right)F(s)\right]$$

$$+\frac{1}{R^2}r(R)\left[(\mu-2)\mu\left(\frac{\mu}{1+\mu}s^2+1\right)+(\mu+3)\mu+\frac{(1+\mu)^2}{\frac{1-f_S^2}{h_R^2}}\right]\sqrt{\frac{(1+\mu)^{\mu}s^2}{\left[(1+\mu)\ -\ s^2\right]^{1+\mu}}}\frac{dF(W)}{dW}$$

$$+2\frac{1}{R}\frac{dr(R)}{dR}\mu\sqrt{\frac{(1+\mu)^{\mu}s^2}{\left[(1+\mu)\ -\ s^2\right]^{1+\mu}}}\left(\frac{\mu}{1+\mu}s^2+1\right)\frac{dF(W)}{dW} \tag{A6}$$

$$+\frac{1}{R^2}r(R)\left(\mu^2\left(\frac{\mu}{1+\mu}s^2+1\right)+\frac{(1+\mu)^2}{\frac{(1+\mu)h_R^2-1}{\mu}s^2}\right)\frac{(1+\mu)^{\mu}s^2}{\left[(1+\mu)\ -\ s^2\right]^{1+\mu}}\frac{d^2F(W)}{dW^2}$$

When inserting the derivatives (C25) and (C28), and after simplification, we arrive at

$$\Delta V = \left(\frac{\mu}{1+\mu}s^2+1\right)\frac{d^2r(R)}{dR^2}F(s)+\left[(\mu+2)-\frac{\mu}{1+\mu}s^2\right]\frac{1}{R}\frac{dr(R)}{dR}F(s)$$

$$+\frac{1}{R^2}r(R)\left[(\mu-2)\mu\left(\frac{\mu}{1+\mu}s^2+1\right)+(\mu+3)\mu+\frac{(1+\mu)^2}{\frac{\mu s^2+(1+\mu)\left(1-s^2\right)}{1+\mu}}\right]$$

$$s\frac{\left[(1+\mu)-s^2\right]}{\left[(1+\mu)+\mu s^2\right]}\frac{dF(s)}{ds}+2\frac{1}{R}\frac{dr(R)}{dR}s\frac{\mu}{1+\mu}\left[(1+\mu)-s^2\right]\frac{dF(s)}{ds} \tag{A7}$$

$$+\mu\left(\mu\frac{1}{1+\mu}\left((1+\mu)+\mu s^2\right)+\frac{(1+\mu)^2}{\frac{(1+\mu)s^2}{\frac{1}{1+\mu}\left(\mu s^2+(1+\mu)\right)}-s^2}\right)$$

$$\left[\frac{\left[(1+\mu)-s^2\right]^2}{\left[(1+\mu)+\mu s^2\right]^2}s^2\frac{d^2F(s)}{ds^2}-\frac{(1+\mu)\left[(1+\mu)-s^2\right]\left[3(1+\mu)+\mu s^2\right]}{\left[(1+\mu)+\mu s^2\right]^3}s^3\frac{dF(s)}{ds}\right]\frac{1}{R^2}r(R)$$

Simplification of some factors containing $\mu$ and $s$ leads to

$$
\begin{aligned}
\Delta V =& \left(\frac{\mu}{1+\mu}s^2+1\right)\frac{d^2r(R)}{dR^2}F(s)+\left[(\mu+2)-\frac{\mu}{1+\mu}s^2\right]\frac{1}{R}\frac{dr(R)}{dR}F(s)\\
&+\frac{1}{R^2}r(R)\left[\frac{\left(\frac{\mu\mu\mu}{1+\mu}s^2-2\frac{\mu\mu}{1+\mu}s^2+2\mu\mu+\mu\right)\left[(1+\mu)-s^2\right]+(1+\mu)^3}{\left[(1+\mu)+\mu s^2\right]}\right]s\frac{dF(s)}{ds}\\
&+2\frac{\mu}{1+\mu}\left[(1+\mu)-s^2\right]s\frac{1}{R}\frac{dr(R)}{dR}\frac{dF(s)}{ds}\\
&+\left(\frac{\mu^2s^2\left((1+\mu)-s^2\right)+(1+\mu)^3}{(1+\mu)}\right)\\
&\left[\frac{(1+\mu)-s^2}{(1+\mu)+\mu s^2}\frac{d^2F(s)}{ds^2}-\frac{(1+\mu)\left[3(1+\mu)+\mu s^2\right]}{\left[(1+\mu)+\mu s^2\right]^2}s\frac{dF(s)}{ds}\right]\frac{1}{R^2}r(R)
\end{aligned}
\tag{A8}
$$

Then, the identical derivatives are put together,

$$
\begin{aligned}
\Delta V =& \left(\frac{\mu}{1+\mu}s^2+1\right)\frac{d^2r(R)}{dR^2}F(s)+\left[(\mu+2)-\frac{\mu}{1+\mu}s^2\right]\frac{1}{R}\frac{dr(R)}{dR}F(s)\\
&+2\frac{\mu}{1+\mu}\left[(1+\mu)-s^2\right]s\frac{1}{R}\frac{dr(R)}{dR}\frac{dF(s)}{ds}-\left(\frac{\mu^2s^2\left((1+\mu)-s^2\right)+(1+\mu)^3}{(1+\mu)}\right)\\
&\frac{(1+\mu)\left[3(1+\mu)+\mu s^2\right]}{\left[(1+\mu)+\mu s^2\right]^2}s\frac{1}{R^2}r(R)\frac{dF(s)}{ds}\\
&+\left[\frac{\left(\frac{\mu\mu\mu}{1+\mu}s^2-2\frac{\mu\mu}{1+\mu}s^2+2\mu\mu+\mu\right)\left[(1+\mu)-s^2\right]+(1+\mu)^3}{\left[(1+\mu)+\mu s^2\right]}\right]s\frac{1}{R^2}r(R)\frac{dF(s)}{ds}\\
&+\frac{\mu^2s^2\left((1+\mu)-s^2\right)+(1+\mu)^3}{(1+\mu)}\frac{(1+\mu)-s^2}{(1+\mu)+\mu s^2}\frac{1}{R^2}r(R)\frac{d^2F(s)}{ds^2}
\end{aligned}
\tag{A9}
$$

their pre-factors are joined, and the relation is simplified:

$$
\begin{aligned}
\Delta V = & \frac{1}{1+\mu}\left((1+\mu)+\mu s^2\right) F(s) \frac{d^2 r(R)}{dR^2} + \left[(\mu+2) - \frac{\mu}{1+\mu} s^2\right] F(s) \frac{1}{R} \frac{dr(R)}{dR} \\
& + 2\frac{\mu}{1+\mu}\left[(1+\mu)-s^2\right] s \frac{dF(s)}{ds} \frac{1}{R} \frac{dr(R)}{dR} + \frac{3\mu s^2 - \left(3\mu^2+5\mu+2\right)}{1+\mu} s \frac{dF(s)}{ds} \frac{r(R)}{R^2} \qquad \text{(A10)} \\
& + \frac{\left[(1+\mu)-s^2\right]\left[(1+\mu)^2-\mu s^2\right]}{1+\mu} \frac{d^2 F(s)}{ds^2} \frac{r(R)}{R^2}
\end{aligned}
$$

The above relation represents the Laplacian in the SOS coordinates expressed in $s$-terms instead of $W$-terms.

**Supplement 1** to P. Strunz: On Laplace equation solution in orthogonal similar oblate spheroidal coordinates

## Laplacian rewriten to *s*-terms without separation of the function

The Laplacian in azimuthaly symmetric case (see the relation (13) of the main text),

$$\Delta V_{\mathrm{N}}(R,s) = \frac{1}{\Im}\left[\frac{\partial}{\partial R}\left(\frac{\Im}{h_R^2}\frac{\partial[V_{\mathrm{N}}(R,s)]}{\partial R}\right)+\frac{\partial}{\partial \mathrm{v}}\left(\frac{\Im}{h_{\mathrm{v}}^2}\frac{\partial[V_{\mathrm{N}}(R,s)]}{\partial \mathrm{v}}\right)\right] \quad , \tag{S1.1}$$

can be modified as follows:

$$\Delta V_{\mathrm{N}}(R,s) = \frac{1}{\Im}\left[\frac{\partial}{\partial R}\left(\frac{\Im}{h_R^2}\frac{\frac{\partial W}{\partial \mathrm{v}}}{\frac{\partial W}{\partial \mathrm{v}}}\frac{\partial[V_{\mathrm{N}}(R,s)]}{\partial R}\right)+\frac{\partial}{\partial \mathrm{v}}\left(\frac{\Im}{h_{\mathrm{v}}^2}\frac{\frac{\partial W}{\partial \mathrm{v}}}{\frac{\partial W}{\partial \mathrm{v}}}\frac{\partial[V_{\mathrm{N}}(R,s)]}{\partial s}\frac{\partial s}{\partial \mathrm{v}}\right)\right] \quad . \tag{S1.2}$$

The product rule leads to

$$\Delta V_{\mathrm{N}}(R,s) = \frac{1}{\Im}\left[\frac{\partial V_{\mathrm{N}}}{\partial R}\frac{\partial W}{\partial \mathrm{v}}\frac{\partial}{\partial R}\left(\frac{\Im}{h_R^2\frac{\partial W}{\partial \mathrm{v}}}\right)+\frac{\Im}{h_R^2\frac{\partial W}{\partial \mathrm{v}}}\frac{\partial W}{\partial \mathrm{v}}\frac{\partial}{\partial R}\left(\frac{\partial V_{\mathrm{N}}}{\partial R}\right)+\frac{\Im}{h_R^2\frac{\partial W}{\partial \mathrm{v}}}\frac{\partial V_{\mathrm{N}}}{\partial R}\frac{\partial}{\partial R}\left(\frac{\partial W}{\partial \mathrm{v}}\right)+\frac{\partial V_{\mathrm{N}}}{\partial s}\frac{\partial s}{\partial \mathrm{v}}\frac{1}{\frac{\partial W}{\partial \mathrm{v}}}\frac{\partial}{\partial \mathrm{v}}\left(\frac{\Im}{h_{\mathrm{v}}^2}\frac{\partial W}{\partial \mathrm{v}}\right)+\frac{\Im}{h_{\mathrm{v}}^2}\frac{\partial W}{\partial \mathrm{v}}\frac{1}{\frac{\partial W}{\partial \mathrm{v}}}\frac{\partial s}{\partial \mathrm{v}}\frac{\partial}{\partial \mathrm{v}}\left(\frac{\partial V_{\mathrm{N}}}{\partial s}\right)+\frac{\Im}{h_{\mathrm{v}}^2}\frac{\partial W}{\partial \mathrm{v}}\frac{\partial V_{\mathrm{N}}}{\partial s}\frac{\partial}{\partial \mathrm{v}}\left(\frac{\partial s}{\partial \mathrm{v}}\frac{1}{\frac{\partial W}{\partial \mathrm{v}}}\right)\right] \quad , \tag{S1.3}$$

and simplification involving Eqs. (A96), (A97), (A16) of [15] for the Jacobian derivatives leads to

$$\Delta V_{\mathrm{N}}(R,s) = \frac{1}{\Im}\left[\frac{\partial V_{\mathrm{N}}}{\partial R}\frac{\partial W}{\partial \mathrm{v}}\frac{\Im}{R\frac{\partial W}{\partial \mathrm{v}}}\left[(\mu+3)-\frac{1+\mu}{h_R^2}\right]+\frac{\Im}{h_R^2}\frac{\partial^2 V_{\mathrm{N}}}{\partial R^2}+\frac{\Im}{h_R^2\frac{\partial W}{\partial \mathrm{v}}}\frac{\partial V_{\mathrm{N}}}{\partial R}\left(\frac{\partial}{\partial \mathrm{v}}\left[\frac{\partial W}{\partial R}\right]\right)+\frac{\partial V_{\mathrm{N}}}{\partial s}\frac{\partial s}{\partial \mathrm{v}}\frac{1}{\frac{\partial W}{\partial \mathrm{v}}}\frac{f_S^2}{W}\frac{\Im}{h_{\mathrm{v}}^2}\left(\frac{\partial W}{\partial \mathrm{v}}\right)^2+\frac{\Im}{h_{\mathrm{v}}^2}\frac{\partial W}{\partial \mathrm{v}}\frac{1}{\frac{\partial W}{\partial \mathrm{v}}}\frac{\partial s}{\partial \mathrm{v}}\frac{\partial^2 V_{\mathrm{N}}}{\partial s^2}\frac{\partial s}{\partial \mathrm{v}}+\frac{\Im}{h_{\mathrm{v}}^2}\frac{\partial W}{\partial \mathrm{v}}\frac{\partial V_{\mathrm{N}}}{\partial s}\left[\frac{1}{\frac{\partial W}{\partial \mathrm{v}}}\frac{\partial}{\partial \mathrm{v}}\left(\frac{\partial s}{\partial \mathrm{v}}\right)+\frac{\partial s}{\partial \mathrm{v}}\frac{\partial}{\partial \mathrm{v}}\left(\frac{1}{\frac{\partial W}{\partial \mathrm{v}}}\right)\right]\right] \quad . \tag{S1.4}$$

Employ the derivative of *W* with respect to *R* from [15], Eq. (10), and do further simplifications:

$$\Delta V_{\mathrm{N}}(R,s) = \frac{1}{\Im}\left[\frac{\partial V_{\mathrm{N}}}{\partial R}\frac{\Im}{R}\left[(\mu+3)-\frac{1+\mu}{h_R^2}\right]+\frac{\Im}{h_R^2}\frac{\partial^2 V_{\mathrm{N}}}{\partial R^2}+\frac{\Im}{h_R^2\frac{\partial W}{\partial \mathrm{v}}}\frac{\partial V_{\mathrm{N}}}{\partial R}\left(\frac{\mu}{R}\frac{\partial W}{\partial \mathrm{v}}\right)+\frac{\partial V_{\mathrm{N}}}{\partial s}\frac{\partial s}{\partial \mathrm{v}}\frac{f_S^2}{W}\frac{\Im}{h_{\mathrm{v}}^2}\frac{\partial W}{\partial \mathrm{v}}+\frac{\Im}{h_{\mathrm{v}}^2}\frac{\partial s}{\partial \mathrm{v}}\frac{\partial^2 V_{\mathrm{N}}}{\partial s^2}\frac{\partial s}{\partial \mathrm{v}}+\frac{\Im}{h_{\mathrm{v}}^2}\frac{\partial W}{\partial \mathrm{v}}\frac{\partial V_{\mathrm{N}}}{\partial s}\left[\frac{1}{\frac{\partial W}{\partial \mathrm{v}}}\frac{\partial^2 s}{\partial \mathrm{v}^2}+\frac{\partial s}{\partial \mathrm{v}}\frac{-\frac{\partial^2 W}{\partial \mathrm{v}^2}}{\left(\frac{\partial W}{\partial \mathrm{v}}\right)^2}\right]\right] \quad . \tag{S1.5}$$

Then, cancel the Jacobian in the denominator and in the numerators:

$$\Delta V_{\mathrm{N}}(R,s) = \left[\frac{\partial V_{\mathrm{N}}}{\partial R}\frac{1}{R}\left[(\mu+3)-\frac{1+\mu}{h_R^2}\right]+\frac{1}{h_R^2}\frac{\partial^2 V_{\mathrm{N}}}{\partial R^2}+\frac{1}{h_R^2}\frac{\partial V_{\mathrm{N}}}{\partial R}\left(\frac{\mu}{R}\right)+\frac{\partial V_{\mathrm{N}}}{\partial s}\frac{\partial s}{\partial \mathrm{v}}\frac{f_S^2}{W}\frac{1}{h_{\mathrm{v}}^2}\frac{\partial W}{\partial \mathrm{v}}+\frac{1}{h_{\mathrm{v}}^2}\frac{\partial^2 V_{\mathrm{N}}}{\partial s^2}\left(\frac{\partial s}{\partial \mathrm{v}}\right)^2+\frac{1}{h_{\mathrm{v}}^2}\frac{\partial V_{\mathrm{N}}}{\partial s}\left[\frac{\partial^2 s}{\partial \mathrm{v}^2}+\frac{\partial s}{\partial \mathrm{v}}\frac{-\frac{\partial^2 W}{\partial \mathrm{v}^2}}{\frac{\partial W}{\partial \mathrm{v}}}\right]\right] . \tag{S1.6}$$

Further employ Eq. (A38) of [15], which can be rewritten to the shape

$$\frac{f_S^2}{h_{\mathrm{v}}^2} = \frac{h_R^2(1+\mu)^2 W^2}{f_C^2 R^2\left(\frac{\partial W}{\partial \mathrm{v}}\right)^2} \quad , \tag{S1.7}$$

and insert it to (S1.6):

$$\Delta V_{\mathrm{N}}(R,s) = \frac{\partial V_{\mathrm{N}}}{\partial R}\frac{1}{R}\left[(\mu+3)-\frac{1+\mu}{h_R^2}\right]+\frac{1}{h_R^2}\frac{\partial^2 V_{\mathrm{N}}}{\partial R^2}+\frac{1}{h_R^2}\frac{\partial V_{\mathrm{N}}}{\partial R}\left(\frac{\mu}{R}\right)+\frac{\partial V_{\mathrm{N}}}{\partial s}\frac{\partial s}{\partial \nu}\frac{1}{W}\frac{h_R^2(1+\mu)^2W^2}{f_C^2R^2\left(\frac{\partial W}{\partial \nu}\right)^2}\frac{\partial W}{\partial \nu}+\frac{h_R^2(1+\mu)^2W^2}{f_S^2f_C^2R^2\left(\frac{\partial W}{\partial \nu}\right)^2}\frac{\partial^2 V_{\mathrm{N}}}{\partial s^2}\left(\frac{\partial s}{\partial \nu}\right)^2+\frac{h_R^2(1+\mu)^2W^2}{f_S^2f_C^2R^2\left(\frac{\partial W}{\partial \nu}\right)^2}\frac{\partial V_{\mathrm{N}}}{\partial s}\left[\frac{\partial^2 s}{\partial \nu^2}-\frac{\partial s}{\partial \nu}\frac{\frac{\partial^2 W}{\partial \nu^2}}{\frac{\partial W}{\partial \nu}}\right]. \tag{S1.8}$$

Mutually cancel some terms in the numerators and denominators, and we get

$$\Delta V_{\mathrm{N}}(R,s) = \frac{\partial V_{\mathrm{N}}}{\partial R}\frac{1}{R}\left[(\mu+3)-\frac{1+\mu}{h_R^2}\right]+\frac{1}{h_R^2}\frac{\partial^2 V_{\mathrm{N}}}{\partial R^2}+\frac{1}{h_R^2}\frac{\partial V_{\mathrm{N}}}{\partial R}\left(\frac{\mu}{R}\right)+\frac{\partial V_{\mathrm{N}}}{\partial s}\frac{\partial s}{\partial \nu}\frac{h_R^2(1+\mu)^2W}{f_C^2R^2\frac{\partial W}{\partial \nu}}+\frac{h_R^2(1+\mu)^2W^2}{f_S^2f_C^2R^2\left(\frac{\partial W}{\partial \nu}\right)^2}\frac{\partial^2 V_{\mathrm{N}}}{\partial s^2}\left(\frac{\partial s}{\partial \nu}\right)^2+\frac{h_R^2(1+\mu)^2W^2}{f_S^2f_C^2R^2\left(\frac{\partial W}{\partial \nu}\right)^2}\frac{\partial V_{\mathrm{N}}}{\partial s}\left[\frac{\partial^2 s}{\partial \nu^2}-\frac{\partial s}{\partial \nu}\frac{\frac{\partial^2 W}{\partial \nu^2}}{\frac{\partial W}{\partial \nu}}\right]. \tag{S1.9}$$

As $\frac{\partial s}{\partial \nu}=\frac{\partial s}{\partial W}\frac{\partial W}{\partial \nu}$, we obtain

$$\Delta V_{\mathrm{N}}(R,s) = \frac{\partial V_{\mathrm{N}}}{\partial R}\frac{1}{R}\left[(\mu+3)-\frac{1+\mu}{h_R^2}\right]+\frac{1}{h_R^2}\frac{\partial^2 V_{\mathrm{N}}}{\partial R^2}+\frac{1}{h_R^2}\frac{\partial V_{\mathrm{N}}}{\partial R}\left(\frac{\mu}{R}\right)+\frac{\partial V_{\mathrm{N}}}{\partial s}\frac{\partial s}{\partial W}\frac{\partial W}{\partial \nu}\frac{h_R^2(1+\mu)^2W}{f_C^2R^2\frac{\partial W}{\partial \nu}}+\frac{h_R^2(1+\mu)^2W^2}{f_S^2f_C^2R^2\left(\frac{\partial W}{\partial \nu}\right)^2}\frac{\partial^2 V_{\mathrm{N}}}{\partial s^2}\left(\frac{\partial s}{\partial W}\frac{\partial W}{\partial \nu}\right)^2+\frac{h_R^2(1+\mu)^2W^2}{f_S^2f_C^2R^2\left(\frac{\partial W}{\partial \nu}\right)^2}\frac{\partial V_{\mathrm{N}}}{\partial s}\left[\frac{\partial}{\partial \nu}\left(\frac{\partial s}{\partial W}\frac{\partial W}{\partial \nu}\right)-\frac{\partial s}{\partial W}\frac{\partial W}{\partial \nu}\frac{\frac{\partial^2 W}{\partial \nu^2}}{\frac{\partial W}{\partial \nu}}\right]. \tag{S1.10}$$

Further, as

$$\frac{\partial}{\partial \nu}\left(\frac{\partial s}{\partial \nu}\right)=\frac{\partial}{\partial \nu}\left(\frac{\partial s}{\partial W}\frac{\partial W}{\partial \nu}\right)=\left(\frac{\partial^2 s}{\partial W^2}\frac{\partial W}{\partial \nu}\frac{\partial W}{\partial \nu}+\frac{\partial s}{\partial W}\frac{\partial^2 W}{\partial \nu^2}\right), \tag{S1.11}$$

we receive the Laplacian in the form

$$\Delta V_{\mathrm{N}}(R,s) = \frac{1}{R}\left[(\mu+3)-\frac{1}{h_R^2}-\cancel{\frac{\mu}{h_R^2}}\right]\frac{\partial V_{\mathrm{N}}}{\partial R}+\frac{1}{h_R^2}\frac{\partial^2 V_{\mathrm{N}}}{\partial R^2}+\cancel{\frac{1}{h_R^2}\left(\frac{\mu}{R}\right)\frac{\partial V_{\mathrm{N}}}{\partial R}}+\frac{\partial s}{\partial W}\frac{h_R^2(1+\mu)^2W}{f_C^2R^2}\frac{\partial V_{\mathrm{N}}}{\partial s}+\frac{h_R^2(1+\mu)^2W^2}{f_S^2f_C^2R^2}\left(\frac{\partial s}{\partial W}\right)^2\frac{\partial^2 V_{\mathrm{N}}}{\partial s^2}+\frac{h_R^2(1+\mu)^2W^2}{f_S^2f_C^2R^2\left(\frac{\partial W}{\partial \nu}\right)^2}\frac{\partial V_{\mathrm{N}}}{\partial s}\left[\left(\frac{\partial^2 s}{\partial W^2}\frac{\partial W}{\partial \nu}\frac{\partial W}{\partial \nu}+\cancel{\frac{\partial s}{\partial W}\frac{\partial^2 W}{\partial \nu^2}}\right)-\cancel{\frac{\partial s}{\partial W}\frac{\partial^2 W}{\partial \nu^2}}\right]. \tag{S1.12}$$

Some terms are mutually cancelled, and then

$$\Delta V_{\mathrm{N}}(R,s) = \left[\frac{1}{h_R^2}\frac{\partial^2 V_{\mathrm{N}}}{\partial R^2}+\frac{h_R^2(1+\mu)^2W^2}{f_S^2f_C^2R^2}\left(\frac{\partial s}{\partial W}\right)^2\frac{\partial^2 V_{\mathrm{N}}}{\partial s^2}\right]+\left\{\frac{1}{R}\left[(\mu+3)-\frac{1}{h_R^2}\right]\frac{\partial V_{\mathrm{N}}}{\partial R}+\frac{\partial s}{\partial W}\frac{h_R^2(1+\mu)^2W}{f_C^2R^2}\frac{\partial V_{\mathrm{N}}}{\partial s}+\frac{h_R^2(1+\mu)^2W^2}{f_S^2f_C^2R^2}\left(\frac{\partial^2 s}{\partial W^2}\right)\frac{\partial V_{\mathrm{N}}}{\partial s}\right\}. \tag{S1.13}$$

Now, proceed with rewriting to *s*-terms, starting from (S1.13). We use the relations Eq. (A93), (A24), (A19), (A87), (A21) and (9) from the article [15].

First, use (A93) of [15] for the *s* derivative with respect to *W*:

$$\Delta V_{\mathrm{N}}(R,s) = \left[\frac{1}{h_R^2}\frac{\partial^2 V_{\mathrm{N}}}{\partial R^2}+\frac{h_R^2(1+\mu)^2W^2}{f_S^2f_C^2R^2}\left(\frac{f_C^2}{W}s\right)^2\frac{\partial^2 V_{\mathrm{N}}}{\partial s^2}\right]+\left\{\frac{1}{R}\left[(\mu+3)-\frac{1}{h_R^2}\right]\frac{\partial V_{\mathrm{N}}}{\partial R}+\frac{f_C^2}{W}s\frac{h_R^2(1+\mu)^2W}{f_C^2R^2}\frac{\partial V_{\mathrm{N}}}{\partial s}+\frac{h_R^2(1+\mu)^2W^2}{f_S^2f_C^2R^2}\frac{\partial}{\partial W}\left(\frac{f_C^2}{W}s\right)\frac{\partial V_{\mathrm{N}}}{\partial s}\right\}. \tag{S1.14}$$

Simplify and perform the remaining derivative with respect to *W*:

$$\Delta V_{\mathrm{N}}(R,s) = \left[\frac{1}{h_R^2}\frac{\partial^2 V_{\mathrm{N}}}{\partial R^2}+\frac{h_R^2(1+\mu)^2}{R^2}\frac{f_C^2}{f_S^2}s^2\frac{\partial^2 V_{\mathrm{N}}}{\partial s^2}\right]+\left\{\frac{1}{R}\left[(\mu+3)-\frac{1}{h_R^2}\right]\frac{\partial V_{\mathrm{N}}}{\partial R}+s\frac{h_R^2(1+\mu)^2}{R^2}\frac{\partial V_{\mathrm{N}}}{\partial s}+\frac{h_R^2(1+\mu)^2W^2}{f_S^2f_C^2R^2}\left(\frac{\partial}{\partial W}\left[\frac{f_C^2}{W}\right]s+\frac{f_C^2}{W}\frac{\partial s}{\partial W}\right)\frac{\partial V_{\mathrm{N}}}{\partial s}\right\}. \tag{S1.15}$$

Use (A24) of [15] for the ratio of generalized sine and cosine, and (A19) of [15] for $h_R^2$, and use (A87) of [15] for the derivative in the last term involving the generalized cosine:

$$\Delta V_{\mathrm{N}}(R,s) = \left[\left(\frac{\mu}{1+\mu}s^2+1\right)\frac{\partial^2 V_{\mathrm{N}}}{\partial R^2} + \frac{\frac{1}{\frac{\mu}{1+\mu}s^2+1}(1+\mu)^2}{R^2}\frac{(1+\mu)-s^2}{(1+\mu)\,s^2}s^2\frac{\partial^2 V_{\mathrm{N}}}{\partial s^2}\right] + \left\{\frac{1}{R}\left[(\mu+3)-\left(\frac{\mu}{1+\mu}s^2+1\right)\right]\frac{\partial V_{\mathrm{N}}}{\partial R} + s\frac{\frac{1}{\frac{\mu}{1+\mu}s^2+1}(1+\mu)^2}{R^2}\frac{\partial V_{\mathrm{N}}}{\partial s} + \frac{(1+\mu)^2 W^2}{\frac{f_S^2}{h_R^2}f_C^2 R^2}\left(-\frac{f_C^2}{W^2}(2h_R^2 f_S^2+1)s + \frac{f_C^2}{W}\frac{f_C^2}{W}s\right)\frac{\partial V_{\mathrm{N}}}{\partial s}\right\} \quad . \tag{S1.16}$$

Further simplify

$$\Delta V_{\mathrm{N}}(R,s) = \left[\left(\frac{\mu}{1+\mu}s^2+1\right)\frac{\partial^2 V_{\mathrm{N}}}{\partial R^2} + \frac{\frac{(1+\mu)}{\mu\,s^2+(1+\mu)}(1+\mu)}{R^2}\frac{(1+\mu)-s^2}{1}\frac{\partial^2 V_{\mathrm{N}}}{\partial s^2}\right] + \left\{\frac{1}{R}\left[(\mu+2)-\frac{\mu}{1+\mu}s^2\right]\frac{\partial V_{\mathrm{N}}}{\partial R} + s\frac{\frac{(1+\mu)}{\mu\,s^2+(1+\mu)}(1+\mu)^2}{R^2}\frac{\partial V_{\mathrm{N}}}{\partial s} + \frac{(1+\mu)^2}{s^2R^2}\left(-(2h_R^2 f_S^2+1)+f_C^2\right)s\frac{\partial V_{\mathrm{N}}}{\partial s}\right\} \quad , \tag{S1.17}$$

and still simplify

$$\Delta V_{\mathrm{N}}(R,s) = \left[\left(\frac{\mu}{1+\mu}s^2+1\right)\frac{\partial^2 V_{\mathrm{N}}}{\partial R^2} + (1+\mu)^2\frac{(1+\mu)-s^2}{\mu\,s^2+(1+\mu)}\frac{1}{R^2}\frac{\partial^2 V_{\mathrm{N}}}{\partial s^2}\right] + \left\{\frac{(1+\mu)(\mu+2)-\mu s^2}{1+\mu}\frac{1}{R}\frac{\partial V_{\mathrm{N}}}{\partial R} + \frac{(1+\mu)^3}{\mu\,s^2+(1+\mu)}\frac{s}{R^2}\frac{\partial V_{\mathrm{N}}}{\partial s} + (1+\mu)^2(-2h_R^2 f_S^2 - 1 + f_C^2)\frac{1}{sR^2}\frac{\partial V_{\mathrm{N}}}{\partial s}\right\} \quad . \tag{S1.18}$$

As

$$(-2h_R^2 f_S^2 - 1 + f_C^2) = (-2h_R^2 f_S^2 - f_S^2) = -f_S^2(2h_R^2+1) = -\frac{f_S^2}{h_R^2}(2h_R^2+1)h_R^2 = -s^2\left(2\frac{1}{\frac{\mu}{1+\mu}s^2+1}+1\right)\frac{1}{\frac{\mu}{1+\mu}s^2+1} = -s^2\left(\frac{2(1+\mu)}{\mu s^2+(1+\mu)}+1\right)\frac{(1+\mu)}{\mu s^2+(1+\mu)} = -s^2\left(\frac{2(1+\mu)+\mu s^2+(1+\mu)}{\mu s^2+(1+\mu)}\right)\frac{(1+\mu)}{\mu s^2+(1+\mu)} = -(1+\mu)s^2\frac{3(1+\mu)+\mu s^2}{[\mu s^2+(1+\mu)]^2} \quad , \tag{S1.19}$$

then

$$\Delta V_{\mathrm{N}}(R,s) = \left[\left(\frac{\mu}{1+\mu}s^2+1\right)\frac{\partial^2 V_{\mathrm{N}}}{\partial R^2} + (1+\mu)^2\frac{(1+\mu)-s^2}{\mu s^2+(1+\mu)}\frac{1}{R^2}\frac{\partial^2 V_{\mathrm{N}}}{\partial s^2}\right] + \left\{\frac{(1+\mu)(\mu+2)-\mu s^2}{1+\mu}\frac{1}{R}\frac{\partial V_{\mathrm{N}}}{\partial R} + \frac{(1+\mu)^3}{\mu s^2+(1+\mu)}\frac{s}{R^2}\frac{\partial V_{\mathrm{N}}}{\partial s} - (1+\mu)^2(1+\mu)s^2\frac{3(1+\mu)+\mu s^2}{[\mu s^2+(1+\mu)]^2}\frac{1}{sR^2}\frac{\partial V_{\mathrm{N}}}{\partial s}\right\} . \tag{S1.20}$$

Factor out the derivative from the last two terms and obtain

$$\Delta V_{\mathrm{N}}(R,s) = \left[\left(\frac{\mu}{1+\mu}s^2+1\right)\frac{\partial^2 V_{\mathrm{N}}}{\partial R^2} + (1+\mu)^2\frac{(1+\mu)-s^2}{\mu s^2+(1+\mu)}\frac{1}{R^2}\frac{\partial^2 V_{\mathrm{N}}}{\partial s^2}\right] + \left\{\frac{(1+\mu)(\mu+2)-\mu s^2}{1+\mu}\frac{1}{R}\frac{\partial V_{\mathrm{N}}}{\partial R} + (1+\mu)^3\left[\frac{1}{\mu s^2+(1+\mu)} - \frac{3(1+\mu)+\mu s^2}{[\mu s^2+(1+\mu)]^2}\right]\frac{s}{R^2}\frac{\partial V_{\mathrm{N}}}{\partial s}\right\} \quad , \tag{S1.21}$$

and still further

$$\Delta V_{\mathrm{N}}(R,s) = \left[\left(\frac{\mu}{1+\mu}s^2+1\right)\frac{\partial^2 V_{\mathrm{N}}}{\partial R^2} + (1+\mu)^2\frac{(1+\mu)-s^2}{\mu s^2+(1+\mu)}\frac{1}{R^2}\frac{\partial^2 V_{\mathrm{N}}}{\partial s^2}\right] + \left[\frac{(1+\mu)(\mu+2)-\mu s^2}{1+\mu}\frac{1}{R}\frac{\partial V_{\mathrm{N}}}{\partial R} + (1+\mu)^3\frac{\mu s^2+(1+\mu)-3(1+\mu)-\mu s^2}{[\mu s^2+(1+\mu)]^2}\frac{s}{R^2}\frac{\partial V_{\mathrm{N}}}{\partial s}\right] \quad . \tag{S1.22}$$

The final algebraic manipulations lead to

$$\Delta V_{\mathrm{N}}(R,s) = \left[\frac{\mu s^2+(1+\mu)}{1+\mu}\frac{\partial^2 V_{\mathrm{N}}}{\partial R^2} + (1+\mu)^2\frac{(1+\mu)-s^2}{\mu s^2+(1+\mu)}\frac{1}{R^2}\frac{\partial^2 V_{\mathrm{N}}}{\partial s^2}\right] + \left[\frac{(1+\mu)(\mu+2)-\mu s^2}{1+\mu}\frac{1}{R}\frac{\partial V_{\mathrm{N}}}{\partial R} - 2(1+\mu)^4\frac{s}{[\mu s^2+(1+\mu)]^2}\frac{1}{R^2}\frac{\partial V_{\mathrm{N}}}{\partial s}\right] . \tag{S1.23}$$

On the right side, there is the Laplacian expressed in *R* and *s*-terms.

It is also possible to rewrite the derivatives in (S1.23) from the form “with respect to *s*” to the form “with respect to *ν*”. The following relations for the derivatives are to be employed for this purpose ((A93), (A21) and (9) relations of the article [15] are used):

$$\frac{\partial V_{\mathrm{N}}}{\partial s}=\frac{\frac{\partial V_{\mathrm{N}}}{\partial \nu}}{\frac{\partial s}{\partial \nu}}=\frac{\partial V_{\mathrm{N}}}{\partial \nu}\frac{1}{\frac{\partial s}{\partial W}\frac{\partial W}{\partial \nu}}=\frac{\partial V_{\mathrm{N}}}{\partial \nu}\frac{1}{\frac{f_C^2}{W}s\,\frac{1+\mu\sin^2\nu}{\sin\nu\cos\nu}W}=\frac{\partial V_{\mathrm{N}}}{\partial \nu}\frac{1}{f_C^2 s\,\frac{1+\mu\sin^2\nu}{\sin\nu\cos\nu}}=\frac{\partial V_{\mathrm{N}}}{\partial \nu}\frac{1}{\frac{1-\frac{1}{1+\mu}s^2}{\frac{\mu}{1+\mu}s^2+1}s\,\frac{1+\mu\sin^2\nu}{\sin\nu\cos\nu}}=$$
$$\frac{\partial V_{\mathrm{N}}}{\partial \nu}\frac{1}{\frac{(1+\mu)-s^2}{\mu s^2+(1+\mu)}s\,\frac{1+\mu\sin^2\nu}{\sin\nu\cos\nu}}=\frac{\partial V_{\mathrm{N}}}{\partial \nu}\frac{[\mu s^2+(1+\mu)]\sin\nu\cos\nu}{[(1+\mu)-s^2]s\,(1+\mu\sin^2\nu)}\quad . \qquad \text{(S1.24)}$$

The second derivative is

$$\frac{\partial^2 V_{\mathrm{N}}}{\partial s^2}=\frac{\frac{\partial^2 V_{\mathrm{N}}}{\partial \nu^2}-\frac{\partial V_{\mathrm{N}}}{\partial s}\frac{\partial^2 s}{\partial \nu^2}}{\left(\frac{\partial s}{\partial \nu}\right)^2}=\frac{\frac{\partial^2 V_{\mathrm{N}}}{\partial \nu^2}-\frac{\frac{\partial V_{\mathrm{N}}}{\partial \nu}}{\frac{\partial s}{\partial \nu}}\frac{\partial^2 s}{\partial \nu^2}}{\left(\frac{\partial s}{\partial \nu}\right)^2}=\left[\frac{\partial^2 V_{\mathrm{N}}}{\partial \nu^2}-\frac{\frac{\partial V_{\mathrm{N}}}{\partial \nu}}{\frac{\partial s}{\partial W}\frac{\partial W}{\partial \nu}}\frac{\partial}{\partial \nu}\left(\frac{\partial s}{\partial \nu}\right)\right]\frac{1}{\left(\frac{\partial s}{\partial W}\frac{\partial W}{\partial \nu}\right)^2}=$$
$$\left[\frac{\partial^2 V_{\mathrm{N}}}{\partial \nu^2}-\frac{\frac{\partial V_{\mathrm{N}}}{\partial \nu}}{\frac{\partial s}{\partial W}\frac{\partial W}{\partial \nu}}\frac{\partial}{\partial \nu}\left(\frac{\partial s}{\partial W}\frac{\partial W}{\partial \nu}\right)\right]\frac{1}{\left(\frac{\partial s}{\partial W}\frac{\partial W}{\partial \nu}\right)^2}=\left[\frac{\partial^2 V_{\mathrm{N}}}{\partial \nu^2}-\frac{\frac{\partial V_{\mathrm{N}}}{\partial \nu}}{\frac{\partial s}{\partial W}\frac{\partial W}{\partial \nu}}\left(\frac{\partial}{\partial \nu}\left[\frac{\partial s}{\partial W}\right]\frac{\partial W}{\partial \nu}+\frac{\partial s}{\partial W}\frac{\partial}{\partial \nu}\left[\frac{\partial W}{\partial \nu}\right]\right)\right]\frac{1}{\left(\frac{\partial s}{\partial W}\frac{\partial W}{\partial \nu}\right)^2}=$$
$$\left[\frac{\partial^2 V_{\mathrm{N}}}{\partial \nu^2}-\frac{\frac{\partial V_{\mathrm{N}}}{\partial \nu}}{\frac{\partial s}{\partial W}\frac{\partial W}{\partial \nu}}\left(\frac{\partial}{\partial W}\left[\frac{f_C^2}{W}s\right]\left(\frac{\partial W}{\partial \nu}\right)^2+\frac{\partial s}{\partial W}\frac{\partial^2 W}{\partial \nu^2}\right)\right]\frac{1}{\left(\frac{\partial s}{\partial W}\frac{\partial W}{\partial \nu}\right)^2}=\left[\frac{\partial^2 V_{\mathrm{N}}}{\partial \nu^2}\frac{1}{\left(\frac{\partial s}{\partial W}\frac{\partial W}{\partial \nu}\right)^2}-\frac{\partial V_{\mathrm{N}}}{\partial \nu}\frac{1}{\left(\frac{\partial s}{\partial W}\frac{\partial W}{\partial \nu}\right)^3}\left(\left[\frac{\partial}{\partial W}\left(\frac{f_C^2}{W}\right)s+\right.\right.\right.$$
$$\left.\left.\left.\frac{f_C^2}{W}\frac{\partial s}{\partial W}\right]\left(\frac{\partial W}{\partial \nu}\right)^2+\frac{\partial s}{\partial W}\frac{\partial^2 W}{\partial \nu^2}\right)\right]\quad . \qquad \text{(S1.25)}$$

When we employ the relation (9) of [15] for $W$ derivative $\frac{\partial W}{\partial \nu}$, the second derivative $\frac{\partial^2 W}{\partial \nu^2}$ can be writtten as

$$\frac{\partial^2 W}{\partial \nu^2}=\frac{\partial}{\partial \nu}\left[\frac{\partial W}{\partial \nu}\right]=\frac{\partial}{\partial \nu}\left[\frac{1+\mu\sin^2\nu}{\sin\nu\cos\nu}W\right]=\frac{\partial}{\partial \nu}\left[\frac{1+\mu\sin^2\nu}{\sin\nu\cos\nu}\right]W+\frac{1+\mu\sin^2\nu}{\sin\nu\cos\nu}\frac{\partial W}{\partial \nu}=\left[\frac{(1+\mu)}{\cos^2\nu}-\frac{1}{\sin^2\nu}\right]W+$$
$$\frac{1+\mu\sin^2\nu}{\sin\nu\cos\nu}\frac{1+\mu\sin^2\nu}{\sin\nu\cos\nu}W=\left\{\frac{(1+\mu)}{\cos^2\nu}-\frac{1}{\sin^2\nu}+\frac{(1+\mu\sin^2\nu)^2}{\sin^2\nu\cos^2\nu}\right\}W=\left[\frac{(1+\mu)(2+\mu)}{\cos^2\nu}-\mu^2\right]W\quad . \qquad \text{(S1.26)}$$

Further, we use the above derivative, and also (A87), (A93) relations of [15], for the following algebraic changes

$$\frac{\partial^2 V_{\mathrm{N}}}{\partial s^2}=\left[\frac{\partial^2 V_{\mathrm{N}}}{\partial \nu^2}\frac{1}{\left(\frac{\partial s}{\partial W}\frac{\partial W}{\partial \nu}\right)^2}-\frac{\partial V_{\mathrm{N}}}{\partial \nu}\left(\left[-\frac{f_C^2}{W^2}(2h_R^2f_S^2+1)s+\frac{f_C^2}{W}\frac{\partial s}{\partial W}\right]\frac{1}{\left(\frac{\partial s}{\partial W}\right)^3\frac{\partial W}{\partial \nu}}+\frac{\frac{\partial^2 W}{\partial \nu^2}}{\left(\frac{\partial s}{\partial W}\right)^2\left(\frac{\partial W}{\partial \nu}\right)^3}\right)\right]=$$
$$\left[\frac{\partial^2 V_{\mathrm{N}}}{\partial \nu^2}\frac{1}{\left(\frac{f_C^2}{W}s\frac{\partial W}{\partial \nu}\right)^2}-\frac{\partial V_{\mathrm{N}}}{\partial \nu}\left(\left[-\frac{1}{W}(2h_R^2f_S^2+1)s+\frac{f_C^2}{W}s\right]\frac{f_C^2}{W}\frac{1}{\left(\frac{f_C^2}{W}s\right)^3\frac{\partial W}{\partial \nu}}+\frac{\frac{\partial^2 W}{\partial \nu^2}}{\left(\frac{f_C^2}{W}s\right)^2\left(\frac{\partial W}{\partial \nu}\right)^3}\right)\right]=\left[\frac{\partial^2 V_{\mathrm{N}}}{\partial \nu^2}\frac{1}{\left(\frac{f_C^2}{W}s\frac{\partial W}{\partial \nu}\right)^2}-\right.$$
$$\left.\frac{\partial V_{\mathrm{N}}}{\partial \nu}\left([-2h_R^2f_S^2-1+f_C^2]\frac{s}{W}\frac{1}{\left(\frac{f_C^2}{W}\right)^2s^3\frac{\partial W}{\partial \nu}}+\frac{\frac{\partial^2 W}{\partial \nu^2}}{\left(\frac{f_C^2}{W}\right)^2s^2\left(\frac{\partial W}{\partial \nu}\right)^3}\right)\right]\quad . \qquad \text{(S1.27)}$$

Use (S1.19) and get

$$\frac{\partial^2 V_{\mathrm{N}}}{\partial s^2}=\left[\frac{\partial^2 V_{\mathrm{N}}}{\partial \nu^2}\frac{1}{\left(\frac{f_C^2}{W}s\frac{\partial W}{\partial \nu}\right)^2}-\frac{\partial V_{\mathrm{N}}}{\partial \nu}\left(-(1+\mu)s^2\frac{3(1+\mu)+\mu s^2}{[\mu s^2+(1+\mu)]^2}\frac{s}{W}\frac{1}{\left(\frac{f_C^2}{W}\right)^2s^3\frac{\partial W}{\partial \nu}}+\frac{\frac{\partial^2 W}{\partial \nu^2}}{\left(\frac{f_C^2}{W}\right)^2s^2\left(\frac{\partial W}{\partial \nu}\right)^3}\right)\right]\quad , \qquad \text{(S1.28)}$$

which can be simplified to

$$\frac{\partial^2 V_\mathrm{N}}{\partial s^2} = \left[\frac{\partial^2 V_\mathrm{N}}{\partial \nu^2}\frac{1}{\left(\frac{f_C^2}{W}s\frac{\partial W}{\partial \nu}\right)^2} - \frac{\partial V_\mathrm{N}}{\partial \nu}\frac{1}{\left(f_C^2\right)^2}\left(-(1+\mu)\frac{3(1+\mu)+\mu s^2}{[\mu s^2+(1+\mu)]^2}\frac{1}{\frac{1}{W}\frac{\partial W}{\partial \nu}} + \frac{\frac{\partial^2 W}{\partial \nu^2}}{s^2\frac{1}{W^2}\left(\frac{\partial W}{\partial \nu}\right)^3}\right)\right] \quad . \qquad \text{(S1.29)}$$

Now use (S1.26) and the relation (9) of [15] to get

$$\frac{\partial^2 V_\mathrm{N}}{\partial s^2} = \left[\frac{\partial^2 V_\mathrm{N}}{\partial \nu^2}\frac{1}{\left(\frac{f_C^2}{W}s\frac{1+\mu\sin^2\nu}{\sin\nu\cos\nu}W\right)^2} - \frac{\partial V_\mathrm{N}}{\partial \nu}\frac{1}{\left(f_C^2\right)^2}\left(-(1+\mu)\frac{3(1+\mu)+\mu s^2}{[\mu s^2+(1+\mu)]^2}\frac{1}{\frac{1}{W}\frac{1+\mu\sin^2\nu}{\sin\nu\cos\nu}W} + \frac{\left[\frac{(1+\mu)(2+\mu)}{\cos^2\nu}-\mu^2\right]W}{s^2\frac{1}{W^2}\left(\frac{1+\mu\sin^2\nu}{\sin\nu\cos\nu}W\right)^3}\right)\right] =$$

$$\left[\frac{\partial^2 V_\mathrm{N}}{\partial \nu^2}\frac{1}{\left(f_C^2 s\frac{1+\mu\sin^2\nu}{\sin\nu\cos\nu}\right)^2} - \frac{\partial V_\mathrm{N}}{\partial \nu}\frac{1}{\left(f_C^2\right)^2}\left(-(1+\mu)\frac{3(1+\mu)+\mu s^2}{[\mu s^2+(1+\mu)]^2}\frac{\sin\nu\cos\nu}{(1+\mu\sin^2\nu)} + \frac{\left[\frac{(1+\mu)(2+\mu)-\mu^2\cos^2\nu}{\cos^2\nu}\right]}{s^2\left(\frac{1+\mu\sin^2\nu}{\sin\nu\cos\nu}\right)^3}\right)\right] =$$

$$\frac{1}{\left(f_C^2\right)^2}\left[\frac{\partial^2 V_\mathrm{N}}{\partial \nu^2}\frac{\sin^2\nu\cos^2\nu}{s^2(1+\mu\sin^2\nu)^2} - \frac{\partial V_\mathrm{N}}{\partial \nu}\left(-(1+\mu)\frac{3(1+\mu)+\mu s^2}{[\mu s^2+(1+\mu)]^2}\frac{\sin\nu\cos\nu}{(1+\mu\sin^2\nu)} + \frac{(1+\mu)(2+\mu)-\mu^2\cos^2\nu}{\cos^2\nu}\frac{\sin^3\nu\cos^3\nu}{s^2(1+\mu\sin^2\nu)^3}\right)\right] =$$

$$\frac{1}{\left(f_C^2\right)^2}\left[\frac{\partial^2 V_\mathrm{N}}{\partial \nu^2}\frac{\sin^2\nu\cos^2\nu}{s^2(1+\mu\sin^2\nu)^2} + \frac{\partial V_\mathrm{N}}{\partial \nu}\left((1+\mu)\frac{3(1+\mu)+\mu s^2}{[\mu s^2+(1+\mu)]^2}\frac{\sin\nu\cos\nu}{(1+\mu\sin^2\nu)} - \frac{1}{s^2}\frac{[(1+\mu)(2+\mu)-\mu^2\cos^2\nu]\sin^3\nu\cos\nu}{(1+\mu\sin^2\nu)^3}\right)\right] \quad . \qquad \text{(S1.30)}$$

Still express the generalized cosine in front of the bracket by means of *s* (using (A21) of [15]):

$$\frac{\partial^2 V_\mathrm{N}}{\partial s^2} = \frac{1}{\left(\frac{(1+\mu)-s^2}{\mu s^2+(1+\mu)}\right)^2}\left[\frac{\partial^2 V_\mathrm{N}}{\partial \nu^2}\frac{\sin^2\nu\cos^2\nu}{s^2(1+\mu\sin^2\nu)^2} + \frac{\partial V_\mathrm{N}}{\partial \nu}\left((1+\mu)\frac{3(1+\mu)+\mu s^2}{[\mu s^2+(1+\mu)]^2}\frac{\sin\nu\cos\nu}{(1+\mu\sin^2\nu)} - \frac{1}{s^2}\frac{[(1+\mu)(2+\mu)-\mu^2\cos^2\nu]\sin^3\nu\cos\nu}{(1+\mu\sin^2\nu)^3}\right)\right] =$$

$$\left[\frac{\partial^2 V_\mathrm{N}}{\partial \nu^2}\frac{[\mu s^2+(1+\mu)]^2}{\left((1+\mu)-s^2\right)^2}\frac{\sin^2\nu\cos^2\nu}{s^2(1+\mu\sin^2\nu)^2} + \frac{\partial V_\mathrm{N}}{\partial \nu}\left((1+\mu)\frac{3(1+\mu)+\mu s^2}{\left((1+\mu)-s^2\right)^2}\frac{\sin\nu\cos\nu}{(1+\mu\sin^2\nu)} - \frac{[\mu s^2+(1+\mu)]^2}{s^2\left((1+\mu)-s^2\right)^2}\frac{[(1+\mu)(2+\mu)-\mu^2\cos^2\nu]\sin^3\nu\cos\nu}{(1+\mu\sin^2\nu)^3}\right)\right] \quad . \qquad \text{(S1.31)}$$

As

$$[(1+\mu)(2+\mu)-\mu^2\cos^2\nu] = [(2+\mu+2\mu+\mu\mu)-\mu^2(1-\sin^2\nu)] = [2+3\mu+\mu^2-\mu^2+\mu^2\sin^2\nu] = [2+3\mu+\mu^2\sin^2\nu] \quad , \qquad \text{(S1.32)}$$

then

$$\frac{\partial^2 V_\mathrm{N}}{\partial s^2} = \left[\frac{\partial^2 V_\mathrm{N}}{\partial \nu^2}\frac{[\mu s^2+(1+\mu)]^2}{s^2\left((1+\mu)-s^2\right)^2}\frac{\sin^2\nu\cos^2\nu}{(1+\mu\sin^2\nu)^2} + \frac{\partial V_\mathrm{N}}{\partial \nu}\left((1+\mu)\frac{3(1+\mu)+\mu s^2}{\left((1+\mu)-s^2\right)^2}\frac{\sin\nu\cos\nu}{(1+\mu\sin^2\nu)} - \frac{[\mu s^2+(1+\mu)]^2}{s^2\left((1+\mu)-s^2\right)^2}\frac{[2+3\mu+\mu^2\sin^2\nu]\sin^3\nu\cos\nu}{(1+\mu\sin^2\nu)^3}\right)\right] \quad . \qquad \text{(S1.33)}$$

Finally, we input both derivatives, the first and the second, to the Laplacian relation (S1.23), resulting in

$$\Delta V_{\mathrm{N}}(R,\nu)=\left[\frac{\mu s^2+(1+\mu)}{1+\mu}\frac{\partial^2 V_{\mathrm{N}}}{\partial R^2}+(1+\mu)^2\frac{(1+\mu)-s^2}{\mu s^2+(1+\mu)}\frac{1}{R^2}\left[\frac{\partial^2 V_{\mathrm{N}}}{\partial \nu^2}\frac{\left[\mu s^2+(1+\mu)\right]^2}{s^2\left((1+\mu)-s^2\right)^2}\frac{\sin^2\nu\cos^2\nu}{(1+\mu\sin^2\nu)^2}+\frac{\partial V_{\mathrm{N}}}{\partial \nu}\left((1+\mu)\frac{3(1+\mu)+\mu s^2}{\left((1+\mu)-s^2\right)^2}\frac{\sin\nu\cos\nu}{(1+\mu\sin^2\nu)}-\frac{\left[\mu s^2+(1+\mu)\right]^2}{s^2\left((1+\mu)-s^2\right)^2}\frac{\left[2+3\mu+\mu^2\sin^2\nu\right]\sin^3\nu\cos\nu}{(1+\mu\sin^2\nu)^3}\right)\right]\right]+\left[\frac{(1+\mu)(\mu+2)-\mu s^2}{1+\mu}\frac{1}{R}\frac{\partial U_{\mathrm{T}}}{\partial R}-2(1+\mu)^4\frac{s}{\left[\mu s^2+(1+\mu)\right]^2}\frac{1}{R^2}\frac{\partial U_{\mathrm{T}}}{\partial \nu}\frac{\left[\mu s^2+(1+\mu)\right]\sin\nu\cos\nu}{\left[(1+\mu)-s^2\right]s(1+\mu\sin^2\nu)}\right] \quad . \tag{S1.34}$$

Now, combine the same kind and order derivatives,

$$\Delta V_{\mathrm{N}}(R,\nu)=\left[\frac{\mu s^2+(1+\mu)}{1+\mu}\frac{\partial^2 V_{\mathrm{N}}}{\partial R^2}+\left[\frac{1}{R^2}\frac{\partial^2 V_{\mathrm{N}}}{\partial \nu^2}(1+\mu)^2\frac{(1+\mu)-s^2}{\mu s^2+(1+\mu)}\frac{\left[\mu s^2+(1+\mu)\right]^2}{s^2\left((1+\mu)-s^2\right)^2}\frac{\sin^2\nu\cos^2\nu}{(1+\mu\sin^2\nu)^2}+\frac{1}{R^2}\frac{\partial V_{\mathrm{N}}}{\partial \nu}\left((1+\mu)^2\frac{(1+\mu)-s^2}{\mu s^2+(1+\mu)}(1+\mu)\frac{3(1+\mu)+\mu s^2}{\left((1+\mu)-s^2\right)^2}\frac{\sin\nu\cos\nu}{(1+\mu\sin^2\nu)}-(1+\mu)^2\frac{(1+\mu)-s^2}{\mu s^2+(1+\mu)}\frac{\left[\mu s^2+(1+\mu)\right]^2}{s^2\left((1+\mu)-s^2\right)^2}\frac{\left[2+3\mu+\mu^2\sin^2\nu\right]\sin^3\nu\cos\nu}{(1+\mu\sin^2\nu)^3}\right)-2(1+\mu)^4\frac{s}{\left[\mu s^2+(1+\mu)\right]^2}\frac{1}{R^2}\frac{\partial V_{\mathrm{N}}}{\partial \nu}\frac{\left[\mu s^2+(1+\mu)\right]\sin\nu\cos\nu}{\left[(1+\mu)-s^2\right]s(1+\mu\sin^2\nu)}\right]\right]+\left[\frac{(1+\mu)(\mu+2)-\mu s^2}{1+\mu}\frac{1}{R}\frac{\partial V_{\mathrm{N}}}{\partial R}\right] \quad , \tag{S1.35}$$

and simplify:

$$\Delta V_{\mathrm{N}}(R,\nu)=\left[\frac{\mu s^2+(1+\mu)}{1+\mu}\frac{\partial^2 V_{\mathrm{N}}}{\partial R^2}+\left[\frac{1}{R^2}\frac{\partial^2 V_{\mathrm{N}}}{\partial \nu^2}\frac{(1+\mu)^2}{s^2}\frac{\mu s^2+(1+\mu)}{(1+\mu)-s^2}\frac{\sin^2\nu\cos^2\nu}{(1+\mu\sin^2\nu)^2}+\frac{1}{R^2}\frac{\partial V_{\mathrm{N}}}{\partial \nu}\left((1+\mu)^3\frac{1}{\mu s^2+(1+\mu)}\frac{3(1+\mu)+\mu s^2}{(1+\mu)-s^2}\frac{\sin\nu\cos\nu}{(1+\mu\sin^2\nu)}-(1+\mu)^2\frac{\mu s^2+(1+\mu)}{s^2\left((1+\mu)-s^2\right)}\frac{\left[2+3\mu+\mu^2\sin^2\nu\right]\sin^3\nu\cos\nu}{(1+\mu\sin^2\nu)^3}-2(1+\mu)^4\frac{1}{\left[\mu s^2+(1+\mu)\right]}\frac{1}{\left[(1+\mu)-s^2\right]}\frac{\sin\nu\cos\nu}{(1+\mu\sin^2\nu)}\right)\right]\right]+\left[\frac{(1+\mu)(\mu+2)-\mu s^2}{1+\mu}\frac{1}{R}\frac{\partial V_{\mathrm{N}}}{\partial R}\right] \quad . \tag{S1.36}$$

Further simplification leads to

$$\Delta V_{\mathrm{N}}(R,\nu)=\left[\frac{\mu s^2+(1+\mu)}{1+\mu}\frac{\partial^2 V_{\mathrm{N}}}{\partial R^2}+\left[\frac{1}{R^2}\frac{\partial^2 V_{\mathrm{N}}}{\partial \nu^2}\frac{(1+\mu)^2}{s^2}\frac{\mu s^2+(1+\mu)}{(1+\mu)-s^2}\frac{\sin^2\nu\cos^2\nu}{(1+\mu\sin^2\nu)^2}+\frac{1}{R^2}\frac{\partial V_{\mathrm{N}}}{\partial \nu}\frac{1}{\left[(1+\mu)-s^2\right]}\left((1+\mu)^3\frac{3(1+\mu)+\mu s^2}{\mu s^2+(1+\mu)}\frac{\sin\nu\cos\nu}{(1+\mu\sin^2\nu)}-(1+\mu)^3\frac{2(1+\mu)}{\left[\mu s^2+(1+\mu)\right]}\frac{\sin\nu\cos\nu}{(1+\mu\sin^2\nu)}-(1+\mu)^2\frac{\mu s^2+(1+\mu)}{s^2}\frac{\left[2+3\mu+\mu^2\sin^2\nu\right]\sin^3\nu\cos\nu}{(1+\mu\sin^2\nu)^3}\right)\right]\right]+\left[\frac{(1+\mu)(\mu+2)-\mu s^2}{1+\mu}\frac{1}{R}\frac{\partial V_{\mathrm{N}}}{\partial R}\right] \quad . \tag{S1.37}$$

As

$$\left((1+\mu)^3\frac{3(1+\mu)+\mu s^2}{\mu s^2+(1+\mu)}\frac{\sin\nu\cos\nu}{(1+\mu\sin^2\nu)}-(1+\mu)^3\frac{2(1+\mu)}{\left[\mu s^2+(1+\mu)\right]}\frac{\sin\nu\cos\nu}{(1+\mu\sin^2\nu)}-(1+\mu)^2\frac{\mu s^2+(1+\mu)}{s^2}\frac{\left[2+3\mu+\mu^2\sin^2\nu\right]\sin^3\nu\cos\nu}{(1+\mu\sin^2\nu)^3}\right)=\left((1+\mu)^3\left[\frac{3(1+\mu)+\mu s^2}{\mu s^2+(1+\mu)}-\frac{2(1+\mu)}{\left[\mu s^2+(1+\mu)\right]}\right]\frac{\sin\nu\cos\nu}{(1+\mu\sin^2\nu)}-(1+\mu)^2\frac{\mu s^2+(1+\mu)}{s^2}\frac{\left[2+3\mu+\mu^2\sin^2\nu\right]\sin^3\nu\cos\nu}{(1+\mu\sin^2\nu)^3}\right)=$$

$$\left((1+\mu)^3 \frac{(1+\mu)+\mu s^2}{\mu s^2+(1+\mu)} \frac{\sin \nu \cos \nu}{(1+\mu \sin^2 \nu)} - (1+\mu)^2 \frac{(1+\mu)+\mu s^2}{s^2} \frac{\left[2+3\mu+\mu^2 \sin^2 \nu\right] \sin^3 \nu \cos \nu}{(1+\mu \sin^2 \nu)^3}\right) =$$

$$(1+\mu)^2 \frac{\sin \nu \cos \nu}{(1+\mu \sin^2 \nu)} \left((1+\mu) - \frac{(1+\mu)+\mu s^2}{s^2} \frac{\left[2+3\mu+\mu^2 \sin^2 \nu\right] \sin^2 \nu}{(1+\mu \sin^2 \nu)^2}\right) \quad , \tag{S1.38}$$

the final relation for the Laplacian in the SOS coordinates with the $V_{\mathrm{N}}$ derivatives expressed in the form "with respect to $R$ or to $\nu$" is

$$\Delta V_{\mathrm{N}}(R, \nu) =$$

$$\frac{\mu s^2+(1+\mu)}{1+\mu} \frac{\partial^2 V_{\mathrm{N}}}{\partial R^2} + \frac{(1+\mu)(\mu+2)-\mu s^2}{1+\mu} \frac{1}{R} \frac{\partial V_{\mathrm{N}}}{\partial R} + \frac{(1+\mu)^2}{s^2} \frac{\mu s^2+(1+\mu)}{(1+\mu)-s^2} \frac{\sin^2 \nu \cos^2 \nu}{(1+\mu \sin^2 \nu)^2} \frac{1}{R^2} \frac{\partial^2 V_{\mathrm{N}}}{\partial \nu^2} +$$

$$\frac{(1+\mu)^2}{[(1+\mu)-s^2]} \frac{\sin \nu \cos \nu}{(1+\mu \sin^2 \nu)} \left((1+\mu) - \frac{(1+\mu)+\mu s^2}{s^2} \frac{\left[2+3\mu+\mu^2 \sin^2 \nu\right] \sin^2 \nu}{(1+\mu \sin^2 \nu)^2}\right) \frac{1}{R^2} \frac{\partial V_{\mathrm{N}}}{\partial \nu} \quad . \tag{S1.39}$$

**Supplement 2** to P. Strunz: On Laplace equation solution in orthogonal similar oblate spheroidal coordinates

## Check of the relation leading to the SOS and solid harmonics equivalence for degree up to 6.

Here, a check if we can obtain the coefficients of the standard Legendree polynomials $a_{n,n-2L}$ from the coefficients of the generalized Legendree polynomials $A_{n,n-2N}$ according to the relation (56) of the main text,

$$\sum_{N=L}^{[n/2]} A_{n,n-2N}\mu^{N-L}(1+\mu)^{n-2N}\binom{N}{L} = \frac{1}{2^n}(-1)^L\binom{n}{L}\binom{2n-2L}{n} = a_{n,n-2L} \quad , \qquad \text{(S2.1)}$$

is carried out. The check is carried out up to $n$=6 in **Table 1**. Corresponding WolframAlpha scripts facilitating the comparison are reported as well below the table. **Table 1** shows $A_{n,n-2N}$ coeficients taken from the known generalized Legendre polynomials [15], which are then used for calculation of terms in the left-hand sum of (S2.1) for various values of $L$ and $N$ (also shown in **Table 1**). The sums of the terms with the same $L$ are shown in the right-hand column of **Table 1**. The results are always equal to $a_{n,n-2L}$ coefficients of the standard Legendre polynomials, which strongly indicates that the physical shape of the harmonic functions in SOS coordinates is the same as in the spherical coordinates.

**Table 1.** $A_{n,n-2N}$ coeficients and terms in the left-hand sum in (S2.1) for various values of $L$ and $N$ taken from the known generalized Legendre polynomials [15]. Further, for particular $L$ and $N$, the terms in the left-hand side of (S2.1) are calculated and recorded. Then, the sum of the terms with the same $L$ are shown in the right-hand column of the table. It can be seen that they are equal to $a_{n,n-2L}$ coefficients of the standard Legendre polynomials in all examined cases. Note also, that the formula (S2.1) works for both even as well as for odd $n$.

| $n$ | $N$ | 0 | 1 | 2 | 3 | Sum over $N$, equal to $a_{n,n-2L}$ in all cases |
|---|---|---|---|---|---|---|
| 0 | $n$-2$N$ | 0 | | | | |
| | $A_{n,n-2N}$ | 1 | | | | |
| | $L$=0 | 1 | | | | $1 = a_{0,0}$ |
| 1 | $n$-2$N$ | 1 | | | | |
| | $A_{n,n-2N}$ | $\frac{1}{1+\mu}$ | | | | |
| | $L$=0 | 1 | | | | $1 = a_{1,1}$ |
| 2 | $n$-2$N$ | 2 | 0 | | | |
| | $A_{n,n-2N}$ | $\frac{1}{(1+\mu)^2}\cdot\frac{1}{2}\cdot(\mu+3)$ | $-\frac{1}{(1+\mu)^2}\cdot\frac{1}{2}\cdot(1+\mu)^2$ | | | |
| | $L$=0 | $\frac{1}{2}\cdot(\mu+3)$ | $-\frac{1}{2}\cdot\mu$ | | | $\frac{3}{2} = a_{2,2}$ |
| | $L$=1 | 0 | $-\frac{1}{2}$ | | | $-\frac{1}{2} = a_{2,0}$ |
| 3 | $n$-2$N$ | 3 | 1 | | | |
| | $A_{n,n\text{-}2N}$ | $\frac{1}{(1+\mu)^3}\cdot\frac{1}{2}\cdot(3\mu+5)$ | $-\frac{1}{(1+\mu)^3}\cdot\frac{1}{2}\cdot 3(1+\mu)^2$ | | | |
| | $L$=0 | $\frac{1}{2}\cdot(3\mu+5)$ | $-3\mu\cdot\frac{1}{2}$ | | | $\frac{5}{2}=a_{3,3}$ |
| | $L$=1 | 0 | $-3\cdot\frac{1}{2}$ | | | $-\frac{3}{2}=a_{3,1}$ |
| 4 | $n$-2$N$ | 4 | 2 | 0 | | |
| | $A_{n,n\text{-}2N}$ | $\frac{1}{(1+\mu)^4}\cdot\frac{1}{2^3}\cdot(3\mu^2+30\mu+35)$ | $-\frac{1}{(1+\mu)^4}\cdot\frac{1}{2^3}\cdot 2\cdot 3(\mu+5)(1+\mu)^2$ | $\frac{1}{(1+\mu)^4}\cdot\frac{1}{2^3}\cdot 3(1+\mu)^4$ | | |
| | $L$=0 | $\frac{1}{8}\cdot(3\mu^2+30\mu+35)$ | $-3\cdot\frac{1}{4}\mu(\mu+5)$ | $3\mu^2\cdot\frac{1}{8}$ | | $\frac{35}{8}=a_{4,4}$ |
| | $L$=1 | 0 | $-3\cdot\frac{1}{4}(\mu+5)$ | $3\cdot\frac{1}{4}\mu$ | | $-\frac{15}{4}=a_{4,2}$ |
| | $L$=2 | 0 | 0 | $3\cdot\frac{1}{8}$ | | $\frac{3}{8}=a_{4,0}$ |
| 5 | $n$-2$N$ | 5 | 3 | 1 | | |
| | $A_{n,n\text{-}2N}$ | $\frac{1}{(1+\mu)^5}\cdot\frac{1}{2^3}\cdot(15\mu^2+70\mu+63)$ | $-\frac{1}{(1+\mu)^5}\cdot\frac{1}{2^3}\cdot 10(3\mu+7)(1+\mu)^2$ | $\frac{1}{(1+\mu)^5}\cdot\frac{1}{2^3}\cdot 15(1+\mu)^4$ | | |
| | $L$=0 | $\frac{1}{8}(15\mu^2+70\mu+63)$ | $-5\cdot\frac{1}{4}\mu(3\mu+7)$ | $15\mu^2\cdot\frac{1}{8}$ | | $\frac{63}{8}=a_{5,5}$ |
| | $L$=1 | 0 | $-5\cdot\frac{1}{4}(3\mu+7)$ | $15\mu\cdot\frac{1}{4}$ | | $-\frac{35}{4}=a_{5,3}$ |
| | $L$=2 | 0 | 0 | $15\cdot\frac{1}{8}$ | | $\frac{15}{8}=a_{5,1}$ |
| 6 | $n$-2$N$ | 6 | 4 | 2 | 0 | |
| | $A_{n,n\text{-}2N}$ | $\frac{1}{(1+\mu)^6}\cdot\frac{1}{2^4}\cdot(5\mu^3+105\mu^2+315\mu+231)$ | $-\frac{1}{(1+\mu)^6}\cdot\frac{1}{2^4}\cdot 5(3\mu^2+42\mu+63)(1+\mu)^2$ | $\frac{1}{(1+\mu)^6}\cdot\frac{1}{2^4}\cdot 3\cdot 5(\mu+7)(1+\mu)^4$ | $-\frac{1}{(1+\mu)^6}\cdot\frac{1}{2^4}\cdot 5(1+\mu)^6$ | |
| | $L$=0 | $\frac{1}{16}\cdot(5\mu^3+105\mu^2+315\mu+231)$ | $-\frac{15}{16}\mu(\mu^2+14\mu+21)$ | $\frac{15}{16}\mu^2(\mu+7)$ | $-\frac{5}{16}\mu^3$ | $\frac{231}{16}=a_{6,6}$ |
| | $L$=1 | 0 | $-\frac{15}{16}(\mu^2+14\mu+21)$ | $\frac{15}{8}\mu(\mu+7)$ | $-\frac{15}{16}\mu^2$ | $-\frac{315}{16}=a_{6,4}$ |
| | $L$=2 | 0 | 0 | $\frac{15}{16}(\mu+7)$ | $-\frac{15}{16}\mu$ | $\frac{105}{16}=a_{6,2}$ |
| | $L$=3 | 0 | 0 | 0 | $-\frac{5}{16}$ | $-\frac{5}{16}=a_{6,0}$ |

WolframAlpha scripts used for the calculations in **Table 1**:

n=0, L=0

evaluate (A*m^(N-L) *(1+m)^(n-2N) *(bin(N,L)) , n=0, L=0, N=0, A=1 )

n=1, L=0

evaluate A*m^(N-L) *(1+m)^(n-2N) *(bin(N,L)), n=1, L=0, N=0, A=1/(1+m)

n=2, L=0

evaluate A*m^(N-L) *(1+m)^(n-2N) *(bin(N,L)), n=2, L=0, N=0, A=(1/2)*(μ+3)/(1+m)^2

evaluate A*m^(N-L) *(1+m)^(n-2N) *(bin(N,L)), n=2, L=0, N=1, A=-(1/2)*(1+m)^2/(1+m)^2

n=2, L=1

evaluate A*m^(N-L) *(1+m)^(n-2N) *(bin(N,L)), n=2, L=1, N=0, A=(1/2)*(3+m)/(1+m)^2
evaluate A*m^(N-L) *(1+m)^(n-2N) *(bin(N,L)), n=2, L=1, N=1, A=-(1/2)*(1+m)^2/(1+m)^2
n=3, L=0
evaluate A*m^(N-L) *(1+m)^(n-2N) *(bin(N,L)), n=3, L=0, N=0, A=(1/2)*(3m+5)/(1+m)^3
evaluate A*m^(N-L) *(1+m)^(n-2N) *(bin(N,L)), n=3, L=0, N=1, A=-(1/2)*3*(1+m)^2/(1+m)^3
n=3, L=1
evaluate A*m^(N-L) *(1+m)^(n-2N) *(bin(N,L)), n=3, L=1, N=0, A=(1/2)*(3m+5)/(1+m)^3
evaluate A*m^(N-L) *(1+m)^(n-2N) *(bin(N,L)), n=3, L=1, N=1, A=-(1/2)*3*(1+m)^2/(1+m)^3
n=4, L=0
evaluate A*m^(N-L) *(1+m)^(n-2N) *(bin(N,L)), n=4, L=0, N=0, A=(1/2^3)*(3m^2+30m+35) /(1+m)^4
evaluate A*m^(N-L) *(1+m)^(n-2N) *(bin(N,L)), n=4, L=0, N=1, A=-(1/2^3)*2*3(m+5)*(1+m)^2/(1+m)^4
evaluate A*m^(N-L) *(1+m)^(n-2N) *(bin(N,L)), n=4, L=0, N=2, A=(1/2^3)*3*(1+m)^4/(1+m)^4
n=4, L=1
evaluate A*m^(N-L) *(1+m)^(n-2N) *(bin(N,L)), n=4, L=1, N=0, A=(1/2^3)*(3m^2+30m+35) /(1+m)^4
evaluate A*m^(N-L) *(1+m)^(n-2N) *(bin(N,L)), n=4, L=1, N=1, A=-(1/2^3)*2*3(m+5)*(1+m)^2/(1+m)^4
evaluate A*m^(N-L) *(1+m)^(n-2N) *(bin(N,L)), n=4, L=1, N=2, A=(1/2^3)*3*(1+m)^4/(1+m)^4
n=4, L=2
evaluate A*m^(N-L) *(1+m)^(n-2N) *(bin(N,L)), n=4, L=2, N=0, A=(1/2^3)*(3m^2+30m+35) /(1+m)^4
evaluate A*m^(N-L) *(1+m)^(n-2N) *(bin(N,L)), n=4, L=2, N=1, A=-(1/2^3)*2*3(m+5)*(1+m)^2/(1+m)^4
evaluate A*m^(N-L) *(1+m)^(n-2N) *(bin(N,L)), n=4, L=2, N=2, A=(1/2^3)*3*(1+m)^4/(1+m)^4
n=5, L=0
evaluate A*m^(N-L) *(1+m)^(n-2N) *(bin(N,L)), n=5, L=0, N=0, A=(1/2^3)*(15m^2+70m+63) /(1+m)^5
evaluate A*m^(N-L) *(1+m)^(n-2N) *(bin(N,L)), n=5, L=0, N=1, A=-(1/2^3)*10(3m+7)*(1+m)^2/(1+m)^5
evaluate A*m^(N-L) *(1+m)^(n-2N) *(bin(N,L)), n=5, L=0, N=2, A=(1/2^3)*15*(1+m)^4/(1+m)^5
n=5, L=1
evaluate A*m^(N-L) *(1+m)^(n-2N) *(bin(N,L)), n=5, L=1, N=0, A=(1/2^3)*(15m^2+70m+63) /(1+m)^5
evaluate A*m^(N-L) *(1+m)^(n-2N) *(bin(N,L)), n=5, L=1, N=1, A=-(1/2^3)*10(3m+7)*(1+m)^2/(1+m)^5
evaluate A*m^(N-L) *(1+m)^(n-2N) *(bin(N,L)), n=5, L=1, N=2, A=(1/2^3)*15*(1+m)^4/(1+m)^5
n=5, L=2
evaluate A*m^(N-L) *(1+m)^(n-2N) *(bin(N,L)), n=5, L=2, N=0, A=(1/2^3)*(15m^2+70m+63) /(1+m)^5
evaluate A*m^(N-L) *(1+m)^(n-2N) *(bin(N,L)), n=5, L=2, N=1, A=-(1/2^3)*10(3m+7)*(1+m)^2/(1+m)^5
evaluate A*m^(N-L) *(1+m)^(n-2N) *(bin(N,L)), n=5, L=2, N=2, A=(1/2^3)*15*(1+m)^4/(1+m)^5
n=6, L=0
evaluate A*m^(N-L) *(1+m)^(n-2N) *(bin(N,L)), n=6, L=0, N=0, A=(1/2^4)*( 5m^3+105m^2+315m+231) /(1+m)^6
evaluate A*m^(N-L) *(1+m)^(n-2N) *(bin(N,L)), n=6, L=0, N=1, A=-(1/2^4)*5(3m^2+42m+63)*(1+m)^2/(1+m)^6
evaluate A*m^(N-L) *(1+m)^(n-2N) *(bin(N,L)), n=6, L=0, N=2, A=(1/2^4)*3*5*(m+7) (1+m)^4/(1+m)^6
evaluate A*m^(N-L) *(1+m)^(n-2N) *(bin(N,L)), n=6, L=0, N=3, A=-(1/2^4)*5*(1+m)^6/(1+m)^6
n=6, L=1
evaluate A*m^(N-L) *(1+m)^(n-2N) *(bin(N,L)), n=6, L=1, N=0, A=(1/2^4)*( 5m^3+105m^2+315m+231) /(1+m)^6
evaluate A*m^(N-L) *(1+m)^(n-2N) *(bin(N,L)), n=6, L=1, N=1, A=-(1/2^4)*5(3m^2+42m+63)*(1+m)^2/(1+m)^6
evaluate A*m^(N-L) *(1+m)^(n-2N) *(bin(N,L)), n=6, L=1, N=2, A=(1/2^4)*3*5*(m+7) (1+m)^4/(1+m)^6
evaluate A*m^(N-L) *(1+m)^(n-2N) *(bin(N,L)), n=6, L=1, N=3, A=-(1/2^4)*5*(1+m)^6/(1+m)^6
n=6, L=2
evaluate A*m^(N-L) *(1+m)^(n-2N) *(bin(N,L)), n=6, L=2, N=0, A=(1/2^4)*( 5m^3+105m^2+315m+231) /(1+m)^6
evaluate A*m^(N-L) *(1+m)^(n-2N) *(bin(N,L)), n=6, L=2, N=1, A=-(1/2^4)*5(3m^2+42m+63)*(1+m)^2/(1+m)^6
evaluate A*m^(N-L) *(1+m)^(n-2N) *(bin(N,L)), n=6, L=2, N=2, A=(1/2^4)*3*5*(m+7) (1+m)^4/(1+m)^6
evaluate A*m^(N-L) *(1+m)^(n-2N) *(bin(N,L)), n=6, L=2, N=3, A=-(1/2^4)*5*(1+m)^6/(1+m)^6
n=6, L=3
evaluate A*m^(N-L) *(1+m)^(n-2N) *(bin(N,L)), n=6, L=3, N=0, A=(1/2^4)*( 5m^3+105m^2+315m+231) /(1+m)^6
evaluate A*m^(N-L) *(1+m)^(n-2N) *(bin(N,L)), n=6, L=3, N=1, A=-(1/2^4)*5(3m^2+42m+63)*(1+m)^2/(1+m)^6
evaluate A*m^(N-L) *(1+m)^(n-2N) *(bin(N,L)), n=6, L=3, N=2, A=(1/2^4)*3*5*(m+7) (1+m)^4/(1+m)^6
evaluate A*m^(N-L) *(1+m)^(n-2N) *(bin(N,L)), n=6, L=3, N=3, A=-(1/2^4)*5*(1+m)^6/(1+m)^6

**Supplement 3** to P. Strunz: On Laplace equation solution in orthogonal similar oblate spheroidal coordinates

## Coefficients of the generalized Legendre polynomials

Relation (56) of the main text,

$$\sum_{N=L}^{[n/2]} A_{n,n-2N}\mu^{N-L}(1+\mu)^{n-2N}\binom{N}{L} = \frac{1}{2^n}(-1)^L\binom{n}{L}\binom{2n-2L}{n} = a_{n,n-2L} \quad , \tag{S3.1}$$

can be employed for determination of the coefficients $A_{n,n\text{-}2N}$ and, consequently, for expressing the generalized Legendre polynomials as a sum of monomials (similarly as the expressions for the standard Legendre polynomials derived using Rodrigues formula, see e.g. [17], pages 38-39). It would facilitate direct determination of $A_{n,n-2N}$ coefficients for any *n* without the use of the recurrence formula.

The relation (S3.1) is to be used separately for *n* even and for *n* odd.

*Index n even*

First, when $L$=$n$/2 in (S3.1), then the sum in (S3.1) is composed of one term only and thus:

$$\sum_{N=n/2}^{[n/2]} A_{n,n-2N}\mu^{N-n/2}(1+\mu)^{n-2N}\binom{N}{n/2} = \frac{1}{2^n}(-1)^{n/2}\binom{n}{n/2}\binom{2n-2n/2}{n} = a_{n,n-2n/2} \; . \tag{S3.2}$$

Then

$$A_{n,n-2n/2}\mu^{n/2-n/2}(1+\mu)^{n-2n/2}\binom{n/2}{n/2} = \frac{1}{2^n}(-1)^{n/2}\binom{n}{n/2}\binom{2n-2n/2}{n} = a_{n,n-2n/2} \quad , \tag{S3.3}$$

$$A_{n,0}\mu^0(1+\mu)^0\cdot 1 = \frac{1}{2^n}(-1)^{n/2}\binom{n}{n/2}\binom{n}{n} = a_{n,0} \quad , \tag{S3.4}$$

$$A_{n,0} = \frac{1}{2^n}(-1)^{n/2}\binom{n}{n/2} = a_{n,0} \quad . \tag{S3.5}$$

Second, when $L$=$n$/2–1, then the sum is composed of two terms only and thus:

$$\sum_{N=\frac{n}{2}-1}^{[n/2]} A_{n,n-2N}\mu^{N-\left(\frac{n}{2}-1\right)}(1+\mu)^{n-2N}\binom{N}{\left(\frac{n}{2}-1\right)} = \frac{1}{2^n}(-1)^{\left(\frac{n}{2}-1\right)}\binom{n}{\left(\frac{n}{2}-1\right)}\binom{2n-2\left(\frac{n}{2}-1\right)}{n} = a_{n,n-2\left(\frac{n}{2}-1\right)} \quad , \tag{S3.6}$$

$$A_{n,n-2\left(\frac{n}{2}-1\right)}\mu^{\left(\frac{n}{2}-1\right)-\left(\frac{n}{2}-1\right)}(1+\mu)^{n-2\left(\frac{n}{2}-1\right)}\binom{\left(\frac{n}{2}-1\right)}{\left(\frac{n}{2}-1\right)} + A_{n,n-2\frac{n}{2}}\mu^{\frac{n}{2}-\left(\frac{n}{2}-1\right)}(1+\mu)^{n-2\frac{n}{2}}\binom{\frac{n}{2}}{\left(\frac{n}{2}-1\right)} = \frac{1}{2^n}(-1)^{\left(\frac{n}{2}-1\right)}\binom{n}{\left(\frac{n}{2}-1\right)}\binom{n+2}{n} = a_{n,2} \, , \tag{S3.7}$$

$$A_{n,2}\mu^0(1+\mu)^2\cdot 1 + A_{n,0}\mu^1(1+\mu)^0\binom{\frac{n}{2}}{\frac{n}{2}-1} = \frac{1}{2^n}(-1)^{\left(\frac{n}{2}-1\right)}\binom{n}{\left(\frac{n}{2}-1\right)}\binom{n+2}{n} = a_{n,2} \quad , \tag{S3.8}$$

Insert the above derived $A_{n,0}$ ,

$$A_{n,2}\mu^0(1+\mu)^2 + \frac{1}{2^n}(-1)^{n/2}\binom{n}{n/2}\mu^1(1+\mu)^0\binom{\frac{n}{2}}{\frac{n}{2}-1} = \frac{1}{2^n}(-1)^{\left(\frac{n}{2}-1\right)}\binom{n}{\left(\frac{n}{2}-1\right)}\binom{n+2}{n} = a_{n,2} \quad , \tag{S3.9}$$

and separate $A_{n,2}$:

$$A_{n,2} = \frac{1}{(1+\mu)^2}\left[\frac{1}{\mu^0 2^n}(-1)^{\left(\frac{n}{2}-1\right)}\binom{n}{\left(\frac{n}{2}-1\right)}\binom{n+2}{n} + \frac{1}{2^n}(-1)^{\frac{n}{2}-1}\binom{n}{n/2}\frac{\mu^1}{\mu^0}(1+\mu)^0\binom{\frac{n}{2}}{\frac{n}{2}-1}\right] \quad , \tag{S3.10}$$

Several subsequent steps lead to

$$A_{n,2} = \frac{1}{(1+\mu)^2}\frac{1}{2^n}(-1)^{\left(\frac{n}{2}-1\right)}\left[\binom{n}{\frac{n}{2}-1}\binom{n+2}{n} + \binom{n}{n/2}\binom{\frac{n}{2}}{\frac{n}{2}-1}\mu^1(1+\mu)^0\right] . \quad \text{(S3.11)}$$

Third, when *L*=*n*/2–2, then the sum is composed of three terms:

$$\sum_{N=\frac{n}{2}-2}^{[n/2]} A_{n,n-2N}\mu^{N-\left(\frac{n}{2}-2\right)}(1+\mu)^{n-2N}\binom{N}{\left(\frac{n}{2}-2\right)} = \frac{1}{2^n}(-1)^{\left(\frac{n}{2}-2\right)}\binom{n}{\left(\frac{n}{2}-2\right)}\binom{2n-2\left(\frac{n}{2}-2\right)}{n} = a_{n,n-2\left(\frac{n}{2}-2\right)} , \quad \text{(S3.12)}$$

that is

$$A_{n,n-2\left(\frac{n}{2}-2\right)}\mu^{\left(\frac{n}{2}-2\right)-\left(\frac{n}{2}-2\right)}(1+\mu)^{n-2\left(\frac{n}{2}-2\right)}\binom{\left(\frac{n}{2}-2\right)}{\left(\frac{n}{2}-2\right)} + A_{n,n-2\left(\frac{n}{2}-1\right)}\mu^{\left(\frac{n}{2}-1\right)-\left(\frac{n}{2}-2\right)}(1+\mu)^{n-2\left(\frac{n}{2}-1\right)}\binom{\left(\frac{n}{2}-1\right)}{\left(\frac{n}{2}-2\right)} + A_{n,n-2\frac{n}{2}}\mu^{\frac{n}{2}-\left(\frac{n}{2}-2\right)}(1+\mu)^{n-2\frac{n}{2}}\binom{\frac{n}{2}}{\left(\frac{n}{2}-2\right)} = \frac{1}{2^n}(-1)^{\left(\frac{n}{2}\right)}\binom{n}{\left(\frac{n}{2}-2\right)}\binom{n+4}{n} = a_{n,4} . \quad \text{(S3.13)}$$

After simplification

$$A_{n,4}\mu^0(1+\mu)^4\cdot 1 + A_{n,2}\mu^1(1+\mu)^2\binom{\frac{n}{2}-1}{\frac{n}{2}-2} + A_{n,0}\mu^2(1+\mu)^0\binom{\frac{n}{2}}{\frac{n}{2}-2} = \frac{1}{2^n}(-1)^{\left(\frac{n}{2}\right)}\binom{n}{\left(\frac{n}{2}-2\right)}\binom{n+4}{n} = a_{n,4} , \quad \text{(S3.14)}$$

Insert the above derived $A_{n,0}$ and $A_{n,2}$ ,

$$A_{n,4}\mu^0(1+\mu)^4 + \frac{1}{(1+\mu)^2}\frac{1}{2^n}(-1)^{\left(\frac{n}{2}-1\right)}\left[\binom{n}{\frac{n}{2}-1}\binom{n+2}{n} + \binom{n}{n/2}\binom{\frac{n}{2}}{\frac{n}{2}-1}\mu^1(1+\mu)^0\right]\mu^1(1+\mu)^2\binom{\frac{n}{2}-1}{\frac{n}{2}-2} + \frac{1}{2^n}(-1)^{n/2}\binom{n}{n/2}\mu^2(1+\mu)^0\binom{\frac{n}{2}}{\frac{n}{2}-2} = \frac{1}{2^n}(-1)^{\left(\frac{n}{2}\right)}\binom{n}{\left(\frac{n}{2}-2\right)}\binom{n+4}{n} = a_{n,4} , \quad \text{(S3.15)}$$

and separate $A_{n,4}$:

$$A_{n,4} = \frac{1}{(1+\mu)^4}\left\{\frac{1}{\mu^0 2^n}(-1)^{\left(\frac{n}{2}\right)}\binom{n}{\left(\frac{n}{2}-2\right)}\binom{n+4}{n} - \frac{1}{(1+\mu)^2}\frac{1}{2^n}(-1)^{\left(\frac{n}{2}-1\right)}\left[\frac{1}{\mu^0}\binom{n}{\frac{n}{2}-1}\binom{n+2}{n} + \binom{n}{n/2}\binom{\frac{n}{2}}{\frac{n}{2}-1}\frac{\mu^1}{\mu^0}(1+\mu)^0\right]\mu^1(1+\mu)^2\binom{\frac{n}{2}-1}{\frac{n}{2}-2} - \frac{1}{2^n}(-1)^{n/2}\binom{n}{n/2}\frac{\mu^2}{\mu^0}(1+\mu)^0\binom{\frac{n}{2}}{\frac{n}{2}-2}\right\} . \quad \text{(S3.16)}$$

Several subsequent steps lead to

$$A_{n,4} = \frac{1}{(1+\mu)^4}\frac{1}{2^n}(-1)^{\left(\frac{n}{2}\right)}\left\{\binom{n}{\frac{n}{2}-2}\binom{n+4}{n} - \frac{1}{(1+\mu)^2}(-1)^{(-1)}\left[\binom{n}{\frac{n}{2}-1}\binom{n+2}{n} + \binom{n}{n/2}\binom{\frac{n}{2}}{\frac{n}{2}-1}\mu^1(1+\mu)^0\right]\mu^1(1+\mu)^2\binom{\frac{n}{2}-1}{\frac{n}{2}-2} - \binom{n}{n/2}\mu^2(1+\mu)^0\binom{\frac{n}{2}}{\frac{n}{2}-2}\right\} , \quad \text{(S3.17)}$$

$$A_{n,4} = \frac{1}{(1+\mu)^4}\frac{1}{2^n}(-1)^{\left(\frac{n}{2}\right)}\left\{\binom{n}{\frac{n}{2}-2}\binom{n+4}{n} + \left[\binom{n}{\frac{n}{2}-1}\binom{n+2}{n}\binom{\frac{n}{2}-1}{\frac{n}{2}-2}\mu^1 + \binom{n}{n/2}\binom{\frac{n}{2}}{\frac{n}{2}-1}\binom{\frac{n}{2}-1}{\frac{n}{2}-2}\mu^2(1+\mu)^0\right] - \binom{n}{n/2}\mu^2(1+\mu)^0\binom{\frac{n}{2}}{\frac{n}{2}-2}\right\} , \quad \text{(S3.18)}$$

$$A_{n,4} = \frac{1}{(1+\mu)^4}\frac{1}{2^n}(-1)^{\left(\frac{n}{2}\right)}\left\{\binom{n}{\frac{n}{2}-2}\binom{n+4}{n} + \binom{n}{\frac{n}{2}-1}\binom{n+2}{n}\binom{\frac{n}{2}-1}{\frac{n}{2}-2}\mu^1 + \binom{n}{n/2}\left[\binom{\frac{n}{2}}{\frac{n}{2}-1}\binom{\frac{n}{2}-1}{\frac{n}{2}-2} - \binom{\frac{n}{2}}{\frac{n}{2}-2}\right]\mu^2(1+\mu)^0\right\} . \quad \text{(S3.19)}$$

As

$$\binom{n}{n/2}\left[\binom{\frac{n}{2}}{\frac{n}{2}-1}\binom{\frac{n}{2}-1}{\frac{n}{2}-2} - \binom{\frac{n}{2}}{\frac{n}{2}-2}\right] = \binom{n}{n/2}\left[\frac{\frac{n}{2}-\left(\frac{n}{2}-1\right)+1}{\frac{n}{2}-1}\binom{\frac{n}{2}}{\frac{n}{2}-2}\binom{\frac{n}{2}-1}{\frac{n}{2}-2} - \binom{\frac{n}{2}}{\frac{n}{2}-2}\right] = \binom{n}{n/2}\binom{\frac{n}{2}}{\frac{n}{2}-2}\left[\frac{\frac{n}{2}-\left(\frac{n}{2}-1\right)+1}{\frac{n}{2}-1}\left(\frac{n}{2}-1\right) - 1\right] = \binom{n}{n/2}\binom{\frac{n}{2}}{\frac{n}{2}-2} , \quad \text{(S3.20)}$$

then

$$A_{n,4} = \frac{1}{(1+\mu)^4}\frac{1}{2^n}(-1)^{\left(\frac{n}{2}\right)}\left\{\binom{n}{\frac{n}{2}-2}\binom{n+4}{n} + \binom{n}{\frac{n}{2}-1}\binom{n+2}{n}\binom{\frac{n}{2}-1}{\frac{n}{2}-2}\mu^1 + \binom{n}{n/2}\binom{n}{n}\binom{\frac{n}{2}}{\frac{n}{2}-2}\mu^2(1+\mu)^0\right\} . \quad \text{(S3.21)}$$

Now, slightly modify the look of the all three found coefficient expressions (by multiplying some terms by binomial coefficients which are always equal to 1):

$$A_{n,4}=\frac{1}{(1+\mu)^4}\frac{1}{2^n}(-1)^{\left(\frac{n}{2}\right)}\left\{\binom{n}{\frac{n}{2}-2}\binom{n+4}{n}\binom{\frac{n}{2}-2}{\frac{n}{2}-2}+\binom{n}{\frac{n}{2}-1}\binom{n+2}{n}\binom{\frac{n}{2}-1}{\frac{n}{2}-2}\mu^1+\binom{n}{n/2}\binom{n}{n}\binom{\frac{n}{2}}{\frac{n}{2}-2}\mu^2(1+\mu)^0\right\},\tag{S3.22}$$

$$A_{n,2}=\frac{1}{(1+\mu)^2}\frac{1}{2^n}(-1)^{\left(\frac{n}{2}-1\right)}\left[\binom{n}{\frac{n}{2}-1}\binom{n+2}{n}\binom{\frac{n}{2}-1}{\frac{n}{2}-1}+\binom{n}{n/2}\binom{n}{n}\binom{\frac{n}{2}}{\frac{n}{2}-1}\mu^1(1+\mu)^0\right],\tag{S3.23}$$

$$A_{n,0}=\frac{1}{(1+\mu)^0}\frac{1}{2^n}(-1)^{n/2}\binom{n}{n/2}\binom{n}{n}\binom{\frac{n}{2}}{\frac{n}{2}}.\tag{S3.24}$$

This shape of the found coefficients facilitates generalization of the coefficient expression by the sum of three binomial coefficients. The attempt leads to the generalization

$$A_{n,m}=\frac{1}{(1+\mu)^m}\frac{1}{2^n}(-1)^{\left(\frac{n-m}{2}\right)}\left\{\sum_{k=m/2}^{0}\binom{n}{\frac{n}{2}-k}\binom{n+2k}{n}\binom{\frac{n}{2}-k}{\frac{n}{2}-\frac{m}{2}}\mu^{\frac{m}{2}-k}\right\}.\tag{S3.25}$$

Using binomial identity

$$\binom{n_W}{h_W+k_W}\binom{h_W+k_W}{h_W}=\binom{n_W}{h_W}\binom{n_W-h_W}{k_W}\tag{S3.26}$$

for the first and the third binomial coefficients (with $n=n_W$, $\frac{n}{2}-k=h_W+k_W$, $\frac{n}{2}-\frac{m}{2}=h_W$), we get

$$\binom{n}{\frac{n}{2}-k}\binom{\frac{n}{2}-k}{\frac{n}{2}-\frac{m}{2}}=\binom{n_W}{h_W+k_W}\binom{h_W+k_W}{h_W}=\binom{n_W}{h_W}\binom{n_W-h_W}{k_W}=\binom{n}{\frac{n}{2}-\frac{m}{2}}\binom{n-\left(\frac{n}{2}-\frac{m}{2}\right)}{\frac{n}{2}-k-\left(\frac{n}{2}-\frac{m}{2}\right)}=\binom{n}{\frac{n-m}{2}}\binom{\frac{n+m}{2}}{\frac{m}{2}-k}.\tag{S3.27}$$

This operation moved $k$ index (over which counting is carried out in (S3.25)) to only one of the binomial coefficients. Therefore, one binomial coeficient can be moved out of the sum using (S3.27) identity. When also changing the order of counting, we obtain from (S3.25)

$$A_{n,m}=\frac{1}{(1+\mu)^m}\frac{1}{2^n}(-1)^{\frac{n-m}{2}}\binom{n}{\frac{n-m}{2}}\sum_{k=0}^{\frac{m}{2}}\binom{n+2k}{n}\binom{\frac{n+m}{2}}{\frac{m}{2}-k}\mu^{\left(\frac{m}{2}-k\right)}.\tag{S3.28}$$

Test of this relation has been successfully performed using WolframAlpha script, see
1/(1+µ)^m *1/2^n *(-1)^((n-m)/2) *(bin(n,(n-m)/2))*[SUM (bin(n+2k,n)*bin((n+m)/2,m/2-k)*µ^(m/2-k)), k=0 to m/2], n=2, m=2 ,
for a number of low-index $A_{n,m}$. The relation (S3.28) correctly reproduces, for even numbers $n$, the previously found coefficients of generalized Legendre polynomials [15].

*Index n odd*

$L$ varies from $[n/2]$ to 0 in this case. First, when $L=[n/2]$, i.e. $L=(n-1)/2$ for $n$ being odd, then the sum in (S3.1) is

$$\sum_{N=[n/2]}^{[n/2]}A_{n,n-2N}\mu^{N-[n/2]}(1+\mu)^{n-2N}\binom{N}{[n/2]}=\frac{1}{2^n}(-1)^{[n/2]}\binom{n}{[n/2]}\binom{2n-2[n/2]}{n}=a_{n,n-2[n/2]},\tag{S3.29}$$

$$\sum_{N=(n-1)/2}^{[n/2]}A_{n,n-2N}\mu^{N-(n-1)/2}(1+\mu)^{n-2N}\binom{N}{(n-1)/2}=\frac{1}{2^n}(-1)^{(n-1)/2}\binom{n}{(n-1)/2}\binom{2n-2(n-1)/2}{n}=a_{n,n-2(n-1)/2}.\tag{S3.30}$$

The sum is composed of one term only. Then

$$A_{n,n-2[n/2]}\mu^{[n/2]-[n/2]}(1+\mu)^{n-2[n/2]}\binom{[n/2]}{[n/2]}=\frac{1}{2^n}(-1)^{[n/2]}\binom{n}{[n/2]}\binom{2n-2[n/2]}{n}=a_{n,n-2(n-1)/2},\tag{S3.31}$$

$$A_{n,n-2(n-1)/2}\mu^{(n-1)/2-(n-1)/2}(1+\mu)^{n-2(n-1)/2}\binom{(n-1)/2}{(n-1)/2} = \frac{1}{2^n}(-1)^{(n-1)/2}\binom{n}{(n-1)/2}\binom{2n-(n-1)}{n} = a_{n,n-2(n-1)/2} \quad , \quad \text{(S3.32)}$$

$$A_{n,1}\mu^0(1+\mu)^{n-2[n/2]}\cdot 1 = \frac{1}{2^n}(-1)^{[n/2]}\binom{n}{[n/2]}\binom{2n-2[n/2]}{n} = a_{n,1} \quad , \quad \text{(S3.33)}$$

$$A_{n,1}\mu^0(1+\mu)^1\cdot 1 = \frac{1}{2^n}(-1)^{(n-1)/2}\binom{n}{(n-1)/2}\binom{n+1}{n} = a_{n,1} \quad , \quad \text{(S3.34)}$$

For $n$ odd, $n$–2[$n$/2] is equal to 1, 2$n$–2[$n$/2] is equal to $n$+1. We thus obtain

$$A_{n,1} = \frac{1}{(1+\mu)^1}\frac{1}{\mu^0 2^n}(-1)^{(n-1)/2}\binom{n}{(n-1)/2}\binom{n+1}{n} \quad , \quad \text{(S3.35)}$$

and finally

$$A_{n,1} = \frac{1}{(1+\mu)^1}\frac{1}{\mu^0 2^n}(-1)^{(n-1)/2}\binom{n}{(n-1)/2}(n+1) . \quad \text{(S3.36)}$$

Second, when $L$=[$n$/2]–1, i.e. $L$=($n$–1)/2–1=($n$–3)/2 for $n$ being odd, then the sum in (S3.1) is composed of two terms only and thus:

$$\sum_{N=[n/2]-1}^{[n/2]} A_{n,n-2N}\mu^{N-([n/2]-1)}(1+\mu)^{n-2N}\binom{N}{([n/2]-1)} = \frac{1}{2^n}(-1)^{([n/2]-1)}\binom{n}{([n/2]-1)}\binom{2n-2([n/2]-1)}{n} = a_{n,n-2([n/2]-1)} \quad , \quad \text{(S3.37)}$$

$$\sum_{N=(n-3)/2}^{[n/2]} A_{n,n-2N}\mu^{N-(n-3)/2}(1+\mu)^{n-2N}\binom{N}{(n-3)/2} = \frac{1}{2^n}(-1)^{(n-3)/2}\binom{n}{(n-3)/2}\binom{2n-2(n-3)/2}{n} = a_{n,n-2(n-3)/2} \quad , \quad \text{(S3.38)}$$

$$A_{n,n-2(n-3)/2}\mu^{(n-3)/2-(n-3)/2}(1+\mu)^{n-2(n-3)/2}\binom{(n-3)/2}{(n-3)/2} + A_{n,n-2(n-1)/2}\mu^{(n-1)/2-(n-3)/2}(1+\mu)^{n-2(n-1)/2}\binom{(n-1)/2}{(n-3)/2} = \frac{1}{2^n}(-1)^{(n-3)/2}\binom{n}{(n-3)/2}\binom{n+3}{n} = a_{n,3} \quad , \quad \text{(S3.39)}$$

$$A_{n,3}\mu^0(1+\mu)^3\binom{(n-3)/2}{(n-3)/2} + A_{n,1}\mu^1(1+\mu)^1\binom{(n-1)/2}{(n-3)/2} = \frac{1}{2^n}(-1)^{(n-3)/2}\binom{n}{(n-3)/2}\binom{n+3}{n} = a_{n,3} \quad . \quad \text{(S3.40)}$$

Insert the alreadey derived $A_{n,1}$:

$$A_{n,3}\mu^0(1+\mu)^3\binom{(n-3)/2}{(n-3)/2} + \frac{1}{(1+\mu)^1}\frac{1}{\mu^0 2^n}(-1)^{(n-1)/2}\binom{n}{(n-1)/2}(n+1)\mu^1(1+\mu)^1\binom{(n-1)/2}{(n-3)/2} = \frac{1}{2^n}(-1)^{(n-3)/2}\binom{n}{(n-3)/2}\binom{n+3}{n} = a_{n,3} \quad , \quad \text{(S3.41)}$$

$$A_{n,3}\cdot 1 = \frac{1}{\mu^0(1+\mu)^3}\left[\frac{1}{2^n}(-1)^{(n-3)/2}\binom{n}{(n-3)/2}\binom{n+3}{n} - \frac{1}{2^n}(-1)^{\frac{(n-3)}{2}+1}\binom{n}{(n-1)/2}\binom{(n-1)/2}{(n-3)/2}(n+1)\mu^1\right] \quad , \quad \text{(S3.42)}$$

$$A_{n,3} = \frac{1}{\mu^0(1+\mu)^3}\frac{1}{2^n}(-1)^{(n-3)/2}\left[\binom{n}{(n-3)/2}\binom{n+3}{n} - (-1)^1\binom{n}{(n-1)/2}\binom{(n-1)/2}{(n-3)/2}(n+1)\mu^1\right] , \quad \text{(S3.43)}$$

$$A_{n,3} = \frac{1}{(1+\mu)^3}\frac{1}{2^n}(-1)^{\frac{(n-1)}{2}-1}\left[\binom{n}{(n-3)/2}\binom{n+3}{n} + \binom{n}{(n-1)/2}\binom{(n-1)/2}{(n-3)/2}(n+1)\mu^1\right] \quad . \quad \text{(S3.44)}$$

Third, when $L$=[$n$/2]–2, i.e. $L$=($n$–1)/2–2=($n$–5)/2 for $n$ being odd, then the sum in (S3.1) is composed of three terms:

$$\sum_{N=[n/2]-2}^{[n/2]} A_{n,n-2N}\mu^{N-([n/2]-2)}(1+\mu)^{n-2N}\binom{N}{([n/2]-2)} = \frac{1}{2^n}(-1)^{([n/2]-2)}\binom{n}{([n/2]-2)}\binom{2n-2([n/2]-2)}{n} = a_{n,n-2([n/2]-2)} \quad , \quad \text{(S3.45)}$$

$$\sum_{N=(n-5)/2}^{[n/2]} A_{n,n-2N}\mu^{N-(n-5)/2}(1+\mu)^{n-2N}\binom{N}{(n-5)/2} = \frac{1}{2^n}(-1)^{(n-5)/2}\binom{n}{(n-5)/2}\binom{2n-(n-5)}{n} = a_{n,n-(n-5)} \quad , \tag{S3.46}$$

$$A_{n,n-2(n-5)/2}\mu^{(n-5)/2-(n-5)/2}(1+\mu)^{n-2(n-5)/2}\binom{(n-5)/2}{(n-5)/2} + A_{n,n-2(n-3)/2}\mu^{(n-3)/2-(n-5)/2}(1+\mu)^{n-2(n-3)/2}\binom{(n-3)/2}{(n-5)/2} + A_{n,n-2(n-1)/2}\mu^{(n-1)/2-(n-5)/2}(1+\mu)^{n-2(n-1)/2}\binom{(n-1)/2}{(n-5)/2} = \frac{1}{2^n}(-1)^{(n-5)/2}\binom{n}{(n-5)/2}\binom{n+5}{n} = a_{n,5} \quad , \tag{S3.47}$$

$$A_{n,5}\mu^0(1+\mu)^5\cdot 1 + A_{n,3}\mu^1(1+\mu)^3\binom{(n-3)/2}{(n-5)/2} + A_{n,1}\mu^2(1+\mu)^1\binom{(n-1)/2}{(n-5)/2} = \frac{1}{2^n}(-1)^{(n-5)/2}\binom{n}{(n-5)/2}\binom{n+5}{n} = a_{n,5} \quad , \tag{S3.48}$$

$$A_{n,5} = \frac{1}{\mu^0(1+\mu)^5}\left[\frac{1}{2^n}(-1)^{(n-5)/2}\binom{n}{(n-5)/2}\binom{n+5}{n} - A_{n,3}\mu^1(1+\mu)^3\binom{(n-3)/2}{(n-5)/2} - A_{n,1}\mu^2(1+\mu)^1\binom{(n-1)/2}{(n-5)/2}\right] \quad . \tag{S3.49}$$

Insert the already derived $A_{n,1}$, $A_{n,3}$:

$$A_{n,5} = \frac{1}{\mu^0(1+\mu)^5}\left\{\frac{1}{2^n}(-1)^{(n-5)/2}\binom{n}{(n-5)/2}\binom{n+5}{n} - \frac{1}{(1+\mu)^3}\frac{1}{2^n}(-1)^{\frac{(n-1)}{2}-1}\left[\binom{n}{(n-3)/2}\binom{n+3}{n} + \binom{n}{(n-1)/2}\binom{(n-1)/2}{(n-3)/2}(n+1)\mu^1\right]\mu^1(1+\mu)^3\binom{(n-3)/2}{(n-5)/2} - \frac{1}{(1+\mu)^1}\frac{1}{\mu^0 2^n}(-1)^{(n-1)/2}\binom{n}{(n-1)/2}(n+1)\mu^2(1+\mu)^1\binom{(n-1)/2}{(n-5)/2}\right\} \quad , \tag{S3.50}$$

$$A_{n,5} = \frac{1}{(1+\mu)^5}\frac{1}{2^n}(-1)^{(n-5)/2}\left\{\binom{n}{(n-5)/2}\binom{n+5}{n} - (-1)^1\left[\binom{n}{(n-3)/2}\binom{n+3}{n} + \binom{n}{(n-1)/2}\binom{(n-1)/2}{(n-3)/2}(n+1)\mu^1\right]\mu^1\binom{(n-3)/2}{(n-5)/2} - (-1)^2\binom{n}{(n-1)/2}(n+1)\mu^2\binom{(n-1)/2}{(n-5)/2}\right\} \quad , \tag{S3.51}$$

$$A_{n,5} = \frac{1}{(1+\mu)^5}\frac{1}{2^n}(-1)^{\frac{(n-1)}{2}-2}\left\{\binom{n}{(n-5)/2}\binom{n+5}{n} + \left[\binom{n}{(n-3)/2}\binom{n+3}{n}\mu^1\binom{(n-3)/2}{(n-5)/2} + \binom{n}{(n-1)/2}\binom{(n-1)/2}{(n-3)/2}(n+1)\mu^1\mu^1\binom{(n-3)/2}{(n-5)/2}\right] - \binom{n}{(n-1)/2}(n+1)\mu^2\binom{(n-1)/2}{(n-5)/2}\right\} \quad , \tag{S3.52}$$

$$A_{n,5} = \frac{1}{(1+\mu)^5}\frac{1}{2^n}(-1)^{\frac{(n-1)}{2}-2}\left\{\binom{n}{(n-5)/2}\binom{n+5}{n} + \binom{n}{(n-3)/2}\binom{n+3}{n}\binom{(n-3)/2}{(n-5)/2}\mu^1 + \left[\binom{n}{(n-1)/2}\binom{(n-1)/2}{(n-3)/2}(n+1)\binom{(n-3)/2}{(n-5)/2}\mu^2 - \binom{n}{(n-1)/2}(n+1)\binom{(n-1)/2}{(n-5)/2}\mu^2\right]\right\} \quad , \tag{S3.53}$$

$$A_{n,5} = \frac{1}{(1+\mu)^5}\frac{1}{2^n}(-1)^{\frac{(n-1)}{2}-2}\left\{\binom{n}{(n-5)/2}\binom{n+5}{n} + \binom{n}{(n-3)/2}\binom{n+3}{n}\binom{(n-3)/2}{(n-5)/2}\mu^1 + \binom{n}{(n-1)/2}(n+1)\left[\binom{(n-1)/2}{(n-3)/2}\binom{(n-3)/2}{(n-5)/2} - \binom{(n-1)/2}{(n-5)/2}\right]\mu^2\right\}. \tag{S3.54}$$

As (thanks to the basic binomial identity applied onto the first binomial coefficient in angle brackets)

$$\left[\binom{(n-1)/2}{(n-3)/2}\binom{(n-3)/2}{(n-5)/2}-\binom{(n-1)/2}{(n-5)/2}\right]=\left[\frac{(n-1)/2-(n-3)/2+1}{(n-3)/2}\binom{(n-1)/2}{\frac{(n-3)}{2}-1}\binom{(n-3)/2}{(n-5)/2}-\binom{(n-1)/2}{(n-5)/2}\right]=$$

$$\left[\frac{2}{(n-3)/2}\binom{(n-1)/2}{\frac{(n-5)}{2}}\binom{\frac{(n-5)}{2}+1}{(n-5)/2}-\binom{(n-1)/2}{(n-5)/2}\right]=\binom{(n-1)/2}{(n-5)/2}\left[\frac{2}{(n-3)/2}\left(\frac{(n-5)}{2}+1\right)-1\right]=$$

$$\binom{(n-1)/2}{\frac{(n-1)}{2}-2}\left[\frac{2}{(n-3)/2}\left(\frac{(n-3)}{2}\right)-1\right]=\binom{(n-1)/2}{\frac{(n-1)}{2}-2}\quad, \tag{S3.55}$$

then

$$A_{n,5}=\frac{1}{(1+\mu)^5}\frac{1}{2^n}(-1)^{\frac{(n-1)}{2}-2}\left\{\binom{n}{\frac{(n-1)}{2}-2}\binom{n+5}{n}+\binom{n}{\frac{(n-1)}{2}-1}\binom{n+3}{n}\binom{\frac{(n-1)}{2}-1}{\frac{(n-1)}{2}-2}\mu^1+\binom{n}{(n-1)/2}(n+1)\binom{(n-1)/2}{\frac{(n-1)}{2}-2}\mu^2\right\}\quad. \tag{S3.56}$$

Implement the floor function $[n/2]=(n-1)/2$ (for $n$ being odd) and consider that $n+1=\binom{n+1}{n}$:

$$A_{n,5}=\frac{1}{(1+\mu)^5}\frac{1}{2^n}(-1)^{[n/2]-2}\left\{\binom{n}{[n/2]-2}\binom{n+5}{n}\binom{[n/2]-2}{[n/2]-2}+\binom{n}{[n/2]-1}\binom{n+3}{n}\binom{[n/2]-1}{[n/2]-2}\mu^1+\binom{n}{[n/2]}\binom{n+1}{n}\binom{[n/2]}{[n/2]-2}\mu^2\right\}\quad. \tag{S3.57}$$

Implement the floor function $[n/2]$ also to the previously found coefficients $A_{n,3}$ and $A_{n,1}$ (for $n$ being odd), and also slightly modify the look of the found coefficient expressions (by multiplying some terms by binomial coefficients always equal to 1 as well as using $n+1=\binom{n+1}{n}$). Then (see (S3.44))

$$A_{n,3}=\frac{1}{(1+\mu)^3}\frac{1}{2^n}(-1)^{[n/2]-1}\left[\binom{n}{[n/2]-1}\binom{n+3}{n}\binom{[n/2]-1}{[n/2]-1}+\binom{n}{[n/2]}\binom{n+1}{n}\binom{[n/2]}{[n/2]-1}\mu^1\right], \tag{S3.58}$$

and (see (S3.35))

$$A_{n,1}=\frac{1}{(1+\mu)^1}\frac{1}{2^n}(-1)^{[n/2]}\binom{n}{[n/2]}\binom{n+1}{n}\binom{[n/2]}{[n/2]}\quad. \tag{S3.59}$$

This shape of the found coefficients facilitates the generalization of the coefficient expression by the sum of a product of three binomial coefficients also for the present case when $n$ is odd number. The attempt leads to the generalization

$$A_{n,m}=\frac{1}{(1+\mu)^m}\frac{1}{2^n}(-1)^{\left(\frac{n-m}{2}\right)}\left\{\sum_{k=[m/2]}^{0}\binom{n}{[n/2]-k}\binom{n+2k+m-2[m/2]}{n}\binom{[n/2]-k}{\frac{n-m}{2}}\mu^{[m/2]-k}\right\}\quad. \tag{S3.60}$$

*Generalization*

The factor $m-2[m/2]$ in the middle binomial coefficient of (S3.60) is present to give 1 for $m$ being odd, but also to give 0 for m being even (for later joining of both cases).

Using binomial identity (S3.26) for the first and the third binomial coefficients in (S3.60) (with $n=n_W$, $[n/2]-k=h_W+k_W, \frac{n-m}{2}=h_W$) we get

$$\binom{n}{[n/2]-k}\binom{[n/2]-k}{\frac{n-m}{2}}=\binom{n_W}{h_W+k_W}\binom{h_W+k_W}{h_W}=\binom{n_W}{h_W}\binom{n_W-h_W}{k_W}=\binom{n}{\frac{n-m}{2}}\binom{n-\frac{n-m}{2}}{[n/2]-k-\frac{n-m}{2}}=$$

$$\binom{n}{\frac{n-m}{2}}\binom{\frac{n+m}{2}}{[n/2]-k-([n/2]-[m/2])}=\binom{n}{\frac{n-m}{2}}\binom{\frac{n+m}{2}}{[m/2]-k}\quad, \tag{S3.61}$$

where we used the relation ($n$–$m$)/2=[$n$/2]–[$m$/2], which is valid when both $n$, $m$ are odd and also when both $n$, $m$ are even. Then

$$A_{n,m}=\frac{1}{(1+\mu)^m}\frac{1}{2^n}(-1)^{\frac{n-m}{2}}\binom{n}{\frac{n-m}{2}}\sum_{k=0}^{[m/2]}\binom{\frac{n+m}{2}}{[m/2]-k}\binom{n+m-2([m/2]-k)}{n}\mu^{[m/2]-k} \quad . \tag{S3.62}$$

Note, that for $\mu$=0 (i.e. spherical coordinates case), the only non-zero term in (S3.62) is the one with $k$=[$m$/2], and thus

$$A_{n,m}=\frac{1}{2^n}(-1)^{\frac{n-m}{2}}\binom{n}{\frac{n-m}{2}}\binom{\frac{n+m}{2}}{[m/2]-[m/2]}\binom{n+m-2([m/2]-[m/2])}{n}=\frac{1}{2^n}(-1)^{\frac{n-m}{2}}\binom{n}{\frac{n-m}{2}}\binom{n+m}{n} \quad , \tag{S3.63}$$

which, with substitution $L$=($n$–$m$)/2, leads to

$$A_{n,n-2L}=\frac{1}{2^n}(-1)^L\binom{n}{L}\binom{n+n-2L}{n}=\frac{1}{2^n}(-1)^L\binom{n}{L}\binom{2n-2L}{n}=a_{n,n-2L} \quad . \tag{S3.64}$$

Cf. (45) of the main text to confirm that, in the case $\mu$=0, the formula reduces to the well known formula for calculation of coefficients of Legendre polynomials. The above substitution (keeping $L$ non-negative integer) also removes possibility of existence of coefficients with mixed $n$ and $m$, i.e. with odd $n$ and even $m$, or vice versa, which are in reality always zero.

(S3.62) estimated formula for generalized Legendre polynomials coefficients can be still modified with the substitution $j$=[$m$/2]–$k$, which leads to

$$A_{n,m}=\frac{1}{(1+\mu)^m}\frac{1}{2^n}(-1)^{\frac{n-m}{2}}\binom{n}{\frac{n-m}{2}}\sum_{j=0}^{[m/2]}\binom{\frac{n+m}{2}}{j}\binom{n+m-2j}{n}\mu^j \quad . \tag{S3.65}$$

This relation then should work for even (see (S3.28) and perform in it substitution $m$/2-$k$=$j$) as well as for odd values of $n$. Also of advantage is the substitution $N$=($n$–$m$)/2, $N$ being a non-negative integer, which leads to

$$A_{n,n-2N}=\frac{1}{(1+\mu)^{n-2N}}\frac{1}{2^n}(-1)^N\binom{n}{N}\sum_{j=0}^{[(n-2N)/2]}\binom{\frac{n+n-2N}{2}}{j}\binom{n+n-2N-2j}{n}\mu^j=$$

$$\frac{1}{(1+\mu)^{n-2N}}\frac{1}{2^n}(-1)^N\binom{n}{N}\sum_{j=0}^{\left[\frac{n}{2}\right]-N}\binom{n-N}{j}\binom{2(n-N-j)}{n}\mu^j \quad , \tag{S3.66}$$

and removes a possibility of existence of coefficients with mixed parity, i.e. with odd polynomial degree $n$ and even power $m$ of $s$ term, or vice versa, which are in reality always zero.

(S3.66) formula can be implemented into (43) of the main text,

$$P_n^{SI}(s)=\sum_{N=0}^{[n/2]}s^{n-2N}A_{n,n-2N} \quad , \tag{S3.67}$$

and we obtain finally the formula for the generalized Legendre polynomial (without the need of recurrence) in the form

$$P_n^{SI}(s)=\frac{1}{2^n}\sum_{N=0}^{[n/2]}s^{n-2N}\frac{(-1)^N}{(1+\mu)^{n-2N}}\binom{n}{N}\sum_{j=0}^{\left[\frac{n}{2}\right]-N}\binom{n-N}{j}\binom{2(n-N-j)}{n}\mu^j \quad . \tag{S3.68}$$

This formula comes from the generalization of coefficients found for the particular low-index cases. Therefore, of course, it has to be tested in the Laplace equation in SOS coordinates if the equation is fulfilled when using the generalized Legendre polynomials with the coefficients given by (S3.68).

**Supplement 4** to P. Strunz: On Laplace equation solution in orthogonal similar oblate spheroidal coordinates

## Test of the angular part of the Laplace equation with a model function in the exterior space

The angular part of the Laplace equation (relation (20) in the main text), i.e. the generalized Legendre equation

$$[(1+\mu)-s^2][(1+\mu)^2-\mu s^2]\frac{d^2F(s)}{ds^2} + s[-(3\mu+2)(1+\mu)+2\mu(1+\mu)K_d+\mu(3-2K_d)s^2]\frac{dF(s)}{ds} + K_d[(K_d-2)\mu s^2+(1+\mu)K_d+(1+\mu)^2]F(s) = 0, \qquad \text{(S4.1)}$$

is to be tested in the exterior points with a model function

$$F_n^{SE}(s) = \frac{(1+\mu)^{2n+1}}{\left(\sqrt{(1+\mu)^2-\mu s^2}\right)^{2n+1}} P_n^{SI}(s) = \left(\frac{1+\mu}{\sqrt{(1+\mu)^2-\mu s^2}}\right)^{2n+1} P_n^{SI}(s) . \qquad \text{(S4.2)}$$

Note, that $K_d$ is negative, particularly $K_d$=–1–$n$, $n$ being integer and $n$≥0, for the prospective exterior space solution $F_n^{SE}$. When we set $F$=$F_n^{SE}$ and $K_d$=–1–$n$ in (S4.1), we receive

$$[(1+\mu)-s^2][(1+\mu)^2-\mu s^2]\frac{d^2F_n^{SE}(s)}{ds^2} + s[-(3\mu+2)(1+\mu)+2\mu(1+\mu)(-1-n)+\mu(3-2(-1-n))s^2]\frac{dF_n^{SE}(s)}{ds} + (-1-n)[((-1-n)-2)\mu s^2+(1+\mu)(-1-n)+(1+\mu)^2]F_n^{SE}(s) = 0 \quad , \qquad \text{(S4.3)}$$

and thus

$$[(1+\mu)-s^2][(1+\mu)^2-\mu s^2]\frac{d^2F_n^{SE}(s)}{ds^2} + s[-(1+\mu)[(5+2n)\mu+2]+(5+2n)\mu s^2]\frac{dF_n^{SE}(s)}{ds} + (1+n)[(n+3)\mu s^2+(1+\mu)(n-\mu)]F_n^{SE}(s) = 0 \quad . \qquad \text{(S4.4)}$$

Now, we calculate the derivatives of $F_n^{SE}$. The first derivative is

$$\frac{dF_n^{SE}(s)}{ds} = \frac{d}{ds}\left[\frac{(1+\mu)^{2n+1}}{\left(\sqrt{(1+\mu)^2-\mu s^2}\right)^{2n+1}}\right]P_n^{SI}(s) + \frac{(1+\mu)^{2n+1}}{\left(\sqrt{(1+\mu)^2-\mu s^2}\right)^{2n+1}}\frac{dP_n^{SI}(s)}{ds} = \left[\frac{(2n+1)(1+\mu)^{2n+1}\mu s}{((1+\mu)^2-\mu s^2)^{n+\frac{3}{2}}}\right]P_n^{SI}(s) + \frac{(1+\mu)^{2n+1}}{((1+\mu)^2-\mu s^2)^{n+\frac{1}{2}}}\frac{dP_n^{SI}(s)}{ds} \quad . \qquad \text{(S4.5)}$$

The second derivative is

$$\frac{d^2F_n^{SE}(s)}{ds^2} = \frac{d}{ds}\left[\frac{dF_n^{SE}(s)}{ds}\right] = \frac{d}{ds}\left[\frac{(2n+1)(1+\mu)^{2n+1}\mu s}{((1+\mu)^2-\mu s^2)^{n+\frac{3}{2}}}P_n^{SI}(s)\right] + \frac{d}{ds}\left[\frac{(1+\mu)^{2n+1}}{((1+\mu)^2-\mu s^2)^{n+\frac{1}{2}}}\frac{dP_n^{SI}(s)}{ds}\right] =$$
$$\frac{d}{ds}\left[\frac{(2n+1)(1+\mu)^{2n+1}\mu s}{((1+\mu)^2-\mu s^2)^{n+\frac{3}{2}}}\right]P_n^{SI}(s) + \frac{(2n+1)(1+\mu)^{2n+1}\mu s}{((1+\mu)^2-\mu s^2)^{n+\frac{3}{2}}}\frac{dP_n^{SI}(s)}{ds} + \frac{d}{ds}\left[\frac{(1+\mu)^{2n+1}}{((1+\mu)^2-\mu s^2)^{n+\frac{1}{2}}}\right]\frac{dP_n^{SI}(s)}{ds} +$$
$$\frac{(1+\mu)^{2n+1}}{((1+\mu)^2-\mu s^2)^{n+\frac{1}{2}}}\left[\frac{d^2P_n^{SI}(s)}{ds^2}\right] = \left[\frac{(2n+1)\mu(1+\mu)^{2n+1}}{((1+\mu)^2-\mu s^2)^{n+\frac{3}{2}}} + \frac{2\mu^2s^2\left(n+\frac{3}{2}\right)(2n+1)(1+\mu)^{2n+1}}{((1+\mu)^2-\mu s^2)^{n+\frac{5}{2}}}\right]P_n^{SI}(s) + \frac{(2n+1)(1+\mu)^{2n+1}\mu s}{((1+\mu)^2-\mu s^2)^{n+\frac{3}{2}}}\frac{dP_n^{SI}(s)}{ds} +$$
$$\left[\frac{(2n+1)(1+\mu)^{2n+1}\mu s}{((1+\mu)^2-\mu s^2)^{n+\frac{3}{2}}}\right]\frac{dP_n^{SI}(s)}{ds} + \frac{(1+\mu)^{2n+1}}{((1+\mu)^2-\mu s^2)^{n+\frac{1}{2}}}\left[\frac{d^2P_n^{SI}(s)}{ds^2}\right] = \left[\frac{(2n+1)(1+\mu)^{2n+1}\mu}{((1+\mu)^2-\mu s^2)^{n+\frac{3}{2}}} + \frac{(1+\mu)^{2n+1}(2n+3)(2n+1)\mu^2s^2}{((1+\mu)^2-\mu s^2)^{n+\frac{5}{2}}}\right]P_n^{SI}(s) +$$
$$2\frac{(2n+1)(1+\mu)^{2n+1}\mu s}{((1+\mu)^2-\mu s^2)^{n+\frac{3}{2}}}\frac{dP_n^{SI}(s)}{ds} + \frac{(1+\mu)^{2n+1}}{((1+\mu)^2-\mu s^2)^{n+\frac{1}{2}}}\left[\frac{d^2P_n^{SI}(s)}{ds^2}\right] \quad . \qquad \text{(S4.6)}$$

By inserting the derivatives and the model function itself to (S4.4), we get

$$[(1+\mu)-s^2][(1+\mu)^2-\mu s^2]\left\{\left[\frac{(2n+1)(1+\mu)^{2n+1}\mu}{((1+\mu)^2-\mu s^2)^{n+\frac{3}{2}}}+\frac{(1+\mu)^{2n+1}(2n+3)(2n+1)\mu^2 s^2}{((1+\mu)^2-\mu s^2)^{n+\frac{5}{2}}}\right]P_n^{SI}(s)+ 2\frac{(2n+1)(1+\mu)^{2n+1}\mu s}{((1+\mu)^2-\mu s^2)^{n+\frac{3}{2}}}\frac{d\,P_n^{SI}(s)}{ds}+\frac{(1+\mu)^{2n+1}}{((1+\mu)^2-\mu s^2)^{n+\frac{1}{2}}}\left[\frac{d^2\,P_n^{SI}(s)}{ds^2}\right]\right\}+ s[-(1+\mu)[(5+2n)\mu+2]+(5+2n)\mu s^2]\left\{\left[\frac{(2n+1)(1+\mu)^{2n+1}\mu s}{((1+\mu)^2-\mu s^2)^{n+\frac{3}{2}}}\right]P_n^{SI}(s)+\frac{(1+\mu)^{2n+1}}{((1+\mu)^2-\mu s^2)^{n+\frac{1}{2}}}\frac{d\,P_n^{SI}(s)}{ds}\right\}+(1+ n)[(n+3)\mu s^2+(1+\mu)(n-\mu)]\frac{(1+\mu)^{2n+1}}{((1+\mu)^2-\mu s^2)^{n+\frac{1}{2}}}P_n^{SI}(s)=0\,. \quad \text{(S4.7)}$$

Expanding

$$(1+\mu)^{2n+1}\left\{[(1+\mu)-s^2][(1+\mu)^2-\mu s^2]\left[\frac{(2n+1)\mu}{((1+\mu)^2-\mu s^2)^{n+\frac{3}{2}}}+\frac{(2n+3)(2n+1)\mu^2 s^2}{((1+\mu)^2-\mu s^2)^{n+\frac{5}{2}}}\right]P_n^{SI}(s)+2[(1+\mu)- s^2][(1+\mu)^2-\mu s^2]\frac{(2n+1)\mu s}{((1+\mu)^2-\mu s^2)^{n+\frac{3}{2}}}\frac{d\,P_n^{SI}(s)}{ds}+[(1+\mu)-s^2][(1+\mu)^2-\mu s^2]\frac{1}{((1+\mu)^2-\mu s^2)^{n+\frac{1}{2}}}\left[\frac{d^2\,P_n^{SI}(s)}{ds^2}\right]\right\}+ (1+\mu)^{2n+1}\left\{s[-(1+\mu)[(5+2n)\mu+2]+(5+2n)\mu s^2]\left[\frac{(2n+1)\mu s}{((1+\mu)^2-\mu s^2)^{n+\frac{3}{2}}}\right]P_n^{SI}(s)+s[-(1+\mu)[(5+2n)\mu+2]+ (5+2n)\mu s^2]\frac{1}{((1+\mu)^2-\mu s^2)^{n+\frac{1}{2}}}\frac{d\,P_n^{SI}(s)}{ds}\right\}+ (1+\mu)^{2n+1}(1+n)[(n+3)\mu s^2+(1+\mu)(n-\mu)]\frac{1}{((1+\mu)^2-\mu s^2)^{n+\frac{1}{2}}}P_n^{SI}(s)=0\quad . \quad \text{(S4.8)}$$

Canceling some terms in the numerators and denominators

$$(1+\mu)^{2n+1}\left\{[(1+\mu)-s^2]\left[\frac{(2n+1)\mu}{((1+\mu)^2-\mu s^2)^{n+\frac{1}{2}}}+\frac{(2n+3)(2n+1)\mu^2 s^2}{((1+\mu)^2-\mu s^2)^{n+\frac{3}{2}}}\right]P_n^{SI}(s)+ 2[(1+\mu)-s^2]\frac{(2n+1)\mu s}{((1+\mu)^2-\mu s^2)^{n+\frac{1}{2}}}\frac{d\,P_n^{SI}(s)}{ds}+[(1+\mu)-s^2][(1+\mu)^2-\mu s^2]\frac{1}{((1+\mu)^2-\mu s^2)^{n+\frac{1}{2}}}\left[\frac{d^2\,P_n^{SI}(s)}{ds^2}\right]\right\}+ (1+\mu)^{2n+1}\left\{s[-(1+\mu)[(5+2n)\mu+2]+(5+2n)\mu s^2]\left[\frac{(2n+1)\mu s}{((1+\mu)^2-\mu s^2)^{n+\frac{3}{2}}}\right]P_n^{SI}(s)+s[-(1+\mu)[(5+2n)\mu+2]+ (5+2n)\mu s^2]\frac{1}{((1+\mu)^2-\mu s^2)^{n+\frac{1}{2}}}\frac{d\,P_n^{SI}(s)}{ds}\right\}+ (1+\mu)^{2n+1}(1+n)[(n+3)\mu s^2+(1+\mu)(n-\mu)]\frac{1}{((1+\mu)^2-\mu s^2)^{n+\frac{1}{2}}}P_n^{SI}(s)=0\quad . \quad \text{(S4.9)}$$

Multiplying by $((1+\mu)^2-\mu s^2)^{n+\frac{1}{2}}$:

$$(1+\mu)^{2n+1}\left\{[(1+\mu)-s^2]\left[\frac{(2n+1)\mu}{1}+\frac{(2n+3)(2n+1)\mu^2 s^2}{((1+\mu)^2-\mu s^2)}\right]P_n^{SI}(s)+2[(1+\mu)-s^2]\frac{(2n+1)\mu s}{1}\frac{d\,P_n^{SI}(s)}{ds}+ [(1+\mu)-s^2][(1+\mu)^2-\mu s^2]\left[\frac{d^2\,P_n^{SI}(s)}{ds^2}\right]\right\}+ (1+\mu)^{2n+1}\left\{s[-(1+\mu)[(5+2n)\mu+2]+(5+2n)\mu s^2]\left[\frac{(2n+1)\mu s}{((1+\mu)^2-\mu s^2)}\right]P_n^{SI}(s)+s[-(1+\mu)[(5+2n)\mu+2]+ (5+2n)\mu s^2]\frac{d\,P_n^{SI}(s)}{ds}\right\}+(1+\mu)^{2n+1}(1+n)[(n+3)\mu s^2+(1+\mu)(n-\mu)]P_n^{SI}(s)=0\quad . \quad \text{(S4.10)}$$

Near the terms of the same $P_n^{SI}$-derivative degree:

$$(1+\mu)^{2n+1}\Big\{+[(1+\mu)-s^2][(1+\mu)^2-\mu s^2]\left[\frac{d^2P_n^{SI}(s)}{ds^2}\right]+2[(1+\mu)-s^2]\frac{(2n+1)\mu s}{1}\frac{dP_n^{SI}(s)}{ds}+s[-(1+\mu)[(5+2n)\mu+2]+(5+2n)\mu s^2]\frac{dP_n^{SI}(s)}{ds}+[(1+\mu)-s^2]\left[\frac{(2n+1)\mu}{1}+\frac{(2n+3)(2n+1)\mu^2s^2}{((1+\mu)^2-\mu s^2)}\right]P_n^{SI}(s)+s[-(1+\mu)[(5+2n)\mu+2]+(5+2n)\mu s^2]\left[\frac{(2n+1)\mu s}{((1+\mu)^2-\mu s^2)}\right]P_n^{SI}(s)+(1+n)[(n+3)\mu s^2+(1+\mu)(n-\mu)]P_n^{SI}(s)\Big\}=0 \quad . \qquad (S4.11)$$

Divide by the common pre-factor and put together the terms of the same $P_n^{SI}$-derivative degree:

$$[(1+\mu)-s^2][(1+\mu)^2-\mu s^2]\left[\frac{d^2P_n^{SI}(s)}{ds^2}\right]+\Big\{2[(1+\mu)-s^2]\frac{(2n+1)\mu s}{1}+s[-(1+\mu)[(5+2n)\mu+2]+(5+2n)\mu s^2]\Big\}\frac{dP_n^{SI}(s)}{ds}+\Big\{[(1+\mu)-s^2]\left[\frac{(2n+1)\mu}{1}+\frac{(2n+3)(2n+1)\mu^2s^2}{((1+\mu)^2-\mu s^2)}\right]+s[-(1+\mu)[(5+2n)\mu+2]+(5+2n)\mu s^2]\left[\frac{(2n+1)\mu s}{((1+\mu)^2-\mu s^2)}\right]+(1+n)[(n+3)\mu s^2+(1+\mu)(n-\mu)]\Big\}P_n^{SI}(s)=0 \quad . \qquad (S4.12)$$

Multiply by $((1+\mu)^2-\mu s^2)$:

$$[(1+\mu)-s^2][(1+\mu)^2-\mu s^2]^2\left[\frac{d^2P_n^{SI}(s)}{ds^2}\right]+\Big\{2(2n+1)[(1+\mu)-s^2]((1+\mu)^2-\mu s^2)\mu s+s((1+\mu)^2-\mu s^2)[-(1+\mu)[(5+2n)\mu+2]+(5+2n)\mu s^2]\Big\}\frac{dP_n^{SI}(s)}{ds}+\Big\{[(1+\mu)-s^2][((1+\mu)^2-\mu s^2)(2n+1)\mu+(2n+3)(2n+1)\mu^2s^2]+s[-(1+\mu)[(5+2n)\mu+2]+(5+2n)\mu s^2](2n+1)\mu s+(1+n)((1+\mu)^2-\mu s^2)[(n+3)\mu s^2+(1+\mu)(n-\mu)]\Big\}P_n^{SI}(s)=0 \quad . \qquad (S4.13)$$

Then (S4.13) becomes

$$[(1+\mu)-s^2][(1+\mu)^2-\mu s^2]^2\frac{d^2P_n^{SI}(s)}{ds^2}+s[(1+\mu)^2-\mu s^2]\{(-2n+3)\mu s^2+[(2n-3)\mu^2+(2n-5)\mu-2]\}\frac{dP_n^{SI}(s)}{ds}+\{-n(n-2)\mu^2s^4+n[n\mu(1+\mu)-3(1+\mu)^2]\mu s^2+n(\mu+n+1)(1+\mu)^3\}P_n^{SI}(s)=0, \qquad (S4.14)$$

and still simplified it reads

$$[(1+\mu)-s^2][(1+\mu)^2-\mu s^2]^2\frac{d^2P_n^{SI}(s)}{ds^2}+[(1+\mu)^2-\mu s^2]s\{(-2n+3)\mu s^2+[(2n-3)\mu^2+(2n-5)\mu-2]\}\frac{dP_n^{SI}(s)}{ds}+n\{-(n-2)\mu^2s^4+[n\mu-3(1+\mu)](1+\mu)\mu s^2+(\mu+n+1)(1+\mu)^3\}P_n^{SI}(s)=0 \quad . \qquad (S4.15)$$

(S4.15) is our basic equation for exterior space, and it is to be tested that it is fulfilled for $P_n^{SI}$ for $n\geq0$. It can be done one by one, starting from $n$=0 and increasing incrementally $n$. However, as there is infinite number of polynomials $P_n^{SI}$, this would not be a total proof.

Then, another approach can be tested. We can employ the formula (58) from the main text, which expresses the generalized Legendre polynomials as a sum of monomials:

$$P_n^{SI}(s)=\sum_{N=0}^{[n/2]}s^{n-2N}A_{n,n-2N} \quad , \qquad (S4.16)$$

with (see (57) from the main text)

$$A_{n,n-2N}=\frac{1}{2^n}\frac{(-1)^N}{(1+\mu)^{n-2N}}\binom{n}{N}\sum_{j=0}^{\left[\frac{n}{2}\right]-N}\binom{n-N}{j}\binom{2(n-N-j)}{n}\mu^j \quad , \qquad (S4.17)$$

where $[n/2]$ means the floor function of $n/2$. We can calculate the derivatives of $P_n^{SI}$ relatively easily, as those are polynomials. The first derivative is

$$\frac{dP_n^{SI}(s)}{ds}=\frac{d}{ds}\left[\sum_{N=0}^{[n/2]}s^{n-2N}A_{n,n-2N}\right]=\sum_{N=0}^{[n/2]}(n-2N)A_{n,n-2N}s^{n-2N-1}, \qquad (S4.18)$$

and the second derivative is

$$\frac{d^2 P_n^{SI}(s)}{ds^2} = \frac{d}{ds}\left[\frac{d P_n^{SI}(s)}{ds}\right] = \frac{d}{ds}\left[\sum_{N=0}^{[n/2]}(n-2N)A_{n,n-2N}s^{n-2N-1}\right] = \sum_{N=0}^{[n/2]}(n-2N)(n-2N-1)A_{n,n-2N}s^{n-2N-2} \quad . \qquad \text{(S4.19)}$$

Further, we insert the function and its derivatives to (S4.15) and we get

$$[(1+\mu)-s^2][(1+\mu)^2-\mu s^2]^2\left[\sum_{N=0}^{[n/2]}(n-2N)(n-2N-1)A_{n,n-2N}s^{n-2N-2}\right] + ((1+\mu)^2-\mu s^2)s\{(-2n+3)\mu s^2+(2n-3)\mu^2+(2n-5)\mu-2\}\left[\sum_{N=0}^{[n/2]}(n-2N)A_{n,n-2N}s^{n-2N-1}\right] + n\left[-(n-2)\mu^2 s^4 + \left(n\mu-3(1+\mu)\right)(1+\mu)\mu s^2 + (\mu+n+1)(1+\mu)^3\right]\left[\sum_{N=0}^{[n/2]} s^{n-2N}A_{n,n-2N}\right] = 0 \quad . \qquad \text{(S4.20)}$$

The sum in the second term is expanded by *s* (which is in front of it in (S4.20)), and – further – the pre-factors are expanded:

$$[(1+\mu)-s^2][(1+\mu)^4-2(1+\mu)^2\mu s^2+\mu^2 s^4]\left[\sum_{N=0}^{[n/2]}(n-2N)(n-2N-1)A_{n,n-2N}s^{n-2N-2}\right] + \{(-2n+3)(1+\mu)^2\mu s^2+(2n-3)(1+\mu)^2\mu^2+(2n-5)(1+\mu)^2\mu-2(1+\mu)^2-(-2n+3)\mu s^2\mu s^2-(2n-3)\mu^2\mu s^2-(2n-5)\mu\mu s^2+2\mu s^2\}\left[\sum_{N=0}^{[n/2]}(n-2N)A_{n,n-2N}s^{n-2N}\right] + \left[-n(n-2)\mu^2 s^4 + n\left(n\mu-3(1+\mu)\right)(1+\mu)\mu s^2 + n(\mu+n+1)(1+\mu)^3\right]\left[\sum_{N=0}^{[n/2]} s^{n-2N}A_{n,n-2N}\right] = 0 \quad . \qquad \text{(S4.21)}$$

Further expansion and starting to put together the terms in the pre-factors with the same power in *s* leads to

$$[(1+\mu)^5-2(1+\mu)^3\mu s^2+(1+\mu)\mu^2 s^4-(1+\mu)^4 s^2+2(1+\mu)^2\mu s^4-\mu^2 s^6]\left[\sum_{N=0}^{[n/2]}(n-2N)(n-2N-1)A_{n,n-2N}s^{n-2N-2}\right] + \{(-2n+3)(1+\mu)^2\mu s^2+(2n-3)(1+\mu)^2\mu^2+(2n-5)(1+\mu)^2\mu-2(1+\mu)^2-(-2n+3)\mu^2 s^4-(2n-3)\mu^2\mu s^2-(2n-5)\mu\mu s^2+2\mu s^2\}\left[\sum_{N=0}^{[n/2]}(n-2N)A_{n,n-2N}s^{n-2N}\right] + \left[-n(n-2)\mu^2 s^4 + n\left(n\mu-3(1+\mu)\right)(1+\mu)\mu s^2 + n(\mu+n+1)(1+\mu)^3\right]\left[\sum_{N=0}^{[n/2]} s^{n-2N}A_{n,n-2N}\right] = 0 \quad . \qquad \text{(S4.22)}$$

Put together the terms in the pre-factors with the same power in *s*:

$$[(1+\mu)^5+[-2\mu-(1+\mu)](1+\mu)^3 s^2+[\mu+2(1+\mu)](1+\mu)\mu s^4-\mu^2 s^6]\left[\sum_{N=0}^{[n/2]}(n-2N)(n-2N-1)A_{n,n-2N}s^{n-2N-2}\right] + \{(2n-3)\mu^2 s^4+(-2n+3)(1+\mu)^2\mu s^2-(2n-3)\mu^2\mu s^2-(2n-5)\mu\mu s^2+2\mu s^2+(2n-3)(1+\mu)^2\mu^2+(2n-5)(1+\mu)^2\mu-2(1+\mu)^2\}\left[\sum_{N=0}^{[n/2]}(n-2N)A_{n,n-2N}s^{n-2N}\right] + \left[-n(n-2)\mu^2 s^4 + n\left(n\mu-3(1+\mu)\right)(1+\mu)\mu s^2 + n(\mu+n+1)(1+\mu)^3\right]\left[\sum_{N=0}^{[n/2]} s^{n-2N}A_{n,n-2N}\right] = 0 \quad . \qquad \text{(S4.23)}$$

Continue with putting together the terms in the pre-factors with the same power in *s*, and reorganizing:

$$[-\mu^2 s^6+(3\mu+2)(1+\mu)\mu s^4-(3\mu+1)(1+\mu)^3 s^2+(1+\mu)^5]\left[\sum_{N=0}^{[n/2]}(n-2N)(n-2N-1)A_{n,n-2N}s^{n-2N-2}\right] + \{(2n-3)\mu^2 s^4+[(-2n+3)(1+\mu)^2-(2n-3)\mu^2-(2n-5)\mu+2]\mu s^2+[(2n-3)\mu^2+(2n-5)\mu-2](1+\mu)^2\}\left[\sum_{N=0}^{[n/2]}(n-2N)A_{n,n-2N}s^{n-2N}\right] + \left[-n(n-2)\mu^2 s^4 + n\left(n\mu-3(1+\mu)\right)(1+\mu)\mu s^2 + n(\mu+n+1)(1+\mu)^3\right]\left[\sum_{N=0}^{[n/2]} s^{n-2N}A_{n,n-2N}\right] = 0 \quad . \qquad \text{(S4.24)}$$

Still continue with algebraic simplification:

$$[-\mu^2 s^6+(3\mu+2)(1+\mu)\mu s^4-(3\mu+1)(1+\mu)^3 s^2+(1+\mu)^5]\left[\sum_{N=0}^{[n/2]}(n-2N)(n-2N-1)A_{n,n-2N}s^{n-2N-2}\right] + \{(2n-3)\mu^2 s^4+[2(3-2n)\mu^2+(11-6n)\mu+(5-2n)]\mu s^2+[(2n-3)\mu^2+(2n-5)\mu-2](1+\mu)^2\}\left[\sum_{N=0}^{[n/2]}(n-2N)A_{n,n-2N}s^{n-2N}\right] + \left[-n(n-2)\mu^2 s^4 + n\left(n\mu-3(1+\mu)\right)(1+\mu)\mu s^2 + n(\mu+n+1)(1+\mu)^3\right]\left[\sum_{N=0}^{[n/2]} s^{n-2N}A_{n,n-2N}\right] = 0 \quad . \qquad \text{(S4.25)}$$

Now expand by multiplication by the sums:

$$\left[-\mu^2 s^6\sum_{N=0}^{[n/2]}(n-2N)(n-2N-1)A_{n,n-2N}s^{n-2N-2} + (3\mu+2)(1+\mu)\mu s^4\sum_{N=0}^{[n/2]}(n-2N)(n-2N-1)A_{n,n-2N}s^{n-2N-2} - (3\mu+1)(1+\mu)^3 s^2\sum_{N=0}^{[n/2]}(n-2N)(n-2N-1)A_{n,n-2N}s^{n-2N-2} + (1+\mu)^5\sum_{N=0}^{[n/2]}(n-2N)(n-2N-1)A_{n,n-2N}s^{n-2N-2}\right] + \left\{(2n-3)\mu^2 s^4\sum_{N=0}^{[n/2]}(n-2N)A_{n,n-2N}s^{n-2N} + [2(3-2n)\mu^2+(11-6n)\mu+(5-2n)]\mu s^2\sum_{N=0}^{[n/2]}(n-2N)A_{n,n-2N}s^{n-2N} + [(2n-3)\mu^2+(2n-5)\mu-2](1+\mu)^2\sum_{N=0}^{[n/2]}(n-2N)A_{n,n-2N}s^{n-2N}\right\} + \left[-n(n-2)\mu^2 s^4\sum_{N=0}^{[n/2]} s^{n-2N}A_{n,n-2N} + n\left(n\mu-3(1+\mu)\right)(1+\mu)\mu s^2\sum_{N=0}^{[n/2]} s^{n-2N}A_{n,n-2N} + n(\mu+n+1)(1+\mu)^3\sum_{N=0}^{[n/2]} s^{n-2N}A_{n,n-2N}\right] = 0 \quad . \qquad \text{(S4.26)}$$

Further continue with expansion process, and also multiply by $s^2$ either inside the sum (the first part) or outside the sum (the second and the third part):

$$\left[-\mu^2 s^6 \sum_{N=0}^{[n/2]}(n-2N)(n-2N-1)A_{n,n-2N}s^{n-2N} + (3\mu+2)(1+\mu)\mu s^4 \sum_{N=0}^{[n/2]}(n-2N)(n-2N-1)A_{n,n-2N}s^{n-2N} - (3\mu+1)(1+\mu)^3 s^2 \sum_{N=0}^{[n/2]}(n-2N)(n-2N-1)A_{n,n-2N}s^{n-2N} + (1+\mu)^5 \sum_{N=0}^{[n/2]}(n-2N)(n-2N-1)A_{n,n-2N}s^{n-2N}\right] + \left\{(2n-3)\mu^2 s^6 \sum_{N=0}^{[n/2]}(n-2N)A_{n,n-2N}s^{n-2N} + [2(3-2n)\mu^2+(11-6n)\mu+(5-2n)]\mu s^4 \sum_{N=0}^{[n/2]}(n-2N)A_{n,n-2N}s^{n-2N} + [(2n-3)\mu^2+(2n-5)\mu-2](1+\mu)^2 s^2 \sum_{N=0}^{[n/2]}(n-2N)A_{n,n-2N}s^{n-2N}\right\} + \left[-n(n-2)\mu^2 s^6 \sum_{N=0}^{[n/2]} A_{n,n-2N}s^{n-2N} + n(n\mu-3(1+\mu))(1+\mu)\mu s^4 \sum_{N=0}^{[n/2]} A_{n,n-2N}s^{n-2N} + n(\mu+n+1)(1+\mu)^3 s^2 \sum_{N=0}^{[n/2]} A_{n,n-2N}s^{n-2N}\right] = 0 \quad . \qquad (S4.27)$$

Put the pre-factors inside the sums:

$$\left[\sum_{N=0}^{[n/2]} -\mu^2(n-2N)(n-2N-1)A_{n,n-2N}s^6 s^{n-2N} + \sum_{N=0}^{[n/2]}(3\mu+2)(1+\mu)\mu(n-2N)(n-2N-1)A_{n,n-2N}s^4 s^{n-2N} - \sum_{N=0}^{[n/2]}(3\mu+1)(1+\mu)^3(n-2N)(n-2N-1)A_{n,n-2N}s^2 s^{n-2N} + \sum_{N=0}^{[n/2]}(1+\mu)^5(n-2N)(n-2N-1)A_{n,n-2N}s^{n-2N}\right] + \left\{\sum_{N=0}^{[n/2]}(2n-3)\mu^2(n-2N)A_{n,n-2N}s^6 s^{n-2N} + \sum_{N=0}^{[n/2]}[2(3-2n)\mu^2+(11-6n)\mu+(5-2n)]\mu(n-2N)A_{n,n-2N}s^4 s^{n-2N} + \sum_{N=0}^{[n/2]}[(2n-3)\mu^2+(2n-5)\mu-2](1+\mu)^2(n-2N)A_{n,n-2N}s^2 s^{n-2N}\right\} + \left[\sum_{N=0}^{[n/2]} -n(n-2)\mu^2 A_{n,n-2N}s^6 s^{n-2N} + \sum_{N=0}^{[n/2]} n(n\mu-3(1+\mu))(1+\mu)\mu A_{n,n-2N}s^4 s^{n-2N} + \sum_{N=0}^{[n/2]} n(\mu+n+1)(1+\mu)^3 A_{n,n-2N}s^2 s^{n-2N}\right] = 0 \quad . \qquad (S4.28)$$

The term $\sum_{N=0}^{[n/2]}(1+\mu)^5(n-2N)(n-2N-1)A_{n,n-2N}s^{n-2N}$ is unique, as it has seemingly no counterpart with which its lowest-power-in-*s* term could fully re-combine to give zero as the result. Nevertheless, it can be rewritten, with a help of the substitution *N*=*M*–1, to

$$\sum_{N=0}^{[n/2]}(1+\mu)^5(n-2N)(n-2N-1)A_{n,n-2N}s^{n-2N} = \sum_{N=0}^{[n/2]}(1+\mu)^5(n-2N)(n-2N-1)A_{n,n-2N}s^2 s^{n-2N-2} = \sum_{M=1}^{[n/2]+1}(1+\mu)^5(n-2(M-1))(n-2(M-1)-1)A_{n,n-2(M-1)}s^2 s^{n-2(M-1)-2} = \sum_{M=1}^{[n/2]+1}(1+\mu)^5(n-2M+2)(n-2M+1)A_{n,n-2M+2}s^2 s^{n-2M} = \sum_{M=1}^{[n/2]}(1+\mu)^5(n-2M+2)(n-2M+1)A_{n,n-2M+2}s^2 s^{n-2M} + (1+\mu)^5(n-2([n/2]+1)+2)(n-2([n/2]+1)+1)A_{n,n-2([n/2]+1)+2}s^2 s^{n-2([n/2]+1)} = \sum_{M=0}^{[n/2]}(1+\mu)^5(n-2M+2)(n-2M+1)A_{n,n-2M+2}s^2 s^{n-2M} - (1+\mu)^5(n-2\cdot 0+2)(n-2\cdot 0+1)A_{n,n-2\cdot 0+2}s^2 s^{n-2\cdot 0} + (1+\mu)^5(n-2[n/2])(n-2[n/2]-1)A_{n,n-2[n/2]}s^2 s^{n-2[n/2]-2} \quad . \qquad (S4.29)$$

The very last term is always zero, as for *n* odd is $(n-2[n/2]-1)$ equal to zero, whereas $(n-2[n/2])$ is equal to zero for *n* even. Then (S4.29) equals to

$$\sum_{M=0}^{[n/2]}(1+\mu)^5(n-2M+2)(n-2M+1)A_{n,n-2M+2}s^2 s^{n-2M} - (1+\mu)^5(n+2)(n+1)A_{n,n+2}s^2 s^n \quad . \qquad (S4.30)$$

The last term is zero, because $A_{n,n+2}$ is zero, as this term does not exist in the generalized Legendre polynomial of degree *n*. Thus the special term in (S4.28) can be written equivalently in the following shape (where the index of the sum was renamed from *M* to *N*):

$$\sum_{N=0}^{[n/2]}(1+\mu)^5(n-2N+2)(n-2N+1)A_{n,n-2N+2}s^2 s^{n-2N} \quad . \qquad (S4.31)$$

After inserting it to (S4.28), we obtain

$$\left[\sum_{N=0}^{[n/2]} -\mu^2(n-2N)(n-2N-1)A_{n,n-2N}s^6 s^{n-2N} + \sum_{N=0}^{[n/2]}(3\mu+2)(1+\mu)\mu(n-2N)(n-2N-1)A_{n,n-2N}s^4 s^{n-2N} - \sum_{N=0}^{[n/2]}(3\mu+1)(1+\mu)^3(n-2N)(n-2N-1)A_{n,n-2N}s^2 s^{n-2N} + \sum_{N=0}^{[n/2]}(1+\mu)^5(n-2N+2)(n-2N+1)A_{n,n-2N+2}s^2 s^{n-2N}\right] + \left\{\sum_{N=0}^{[n/2]}(2n-3)\mu^2(n-2N)A_{n,n-2N}s^6 s^{n-2N} + \sum_{N=0}^{[n/2]}[2(3-2n)\mu^2+(11-6n)\mu+(5-2n)]\mu(n-2N)A_{n,n-2N}s^4 s^{n-2N} + \sum_{N=0}^{[n/2]}[(2n-3)\mu^2+(2n-5)\mu-2](1+\mu)^2(n-2N)A_{n,n-2N}s^2 s^{n-2N}\right\} + \left[\sum_{N=0}^{[n/2]} -n(n-2)\mu^2 A_{n,n-2N}s^6 s^{n-2N} + \sum_{N=0}^{[n/2]} n(n\mu-3(1+\mu))(1+\mu)\mu A_{n,n-2N}s^4 s^{n-2N} + \sum_{N=0}^{[n/2]} n(\mu+n+1)(1+\mu)^3 A_{n,n-2N}s^2 s^{n-2N}\right] = 0 \quad . \qquad (S4.32)$$

Now, the terms with the same exponents of *s* are put together:

$$\left[\sum_{N=0}^{[n/2]} -\mu^2(n-2N)(n-2N-1)A_{n,n-2N}s^6 s^{n-2N} + \sum_{N=0}^{[n/2]}(2n-3)\mu^2(n-2N)A_{n,n-2N}s^6 s^{n-2N} + \sum_{N=0}^{[n/2]} -n(n-2)\mu^2 A_{n,n-2N}s^6 s^{n-2N} + \sum_{N=0}^{[n/2]}(3\mu+2)(1+\mu)\mu(n-2N)(n-2N-1)A_{n,n-2N}s^4 s^{n-2N} + \sum_{N=0}^{[n/2]}[2(3-2n)\mu^2+(11-6n)\mu+(5-2n)]\mu(n-2N)A_{n,n-2N}s^4 s^{n-2N} + \sum_{N=0}^{[n/2]} n(n\mu-3(1+\mu))(1+\mu)\mu A_{n,n-2N}s^4 s^{n-2N} - \sum_{N=0}^{[n/2]}(3\mu+1)(1+\mu)^3(n-2N)(n-2N-1)A_{n,n-2N}s^2 s^{n-2N} + \sum_{N=0}^{[n/2]}(1+\mu)^5(n-2N+2)(n-2N+1)A_{n,n-2N+2}s^2 s^{n-2N}\right] + \left\{+\sum_{N=0}^{[n/2]}[(2n-3)\mu^2+(2n-5)\mu-2](1+\mu)^2(n-2N)A_{n,n-2N}s^2 s^{n-2N}\right\} + \left[+\sum_{N=0}^{[n/2]} n(\mu+n+1)(1+\mu)^3 A_{n,n-2N}s^2 s^{n-2N}\right] = 0 \quad , \qquad (S4.33)$$

$$\sum_{N=0}^{[n/2]}\left[-\mu^2(n-2N)(n-2N-1)A_{n,n-2N} + (2n-3)\mu^2(n-2N)A_{n,n-2N} - n(n-2)\mu^2 A_{n,n-2N}\right]s^6 s^{n-2N} + \sum_{N=0}^{[n/2]}\left[(3\mu+2)(1+\mu)\mu(n-2N)(n-2N-1)A_{n,n-2N} + [2(3-2n)\mu^2+(11-6n)\mu+(5-2n)]\mu(n-2N)A_{n,n-2N} + n(n\mu-\right.$$

$$3(1+\mu))(1+\mu)\mu A_{n,n-2N}]s^4 s^{n-2N} + \sum_{N=0}^{[n/2]}[-(3\mu+1)(1+\mu)^3(n-2N)(n-2N-1)A_{n,n-2N} + (1+\mu)^5(n-2N+2)(n-2N+1)A_{n,n-2N+2} + [(2n-3)\mu^2+(2n-5)\mu-2](1+\mu)^2(n-2N)A_{n,n-2N} + n(\mu+n+1)(1+\mu)^3 A_{n,n-2N}]s^2 s^{n-2N} = 0 \quad , \qquad \text{(S4.34)}$$

$$\sum_{N=0}^{[n/2]}[-(n-2N)(n-2N-1)+(2n-3)(n-2N)-n(n-2)]\mu^2 A_{n,n-2N}s^6 s^{n-2N} + \sum_{N=0}^{[n/2]}\{(3\mu+2)(1+\mu)(n-2N)(n-2N-1)+[2(3-2n)\mu^2+(11-6n)\mu+(5-2n)](n-2N)+n(n\mu-3(1+\mu))(1+\mu)\}\mu A_{n,n-2N}s^4 s^{n-2N} + \sum_{N=0}^{[n/2]}\{-(3\mu+1)(1+\mu)(n-2N)(n-2N-1)A_{n,n-2N}+[(2n-3)\mu^2+(2n-5)\mu-2](n-2N)A_{n,n-2N}+n(\mu+n+1)(1+\mu)A_{n,n-2N}+(1+\mu)^3(n-2N+2)(n-2N+1)A_{n,n-2N+2}\}(1+\mu)^2 s^2 s^{n-2N} = 0 \quad . \qquad \text{(S4.35)}$$

Partial expansions

$[-(n-2N)(n-2N-1)+(2n-3)(n-2N)-n(n-2)] = -4N(N-1)$,

and

$\{(3\mu+2)(1+\mu)(n-2N)(n-2N-1)+[2(3-2n)\mu^2+(11-6n)\mu+(5-2n)](n-2N)+n(n\mu-3(1+\mu))(1+\mu)\} = -4\mu^2 nN - 8\mu nN - 4nN + 12\mu^2N^2 + 20\mu N^2 + 8N^2 - 6\mu^2 N - 12\mu N - 6N = +12\mu^2N^2 - 4\mu^2 nN - 6\mu^2 N + 20\mu N^2 - 8\mu nN - 12\mu N + 8N^2 - 4nN - 6N = 2(6N-2n-3)N\mu^2 + 4(5N-2n-3)N\mu + 2(4N-2n-3)N$ ~~$= 2(6N)\mu^2 N + 4(5N)\mu N + 2(4N)N + 2(-2n-3)\mu^2 N + 4(-2n-3)\mu N + 2(-2n-3)N = 12\mu^2N^2 + 20\mu N^2 + 8N^2 + 2(-2n-3)N[\mu^2+2\mu+1] = (3\mu^2+5\mu+2)4N^2 + 2(-2n-3)N(1+\mu)^2 = 4N^2(3\mu+2)(1+\mu) - 2(2n+3)N(1+\mu)^2$~~,

and

$\{[-(3\mu+1)(1+\mu)(n-2N)(n-2N-1)+[(2n-3)\mu^2+(2n-5)\mu-2](n-2N)+n(\mu+n+1)(1+\mu)]\} = -\mu^2n^2 - \mu n^2 + \mu^2 n + \mu n + 8\mu^2 nN + 12\mu nN + 4nN - 12\mu^2N^2 - 16\mu N^2 - 4N^2 + 2\mu N + 2N = -\mu^2n^2 + \mu^2 n + 8\mu^2 nN - 12\mu^2 N^2 - \mu n^2 + \mu n + 12\mu nN - 16\mu N^2 + 2\mu N + 4nN - 4N^2 + 2N = (-n^2+n+8nN-12N^2)\mu^2 + (-n^2+n+12nN-16N^2+2N)\mu + 4nN - 4N^2 + 2N = ((1-n)n+4(2n-3N)N)\mu^2 + ((1-n)n+2(6n-8N+1)N)\mu + 2(2n-2N+1)N$,

are applied. Then

$$\sum_{N=0}^{[n/2]}[-4N(N-1)]\mu^2 A_{n,n-2N}s^6 s^{n-2N} + \sum_{N=0}^{[n/2]}\{2(6N-2n-3)N\mu^2+4(5N-2n-3)N\mu+2(4N-2n-3)N\}\mu A_{n,n-2N}s^4 s^{n-2N} + \sum_{N=0}^{[n/2]}\{[((1-n)n+4(2n-3N)N)\mu^2+((1-n)n+2(6n-8N+1)N)\mu+2(2n-2N+1)N]A_{n,n-2N}+(1+\mu)^3(n-2N+2)(n-2N+1)A_{n,n-2N+2}\}(1+\mu)^2 s^2 s^{n-2N} = 0 \quad . \qquad \text{(S4.36)}$$

It can be seen, that the highest power in *s* term is present in the first sum (the one for *N*=0), but it is zero, as the terms in the first sum are proportional to *N*.

It can be also seen, that the lowest power in *s* term is present in the third sum (the one for *N*=[*n*/2]), and it is proportional to

$$\left\{\left[\left((1-n)n+4\left(2n-3\left[\tfrac{n}{2}\right]\right)\left[\tfrac{n}{2}\right]\right)\mu^2+\left((1-n)n+2\left(6n-8\left[\tfrac{n}{2}\right]+1\right)\left[\tfrac{n}{2}\right]\right)\mu+2\left(2n-2\left[\tfrac{n}{2}\right]+1\right)\left[\tfrac{n}{2}\right]\right]A_{n,n-2\left[\frac{n}{2}\right]}+(1+\mu)^3\left(n-2\left[\tfrac{n}{2}\right]+2\right)\left(n-2\left[\tfrac{n}{2}\right]+1\right)A_{n,n-2\left[\frac{n}{2}\right]+2}\right\} \quad . \qquad \text{(S4.37)}$$

For *n* being even, it is

$$\{[((n-nn)+(4n-3n)n)\mu^2+((1-n)n+(6n-4n+1)n)\mu+(2n-n+1)n]A_{n,n-n}+(1+\mu)^3(n-n+2)(n-n+1)A_{n,n-n+2}\} = \{[n\mu^2+(n+2)n\mu+(n+1)n]A_{n,0}+(1+\mu)^3 2A_{n,2}\} = \{(1+\mu)n(\mu+n+1)A_{n,0}+(1+\mu)^3 2A_{n,2}\} = (1+\mu)\{n(\mu+n+1)A_{n,0}+2(1+\mu)^2 A_{n,2}\} \quad . \qquad \text{(S4.38)}$$

According to (S4.17),

$$A_{n,0} = A_{n,n-2\left[\frac{n}{2}\right]} = A_{n,n-2\left(\frac{n}{2}\right)} = \frac{1}{2^n}\frac{(-1)^{\left(\frac{n}{2}\right)}}{(1+\mu)^{n-2\left(\frac{n}{2}\right)}}\binom{n}{\left(\frac{n}{2}\right)}\sum_{j=0}^{\left(\frac{n}{2}\right)-\left(\frac{n}{2}\right)}\binom{\left(\frac{n}{2}\right)}{j}\binom{2\left(n-\left(\frac{n}{2}\right)-j\right)}{n}\mu^j = \frac{1}{2^n}\frac{(-1)^{\left(\frac{n}{2}\right)}}{(1+\mu)^0}\binom{n}{\frac{n}{2}}\sum_{j=0}^{0}\binom{\frac{n}{2}}{j}\binom{2\left(\frac{n}{2}-j\right)}{n}\mu^j = \frac{1}{2^n}(-1)^{\frac{n}{2}}\binom{n}{\frac{n}{2}}\sum_{j=0}^{0}\binom{\frac{n}{2}}{j}\binom{(n-2j)}{n}\mu^j = \frac{1}{2^n}(-1)^{\frac{n}{2}}\binom{n}{\frac{n}{2}}\binom{\frac{n}{2}}{0}\binom{(n-2\cdot 0)}{n}\mu^0 = \frac{1}{2^n}(-1)^{\frac{n}{2}}\binom{n}{\frac{n}{2}}\cdot 1\cdot 1\cdot 1 = \frac{(-1)^{\frac{n}{2}}}{2^n}\binom{n}{\frac{n}{2}} \quad , \qquad \text{(S4.39)}$$

and

$$A_{n,2} = A_{n,n-2\left(\frac{n}{2}-1\right)} = \frac{1}{2^n}\frac{(-1)^{\left(\frac{n}{2}-1\right)}}{(1+\mu)^{n-2\left(\frac{n}{2}-1\right)}}\binom{n}{\left(\frac{n}{2}-1\right)}\sum_{j=0}^{\left[\frac{n}{2}\right]-\left(\frac{n}{2}-1\right)}\binom{n-\left(\frac{n}{2}-1\right)}{j}\binom{2\left(n-\left(\frac{n}{2}-1\right)-j\right)}{n}\mu^j = \frac{1}{2^n}\frac{(-1)^{\left(\frac{n}{2}-1\right)}}{(1+\mu)^2}\binom{n}{\frac{n}{2}-1}\sum_{j=0}^{\left(\frac{n}{2}\right)-\left(\frac{n}{2}-1\right)}\binom{\frac{n}{2}+1}{j}\binom{2\left(\frac{n}{2}+1-j\right)}{n}\mu^j = \frac{1}{2^n}\frac{(-1)^{\left(\frac{n}{2}-1\right)}}{(1+\mu)^2}\binom{n}{\frac{n}{2}-1}\sum_{j=0}^{1}\binom{\frac{n}{2}+1}{j}\binom{(n+2-2j)}{n}\mu^j =$$

$$\frac{1}{2^n}\frac{(-1)^{\left(\frac{n}{2}-1\right)}}{(1+\mu)^2}\binom{n}{\frac{n}{2}-1}\left[\binom{\frac{n}{2}+1}{0}\binom{n+2-2\cdot 0}{n}\mu^0+\binom{\frac{n}{2}+1}{1}\binom{n+2-2\cdot 1}{n}\mu^1\right]=\frac{1}{2^n}\frac{(-1)^{\left(\frac{n}{2}-1\right)}}{(1+\mu)^2}\binom{n}{\frac{n}{2}-1}\left[\binom{\frac{n}{2}+1}{0}\binom{n+2}{n}+\binom{\frac{n}{2}+1}{1}\binom{n}{n}\mu\right]=\frac{1}{2^n}\frac{(-1)^{\left(\frac{n}{2}-1\right)}}{(1+\mu)^2}\binom{n}{\frac{n}{2}-1}\left[\binom{n+2}{n}+\binom{\frac{n}{2}+1}{1}\mu\right] \quad . \quad \text{(S4.40)}$$

The lowest order in $s$ term coefficient is thus

$$(1+\mu)\{n(\mu+n+1)A_{n,0}+2(1+\mu)^2A_{n,2}\}=(1+\mu)\left\{n(\mu+n+1)\frac{(-1)^{\frac{n}{2}}}{2^n}\binom{n}{\frac{n}{2}}+2(1+\mu)^2\frac{1}{2^n}\frac{(-1)^{\left(\frac{n}{2}-1\right)}}{(1+\mu)^2}\binom{n}{\frac{n}{2}-1}\left[\binom{n+2}{n}+\binom{\frac{n}{2}+1}{1}\mu\right]\right\}=(1+\mu)\left\{n(\mu+n+1)\frac{(-1)^{\frac{n}{2}}}{2^n}\binom{n}{\frac{n}{2}-1}\frac{n-\frac{n}{2}+1}{\frac{n}{2}}+2(1+\mu)^2\frac{1}{2^n}\frac{(-1)^{\frac{n}{2}}(-1)^{-1}}{(1+\mu)^2}\binom{n}{\frac{n}{2}-1}\left[\frac{1}{2}(n+1)(n+2)+\left(\frac{n}{2}+1\right)\mu\right]\right\}=(1+\mu)\frac{(-1)^{\frac{n}{2}}}{2^n}\binom{n}{\frac{n}{2}-1}\left\{n(\mu+n+1)\frac{n-\frac{n}{2}+1}{\frac{n}{2}}+2(1+\mu)^2\frac{-1}{(1+\mu)^2}\left[\frac{1}{2}(n+1)(n+2)+\left(\frac{n}{2}+1\right)\mu\right]\right\}=(1+\mu)\frac{(-1)^{\frac{n}{2}}}{2^n}\binom{n}{\frac{n}{2}-1}\left\{n(\mu+n+1)\frac{\frac{n}{2}+1}{\frac{n}{2}}-2\left[\frac{1}{2}(n+1)(n+2)+\left(\frac{n}{2}+1\right)\mu\right]\right\}=(1+\mu)\frac{(-1)^{\frac{n}{2}}}{2^n}\binom{n}{\frac{n}{2}-1}\left\{(\mu+n+1)2\left(\frac{n}{2}+1\right)-[(n+1)(n+2)+(n+2)\mu]\right\}=(1+\mu)\frac{(-1)^{\frac{n}{2}}}{2^n}\binom{n}{\frac{n}{2}-1}\{(\mu+n+1)(n+2)-(n+2)[n+1+\mu]\}=0 \quad . \quad \text{(S4.41)}$$

The term (see (S4.38)) is thus zero in this case (i.e. for $n$ being even).

For $n$ being odd, the lowest order in $s$ term coefficient of (S4.36) is:

$$\left\{\left[\left((1-n)n+4\left(2n-3\left(\tfrac{n-1}{2}\right)\right)\left(\tfrac{n-1}{2}\right)\right)\mu^2+\left((1-n)n+2\left(6n-8\left(\tfrac{n-1}{2}\right)+1\right)\left(\tfrac{n-1}{2}\right)\right)\mu+2\left(2n-2\left(\tfrac{n-1}{2}\right)+1\right)\left(\tfrac{n-1}{2}\right)\right]A_{n,n-2\left(\frac{n-1}{2}\right)}+(1+\mu)^3\left(n-2\left(\tfrac{n-1}{2}\right)+2\right)\left(n-2\left(\tfrac{n-1}{2}\right)+1\right)A_{n,n-2\left(\frac{n-1}{2}\right)+2}\right\}=\left\{\left[\left((n-nn)+\left(4n-3(n-1)\right)(n-1)\right)\mu^2+\left((1-n)n+(6n-4(n-1)+1)(n-1)\right)\mu+(2n-(n-1)+1)(n-1)\right]A_{n,n-(n-1)}+(1+\mu)^3(n-(n-1)+2)(n-(n-1)+1)A_{n,n-(n-1)+2}\right\}=\left\{\left[\left((n-nn)+\left(4n-(3n-3)\right)(n-1)\right)\mu^2+\left((n-nn)+(6n-(4n-4)+1)(n-1)\right)\mu+(n+2)(n-1)\right]A_{n,1}+(1+\mu)^3(+1+2)(+1+1)A_{n,3}\right\}=\{[((n-nn)+(n+3)(n-1))\mu^2+((n-nn)+(2n+5)(n-1))\mu+(n+2)(n-1)]A_{n,1}+(1+\mu)^3 6A_{n,3}\}=\{[((n-nn)+(nn+3n-n-3))\mu^2+((n-nn)+(2nn+5n-2n-5))\mu+(n+2)(n-1)]A_{n,1}+(1+\mu)^3 6A_{n,3}\}=\{[3(n-1)\mu^2+(nn+4n-5)\mu+(n+2)(n-1)]A_{n,1}+(1+\mu)^3 6A_{n,3}\}=\left\{\left[\tfrac{1}{4}(1+\mu)(3\mu+2n+1)^2-\tfrac{9}{4}(1+\mu)^3\right]A_{n,1}+(1+\mu)^3 6A_{n,3}\right\}=(1+\mu)\left\{\left[\tfrac{1}{4}(3\mu+2n+1)^2-\tfrac{9}{4}(1+\mu)^2\right]A_{n,1}+6(1+\mu)^2A_{n,3}\right\} \quad . \quad \text{(S4.42)}$$

According to (S4.17),

$$A_{n,1}=A_{n,n-2\left[\frac{n}{2}\right]}=A_{n,n-2\left(\frac{n-1}{2}\right)}=\frac{1}{2^n}\frac{(-1)^{\left(\frac{n-1}{2}\right)}}{(1+\mu)^{n-2\left(\frac{n-1}{2}\right)}}\binom{n}{\left(\frac{n-1}{2}\right)}\sum_{j=0}^{\left(\frac{n-1}{2}\right)-\left(\frac{n-1}{2}\right)}\binom{n-\left(\frac{n-1}{2}\right)}{j}\binom{2\left(n-\left(\frac{n-1}{2}\right)-j\right)}{n}\mu^j=\frac{1}{2^n}\frac{(-1)^{\left(\frac{n-1}{2}\right)}}{(1+\mu)^1}\binom{n}{\frac{n-1}{2}}\sum_{j=0}^{0}\binom{\frac{n}{2}+\frac{1}{2}}{j}\binom{(2n-(n-1)-2j)}{n}\mu^j=\frac{1}{2^n}\frac{(-1)^{\left(\frac{n-1}{2}\right)}}{(1+\mu)^1}\binom{n}{\frac{n-1}{2}}\binom{\frac{n}{2}+\frac{1}{2}}{0}\binom{(2n-(n-1)-2\cdot 0)}{n}\mu^0=\frac{1}{2^n}\frac{(-1)^{\left(\frac{n-1}{2}\right)}}{(1+\mu)}\binom{n}{\frac{n-1}{2}}\cdot 1\cdot\binom{n+1}{n}\cdot 1=\frac{1}{2^n}\frac{(-1)^{\frac{n-1}{2}}}{(1+\mu)}\binom{n}{\frac{n-1}{2}}\binom{n+1}{n}=\frac{1}{2^n}\frac{(-1)^{\frac{n-1}{2}}}{(1+\mu)}\binom{n}{\frac{n-1}{2}-1}\frac{n-\frac{n-1}{2}+1}{\frac{n-1}{2}}(n+1)=\frac{1}{2^n}\frac{(-1)^{\frac{n-1}{2}}}{(1+\mu)}\binom{n}{\frac{n-1}{2}-1}\frac{2n-(n-1)+2}{n-1}(n+1)=\frac{1}{2^n}\frac{(-1)^{\frac{n-1}{2}}}{(1+\mu)}\binom{n}{\frac{n-1}{2}-1}\frac{n+3}{n-1}(n+1) \quad , \quad \text{(S4.43)}$$

and

$$A_{n,3}=A_{n,n-2\left(\frac{n-3}{2}\right)}=\frac{1}{2^n}\frac{(-1)^{\left(\frac{n-3}{2}\right)}}{(1+\mu)^{n-2\left(\frac{n-3}{2}\right)}}\binom{n}{\left(\frac{n-3}{2}\right)}\sum_{j=0}^{\left(\frac{n-1}{2}\right)-\left(\frac{n-3}{2}\right)}\binom{n-\left(\frac{n-3}{2}\right)}{j}\binom{2\left(n-\left(\frac{n-3}{2}\right)-j\right)}{n}\mu^j=\frac{1}{2^n}\frac{(-1)^{\frac{n-1}{2}-1}}{(1+\mu)^3}\binom{n}{\frac{n-1}{2}-1}\sum_{j=0}^{1}\binom{\frac{n}{2}+\frac{1}{2}+1}{j}\binom{2\left(\frac{n}{2}+\frac{1}{2}+1-j\right)}{n}\mu^j=\frac{1}{2^n}\frac{(-1)^{\frac{n-1}{2}-1}}{(1+\mu)^3}\binom{n}{\frac{n-1}{2}-1}\left[\binom{\frac{n}{2}+\frac{1}{2}+1}{0}\binom{2\left(\frac{n}{2}+\frac{1}{2}+1-0\right)}{n}\mu^0+\right.$$

$$\binom{\frac{n}{2}+\frac{1}{2}+1}{1}\binom{2\left(\frac{n}{2}+\frac{1}{2}+1-1\right)}{n}\mu^1\Bigg] = \frac{1}{2^n}\frac{(-1)^{\frac{n-1}{2}-1}}{(1+\mu)^3}\binom{n}{\frac{n-1}{2}-1}\left[1\cdot\binom{n+3}{n}+\left(\frac{n}{2}+\frac{1}{2}+1\right)\binom{n+1}{n}\mu\right] =$$
$$\frac{(-1)^{\frac{n-1}{2}}}{2^n}\frac{(-1)^{-1}}{(1+\mu)^3}\binom{n}{\frac{n-1}{2}-1}\left[\frac{1}{6}(n+1)(n+2)(n+3)+\left(\frac{n}{2}+\frac{3}{2}\right)(n+1)\mu\right] = -\frac{(-1)^{\frac{n-1}{2}}}{2^n}\frac{1}{(1+\mu)^3}\binom{n}{\frac{n-1}{2}-1}(n+$$
$$1)\left[\frac{1}{6}(n+2)(n+3)+\frac{1}{6}3(n+3)\mu\right] = -\frac{(-1)^{\frac{n-1}{2}}}{2^n}\frac{1}{(1+\mu)^3}\binom{n}{\frac{n-1}{2}-1}(n+1)(n+3)\frac{1}{6}[(n+2)+3\mu]\,. \tag{S4.44}$$

Thus, the lowest order in *s* term coefficient (see (S4.42)) is then for *n* odd

$$(1+\mu)\left\{\left[\frac{1}{4}(3\mu+2n+1)^2-\frac{9}{4}(1+\mu)^2\right]A_{n,1}+6(1+\mu)^2A_{n,3}\right\} =$$
$$(1+\mu)\Bigg\{\left[\frac{1}{4}(3\mu+2n+1)^2-\frac{9}{4}(1+\mu)^2\right]\frac{1}{2^n}\frac{(-1)^{\frac{n-1}{2}}}{(1+\mu)}\binom{n}{\frac{n-1}{2}-1}\frac{n+3}{n-1}(n+1) - 6(1+\mu)^2\frac{(-1)^{\frac{n-1}{2}}}{2^n}\frac{1}{(1+\mu)^3}\binom{n}{\frac{n-1}{2}-1}(n+1)(n+$$
$$3)\frac{1}{6}[(n+2)+3\mu]\Bigg\} = (1+\mu)\frac{(-1)^{\frac{n-1}{2}}}{2^n}\frac{1}{(1+\mu)}\binom{n}{\frac{n-1}{2}-1}(n+3)\left\{\left[\frac{1}{4}(3\mu+2n+1)^2-\frac{9}{4}(1+\mu)^2\right]\frac{1}{n-1}(n+1) - 6(1+$$
$$\mu)^2\frac{1}{(1+\mu)^2}(n+1)\frac{1}{6}[(n+2)+3\mu]\right\} = \frac{(-1)^{\frac{n-1}{2}}}{2^n}\binom{n}{\frac{n-1}{2}-1}(n+3)\{[3\mu+n+2](n+1)-(n+1)[(n+2)+3\mu]\} =$$
$$\frac{(-1)^{\frac{n-1}{2}}}{2^n}\binom{n}{\frac{n-1}{2}-1}(n+3)\{0\} = 0 \quad . \tag{S4.45}$$

The coefficient of the lowest order *s* term is zero in this case as well. Then, it is proved that the term in the third sum in (S4.36) with the lowest exponent of *s* is always zero.

Therefore, we can proceed with modifications of (S4.36). We can remove the terms just found to be zero (the first term in the first sum with *N*=0, and the last term in the third sum with *N*=[*n*/2]) by adjusting the range of counting in that sums:

$$\sum_{N=1}^{[n/2]}[-4N(N-1)]\mu^2A_{n,n-2N}s^6s^{n-2N}+$$
$$\sum_{N=0}^{[n/2]}\{2(6N-2n-3)N\mu^2+4(5N-2n-3)N\mu+2(4N-2n-3)N\}\mu A_{n,n-2N}s^4s^{n-2N}+\sum_{N=0}^{[n/2]-1}\{[((1-n)n+$$
$$4(2n-3N)N)\mu^2+((1-n)n+2(6n-8N+1)N)\mu+2(2n-2N+1)N]A_{n,n-2N}+(1+\mu)^3(n-2N+2)(n-2N+$$
$$1)A_{n,n-2N+2}\}(1+\mu)^2s^2s^{n-2N}=0 \quad . \tag{S4.46}$$

Further, divide by $s^2$ and put the same powers of *s* together (but separately in the three sums):

$$\sum_{N=1}^{[n/2]}[-4N(N-1)]\mu^2A_{n,n-2N}s^{n-2N+4}+$$
$$\sum_{N=0}^{[n/2]}\{2(6N-2n-3)N\mu^2+4(5N-2n-3)N\mu+2(4N-2n-3)N\}\mu A_{n,n-2N}s^{n-2N+2}+\sum_{N=0}^{[n/2]-1}\{[((1-n)n+$$
$$4(2n-3N)N)\mu^2+((1-n)n+2(6n-8N+1)N)\mu+2(2n-2N+1)N]A_{n,n-2N}+(1+\mu)^3(n-2N+2)(n-2N+$$
$$1)A_{n,n-2N+2}\}(1+\mu)^2s^{n-2N+0}=0 \quad . \tag{S4.47}$$

Then do an index substitution *M*+1=*N* in the first sum and *K*-1=*N* in the third sum:

$$\sum_{M=0}^{[n/2]-1}[-4(M+1)M]\mu^2A_{n,n-2(M+1)}s^{n-2(M+1)+4}+\sum_{N=0}^{[n/2]}\{2(6N-2n-3)N\mu^2+4(5N-2n-3)N\mu+2(4N-2n-$$
$$3)N\}\mu A_{n,n-2N}s^{n-2N+2}+\sum_{K=1}^{[n/2]}\Big\{\Big[\big((1-n)n+4(2n-3(K-1))(K-1)\big)\mu^2+\big((1-n)n+2(6n-8(K-1)+1)(K-$$
$$1)\big)\mu+2(2n-2(K-1)+1)(K-1)\Big]A_{n,n-2(K-1)}+(1+\mu)^3(n-2(K-1)+2)(n-2(K-1)+1)A_{n,n-2(K-1)+2}\Big\}(1+$$
$$\mu)^2s^{n-2(K-1)+0}=0 \quad . \tag{S4.48}$$

Algebraic simplification leads to

$$\sum_{M=0}^{[n/2]-1}[-4(M+1)M]\mu^2A_{n,n-2M-2}s^{n-2M+2}+$$
$$\sum_{N=0}^{[n/2]}2\{(6N-2n-3)\mu^2+2(5N-2n-3)\mu+(4N-2n-3)\}N\mu A_{n,n-2N}s^{n-2N+2}+\sum_{K=1}^{[n/2]}\Big\{\Big[\big((1-n)n+4(2n-$$
$$3(K-1))(K-1)\big)\mu^2+\big((1-n)n+2(6n-8K+9)(K-1)\big)\mu+2(2n-2K+3)(K-1)\Big]A_{n,n-2K+2}+(1+\mu)^3(n-2K+$$
$$4)(n-2K+3)A_{n,n-2K+4}\Big\}(1+\mu)^2s^{n-2K+2}=0 \quad . \tag{S4.49}$$

The first sum index can be extended up to [n/2], as it results in $A_{n,-2}$ for even *n*, or in $A_{n,-1}$ for odd *n*, which is zero (as it is non-existent). The *M*=0 term in the first sum results clearly in zero (multiplication by *M*), can be thus skipped. The *N*=0 term in the second sum results clearly in zero (multiplication by *N*), can be thus skipped. The *N* and *K* indices in the second and third sum, respectively, are then renamed to *M*:

$$\sum_{M=1}^{[n/2]}[-4(M+1)M]\mu^2 A_{n,n-2M-2}s^{n-2M+2} + \sum_{M=1}^{[n/2]} 2\{(6M-2n-3)\mu^2 + 2(5M-2n-3)\mu + (4M-2n-3)\}M\mu A_{n,n-2M}s^{n-2M+2} + \sum_{M=1}^{[n/2]}\Big\{\Big[\Big((1-n)n + 4\big(2n-3(M-1)\big)(M-1)\Big)\mu^2 + \big((1-n)n + 2(6n-8M+9)(M-1)\big)\mu + 2(2n-2M+3)(M-1)\Big] A_{n,n-2M+2} + (1+\mu)^3(n-2M+4)(n-2M+3)A_{n,n-2M+4}\Big\}(1+\mu)^2 s^{n-2M+2} = 0 \quad . \qquad \text{(S4.50)}$$

Now, all three sums have the same range and the exponent of *s* has the same form. Then, the sums can be composed into one sum:

$$\sum_{M=1}^{[n/2]}\Big[[-4(M+1)M]\mu^2 A_{n,n-2M-2} + 2\{(6M-2n-3)\mu^2 + 2(5M-2n-3)\mu + (4M-2n-3)\}M\mu A_{n,n-2M} + \Big\{\Big[\Big((1-n)n + 4\big(2n-3(M-1)\big)(M-1)\Big)\mu^2 + \big((1-n)n + 2(6n-8M+9)(M-1)\big)\mu + 2(2n-2M+3)(M-1)\Big] A_{n,n-2M+2} + (1+\mu)^3(n-2M+4)(n-2M+3)A_{n,n-2M+4}\Big\}(1+\mu)^2\Big] s^{n-2M+2} = 0 \quad . \qquad \text{(S4.51)}$$

In order the pre-factors in front of all powers of *s* to be zero, the term in the outmost angle bracket has to be zero for every *n*≥0 and for all *M* in the range 1 to [*n*/2]. Thus

$$\Big[[-4(M+1)M]\mu^2 A_{n,n-2M-2} + 2\{(6M-2n-3)\mu^2 + 2(5M-2n-3)\mu + (4M-2n-3)\}M\mu A_{n,n-2M} + \Big\{\Big[\Big((1-n)n + 4\big(2n-3(M-1)\big)(M-1)\Big)\mu^2 + \big((1-n)n + 2(6n-8M+9)(M-1)\big)\mu + 2(2n-2M+3)(M-1)\Big] A_{n,n-2M+2} + (1+\mu)^3(n-2M+4)(n-2M+3)A_{n,n-2M+4}\Big\}(1+\mu)^2\Big] = 0 \quad , \qquad \text{(S4.52)}$$

and – by expanding the last part –

$$-4(M+1)M\mu^2 A_{n,n-2M-2} + 2[(6M-2n-3)\mu^2 + 2(5M-2n-3)\mu + (4M-2n-3)]M\mu A_{n,n-2M} + \{[(1-n)n + 4\big(2n-3(M-1)\big)(M-1)]\mu^2 + [(1-n)n + 2(6n-8M+9)(M-1)]\mu + 2(2n-2M+3)(M-1)\}(1+\mu)^2 A_{n,n-2M+2} + (n-2M+4)(n-2M+3)(1+\mu)^5 A_{n,n-2M+4} = 0 \quad . \qquad \text{(S4.53)}$$

The left-side relation is in fact a polynomial in $\mu$. To be zero, the coefficient by each power of $\mu$ has to be zero.

Now, use (S4.17), for expressing the coefficients $A_{n,m}$ in (S4.53).

With substitution *N*=*M*+1, we obtain from (S4.17)

$$A_{n,n-2N} = A_{n,n-2(M+1)} = A_{n,n-2M-2} = \frac{1}{2^n}\frac{(-1)^{M+1}}{(1+\mu)^{n-2(M+1)}}\binom{n}{M+1}\sum_{j=0}^{\left[\frac{n}{2}\right]-(M+1)}\binom{n-(M+1)}{j}\binom{2(n-(M+1)-j)}{n}\mu^j = -\frac{1}{2^n}\frac{(-1)^M}{(1+\mu)^{n-2M}}\frac{n-M}{M+1}\binom{n}{M}(1+\mu)^2\left[\sum_{j=0}^{\left[\frac{n}{2}\right]-M-1}\binom{n-M-1}{j}\binom{2(n-M-j)-2}{n}\mu^j\right] \quad . \qquad \text{(S4.54)}$$

For substitution *N*=*M*, we obtain

$$A_{n,n-2M} = \frac{1}{2^n}\frac{(-1)^M}{(1+\mu)^{n-2M}}\binom{n}{M}\sum_{j=0}^{\left[\frac{n}{2}\right]-M}\binom{n-M}{j}\binom{2(n-M-j)}{n}\mu^j \quad . \qquad \text{(S4.55)}$$

For substitution *N*=*M*-1, we obtain

$$A_{n,n-2M+2} = \frac{1}{2^n}\frac{(-1)^{M-1}}{(1+\mu)^{n-2(M-1)}}\binom{n}{M-1}\sum_{j=0}^{\left[\frac{n}{2}\right]-(M-1)}\binom{n-(M-1)}{j}\binom{2(n-(M-1)-j)}{n}\mu^j = -\frac{1}{2^n}\frac{(-1)^M}{(1+\mu)^{n-2M}}\frac{M}{n-M+1}\binom{n}{M}\frac{1}{(1+\mu)^2}\left[\sum_{j=0}^{\left[\frac{n}{2}\right]-M+1}\binom{n-M+1}{j}\binom{2(n-M-j)+2}{n}\mu^j\right] \quad . \qquad \text{(S4.56)}$$

For substitution *N*=*M*-2, we obtain

$$A_{n,n-2M+4} = \frac{1}{2^n}\frac{(-1)^{M-2}}{(1+\mu)^{n-2(M-2)}}\binom{n}{M-2}\sum_{j=0}^{\left[\frac{n}{2}\right]-(M-2)}\binom{n-(M-2)}{j}\binom{2(n-(M-2)-j)}{n}\mu^j = \frac{1}{2^n}\frac{(-1)^M}{(1+\mu)^{n-2M+4}}\frac{M-1}{n-(M-1)+1}\binom{n}{M-1}\sum_{j=0}^{\left[\frac{n}{2}\right]-M+2}\binom{n-M+2}{j}\binom{2(n-M-j)+4}{n}\mu^j = \frac{1}{2^n}\frac{(-1)^M}{(1+\mu)^{n-2M}}\frac{M-1}{n-M+2}\frac{M}{n-M+1}\binom{n}{M}\frac{1}{(1+\mu)^4}\sum_{j=0}^{\left[\frac{n}{2}\right]-M+2}\binom{n-M+2}{j}\binom{2(n-M-j)+4}{n}\mu^j \quad . \qquad \text{(S4.57)}$$

Then the individual terms in (S4.53) are as follows.
The term with $A_{n,n-2M-2}$ (using (S4.54)):

$$-4(M+1)M\mu^2 A_{n,n-2M-2} = -4(M+1)M\mu^2 \frac{-1}{2^n}\frac{(-1)^M}{(1+\mu)^{n-2M}}\frac{n-M}{M+1}\binom{n}{M}(1+\mu)^2\left[\sum_{j=0}^{\left[\frac{n}{2}\right]-M-1}\binom{n-M-1}{j}\binom{2(n-M-j)-2}{n}\mu^j\right] = \frac{1}{2^n}\frac{(-1)^M}{(1+\mu)^{n-2M}}\binom{n}{M}4(n-M)M\mu^2(1+2\mu+\mu^2)\left[\sum_{j=0}^{\left[\frac{n}{2}\right]-M-1}\binom{n-M-1}{j}\binom{2(n-M-j)-2}{n}\mu^j\right] = \frac{1}{2^n}\frac{(-1)^M}{(1+\mu)^{n-2M}}\binom{n}{M}M4(n-M)(\mu^2+2\mu^3+\mu^4)\left[\sum_{j=0}^{\left[\frac{n}{2}\right]-M-1}\binom{n-M-1}{j}\binom{2(n-M-j)-2}{n}\mu^j\right] = \frac{1}{2^n}\frac{(-1)^M}{(1+\mu)^{n-2M}}\binom{n}{M}M\left\{\left[\sum_{j=0}^{\left[\frac{n}{2}\right]-M-1}4(n-M)\binom{n-M-1}{j}\binom{2(n-M-j)-2}{n}\mu^{j+2}\right]+\left[\sum_{j=0}^{\left[\frac{n}{2}\right]-M-1}8(n-M)\binom{n-M-1}{j}\binom{2(n-M-j)-2}{n}\mu^{j+3}\right]+\left[\sum_{j=0}^{\left[\frac{n}{2}\right]-M-1}4(n-M)\binom{n-M-1}{j}\binom{2(n-M-j)-2}{n}\mu^{j+4}\right]\right\} \quad . \qquad (S4.58)$$

The term with $A_{n,n-2M}$ (using (S4.55)):

$$2[(6M-2n-3)\mu^2+2(5M-2n-3)\mu+(4M-2n-3)]M\mu A_{n,n-2M} = 2[(6M-2n-3)\mu^2+2(5M-2n-3)\mu+(4M-2n-3)]M\mu\frac{1}{2^n}\frac{(-1)^M}{(1+\mu)^{n-2M}}\binom{n}{M}\left[\sum_{j=0}^{\left[\frac{n}{2}\right]-M}\binom{n-M}{j}\binom{2(n-M-j)}{n}\mu^j\right] = \frac{1}{2^n}\frac{(-1)^M}{(1+\mu)^{n-2M}}\binom{n}{M}M[2(6M-2n-3)\mu^3+4(5M-2n-3)\mu^2+2(4M-2n-3)\mu]\left[\sum_{j=0}^{\left[\frac{n}{2}\right]-M}\binom{n-M}{j}\binom{2(n-M-j)}{n}\mu^j\right] = \frac{1}{2^n}\frac{(-1)^M}{(1+\mu)^{n-2M}}\binom{n}{M}M\left\{2(6M-2n-3)\left[\sum_{j=0}^{\left[\frac{n}{2}\right]-M}\binom{n-M}{j}\binom{2(n-M-j)}{n}\mu^{j+3}\right]+4(5M-2n-3)\left[\sum_{j=0}^{\left[\frac{n}{2}\right]-M}\binom{n-M}{j}\binom{2(n-M-j)}{n}\mu^{j+2}\right]+2(4M-2n-3)\left[\sum_{j=0}^{\left[\frac{n}{2}\right]-M}\binom{n-M}{j}\binom{2(n-M-j)}{n}\mu^{j+1}\right]\right\} \quad . \qquad (S4.59)$$

The term with $A_{n,n-2M+2}$ (using (S4.56)):

$$\{[(1-n)n+4(2n-3(M-1))(M-1)]\mu^2+[(1-n)n+2(6n-8M+9)(M-1)]\mu+2(2n-2M+3)(M-1)\}(1+\mu)^2A_{n,n-2M+2} = \{[(1-n)n+4(2n-3(M-1))(M-1)]\mu^2+[(1-n)n+2(6n-8M+9)(M-1)]\mu+2(2n-2M+3)(M-1)\}(1+\mu)^2\frac{-1}{2^n}\frac{(-1)^M}{(1+\mu)^{n-2M}}\frac{M}{n-M+1}\binom{n}{M}\frac{1}{(1+\mu)^2}\left[\sum_{j=0}^{\left[\frac{n}{2}\right]-M+1}\binom{n-M+1}{j}\binom{2(n-M-j)+2}{n}\mu^j\right] = -\frac{1}{2^n}\frac{(-1)^M}{(1+\mu)^{n-2M}}\binom{n}{M}M\frac{1}{n-M+1}\{[(1-n)n+4(2n-3(M-1))(M-1)]\mu^2+[(1-n)n+2(6n-8M+9)(M-1)]\mu+2(2n-2M+3)(M-1)\}\left[\sum_{j=0}^{\left[\frac{n}{2}\right]-M+1}\binom{n-M+1}{j}\binom{2(n-M-j)+2}{n}\mu^j\right] = -\frac{1}{2^n}\frac{(-1)^M}{(1+\mu)^{n-2M}}\binom{n}{M}M\left\{\frac{(1-n)n+4(2n-3(M-1))(M-1)}{n-M+1}\left[\sum_{j=0}^{\left[\frac{n}{2}\right]-M+1}\binom{n-M+1}{j}\binom{2(n-M-j)+2}{n}\mu^{j+2}\right]+\frac{(1-n)n+2(6n-8M+9)(M-1)}{n-M+1}\left[\sum_{j=0}^{\left[\frac{n}{2}\right]-M+1}\binom{n-M+1}{j}\binom{2(n-M-j)+2}{n}\mu^{j+1}\right]+\frac{2(2n-2M+3)(M-1)}{n-M+1}\left[\sum_{j=0}^{\left[\frac{n}{2}\right]-M+1}\binom{n-M+1}{j}\binom{2(n-M-j)+2}{n}\mu^{j}\right]\right\} \quad . \qquad (S4.60)$$

The term with $A_{n,n-2M+4}$ (using (S4.57)):

$$(n-2M+4)(n-2M+3)(1+\mu)^5A_{n,n-2M+4} = (n-2M+4)(n-2M+3)(1+\mu)^5\frac{1}{2^n}\frac{(-1)^M}{(1+\mu)^{n-2M}}\frac{M-1}{n-M+2}\frac{M}{n-M+1}\binom{n}{M}\frac{1}{(1+\mu)^4}\sum_{j=0}^{\left[\frac{n}{2}\right]-M+2}\binom{n-M+2}{j}\binom{2(n-M-j)+4}{n}\mu^j = \frac{1}{2^n}\frac{(-1)^M}{(1+\mu)^{n-2M}}\binom{n}{M}M(M-1)\frac{(n-2M+4)}{n-M+2}\frac{(n-2M+3)}{n-M+1}(1+\mu)\sum_{j=0}^{\left[\frac{n}{2}\right]-M+2}\binom{n-M+2}{j}\binom{2(n-M-j)+4}{n}\mu^j = \frac{1}{2^n}\frac{(-1)^M}{(1+\mu)^{n-2M}}\binom{n}{M}M\left\{(M-1)\frac{(n-2M+4)}{n-M+2}\frac{(n-2M+3)}{n-M+1}\left[\sum_{j=0}^{\left[\frac{n}{2}\right]-M+2}\binom{n-M+2}{j}\binom{2(n-M-j)+4}{n}\mu^j\right]+(M-1)\frac{(n-2M+4)}{n-M+2}\frac{(n-2M+3)}{n-M+1}\left[\sum_{j=0}^{\left[\frac{n}{2}\right]-M+2}\binom{n-M+2}{j}\binom{2(n-M-j)+4}{n}\mu^{j+1}\right]\right\} \quad . \qquad (S4.61)$$

The common factor for all the terms is

$$\frac{1}{2^n}\frac{(-1)^M}{(1+\mu)^{n-2M}}\binom{n}{M}M \quad . \qquad (S4.62)$$

This pre-factor is zero in the case when $M$=0. In all other cases it is non-zero. As $M$ is in the range 1 to [$n$/2] (see (S4.51)), then the pre-factor (S4.62) is non-zero for all cases of interest. Then, the remaining combined expression has to be zero (which is to be tested). Thus, we will skip the factor (S4.62) in the following derivations. Put together all the terms in (S4.53), i.e. (S4.58), (S4.59), (S4.60) and (S4.61), without the pre-factor discussed above:

$$\left\{\left[\sum_{j=0}^{\left[\frac{n}{2}\right]-M-1} 4(n-M)\binom{n-M-1}{j}\binom{2(n-M-j)-2}{n}\mu^{j+2}\right]+\left[\sum_{j=0}^{\left[\frac{n}{2}\right]-M-1} 8(n-M)\binom{n-M-1}{j}\binom{2(n-M-j)-2}{n}\mu^{j+3}\right]+\left[\sum_{j=0}^{\left[\frac{n}{2}\right]-M-1} 4(n-M)\binom{n-M-1}{j}\binom{2(n-M-j)-2}{n}\mu^{j+4}\right]\right\}+$$

$$\left\{2(6M-2n-3)\left[\sum_{j=0}^{\left[\frac{n}{2}\right]-M}\binom{n-M}{j}\binom{2(n-M-j)}{n}\mu^{j+3}\right]+4(5M-2n-3)\left[\sum_{j=0}^{\left[\frac{n}{2}\right]-M}\binom{n-M}{j}\binom{2(n-M-j)}{n}\mu^{j+2}\right]+2(4M-2n-3)\left[\sum_{j=0}^{\left[\frac{n}{2}\right]-M}\binom{n-M}{j}\binom{2(n-M-j)}{n}\mu^{j+1}\right]\right\}-$$

$$\left\{\frac{(1-n)n+4(2n-3(M-1))(M-1)}{n-M+1}\left[\sum_{j=0}^{\left[\frac{n}{2}\right]-M+1}\binom{n-M+1}{j}\binom{2(n-M-j)+2}{n}\mu^{j+2}\right]+\frac{(1-n)n+2(6n-8M+9)(M-1)}{n-M+1}\left[\sum_{j=0}^{\left[\frac{n}{2}\right]-M+1}\binom{n-M+1}{j}\binom{2(n-M-j)+2}{n}\mu^{j+1}\right]+\frac{2(2n-2M+3)(M-1)}{n-M+1}\left[\sum_{j=0}^{\left[\frac{n}{2}\right]-M+1}\binom{n-M+1}{j}\binom{2(n-M-j)+2}{n}\mu^{j}\right]\right\}+$$

$$\left\{(M-1)\frac{(n-2M+4)}{n-M+2}\frac{(n-2M+3)}{n-M+1}\left[\sum_{j=0}^{\left[\frac{n}{2}\right]-M+2}\binom{n-M+2}{j}\binom{2(n-M-j)+4}{n}\mu^{j}\right]+(M-1)\frac{(n-2M+4)}{n-M+2}\frac{(n-2M+3)}{n-M+1}\left[\sum_{j=0}^{\left[\frac{n}{2}\right]-M+2}\binom{n-M+2}{j}\binom{2(n-M-j)+4}{n}\mu^{j+1}\right]\right\}\quad. \qquad (S4.63)$$

Remind that $M$ is in the range from 1 to $[n/2]$ (see the range in (S4.51)) in (S4.63). The further approach for proving that the expression (S4.63) is zero for any non-negative $n$ and for any $M$ in the abovementioned range is to change the expression algebraically in order to have the same exponent of $\mu$ in the sums.

Substitutions of the indices in the individual sums are done first: $i$=$j$+1 in the first, fifth and seventh term of (S4.63), while $k$=$j$+2 in the second and fourth term, $l$=$j$+3 in the third term, and $q$=$j$-1 in the nineth and tenth terms:

$$\left\{\left[\sum_{i=1}^{\left[\frac{n}{2}\right]-M} 4(n-M)\binom{n-M-1}{i-1}\binom{2(n-M-(i-1))-2}{n}\mu^{(i-1)+2}\right]+\left[\sum_{k=2}^{\left[\frac{n}{2}\right]-M+1} 8(n-M)\binom{n-M-1}{k-2}\binom{2(n-M-(k-2))-2}{n}\mu^{(k-2)+3}\right]+\left[\sum_{l=3}^{\left[\frac{n}{2}\right]-M+2} 4(n-M)\binom{n-M-1}{l-3}\binom{2(n-M-(l-3))-2}{n}\mu^{(l-3)+4}\right]\right\}+$$

$$\left\{2(6M-2n-3)\left[\sum_{k=2}^{\left[\frac{n}{2}\right]-M+2}\binom{n-M}{k-2}\binom{2(n-M-(k-2))}{n}\mu^{(k-2)+3}\right]+4(5M-2n-3)\left[\sum_{i=1}^{\left[\frac{n}{2}\right]-M+1}\binom{n-M}{i-1}\binom{2(n-M-(i-1))}{n}\mu^{(i-1)+2}\right]+2(4M-2n-3)\left[\sum_{j=0}^{\left[\frac{n}{2}\right]-M}\binom{n-M}{j}\binom{2(n-M-j)}{n}\mu^{j+1}\right]\right\}-$$

$$\left\{\frac{(1-n)n+4(2n-3(M-1))(M-1)}{n-M+1}\left[\sum_{i=1}^{\left[\frac{n}{2}\right]-M+2}\binom{n-M+1}{i-1}\binom{2(n-M-(i-1))+2}{n}\mu^{(i-1)+2}\right]+\frac{(1-n)n+2(6n-8M+9)(M-1)}{n-M+1}\left[\sum_{j=0}^{\left[\frac{n}{2}\right]-M+1}\binom{n-M+1}{j}\binom{2(n-M-j)+2}{n}\mu^{j+1}\right]+\frac{2(2n-2M+3)(M-1)}{n-M+1}\left[\sum_{q=-1}^{\left[\frac{n}{2}\right]-M}\binom{n-M+1}{q+1}\binom{2(n-M-(q+1))+2}{n}\mu^{(q+1)}\right]\right\}+$$

$$\left\{(M-1)\frac{(n-2M+4)}{n-M+2}\frac{(n-2M+3)}{n-M+1}\left[\sum_{q=-1}^{\left[\frac{n}{2}\right]-M+1}\binom{n-M+2}{q+1}\binom{2(n-M-(q+1))+4}{n}\mu^{(q+1)}\right]+(M-1)\frac{(n-2M+4)}{n-M+2}\frac{(n-2M+3)}{n-M+1}\left[\sum_{j=0}^{\left[\frac{n}{2}\right]-M+2}\binom{n-M+2}{j}\binom{2(n-M-j)+4}{n}\mu^{j+1}\right]\right\}\quad. \qquad (S4.64)$$

After trivial algebra, it reads

$$\left\{\left[\sum_{i=1}^{\left[\frac{n}{2}\right]-M} 4(n-M)\binom{n-M-1}{i-1}\binom{2(n-M-i)}{n}\mu^{i+1}\right]+\left[\sum_{k=2}^{\left[\frac{n}{2}\right]-M+1} 8(n-M)\binom{n-M-1}{k-2}\binom{2(n-M-k)+2}{n}\mu^{k+1}\right]+\left[\sum_{l=3}^{\left[\frac{n}{2}\right]-M+2} 4(n-M)\binom{n-M-1}{l-3}\binom{2(n-M-l)+4}{n}\mu^{l+1}\right]\right\}+$$

$$\left\{2(6M-2n-3)\left[\sum_{k=2}^{\left[\frac{n}{2}\right]-M+2}\binom{n-M}{k-2}\binom{2(n-M-k)+4}{n}\mu^{k+1}\right]+4(5M-2n-3)\left[\sum_{i=1}^{\left[\frac{n}{2}\right]-M+1}\binom{n-M}{i-1}\binom{2(n-M-i)+2}{n}\mu^{i+1}\right]+2(4M-2n-3)\left[\sum_{j=0}^{\left[\frac{n}{2}\right]-M}\binom{n-M}{j}\binom{2(n-M-j)}{n}\mu^{j+1}\right]\right\}-$$

$$\left\{\frac{(1-n)n+4(2n-3(M-1))(M-1)}{n-M+1}\left[\sum_{i=1}^{\left[\frac{n}{2}\right]-M+2}\binom{n-M+1}{i-1}\binom{2(n-M-i)+4}{n}\mu^{i+1}\right]+\frac{(1-n)n+2(6n-8M+9)(M-1)}{n-M+1}\left[\sum_{j=0}^{\left[\frac{n}{2}\right]-M+1}\binom{n-M+1}{j}\binom{2(n-M-j)+2}{n}\mu^{j+1}\right]+\frac{2(2n-2M+3)(M-1)}{n-M+1}\left[\sum_{q=-1}^{\left[\frac{n}{2}\right]-M}\binom{n-M+1}{q+1}\binom{2(n-M-q)}{n}\mu^{q+1}\right]\right\}+$$

$$\left\{(M-1)\frac{(n-2M+4)}{n-M+2}\frac{(n-2M+3)}{n-M+1}\left[\sum_{q=-1}^{\left[\frac{n}{2}\right]-M+1}\binom{n-M+2}{q+1}\binom{2(n-M-q)+2}{n}\mu^{q+1}\right]+\right.$$

$$\left.(M-1)\frac{(n-2M+4)}{n-M+2}\frac{(n-2M+3)}{n-M+1}\left[\sum_{j=0}^{\left[\frac{n}{2}\right]-M+2}\binom{n-M+2}{j}\binom{2(n-M-j)+4}{n}\mu^{j+1}\right]\right\} \quad . \qquad \text{(S4.65)}$$

Do further simple algebraic manipulations and rename the indices *i*, *k*, *l* and *q* to *j*:

$$\left\{\left[\sum_{j=1}^{\left[\frac{n}{2}\right]-M}4(n-M)\binom{n-M-1}{j-1}\binom{2(n-M-j)}{n}\mu^{j+1}\right]+\left[\sum_{j=2}^{\left[\frac{n}{2}\right]-M+1}8(n-M)\binom{n-M-1}{j-2}\binom{2(n-M-j)+2}{n}\mu^{j+1}\right]+\left[\sum_{j=3}^{\left[\frac{n}{2}\right]-M+2}4(n-M)\binom{n-M-1}{j-3}\binom{2(n-M-j)+4}{n}\mu^{j+1}\right]\right\}+$$

$$\left\{\left[\sum_{j=2}^{\left[\frac{n}{2}\right]-M+2}2(6M-2n-3)\binom{n-M}{j-2}\binom{2(n-M-j)+4}{n}\mu^{j+1}\right]+\left[\sum_{j=1}^{\left[\frac{n}{2}\right]-M+1}4(5M-2n-3)\binom{n-M}{j-1}\binom{2(n-M-j)+2}{n}\mu^{j+1}\right]+\left[\sum_{j=0}^{\left[\frac{n}{2}\right]-M}2(4M-2n-3)\binom{n-M}{j}\binom{2(n-M-j)}{n}\mu^{j+1}\right]\right\}-$$

$$\left\{\left[\sum_{j=1}^{\left[\frac{n}{2}\right]-M+2}\frac{(1-n)n+4(2n-3(M-1))(M-1)}{n-M+1}\binom{n-M+1}{j-1}\binom{2(n-M-j)+4}{n}\mu^{j+1}\right]+\right.$$

$$\left.\left[\sum_{j=0}^{\left[\frac{n}{2}\right]-M+1}\frac{(1-n)n+2(6n-8M+9)(M-1)}{n-M+1}\binom{n-M+1}{j}\binom{2(n-M-j)+2}{n}\mu^{j+1}\right]+\left[\sum_{j=-1}^{\left[\frac{n}{2}\right]-M}\frac{2(2n-2M+3)(M-1)}{n-M+1}\binom{n-M+1}{j+1}\binom{2(n-M-j)}{n}\mu^{j+1}\right]\right\}+$$

$$\left\{\left[\sum_{j=-1}^{\left[\frac{n}{2}\right]-M+1}(M-1)\frac{(n-2M+4)}{n-M+2}\frac{(n-2M+3)}{n-M+1}\binom{n-M+2}{j+1}\binom{2(n-M-j)+2}{n}\mu^{j+1}\right]+\right.$$

$$\left.\left[\sum_{j=0}^{\left[\frac{n}{2}\right]-M+2}(M-1)\frac{(n-2M+4)}{n-M+2}\frac{(n-2M+3)}{n-M+1}\binom{n-M+2}{j}\binom{2(n-M-j)+4}{n}\mu^{j+1}\right]\right\} \quad . \qquad \text{(S4.66)}$$

The sums have now the same power in $\mu$, but the range of *j* index differs for the individual sums. Reduce the sums to the common range *j*=3 to [*n*/2]–*M*, and separate the rest of the terms from the sums:

$$\left\{\left[\sum_{j=3}^{\left[\frac{n}{2}\right]-M}4(n-M)\binom{n-M-1}{j-1}\binom{2(n-M-j)}{n}\mu^{j+1}+4(n-M)\binom{n-M-1}{1-1}\binom{2(n-M-1)}{n}\mu^{1+1}+4(n-M)\binom{n-M-1}{2-1}\binom{2(n-M-2)}{n}\mu^{2+1}\right]+\right.$$

$$\left[\sum_{j=3}^{\left[\frac{n}{2}\right]-M}8(n-M)\binom{n-M-1}{j-2}\binom{2(n-M-j)+2}{n}\mu^{j+1}+8(n-M)\binom{n-M-1}{2-2}\binom{2(n-M-2)+2}{n}\mu^{2+1}+\right.$$

$$\left.8(n-M)\binom{n-M-1}{\left[\frac{n}{2}\right]-M+1-2}\binom{2\left(n-M-\left(\left[\frac{n}{2}\right]-M+1\right)\right)+2}{n}\mu^{\left(\left[\frac{n}{2}\right]-M+1\right)+1}\right]+\left[\sum_{j=3}^{\left[\frac{n}{2}\right]-M}4(n-M)\binom{n-M-1}{j-3}\binom{2(n-M-j)+4}{n}\mu^{j+1}+4(n-M)\right.$$

$$\left.\left.\binom{n-M-1}{\left(\left[\frac{n}{2}\right]-M+1\right)-3}\binom{2\left(n-M-\left(\left[\frac{n}{2}\right]-M+1\right)\right)+4}{n}\mu^{\left(\left[\frac{n}{2}\right]-M+1\right)+1}+4(n-M)\binom{n-M-1}{\left(\left[\frac{n}{2}\right]-M+2\right)-3}\binom{2\left(n-M-\left(\left[\frac{n}{2}\right]-M+2\right)\right)+4}{n}\mu^{\left(\left[\frac{n}{2}\right]-M+2\right)+1}\right]\right\}+$$

$$\left\{\left[\sum_{j=3}^{\left[\frac{n}{2}\right]-M}2(6M-2n-3)\binom{n-M}{j-2}\binom{2(n-M-j)+4}{n}\mu^{j+1}+2(6M-2n-3)\binom{n-M}{2-2}\binom{2(n-M-2)+4}{n}\mu^{2+1}+\right.\right.$$

$$2(6M-2n-3)\binom{n-M}{\left[\frac{n}{2}\right]-M+1-2}\binom{2\left(n-M-\left(\left[\frac{n}{2}\right]-M+1\right)\right)+4}{n}\mu^{\left(\left[\frac{n}{2}\right]-M+1\right)+1}+$$

$$\left.2(6M-2n-3)\binom{n-M}{\left[\frac{n}{2}\right]-M+2-2}\binom{2\left(n-M-\left(\left[\frac{n}{2}\right]-M+2\right)\right)+4}{n}\mu^{\left(\left[\frac{n}{2}\right]-M+2\right)+1}\right]+\left[\sum_{j=3}^{\left[\frac{n}{2}\right]-M}4(5M-2n-3)\binom{n-M}{j-1}\binom{2(n-M-j)+2}{n}\mu^{j+1}+\right.$$

$$4(5M-2n-3)\binom{n-M}{1-1}\binom{2(n-M-1)+2}{n}\mu^{1+1}+4(5M-2n-3)\binom{n-M}{2-1}\binom{2(n-M-2)+2}{n}\mu^{2+1}+$$

$$\left.4(5M-2n-3)\binom{n-M}{\left[\frac{n}{2}\right]-M+1-1}\binom{2\left(n-M-\left(\left[\frac{n}{2}\right]-M+1\right)\right)+2}{n}\mu^{\left(\left[\frac{n}{2}\right]-M+1\right)+1}\right]+\left[\sum_{j=3}^{\left[\frac{n}{2}\right]-M}2(4M-2n-3)\binom{n-M}{j}\binom{2(n-M-j)}{n}\mu^{j+1}+\right.$$

$$\left.\left.2(4M-2n-3)\binom{n-M}{0}\binom{2(n-M-0)}{n}\mu^{0+1}+2(4M-2n-3)\binom{n-M}{1}\binom{2(n-M-1)}{n}\mu^{1+1}+2(4M-2n-3)\binom{n-M}{2}\binom{2(n-M-2)}{n}\mu^{2+1}\right]\right\}-$$

$$\left\{\left[\sum_{j=3}^{\left[\frac{n}{2}\right]-M}\frac{(1-n)n+4(2n-3(M-1))(M-1)}{n-M+1}\binom{n-M+1}{j-1}\binom{2(n-M-j)+4}{n}\mu^{j+1}+\frac{(1-n)n+4(2n-3(M-1))(M-1)}{n-M+1}\binom{n-M+1}{1-1}\binom{2(n-M-1)+4}{n}\mu^{1+1}+\right.\right.$$

$$\frac{(1-n)n+4(2n-3(M-1))(M-1)}{n-M+1}\binom{n-M+1}{2-1}\binom{2(n-M-2)+4}{n}\mu^{2+1}+$$

$$\frac{(1-n)n+4(2n-3(M-1))(M-1)}{n-M+1}\binom{n-M+1}{\left[\frac{n}{2}\right]-M+1-1}\binom{2\left(n-M-\left(\left[\frac{n}{2}\right]-M+1\right)\right)+4}{n}\mu^{\left(\left[\frac{n}{2}\right]-M+1\right)+1}+$$

$$\left.\frac{(1-n)n+4(2n-3(M-1))(M-1)}{n-M+1}\binom{n-M+1}{\left[\frac{n}{2}\right]-M+2-1}\binom{2\left(n-M-\left(\left[\frac{n}{2}\right]-M+2\right)\right)+4}{n}\mu^{\left(\left[\frac{n}{2}\right]-M+2\right)+1}\right]+$$

$$\left[\sum_{j=3}^{\left[\frac{n}{2}\right]-M}\frac{(1-n)n+2(6n-8M+9)(M-1)}{n-M+1}\binom{n-M+1}{j}\binom{2(n-M-j)+2}{n}\mu^{j+1}+\frac{(1-n)n+2(6n-8M+9)(M-1)}{n-M+1}\binom{n-M+1}{0}\binom{2(n-M-0)+2}{n}\mu^{0+1}+\right.$$

$$\frac{(1-n)n+2(6n-8M+9)(M-1)}{n-M+1}\binom{n-M+1}{1}\binom{2(n-M-1)+2}{n}\mu^{1+1}+\frac{(1-n)n+2(6n-8M+9)(M-1)}{n-M+1}\binom{n-M+1}{2}\binom{2(n-M-2)+2}{n}\mu^{2+1}+$$

$$\left.\frac{(1-n)n+2(6n-8M+9)(M-1)}{n-M+1}\binom{n-M+1}{\left[\frac{n}{2}\right]-M+1}\binom{2\left(n-M-\left(\left[\frac{n}{2}\right]-M+1\right)\right)+2}{n}\mu^{\left(\left[\frac{n}{2}\right]-M+1\right)+1}\right]+\left[\sum_{j=3}^{\left[\frac{n}{2}\right]-M}\frac{2(2n-2M+3)(M-1)}{n-M+1}\binom{n-M+1}{j+1}\binom{2(n-M-j)}{n}\mu^{j+1}+\right.$$

$$\frac{2(2n-2M+3)(M-1)}{n-M+1}\binom{n-M+1}{-1+1}\binom{2(n-M--1)}{n}\mu^{-1+1}+\frac{2(2n-2M+3)(M-1)}{n-M+1}\binom{n-M+1}{0+1}\binom{2(n-M-0)}{n}\mu^{0+1}+$$

$$\left.\left.\frac{2(2n-2M+3)(M-1)}{n-M+1}\binom{n-M+1}{1+1}\binom{2(n-M-1)}{n}\mu^{1+1}+\frac{2(2n-2M+3)(M-1)}{n-M+1}\binom{n-M+1}{2+1}\binom{2(n-M-2)}{n}\mu^{2+1}\right]\right\}+$$

$$\left\{\left[\sum_{j=3}^{\left[\frac{n}{2}\right]-M}(M-1)\frac{(n-2M+4)}{n-M+2}\frac{(n-2M+3)}{n-M+1}\binom{n-M+2}{j+1}\binom{2(n-M-j)+2}{n}\mu^{j+1}+(M-1)\frac{(n-2M+4)}{n-M+2}\frac{(n-2M+3)}{n-M+1}\binom{n-M+2}{-1+1}\binom{2(n-M--1)+2}{n}\mu^{-1+1}+\right.\right.$$

$$(M-1)\frac{(n-2M+4)}{n-M+2}\frac{(n-2M+3)}{n-M+1}\binom{n-M+2}{0+1}\binom{2(n-M-0)+2}{n}\mu^{0+1}+(M-1)\frac{(n-2M+4)}{n-M+2}\frac{(n-2M+3)}{n-M+1}\binom{n-M+2}{1+1}\binom{2(n-M-1)+2}{n}\mu^{1+1}+(M-$$

$$1)\frac{(n-2M+4)}{n-M+2}\frac{(n-2M+3)}{n-M+1}\binom{n-M+2}{2+1}\binom{2(n-M-2)+2}{n}\mu^{2+1}+$$

$$\left.(M-1)\frac{(n-2M+4)}{n-M+2}\frac{(n-2M+3)}{n-M+1}\binom{n-M+2}{\left[\frac{n}{2}\right]-M+1+1}\binom{2\left(n-M-\left(\left[\frac{n}{2}\right]-M+1\right)\right)+2}{n}\mu^{\left(\left[\frac{n}{2}\right]-M+1\right)+1}\right]+$$

$$\left[\sum_{j=3}^{\left[\frac{n}{2}\right]-M}(M-1)\frac{(n-2M+4)}{n-M+2}\frac{(n-2M+3)}{n-M+1}\binom{n-M+2}{j}\binom{2(n-M-j)+4}{n}\mu^{j+1}+(M-1)\frac{(n-2M+4)}{n-M+2}\frac{(n-2M+3)}{n-M+1}\binom{n-M+2}{0}\binom{2(n-M-0)+4}{n}\mu^{0+1}+\right.$$

$$(M-1)\frac{(n-2M+4)}{n-M+2}\frac{(n-2M+3)}{n-M+1}\binom{n-M+2}{1}\binom{2(n-M-1)+4}{n}\mu^{1+1}+(M-1)\frac{(n-2M+4)}{n-M+2}\frac{(n-2M+3)}{n-M+1}\binom{n-M+2}{2}\binom{2(n-M-2)+4}{n}\mu^{2+1}+(M-$$

$$1)\frac{(n-2M+4)}{n-M+2}\frac{(n-2M+3)}{n-M+1}\binom{n-M+2}{\left[\frac{n}{2}\right]-M+1}\binom{2\left(n-M-\left(\left[\frac{n}{2}\right]-M+1\right)\right)+4}{n}\mu^{\left[\frac{n}{2}\right]-M+1+1}+$$

$$\left.\left.(M-1)\frac{(n-2M+4)}{n-M+2}\frac{(n-2M+3)}{n-M+1}\binom{n-M+2}{\left[\frac{n}{2}\right]-M+2}\binom{2\left(n-M-\left(\left[\frac{n}{2}\right]-M+2\right)\right)+4}{n}\mu^{\left[\frac{n}{2}\right]-M+2+1}\right]\right\}\quad. \qquad (S4.67)$$

In order to shorten the relations, we split the expression basically to two parts: the part with sums and the part of the individual terms. The two parts has to be zero separately, as they contain different powers of $\mu$. Split and perform simple algebraic manipulations.

The part with the sums is:

$$\left\{\left[\sum_{j=3}^{\left[\frac{n}{2}\right]-M}4(n-M)\binom{n-M-1}{j-1}\binom{2(n-M-j)}{n}\mu^{j+1}\right]+\left[\sum_{j=3}^{\left[\frac{n}{2}\right]-M}8(n-M)\binom{n-M-1}{j-2}\binom{2(n-M-j)+2}{n}\mu^{j+1}\right]+\right.$$

$$\left.\left[\sum_{j=3}^{\left[\frac{n}{2}\right]-M}4(n-M)\binom{n-M-1}{j-3}\binom{2(n-M-j)+4}{n}\mu^{j+1}\right]\right\}+\left\{\left[\sum_{j=3}^{\left[\frac{n}{2}\right]-M}2(6M-2n-3)\binom{n-M}{j-2}\binom{2(n-M-j)+4}{n}\mu^{j+1}\right]+\left[\sum_{j=3}^{\left[\frac{n}{2}\right]-M}4(5M-2n-\right.\right.$$

$$\left.\left.3)\binom{n-M}{j-1}\binom{2(n-M-j)+2}{n}\mu^{j+1}\right]+\left[\sum_{j=3}^{\left[\frac{n}{2}\right]-M}2(4M-2n-3)\binom{n-M}{j}\binom{2(n-M-j)}{n}\mu^{j+1}\right]\right\}-$$

$$\left\{\left[\sum_{j=3}^{\left[\frac{n}{2}\right]-M}\frac{(1-n)n+4(2n-3(M-1))(M-1)}{n-M+1}\binom{n-M+1}{j-1}\binom{2(n-M-j)+4}{n}\mu^{j+1}\right]+\left[\sum_{j=3}^{\left[\frac{n}{2}\right]-M}\frac{(1-n)n+2(6n-8M+9)(M-1)}{n-M+1}\binom{n-M+1}{j}\binom{2(n-M-j)+2}{n}\mu^{j+1}\right]+\right.$$

$$\left.\left[\sum_{j=3}^{\left[\frac{n}{2}\right]-M}\frac{2(2n-2M+3)(M-1)}{n-M+1}\binom{n-M+1}{j+1}\binom{2(n-M-j)}{n}\mu^{j+1}\right]\right\}+\left\{\left[\sum_{j=3}^{\left[\frac{n}{2}\right]-M}(M-1)\frac{(n-2M+4)}{n-M+2}\frac{(n-2M+3)}{n-M+1}\binom{n-M+2}{j+1}\binom{2(n-M-j)+2}{n}\mu^{j+1}\right]+\right.$$

$$\left.\left[\sum_{j=3}^{\left[\frac{n}{2}\right]-M}(M-1)\frac{(n-2M+4)}{n-M+2}\frac{(n-2M+3)}{n-M+1}\binom{n-M+2}{j}\binom{2(n-M-j)+4}{n}\mu^{j+1}\right]\right\}\quad. \qquad (S4.68)$$

The part of the individual terms is:

$$\left\{\left[+4(n-M)\binom{n-M-1}{0}\binom{2(n-M)-2}{n}\mu^2+4(n-M)\binom{n-M-1}{1}\binom{2(n-M)-4}{n}\mu^3\right]+\left[+8(n-M)\binom{n-M-1}{0}\binom{2(n-M)-2}{n}\mu^3+8(n-\right.\right.$$

$$\left.M)\binom{n-M-1}{\left[\frac{n}{2}\right]-M-1}\binom{2\left(n-\left[\frac{n}{2}\right]\right)}{n}\mu^{\left[\frac{n}{2}\right]-M+2}\right]+$$

$$\left.\left[+4(n-M)\binom{n-M-1}{\left[\frac{n}{2}\right]-M-2}\binom{2\left(n-\left[\frac{n}{2}\right]\right)+2}{n}\mu^{\left[\frac{n}{2}\right]-M+2}+4(n-M)\binom{n-M-1}{\left[\frac{n}{2}\right]-M-1}\binom{2\left(n-\left[\frac{n}{2}\right]\right)}{n}\mu^{\left[\frac{n}{2}\right]-M+3}\right]\right\}+$$

$$\left\{\left[+2(6M-2n-3)\binom{n-M}{0}\binom{2(n-M)}{n}\mu^3+2(6M-2n-3)\binom{n-M}{\left[\frac{n}{2}\right]-M-1}\binom{2\left(n-\left(\left[\frac{n}{2}\right]\right)\right)+2}{n}\mu^{\left[\frac{n}{2}\right]-M+2}+\right.\right.$$

$$\left.2(6M-2n-3)\binom{n-M}{\left[\frac{n}{2}\right]-M}\binom{2\left(n-\left[\frac{n}{2}\right]\right)}{n}\mu^{\left[\frac{n}{2}\right]-M+3}\right]+\left[+4(5M-2n-3)\binom{n-M}{0}\binom{2(n-M)}{n}\mu^2+4(5M-2n-3)\binom{n-M}{1}\binom{2(n-M)-2}{n}\mu^3+\right.$$

$$\left.4(5M-2n-3)\binom{n-M}{\left[\frac{n}{2}\right]-M}\binom{2\left(n-\left[\frac{n}{2}\right]\right)}{n}\mu^{\left[\frac{n}{2}\right]-M+2}\right]+\left[+2(4M-2n-3)\binom{n-M}{0}\binom{2(n-M)}{n}\mu^1+2(4M-2n-3)\binom{n-M}{1}\binom{2(n-M)-2}{n}\mu^2+\right.$$

$$\left.\left.2(4M-2n-3)\binom{n-M}{2}\binom{2(n-M)-4}{n}\mu^3\right]\right\}-$$

$$\left\{\left[+\frac{(1-n)n+4(2n-3(M-1))(M-1)}{n-M+1}\binom{n-M+1}{0}\binom{2(n-M)+2}{n}\mu^2+\frac{(1-n)n+4(2n-3(M-1))(M-1)}{n-M+1}\binom{n-M+1}{1}\binom{2(n-M)}{n}\mu^3+\right.\right.$$

$$\left.\frac{(1-n)n+4(2n-3(M-1))(M-1)}{n-M+1}\binom{n-M+1}{\left[\frac{n}{2}\right]-M}\binom{2\left(n-\left[\frac{n}{2}\right]\right)+2}{n}\mu^{\left[\frac{n}{2}\right]-M+2}+\frac{(1-n)n+4(2n-3(M-1))(M-1)}{n-M+1}\binom{n-M+1}{\left[\frac{n}{2}\right]-M+1}\binom{2\left(n-\left[\frac{n}{2}\right]\right)}{n}\mu^{\left[\frac{n}{2}\right]-M+3}\right]+$$

$$\left[+\frac{(1-n)n+2(6n-8M+9)(M-1)}{n-M+1}\binom{n-M+1}{0}\binom{2(n-M)+2}{n}\mu^1+\frac{(1-n)n+2(6n-8M+9)(M-1)}{n-M+1}\binom{n-M+1}{1}\binom{2(n-M)}{n}\mu^2+\right.$$

$$\frac{(1-n)n+2(6n-8M+9)(M-1)}{n-M+1}\binom{n-M+1}{2}\binom{2(n-M)-2}{n}\mu^3+\frac{(1-n)n+2(6n-8M+9)(M-1)}{n-M+1}\binom{n-M+1}{\left[\frac{n}{2}\right]-M+1}\binom{2\left(n-\left[\frac{n}{2}\right]\right)+0}{n}\mu^{\left[\frac{n}{2}\right]-M+2}\Big]+$$

$$\Big[+\frac{2(2n-2M+3)(M-1)}{n-M+1}\binom{n-M+1}{0}\binom{2(n-M)+2}{n}\mu^0+\frac{2(2n-2M+3)(M-1)}{n-M+1}\binom{n-M+1}{1}\binom{2(n-M)}{n}\mu^1+\frac{2(2n-2M+3)(M-1)}{n-M+1}\binom{n-M+1}{2}\binom{2(n-M)-2}{n}\mu^2+$$

$$\frac{2(2n-2M+3)(M-1)}{n-M+1}\binom{n-M+1}{3}\binom{2(n-M)-4}{n}\mu^3\Big]\Big\}+$$

$$\Big\{\Big[+(M-1)\frac{(n-2M+4)}{n-M+2}\frac{(n-2M+3)}{n-M+1}\binom{n-M+2}{0}\binom{2(n-M)+4}{n}\mu^0+(M-1)\frac{(n-2M+4)}{n-M+2}\frac{(n-2M+3)}{n-M+1}\binom{n-M+2}{1}\binom{2(n-M)+2}{n}\mu^1+$$

$$(M-1)\frac{(n-2M+4)}{n-M+2}\frac{(n-2M+3)}{n-M+1}\binom{n-M+2}{2}\binom{2(n-M)}{n}\mu^2+(M-1)\frac{(n-2M+4)}{n-M+2}\frac{(n-2M+3)}{n-M+1}\binom{n-M+2}{3}\binom{2(n-M)-2}{n}\mu^3+$$

$$(M-1)\frac{(n-2M+4)}{n-M+2}\frac{(n-2M+3)}{n-M+1}\binom{n-M+2}{\left[\frac{n}{2}\right]-M+2}\binom{2\left(n-\left[\frac{n}{2}\right]\right)+0}{n}\mu^{\left[\frac{n}{2}\right]-M+2}\Big]+\Big[+(M-1)\frac{(n-2M+4)}{n-M+2}\frac{(n-2M+3)}{n-M+1}\binom{n-M+2}{0}\binom{2(n-M)+4}{n}\mu^1+$$

$$(M-1)\frac{(n-2M+4)}{n-M+2}\frac{(n-2M+3)}{n-M+1}\binom{n-M+2}{1}\binom{2(n-M)+2}{n}\mu^2+(M-1)\frac{(n-2M+4)}{n-M+2}\frac{(n-2M+3)}{n-M+1}\binom{n-M+2}{2}\binom{2(n-M)}{n}\mu^3+$$

$$(M-1)\frac{(n-2M+4)}{n-M+2}\frac{(n-2M+3)}{n-M+1}\binom{n-M+2}{\left[\frac{n}{2}\right]-M+1}\binom{2\left(n-\left[\frac{n}{2}\right]\right)+2}{n}\mu^{\left[\frac{n}{2}\right]-M+2}+(M-1)\frac{(n-2M+4)}{n-M+2}\frac{(n-2M+3)}{n-M+1}\binom{n-M+2}{\left[\frac{n}{2}\right]-M+2}\binom{2\left(n-\left[\frac{n}{2}\right]\right)}{n}\mu^{\left[\frac{n}{2}\right]-M+3}\Big]\Big\}$$

$$. \qquad (S4.69)$$

First, evaluate the part containing the individual terms (i.e. the expression (S4.69)), and afterwards, evaluate also the part with the sums, i.e. (S4.68).

The terms in the part of the individual terms of (S4.69) are grouped according to the exponent of $\mu$:

$$\Big[+4(n-M)\binom{n-M-1}{1}\binom{2(n-M)-4}{n}\mu^3+8(n-M)\binom{n-M-1}{0}\binom{2(n-M)-2}{n}\mu^3+2(6M-2n-3)\binom{n-M}{0}\binom{2(n-M)}{n}\mu^3+4(5M-2n-$$

$$3)\binom{n-M}{1}\binom{2(n-M)-2}{n}\mu^3+2(4M-2n-3)\binom{n-M}{2}\binom{2(n-M)-4}{n}\mu^3-\frac{(1-n)n+4(2n-3(M-1))(M-1)}{n-M+1}\binom{n-M+1}{1}\binom{2(n-M)}{n}\mu^3-$$

$$\frac{(1-n)n+2(6n-8M+9)(M-1)}{n-M+1}\binom{n-M+1}{2}\binom{2(n-M)-2}{n}\mu^3-\frac{2(2n-2M+3)(M-1)}{n-M+1}\binom{n-M+1}{3}\binom{2(n-M)-4}{n}\mu^3+$$

$$(M-1)\frac{(n-2M+4)}{n-M+2}\frac{(n-2M+3)}{n-M+1}\binom{n-M+2}{3}\binom{2(n-M)-2}{n}\mu^3+(M-1)\frac{(n-2M+4)}{n-M+2}\frac{(n-2M+3)}{n-M+1}\binom{n-M+2}{2}\binom{2(n-M)}{n}\mu^3\Big]+$$

$$\Big[4(n-M)\binom{n-M-1}{0}\binom{2(n-M)-2}{n}\mu^2+4(5M-2n-3)\binom{n-M}{0}\binom{2(n-M)}{n}\mu^2+2(4M-2n-3)\binom{n-M}{1}\binom{2(n-M)-2}{n}\mu^2-$$

$$\frac{(1-n)n+4(2n-3(M-1))(M-1)}{n-M+1}\binom{n-M+1}{0}\binom{2(n-M)+2}{n}\mu^2-\frac{(1-n)n+2(6n-8M+9)(M-1)}{n-M+1}\binom{n-M+1}{1}\binom{2(n-M)}{n}\mu^2-$$

$$\frac{2(2n-2M+3)(M-1)}{n-M+1}\binom{n-M+1}{2}\binom{2(n-M)-2}{n}\mu^2+(M-1)\frac{(n-2M+4)}{n-M+2}\frac{(n-2M+3)}{n-M+1}\binom{n-M+2}{2}\binom{2(n-M)}{n}\mu^2+$$

$$(M-1)\frac{(n-2M+4)}{n-M+2}\frac{(n-2M+3)}{n-M+1}\binom{n-M+2}{1}\binom{2(n-M)+2}{n}\mu^2\Big]+$$

$$\Big[+2(4M-2n-3)\binom{n-M}{0}\binom{2(n-M)}{n}\mu^1-\frac{(1-n)n+2(6n-8M+9)(M-1)}{n-M+1}\binom{n-M+1}{0}\binom{2(n-M)+2}{n}\mu^1-\frac{2(2n-2M+3)(M-1)}{n-M+1}\binom{n-M+1}{1}\binom{2(n-M)}{n}\mu^1+$$

$$(M-1)\frac{(n-2M+4)}{n-M+2}\frac{(n-2M+3)}{n-M+1}\binom{n-M+2}{0}\binom{2(n-M)+4}{n}\mu^1+(M-1)\frac{(n-2M+4)}{n-M+2}\frac{(n-2M+3)}{n-M+1}\binom{n-M+2}{1}\binom{2(n-M)+2}{n}\mu^1\Big]+$$

$$\Big[-\frac{2(2n-2M+3)(M-1)}{n-M+1}\binom{n-M+1}{0}\binom{2(n-M)+2}{n}\mu^0+(M-1)\frac{(n-2M+4)}{n-M+2}\frac{(n-2M+3)}{n-M+1}\binom{n-M+2}{0}\binom{2(n-M)+4}{n}\mu^0\Big]+$$

$$\Big[+4(n-M)\binom{n-M-1}{\left[\frac{n}{2}\right]-M-2}\binom{2\left(n-\left[\frac{n}{2}\right]\right)+2}{n}\mu^{\left[\frac{n}{2}\right]-M+2}+8(n-M)\binom{n-M-1}{\left[\frac{n}{2}\right]-M-1}\binom{2\left(n-\left[\frac{n}{2}\right]\right)}{n}\mu^{\left[\frac{n}{2}\right]-M+2}+$$

$$2(6M-2n-3)\binom{n-M}{\left[\frac{n}{2}\right]-M-1}\binom{2\left(n-\left(\left[\frac{n}{2}\right]\right)\right)+2}{n}\mu^{\left[\frac{n}{2}\right]-M+2}+4(5M-2n-3)\binom{n-M}{\left[\frac{n}{2}\right]-M}\binom{2\left(n-\left[\frac{n}{2}\right]\right)}{n}\mu^{\left[\frac{n}{2}\right]-M+2}-$$

$$\frac{(1-n)n+4(2n-3(M-1))(M-1)}{n-M+1}\binom{n-M+1}{\left[\frac{n}{2}\right]-M}\binom{2\left(n-\left[\frac{n}{2}\right]\right)+2}{n}\mu^{\left[\frac{n}{2}\right]-M+2}-\frac{(1-n)n+2(6n-8M+9)(M-1)}{n-M+1}\binom{n-M+1}{\left[\frac{n}{2}\right]-M+1}\binom{2\left(n-\left[\frac{n}{2}\right]\right)+0}{n}\mu^{\left[\frac{n}{2}\right]-M+2}+$$

$$(M-1)\frac{(n-2M+4)}{n-M+2}\frac{(n-2M+3)}{n-M+1}\binom{n-M+2}{\left[\frac{n}{2}\right]-M+2}\binom{2\left(n-\left[\frac{n}{2}\right]\right)+0}{n}\mu^{\left[\frac{n}{2}\right]-M+2}+(M-1)\frac{(n-2M+4)}{n-M+2}\frac{(n-2M+3)}{n-M+1}\binom{n-M+2}{\left[\frac{n}{2}\right]-M+1}\binom{2\left(n-\left[\frac{n}{2}\right]\right)+2}{n}\mu^{\left[\frac{n}{2}\right]-M+2}\Big]+$$

$$\Big[+4(n-M)\binom{n-M-1}{\left[\frac{n}{2}\right]-M-1}\binom{2\left(n-\left[\frac{n}{2}\right]\right)}{n}\mu^{\left[\frac{n}{2}\right]-M+3}+2(6M-2n-3)\binom{n-M}{\left[\frac{n}{2}\right]-M}\binom{2\left(n-\left[\frac{n}{2}\right]\right)}{n}\mu^{\left[\frac{n}{2}\right]-M+3}-$$

$$\frac{(1-n)n+4(2n-3(M-1))(M-1)}{n-M+1}\binom{n-M+1}{\left[\frac{n}{2}\right]-M+1}\binom{2\left(n-\left[\frac{n}{2}\right]\right)}{n}\mu^{\left[\frac{n}{2}\right]-M+3}+(M-1)\frac{(n-2M+4)}{n-M+2}\frac{(n-2M+3)}{n-M+1}\binom{n-M+2}{\left[\frac{n}{2}\right]-M+2}\binom{2\left(n-\left[\frac{n}{2}\right]\right)}{n}\mu^{\left[\frac{n}{2}\right]-M+3}\Big] .$$

$$(S4.70)$$

Factor out the powers of $\mu$, and the trivial binomial coefficients are expanded:

$$\Big[+4(n-M)(n-M-1)\binom{2(n-M)-4}{n}+8(n-M)\binom{2(n-M)-2}{n}+2(6M-2n-3)\binom{2(n-M)}{n}+$$

$$4(5M-2n-3)(n-M)\binom{2(n-M)-2}{n}+2(4M-2n-3)\frac{1}{2}(n-M-1)(n-M)\binom{2(n-M)-4}{n}-\frac{(1-n)n+4(2n-3(M-1))(M-1)}{n-M+1}(n-M+$$

$$1)\binom{2(n-M)}{n}-\frac{(1-n)n+2(6n-8M+9)(M-1)}{n-M+1}\frac{1}{2}(n-M)(n-M+1)\binom{2(n-M)-2}{n}-\frac{2(2n-2M+3)(M-1)}{n-M+1}\frac{1}{6}(n-M+1-2)(n-M+1-$$

$$1)(n-M+1)\binom{2(n-M)-4}{n}+(M-1)\frac{(n-2M+4)}{n-M+2}\frac{(n-2M+3)}{n-M+1}\frac{1}{6}(n-M+2-2)(n-M+2-1)(n-M+2)\binom{2(n-M)-2}{n}+$$

$$(M-1)\frac{(n-2M+4)}{n-M+2}\frac{(n-2M+3)}{n-M+1}\frac{1}{2}(n-M+1)(n-M+2)\binom{2(n-M)}{n}\Big]\mu^3+\Big[4(n-M)\binom{2(n-M)-2}{n}+4(5M-2n-3)\binom{2(n-M)}{n}+$$

$$
\begin{aligned}
&2(4M-2n-3)(n-M)\binom{2(n-M)-2}{n}-\frac{(1-n)n+4(2n-3(M-1))(M-1)}{n-M+1}\binom{2(n-M)+2}{n}-\frac{(1-n)n+2(6n-8M+9)(M-1)}{n-M+1}(n-M+1)\binom{2(n-M)}{n}-\\
&\frac{2(2n-2M+3)(M-1)}{n-M+1}\frac{1}{2}(n-M)(n-M+1)\binom{2(n-M)-2}{n}+(M-1)\frac{(n-2M+4)}{n-M+2}\frac{(n-2M+3)}{n-M+1}\frac{1}{2}(n-M+1)(n-M+2)\binom{2(n-M)}{n}+\\
&(M-1)\frac{(n-2M+4)}{n-M+2}\frac{(n-2M+3)}{n-M+1}(n-M+2)\binom{2(n-M)+2}{n}\Bigg]\mu^2+\Bigg[+2(4M-2n-3)\binom{2(n-M)}{n}-\frac{(1-n)n+2(6n-8M+9)(M-1)}{n-M+1}\binom{2(n-M)+2}{n}-\\
&\frac{2(2n-2M+3)(M-1)}{n-M+1}(n-M+1)\binom{2(n-M)}{n}+(M-1)\frac{(n-2M+4)}{n-M+2}\frac{(n-2M+3)}{n-M+1}\binom{2(n-M)+4}{n}+(M-1)\frac{(n-2M+4)}{n-M+2}\frac{(n-2M+3)}{n-M+1}(n-M+\\
&2)\binom{2(n-M)+2}{n}\Bigg]\mu^1+\Bigg[-\frac{2(2n-2M+3)(M-1)}{n-M+1}\binom{2(n-M)+2}{n}+(M-1)\frac{(n-2M+4)}{n-M+2}\frac{(n-2M+3)}{n-M+1}\binom{2(n-M)+4}{n}\mu^0\Bigg]+\\
&\Bigg[+4(n-M)\binom{n-M-1}{\left[\frac{n}{2}\right]-M-2}\binom{2\left(n-\left[\frac{n}{2}\right]\right)+2}{n}+8(n-M)\binom{n-M-1}{\left[\frac{n}{2}\right]-M-1}\binom{2\left(n-\left[\frac{n}{2}\right]\right)}{n}+2(6M-2n-3)\binom{n-M}{\left[\frac{n}{2}\right]-M-1}\binom{2\left(n-\left[\frac{n}{2}\right]\right)+2}{n}+\\
&4(5M-2n-3)\binom{n-M}{\left[\frac{n}{2}\right]-M}\binom{2\left(n-\left[\frac{n}{2}\right]\right)}{n}-\frac{(1-n)n+4(2n-3(M-1))(M-1)}{n-M+1}\binom{n-M+1}{\left[\frac{n}{2}\right]-M}\binom{2\left(n-\left[\frac{n}{2}\right]\right)+2}{n}-\\
&\frac{(1-n)n+2(6n-8M+9)(M-1)}{n-M+1}\binom{n-M+1}{\left[\frac{n}{2}\right]-M+1}\binom{2\left(n-\left[\frac{n}{2}\right]\right)+0}{n}+(M-1)\frac{(n-2M+4)}{n-M+2}\frac{(n-2M+3)}{n-M+1}\binom{n-M+2}{\left[\frac{n}{2}\right]-M+2}\binom{2\left(n-\left[\frac{n}{2}\right]\right)+0}{n}+\\
&(M-1)\frac{(n-2M+4)}{n-M+2}\frac{(n-2M+3)}{n-M+1}\binom{n-M+2}{\left[\frac{n}{2}\right]-M+1}\binom{2\left(n-\left[\frac{n}{2}\right]\right)+2}{n}\Bigg]\mu^{\left[\frac{n}{2}\right]-M+2}+\Bigg[+4(n-M)\binom{n-M-1}{\left[\frac{n}{2}\right]-M-1}\binom{2\left(n-\left[\frac{n}{2}\right]\right)}{n}+2(6M-2n-\\
&3)\binom{n-M}{\left[\frac{n}{2}\right]-M}\binom{2\left(n-\left[\frac{n}{2}\right]\right)}{n}-\frac{(1-n)n+4(2n-3(M-1))(M-1)}{n-M+1}\binom{n-M+1}{\left[\frac{n}{2}\right]-M+1}\binom{2\left(n-\left[\frac{n}{2}\right]\right)}{n}+\\
&(M-1)\frac{(n-2M+4)}{n-M+2}\frac{(n-2M+3)}{n-M+1}\binom{n-M+2}{\left[\frac{n}{2}\right]-M+2}\binom{2\left(n-\left[\frac{n}{2}\right]\right)}{n}\Bigg]\mu^{\left[\frac{n}{2}\right]-M+3}\ .
\end{aligned}
\tag{S4.71}
$$

The part with the individual terms is split to the different sections containing different powers of $\mu$ in order to simplify the relations. Each relation has to be equal to zero individually:

Power of $\mu$ equal to 3:

$$
\begin{aligned}
&\Bigg[+4(n-M)(n-M-1)\binom{2(n-M)-4}{n}+8(n-M)\binom{2(n-M)-2}{n}+2(6M-2n-3)\binom{2(n-M)}{n}+\\
&4(5M-2n-3)(n-M)\binom{2(n-M)-2}{n}+2(4M-2n-3)\frac{1}{2}(n-M-1)(n-M)\binom{2(n-M)-4}{n}-\frac{(1-n)n+4(2n-3(M-1))(M-1)}{n-M+1}(n-M+\\
&1)\binom{2(n-M)}{n}-\frac{(1-n)n+2(6n-8M+9)(M-1)}{n-M+1}\frac{1}{2}(n-M)(n-M+1)\binom{2(n-M)-2}{n}-\frac{2(2n-2M+3)(M-1)}{n-M+1}\frac{1}{6}(n-M-1)(n-M)(n-M+\\
&1)\binom{2(n-M)-4}{n}+(M-1)\frac{(n-2M+4)}{n-M+2}\frac{(n-2M+3)}{n-M+1}\frac{1}{6}(n-M)(n-M+1)(n-M+2)\binom{2(n-M)-2}{n}+(M-1)\frac{(n-2M+4)}{n-M+2}\frac{(n-2M+3)}{n-M+1}\frac{1}{2}(n-\\
&M+1)(n-M+2)\binom{2(n-M)}{n}\Bigg]\mu^3\ .
\end{aligned}
\tag{S4.72}
$$

Power of $\mu$ equal to 2:

$$
\begin{aligned}
&\Bigg[4(n-M)\binom{2(n-M)-2}{n}+4(5M-2n-3)\binom{2(n-M)}{n}+2(4M-2n-3)(n-M)\binom{2(n-M)-2}{n}-\\
&\frac{(1-n)n+4(2n-3(M-1))(M-1)}{n-M+1}\binom{2(n-M)+2}{n}-\frac{(1-n)n+2(6n-8M+9)(M-1)}{n-M+1}(n-M+1)\binom{2(n-M)}{n}-\frac{2(2n-2M+3)(M-1)}{n-M+1}\frac{1}{2}(n-M)(n-M+\\
&1)\binom{2(n-M)-2}{n}+(M-1)\frac{(n-2M+4)}{n-M+2}\frac{(n-2M+3)}{n-M+1}\frac{1}{2}(n-M+1)(n-M+2)\binom{2(n-M)}{n}+(M-1)\frac{(n-2M+4)}{n-M+2}\frac{(n-2M+3)}{n-M+1}(n-M+\\
&2)\binom{2(n-M)+2}{n}\Bigg]\mu^2\ .
\end{aligned}
\tag{S4.73}
$$

Power of $\mu$ equal to 1:

$$
\begin{aligned}
&\Bigg[+2(4M-2n-3)\binom{2(n-M)}{n}-\frac{(1-n)n+2(6n-8M+9)(M-1)}{n-M+1}\binom{2(n-M)+2}{n}-\frac{2(2n-2M+3)(M-1)}{n-M+1}(n-M+1)\binom{2(n-M)}{n}+(M-\\
&1)\frac{(n-2M+4)}{n-M+2}\frac{(n-2M+3)}{n-M+1}\binom{2(n-M)+4}{n}+(M-1)\frac{(n-2M+4)}{n-M+2}\frac{(n-2M+3)}{n-M+1}(n-M+2)\binom{2(n-M)+2}{n}\Bigg]\mu^1\ .
\end{aligned}
\tag{S4.74}
$$

Power of $\mu$ equal to 0:

$$
\Bigg[-\frac{2(2n-2M+3)(M-1)}{n-M+1}\binom{2(n-M)+2}{n}+(M-1)\frac{(n-2M+4)}{n-M+2}\frac{(n-2M+3)}{n-M+1}\binom{2(n-M)+4}{n}\Bigg]\mu^0\ .
\tag{S4.75}
$$

Power of $\mu$ equal to $\left[\frac{n}{2}\right]-M+2$:

$$
\begin{aligned}
&\Bigg[+4(n-M)\binom{n-M-1}{\left[\frac{n}{2}\right]-M-2}\binom{2\left(n-\left[\frac{n}{2}\right]\right)+2}{n}+8(n-M)\binom{n-M-1}{\left[\frac{n}{2}\right]-M-1}\binom{2\left(n-\left[\frac{n}{2}\right]\right)}{n}+2(6M-2n-3)\binom{n-M}{\left[\frac{n}{2}\right]-M-1}\binom{2\left(n-\left[\frac{n}{2}\right]\right)+2}{n}+\\
&4(5M-2n-3)\binom{n-M}{\left[\frac{n}{2}\right]-M}\binom{2\left(n-\left[\frac{n}{2}\right]\right)}{n}-\frac{(1-n)n+4(2n-3(M-1))(M-1)}{n-M+1}\binom{n-M+1}{\left[\frac{n}{2}\right]-M}\binom{2\left(n-\left[\frac{n}{2}\right]\right)+2}{n}-\\
&\frac{(1-n)n+2(6n-8M+9)(M-1)}{n-M+1}\binom{n-M+1}{\left[\frac{n}{2}\right]-M+1}\binom{2\left(n-\left[\frac{n}{2}\right]\right)+0}{n}+(M-1)\frac{(n-2M+4)}{n-M+2}\frac{(n-2M+3)}{n-M+1}\binom{n-M+2}{\left[\frac{n}{2}\right]-M+2}\binom{2\left(n-\left[\frac{n}{2}\right]\right)+0}{n}+\\
&(M-1)\frac{(n-2M+4)}{n-M+2}\frac{(n-2M+3)}{n-M+1}\binom{n-M+2}{\left[\frac{n}{2}\right]-M+1}\binom{2\left(n-\left[\frac{n}{2}\right]\right)+2}{n}\Bigg]\mu^{\left[\frac{n}{2}\right]-M+2}\ .
\end{aligned}
\tag{S4.76}
$$

Power of $\mu$ equal to $\left[\frac{n}{2}\right]-M+3$:

$$
\begin{aligned}
&\Bigg[+4(n-M)\binom{n-M-1}{\left[\frac{n}{2}\right]-M-1}\binom{2\left(n-\left[\frac{n}{2}\right]\right)}{n}+2(6M-2n-3)\binom{n-M}{\left[\frac{n}{2}\right]-M}\binom{2\left(n-\left[\frac{n}{2}\right]\right)}{n}-\frac{(1-n)n+4(2n-3(M-1))(M-1)}{n-M+1}\binom{n-M+1}{\left[\frac{n}{2}\right]-M+1}\binom{2\left(n-\left[\frac{n}{2}\right]\right)}{n}+\\
&(M-1)\frac{(n-2M+4)}{n-M+2}\frac{(n-2M+3)}{n-M+1}\binom{n-M+2}{\left[\frac{n}{2}\right]-M+2}\binom{2\left(n-\left[\frac{n}{2}\right]\right)}{n}\Bigg]\mu^{\left[\frac{n}{2}\right]-M+3}\ .
\end{aligned}
\tag{S4.77}
$$

In the expressions with the individual terms with specific power of $\mu$, the terms with the same binomial coefficients are put together. Also, the basic binomial identities

$$\binom{m-1}{k-1}=\frac{k}{m}\binom{m}{k} \quad \text{and thus also} \quad \binom{m+1}{k+1}=\frac{m+1}{k+1}\binom{m}{k} \quad , \tag{S4.78}$$

$$\binom{m}{k-1}=\frac{k}{m-k+1}\binom{m}{k} \quad , \tag{S4.79}$$

$$\binom{m+1}{k}=\frac{m+1}{m+1-k}\binom{m}{k} \quad , \tag{S4.80}$$

$$\binom{m-1}{k}=\frac{m-k}{m}\binom{m}{k} \quad , \tag{S4.81}$$

are applied to some of the binomial coefficients.

Power of $\mu$ equal to 3:

$$\Big[+4(n-M)(n-M-1)\binom{2(n-M)-4}{n}+2(4M-2n-3)\frac{1}{2}(n-M-1)(n-M)\binom{2(n-M)-4}{n}-\frac{2(2n-2M+3)(M-1)}{n-M+1}\frac{1}{6}(n-M-1)(n-M)(n-M+1)\binom{2(n-M)-4}{n}+8(n-M)\binom{2(n-M)-2}{n}+4(5M-2n-3)(n-M)\binom{2(n-M)-2}{n}-\frac{(1-n)n+2(6n-8M+9)(M-1)}{n-M+1}\frac{1}{2}(n-M)(n-M+1)\binom{2(n-M)-2}{n}+(M-1)(n-2M+4)(n-2M+3)\frac{1}{6}(n-M)\binom{2(n-M)-2}{n}+2(6M-2n-3)\binom{2(n-M)}{n}-\frac{(1-n)n+4(2n-3(M-1))(M-1)}{n-M+1}(n-M+1)\binom{2(n-M)}{n}+(M-1)\frac{(n-2M+4)}{n-M+2}\frac{(n-2M+3)}{n-M+1}\frac{1}{2}(n-M+1)(n-M+2)\binom{2(n-M)}{n}\Big]\mu^3 . \tag{S4.82}$$

Power of $\mu$ equal to 2:

$$\Big[4(n-M)\binom{2(n-M)-2}{n}+2(4M-2n-3)(n-M)\binom{2(n-M)-2}{n}-\frac{2(2n-2M+3)(M-1)}{n-M+1}\frac{1}{2}(n-M)(n-M+1)\binom{2(n-M)-2}{n}+4(5M-2n-3)\binom{2(n-M)}{n}-\frac{(1-n)n+2(6n-8M+9)(M-1)}{n-M+1}(n-M+1)\binom{2(n-M)}{n}+(M-1)\frac{(n-2M+4)}{n-M+2}\frac{(n-2M+3)}{n-M+1}\frac{1}{2}(n-M+1)(n-M+2)\binom{2(n-M)}{n}-\frac{(1-n)n+4(2n-3(M-1))(M-1)}{n-M+1}\binom{2(n-M)+2}{n}+(M-1)\frac{(n-2M+4)}{n-M+2}\frac{(n-2M+3)}{n-M+1}(n-M+2)\binom{2(n-M)+2}{n}\Big]\mu^2 . \tag{S4.83}$$

Power of $\mu$ equal to 1:

$$\Big[+2(4M-2n-3)\binom{2(n-M)}{n}-\frac{2(2n-2M+3)(M-1)}{n-M+1}(n-M+1)\binom{2(n-M)}{n}-\frac{(1-n)n+2(6n-8M+9)(M-1)}{n-M+1}\binom{2(n-M)+2}{n}+(M-1)\frac{(n-2M+4)}{n-M+2}\frac{(n-2M+3)}{n-M+1}(n-M+2)\binom{2(n-M)+2}{n}+(M-1)\frac{(n-2M+4)}{n-M+2}\frac{(n-2M+3)}{n-M+1}\frac{2(n-M)+3+1}{2(n-M)+3+1-n}\frac{2(n-M)+2+1}{2(n-M)+2+1-n}\binom{2(n-M)+2}{n}\Big]\mu^1 . \tag{S4.84}$$

Power of $\mu$ equal to 0:

$$\Big[-\frac{2(2n-2M+3)(M-1)}{n-M+1}\binom{2(n-M)+2}{n}+(M-1)\frac{(n-2M+4)}{n-M+2}\frac{(n-2M+3)}{n-M+1}\frac{2(n-M)+3+1}{2(n-M)+3+1-n}\frac{2(n-M)+2+1}{2(n-M)+2+1-n}\binom{2(n-M)+2}{n}\Big]\mu^0 . \tag{S4.85}$$

Power of $\mu$ equal to $\left[\frac{n}{2}\right]-M+2$:

$$\Big[+8(n-M)\frac{\left[\frac{n}{2}\right]-M}{n-M}\binom{n-M}{\left[\frac{n}{2}\right]-M}\binom{2\left(n-\left[\frac{n}{2}\right]\right)}{n}+4(5M-2n-3)\binom{n-M}{\left[\frac{n}{2}\right]-M}\binom{2\left(n-\left[\frac{n}{2}\right]\right)}{n}+4(n-M)\frac{\left[\frac{n}{2}\right]-M-1}{n-M}\binom{n-M}{\left[\frac{n}{2}\right]-M-1}\binom{2\left(n-\left[\frac{n}{2}\right]\right)+2}{n}+2(6M-2n-3)\binom{n-M}{\left[\frac{n}{2}\right]-M-1}\binom{2\left(n-\left[\frac{n}{2}\right]\right)+2}{n}+(M-1)\frac{(n-2M+4)}{n-M+2}\frac{(n-2M+3)}{n-M+1}\frac{n-M+1+1}{\left[\frac{n}{2}\right]-M+1}\binom{n-M+1}{\left[\frac{n}{2}\right]-M}\binom{2\left(n-\left[\frac{n}{2}\right]\right)+2}{n}-\frac{(1-n)n+4(2n-3(M-1))(M-1)}{n-M+1}\binom{n-M+1}{\left[\frac{n}{2}\right]-M}\binom{2\left(n-\left[\frac{n}{2}\right]\right)+2}{n}-\frac{(1-n)n+2(6n-8M+9)(M-1)}{n-M+1}\frac{n-M+1}{\left[\frac{n}{2}\right]-M+1}\binom{n-M}{\left[\frac{n}{2}\right]-M}\binom{2\left(n-\left[\frac{n}{2}\right]\right)+0}{n}+(M-1)\frac{(n-2M+4)}{n-M+2}\frac{(n-2M+3)}{n-M+1}\frac{n-M+1+1}{\left[\frac{n}{2}\right]-M+1+1}\frac{n-M+1}{\left[\frac{n}{2}\right]-M+1}\binom{n-M}{\left[\frac{n}{2}\right]-M}\binom{2\left(n-\left[\frac{n}{2}\right]\right)+0}{n}\Big]\mu^{\left[\frac{n}{2}\right]-M+2} . \tag{S4.86}$$

Power of $\mu$ equal to $\left[\frac{n}{2}\right]-M+3$:

$$\Big[+4(n-M)\frac{\left[\frac{n}{2}\right]-M}{n-M}\binom{n-M}{\left[\frac{n}{2}\right]-M}+2(6M-2n-3)\binom{n-M}{\left[\frac{n}{2}\right]-M}-\frac{(1-n)n+4(2n-3(M-1))(M-1)}{n-M+1}\frac{n-M+1}{\left[\frac{n}{2}\right]-M+1}\binom{n-M}{\left[\frac{n}{2}\right]-M}+(M-1)\frac{(n-2M+4)}{n-M+2}\frac{(n-2M+3)}{n-M+1}\frac{n-M+1+1}{\left[\frac{n}{2}\right]-M+1+1}\frac{n-M+1}{\left[\frac{n}{2}\right]-M+1}\binom{n-M}{\left[\frac{n}{2}\right]-M}\Big]\binom{2\left(n-\left[\frac{n}{2}\right]\right)}{n}\mu^{\left[\frac{n}{2}\right]-M+3} . \tag{S4.87}$$

The terms with the same binomial coefficients are put together. Also, the basic binomial identities (S4.78), (S4.79) and (S4.80) are applied to some of the binomial coefficients.

Power of $\mu$ equal to 3:

$$\Big[\Big\{+4+(4M-2n-3)-2(2n-2M+3)(M-1)\frac{1}{6}\Big\}(n-M)(n-M-1)\binom{2(n-M)-4}{n}+\Big\{8+4(5M-2n-3)-\frac{(1-n)n+2(6n-8M+9)(M-1)}{2}+(M-1)(n-2M+4)(n-2M+3)\frac{1}{6}\Big\}(n-M)\binom{2(n-M)-2}{n}+\Big\{2(6M-2n-3)-\frac{(1-n)n+4(2n-3(M-1))(M-1)}{1}+(M-1)(n-2M+4)(n-2M+3)\frac{1}{2}\Big\}\binom{2(n-M)}{n}\Big]\mu^3 . \tag{S4.88}$$

Power of $\mu$ equal to 2:

$$\left[\left\{4+2(4M-2n-3)-(2n-2M+3)(M-1)\right\}(n-M)\binom{2(n-M)-2}{n}+\left\{4(5M-2n-3)-\frac{(1-n)n+2(6n-8M+9)(M-1)}{1}+(M-1)(n-2M+4)(n-2M+3)\frac{1}{2}\right\}\binom{2(n-M)}{n}+\left\{-\frac{(1-n)n+4(2n-3(M-1))(M-1)}{n-M+1}+(M-1)\frac{(n-2M+4)}{1}\frac{(n-2M+3)}{n-M+1}\right\}\binom{2(n-M)+2}{n}\right]\mu^2 \quad . \quad \text{(S4.)89}$$

Power of $\mu$ equal to 1:

$$\left[\left\{+2(4M-2n-3)-\frac{2(2n-2M+3)(M-1)}{n-M+1}(n-M+1)\right\}\binom{2(n-M)}{n}+\left\{-\frac{(1-n)n+2(6n-8M+9)(M-1)}{n-M+1}+(M-1)\frac{(n-2M+4)}{1}\frac{(n-2M+3)}{n-M+1}+(M-1)\frac{(n-2M+4)}{n-M+2}\frac{(n-2M+3)}{n-M+1}\frac{2(n-M+2)}{n-2M+4}\frac{2(n-M)+3}{n-2M+3}\right\}\binom{2(n-M)+2}{n}\right]\mu^1 \quad . \quad \text{(S4.90)}$$

Power of $\mu$ equal to 0:

$$\left[-\frac{2(2n-2M+3)(M-1)}{n-M+1}+(M-1)\frac{(n-2M+4)}{n-M+2}\frac{(n-2M+3)}{n-M+1}\frac{2(n-M+2)}{n-2M+4}\frac{2(n-M)+3}{n-2M+3}\right]\binom{2(n-M)+2}{n}\mu^0 \quad . \quad \text{(S4.91)}$$

Power of $\mu$ equal to $\left[\frac{n}{2}\right]-M+2$:

$$\left[+8\left(\left[\tfrac{n}{2}\right]-M\right)\binom{n-M}{\left[\frac{n}{2}\right]-M}\binom{2\left(n-\left[\frac{n}{2}\right]\right)}{n}+4(5M-2n-3)\binom{n-M}{\left[\frac{n}{2}\right]-M}\binom{2\left(n-\left[\frac{n}{2}\right]\right)}{n}+4\left(\left[\tfrac{n}{2}\right]-M-1\right)\frac{\left[\frac{n}{2}\right]-M}{n-M-\left(\left[\frac{n}{2}\right]-M\right)+1}\binom{n-M}{\left[\frac{n}{2}\right]-M}\binom{2\left(n-\left[\frac{n}{2}\right]\right)+2}{n}+2(6M-2n-3)\frac{\left[\frac{n}{2}\right]-M}{n-M-\left(\left[\frac{n}{2}\right]-M\right)+1}\binom{n-M}{\left[\frac{n}{2}\right]-M}\binom{2\left(n-\left[\frac{n}{2}\right]\right)+2}{n}+(M-1)\frac{(n-2M+4)}{n-M+2}\frac{(n-2M+3)}{n-M+1}\frac{n-M+2}{\left[\frac{n}{2}\right]-M+1}\frac{n-M+1}{n-M+1-\left(\left[\frac{n}{2}\right]-M\right)}\binom{n-M}{\left[\frac{n}{2}\right]-M}\binom{2\left(n-\left[\frac{n}{2}\right]\right)+2}{n}-\frac{(1-n)n+4(2n-3(M-1))(M-1)}{n-M+1}\frac{n-M+1}{n-M+1-\left(\left[\frac{n}{2}\right]-M\right)}\binom{n-M}{\left[\frac{n}{2}\right]-M}\binom{2\left(n-\left[\frac{n}{2}\right]\right)+2}{n}-\frac{(1-n)n+2(6n-8M+9)(M-1)}{\left[\frac{n}{2}\right]-M+1}\binom{n-M}{\left[\frac{n}{2}\right]-M}\binom{2\left(n-\left[\frac{n}{2}\right]\right)+0}{n}+(M-1)\frac{(n-2M+4)}{\left[\frac{n}{2}\right]-M+2}\frac{(n-2M+3)}{\left[\frac{n}{2}\right]-M+1}\binom{n-M}{\left[\frac{n}{2}\right]-M}\binom{2\left(n-\left[\frac{n}{2}\right]\right)+0}{n}\right]\mu^{\left[\frac{n}{2}\right]-M+2} \quad . \quad \text{(S4.92)}$$

Power of $\mu$ equal to $\left[\frac{n}{2}\right]-M+3$:

$$\left[+4\left(\left[\tfrac{n}{2}\right]-M\right)+2(6M-2n-3)-\frac{(1-n)n+4(2n-3(M-1))(M-1)}{\left[\frac{n}{2}\right]-M+1}+(M-1)\frac{(n-2M+4)}{\left[\frac{n}{2}\right]-M+2}\frac{(n-2M+3)}{\left[\frac{n}{2}\right]-M+1}\right]\binom{n-M}{\left[\frac{n}{2}\right]-M}\binom{2\left(n-\left[\frac{n}{2}\right]\right)}{n}\mu^{\left[\frac{n}{2}\right]-M+3} \quad . \quad \text{(S4.93)}$$

The basic binomial identities (S4.80) and (S4.81) are again applied to some of the binomial coefficients.

Power of $\mu$ equal to 3:

$$\left[\left\{1+4M-2n-\frac{1}{3}(2n-2M+3)(M-1)\right\}(n-M)(n-M-1)\frac{2(n-M)-3-n}{2(n-M)-3}\frac{2(n-M)-2-n}{2(n-M)-2}\frac{2(n-M)-1-n}{2(n-M)-1}\frac{2(n-M)-n}{2(n-M)}\binom{2(n-M)}{n}+\left\{-4+20M-8n-\frac{(1-n)n}{2}-(6n-8M+9)(M-1)+\frac{1}{6}(M-1)(n-2M+4)(n-2M+3)\right\}(n-M)\frac{2(n-M)-1-n}{2(n-M)-1}\frac{2(n-M)-n}{2(n-M)}\binom{2(n-M)}{n}+\left\{2(6M-2n-3)-(1-n)n-4(2n-3(M-1))(M-1)+\frac{1}{2}(M-1)(n-2M+4)(n-2M+3)\right\}\binom{2(n-M)}{n}\right]\mu^3 \quad . \quad \text{(S4.94)}$$

Power of $\mu$ equal to 2:

$$\left[\left\{-2+8M-4n-(2n-2M+3)(M-1)\right\}(n-M)\frac{2(n-M)-1-n}{2(n-M)-1}\frac{2(n-M)-n}{2(n-M)}\binom{2(n-M)}{n}+\left\{20M-8n-12-(1-n)n-2(6n-8M+9)(M-1)+\frac{1}{2}(M-1)(n-2M+4)(n-2M+3)\right\}\binom{2(n-M)}{n}+\left\{-\frac{(1-n)n+4(2n-3(M-1))(M-1)}{n-M+1}+(M-1)\frac{(n-2M+4)}{1}\frac{(n-2M+3)}{n-M+1}\right\}\frac{2(n-M)+1+1}{2(n-M)+1+1-n}\frac{2(n-M)+1}{2(n-M)+1-n}\binom{2(n-M)}{n}\right]\mu^2 \quad . \quad \text{(S4.95)}$$

Power of $\mu$ equal to 1:

$$\left[\left\{8M-4n-6-\frac{2(2n-2M+3)(M-1)}{n-M+1}(n-M+1)\right\}\binom{2(n-M)}{n}+\left\{-\frac{(1-n)n+2(6n-8M+9)(M-1)}{n-M+1}+(M-1)(n-2M+4)\frac{(n-2M+3)}{n-M+1}+(M-1)\frac{2}{n-M+1}\frac{2(n-M)+3}{1}\right\}\frac{2(n-M)+1+1}{2(n-M)+1+1-n}\frac{2(n-M)+1}{2(n-M)+1-n}\binom{2(n-M)}{n}\right]\mu^1 \quad . \quad \text{(S4.96)}$$

Power of $\mu$ equal to 0:

$$\left[-\frac{(2n-2M+3)}{1}+\frac{2(n-M)+3}{1}\right]\frac{2}{n-M+1}(M-1)\binom{2(n-M)+2}{n}\mu^0=[0]\frac{2}{n-M+1}(M-1)\binom{2(n-M)+2}{n}\mu^0=0 \quad . \quad \text{(S4.97)}$$

Here, it is clearly seen, that the pre-factor (and thus the whole term) is zero. This expression is thus no more investigated.

Power of $\mu$ equal to $\left[\frac{n}{2}\right]-M+2$:

$$\Bigg[+8\left(\left[\frac{n}{2}\right]-M\right)\binom{2\left(n-\left[\frac{n}{2}\right]\right)}{n}+4(5M-2n-3)\binom{2\left(n-\left[\frac{n}{2}\right]\right)}{n}+\Bigg\{4\left(\left[\frac{n}{2}\right]-M-1\right)\frac{\left[\frac{n}{2}\right]-M}{n-\left[\frac{n}{2}\right]+1}+2(6M-2n-3)\frac{\left[\frac{n}{2}\right]-M}{n-\left[\frac{n}{2}\right]+1}+$$

$$(M-1)\frac{(n-2M+4)}{\left[\frac{n}{2}\right]-M+1}\frac{(n-2M+3)}{n-\left[\frac{n}{2}\right]+1}-\frac{(1-n)n+4(2n-3(M-1))(M-1)}{n-\left[\frac{n}{2}\right]+1}\Bigg\}\binom{2\left(n-\left[\frac{n}{2}\right]\right)+2}{n}+$$

$$\Bigg\{-\frac{(1-n)n+2(6n-8M+9)(M-1)}{\left[\frac{n}{2}\right]-M+1}+(M-1)\frac{(n-2M+4)}{\left[\frac{n}{2}\right]-M+2}\frac{(n-2M+3)}{\left[\frac{n}{2}\right]-M+1}\Bigg\}\binom{2\left(n-\left[\frac{n}{2}\right]\right)+0}{n}\Bigg]\binom{n-M}{\left[\frac{n}{2}\right]-M}\mu^{\left[\frac{n}{2}\right]-M+2} \quad . \qquad \text{(S4.98)}$$

Power of $\mu$ equal to $\left[\frac{n}{2}\right]-M+3$:

$$\Bigg[4\left[\frac{n}{2}\right]+8M-4n-6-\frac{(1-n)n+4(2n-3(M-1))(M-1)}{\left[\frac{n}{2}\right]-M+1}+(M-1)\frac{(n-2M+4)}{\left[\frac{n}{2}\right]-M+2}\frac{(n-2M+3)}{\left[\frac{n}{2}\right]-M+1}\Bigg]\binom{n-M}{\left[\frac{n}{2}\right]-M}\binom{2\left(n-\left[\frac{n}{2}\right]\right)}{n}\mu^{\left[\frac{n}{2}\right]-M+3} \quad .$$

(S4.99)

The basic binomial identities (S4.80) and (S4.81) are once more applied to some of the binomial coefficients.

Power of $\mu$ equal to 3:

$$\Bigg[\Bigg\{1+4M-2n+\left(\frac{2}{3}n-\frac{2}{3}M+1\right)(1-M)\Bigg\}\frac{n-2M-3}{2(n-M)-3}\frac{n-2M-2}{2}\frac{n-2M-1}{2(n-M)-1}\frac{n-2M}{2}+\Bigg\{-4+20M-8n-\frac{(1-n)n}{2}+(6n-8M+$$

$$9)(1-M)+\frac{1}{6}(M-1)(4M^2-4Mn-14M+n^2+7n+12)\Bigg\}\frac{2(n-M)-1-n}{2(n-M)-1}\frac{2(n-M)-n}{2}+\Bigg\{(12M-4n-6)-(1-n)n-$$

$$(8n-12M+12)(M-1)+\frac{1}{2}(M-1)(4M^2-4Mn-14M+n^2+7n+12)\Bigg\}\Bigg]\binom{2(n-M)}{n}\mu^3 \quad . \qquad \text{(S4.100)}$$

Power of $\mu$ equal to 2:

$$\Bigg[\{(2M^2-2Mn+3M-2n+1)\}\frac{2(n-M)-1-n}{2(n-M)-1}\frac{2(n-M)-n}{2}+\Bigg\{20M-9n-12+n^2+(-12n+16M-18)(M-1)+$$

$$(M-1)\left(2M^2-2Mn-7M+\frac{1}{2}n^2+\frac{7}{2}n+6\right)\Bigg\}+$$

$$\Bigg\{-\frac{(1-n)n+(8n-12M+12)(M-1)}{\cancel{n-M+1}}+(M-1)\frac{(n-2M+4)}{1}\frac{(n-2M+3)}{\cancel{n-M+1}}\Bigg\}\frac{2\cancel{(n-M+1)}}{(n-2M)+2}\frac{2(n-M)+1}{(n-2M)+1}\Bigg]\binom{2(n-M)}{n}\mu^2 \quad . \qquad \text{(S4.101)}$$

Power of $\mu$ equal to 1:

$$\Bigg[\Bigg\{8M-4n-6-\frac{2(2n-2M+3)(M-1)}{n-M+1}(n-M+1)\Bigg\}\binom{2(n-M)}{n}+\Bigg\{-\frac{(1-n)n+2(6n-8M+9)(M-1)}{\cancel{n-M+1}}+(M-1)(n-2M+4)\frac{(n-2M+3)}{\cancel{n-M+1}}+$$

$$(M-1)\frac{2}{\cancel{n-M+1}}\frac{2(n-M)+3}{1}\Bigg\}\frac{2\cancel{(n-M+1)}}{(n-2M)+2}\frac{2(n-M)+1}{(n-2M)+1}\binom{2(n-M)}{n}\Bigg]\mu^1 \quad . \qquad \text{(S4.102)}$$

Power of $\mu$ equal to $\left[\frac{n}{2}\right]-M+2$:

$$\Bigg[+8\left(\left[\frac{n}{2}\right]-M\right)\binom{2\left(n-\left[\frac{n}{2}\right]\right)}{n}+4(5M-2n-3)\binom{2\left(n-\left[\frac{n}{2}\right]\right)}{n}+\Bigg\{4\left(\left[\frac{n}{2}\right]-M-1\right)\frac{\left[\frac{n}{2}\right]-M}{n-\left[\frac{n}{2}\right]+1}+2(6M-2n-3)\frac{\left[\frac{n}{2}\right]-M}{n-\left[\frac{n}{2}\right]+1}+$$

$$(M-1)\frac{(n-2M+4)}{\left[\frac{n}{2}\right]-M+1}\frac{(n-2M+3)}{n-\left[\frac{n}{2}\right]+1}-\frac{(1-n)n+4(2n-3(M-1))(M-1)}{n-\left[\frac{n}{2}\right]+1}\Bigg\}\frac{2\left(n-\left[\frac{n}{2}\right]\right)+1+1}{2\left(n-\left[\frac{n}{2}\right]\right)+1+1-n}\frac{2\left(n-\left[\frac{n}{2}\right]\right)+1}{2\left(n-\left[\frac{n}{2}\right]\right)+1-n}\binom{2\left(n-\left[\frac{n}{2}\right]\right)}{n}+$$

$$\Bigg\{-\frac{(1-n)n+2(6n-8M+9)(M-1)}{\left[\frac{n}{2}\right]-M+1}+(M-1)\frac{(n-2M+4)}{\left[\frac{n}{2}\right]-M+2}\frac{(n-2M+3)}{\left[\frac{n}{2}\right]-M+1}\Bigg\}\binom{2\left(n-\left[\frac{n}{2}\right]+0\right)}{n}\Bigg]\binom{n-M}{\left[\frac{n}{2}\right]-M}\mu^{\left[\frac{n}{2}\right]-M+2} \quad . \qquad \text{(S4.103)}$$

Power of $\mu$ equal to $\left[\frac{n}{2}\right]-M+3$:

$$\Bigg[\left(4\left[\frac{n}{2}\right]+8M-4n-6\right)\left(\left[\frac{n}{2}\right]-M+1\right)\left(\left[\frac{n}{2}\right]-M+2\right)-\{(1-n)n+(8n-12M+12)(M-1)\}\left(\left[\frac{n}{2}\right]-M+2\right)+$$

$$(M-1)(4M^2-4Mn-14M+n^2+7n+12)\Bigg]\frac{1}{\left[\frac{n}{2}\right]-M+2}\frac{1}{\left[\frac{n}{2}\right]-M+1}\binom{n-M}{\left[\frac{n}{2}\right]-M}\binom{2\left(n-\left[\frac{n}{2}\right]\right)}{n}\mu^{\left[\frac{n}{2}\right]-M+3} \quad . \qquad \text{(S4.104)}$$

The individual expressions are now sufficiently simple that we can finalize the considerations on their values by expansion of the terms:

Power of $\mu$ equal to 3:

$$\Bigg[\Bigg\{1+4M-2n+\left(\frac{2}{3}n-\frac{2}{3}M+1\right)(1-M)\Bigg\}\frac{n-2M-3}{1}\frac{n-2M-2}{2}\frac{n-2M-1}{1}\frac{n-2M}{2}+\Bigg\{-4+20M-\frac{17}{2}n+\frac{n^2}{2}+(-6n+8M-9)(M-$$

$$1)+(M-1)\left(\frac{2}{3}M^2-\frac{2}{3}Mn-\frac{7}{3}M+\frac{1}{6}n^2+\frac{7}{6}n+2\right)\Bigg\}(2(n-M)-3)\frac{(n-2M)-1}{1}\frac{(n-2M)}{2}+\Bigg\{(12M-5n-6+n^2)+$$

$$(-8n+12M-12)(M-1)+(M-1)\left(2M^2-2Mn-7M+\frac{1}{2}n^2+\frac{7}{2}n+6\right)\Bigg\}(2(n-M)-1)(2(n-M)-$$

$$3)\Bigg]\frac{1}{2(n-M)-1}\frac{1}{2(n-M)-3}\binom{2(n-M)}{n}\mu^3 \quad , \qquad \text{(S4.105)}$$

$$\Big[\cancel{\tfrac{8M^6}{3}} - \cancel{8M^5n} + \tfrac{52M^5}{3} + \cancel{\tfrac{28M^4n^2}{3}} - 44M^4n + \tfrac{130M^4}{3} - \cancel{\tfrac{16M^3n^3}{3}} + \tfrac{128M^3n^2}{3} - \tfrac{266M^3n}{3} + \tfrac{155M^3}{3} + \cancel{\tfrac{3M^2n^4}{2}} - \tfrac{59M^2n^3}{3} + \tfrac{397M^2n^2}{6} - \tfrac{238M^2n}{3} + 29M^2 - \cancel{\tfrac{Mn^5}{6}} + \tfrac{17Mn^4}{4} - \tfrac{64Mn^3}{3} + \tfrac{481Mn^2}{12} - \tfrac{59Mn}{2} + 6M - \cancel{\tfrac{n^5}{3}} + \tfrac{5n^4}{2} - \tfrac{20n^3}{3} + \tfrac{15n^2}{2} - 3n + - \cancel{\tfrac{8M^6}{3}} + \cancel{8M^5n} - \tfrac{76M^5}{3} - \cancel{\tfrac{28M^4n^2}{3}} + 68M^4n - \tfrac{214M^4}{3} + \cancel{\tfrac{16M^3n^3}{3}} - \tfrac{206M^3n^2}{3} + \tfrac{464M^3n}{3} - \tfrac{257M^3}{3} - \cancel{\tfrac{3M^2n^4}{2}} + \tfrac{95M^2n^3}{3} - \tfrac{721M^2n^2}{6} + \tfrac{418M^2n}{3} - 46M^2 + \cancel{\tfrac{Mn^5}{6}} - \tfrac{25Mn^4}{4} + \tfrac{118Mn^3}{3} - \tfrac{883Mn^2}{12} + 49Mn - 9M + \cancel{\tfrac{n^5}{3}} - \tfrac{9n^4}{2} + \tfrac{38n^3}{3} - 13n^2 + \tfrac{9n}{2} + 8M^5 - 24M^4n + 28M^4 + 26M^3n^2 - 66M^3n + 34M^3 - 12M^2n^3 + 54M^2n^2 - 60M^2n + 17M^2 + 2Mn^4 - 18Mn^3 + \tfrac{67Mn^2}{2} - \tfrac{39Mn}{2} + 3M + 2n^4 - 6n^3 + \tfrac{11n^2}{2} - \tfrac{3n}{2}\Big] \frac{1}{2(n-M)-1}\frac{1}{2(n-M)-3}\binom{2(n-M)}{n}\mu^3 \quad , \qquad (S4.106)$$

$$\Big[+\tfrac{155M^3}{3} - \tfrac{59M^2n^3}{3} + \tfrac{397M^2n^2}{6} - \tfrac{238M^2n}{3} + 29M^2 + \tfrac{17Mn^4}{4} - \tfrac{64Mn^3}{3} + \tfrac{481Mn^2}{12} - \tfrac{59Mn}{2} + 6M + \tfrac{5n^4}{2} - \tfrac{20n^3}{3} + \tfrac{15n^2}{2} - 3n + - \tfrac{257M^3}{3} + \tfrac{95M^2n^3}{3} - \tfrac{721M^2n^2}{6} + \tfrac{418M^2n}{3} - 46M^2 - \tfrac{25Mn^4}{4} + \tfrac{118Mn^3}{3} - \tfrac{883Mn^2}{12} + 49Mn - 9M - \tfrac{9n^4}{2} + \tfrac{38n^3}{3} - 13n^2 + \tfrac{9n}{2} + 34M^3 - 12M^2n^3 + 54M^2n^2 - 60M^2n + 17M^2 + 2Mn^4 - 18Mn^3 + \tfrac{67Mn^2}{2} - \tfrac{39Mn}{2} + 3M + 2n^4 - 6n^3 + \tfrac{11n^2}{2} - \tfrac{3n}{2}\Big] \frac{1}{2(n-M)-1}\frac{1}{2(n-M)-3}\binom{2(n-M)}{n}\mu^3 \quad , \qquad (S4.107)$$

$$\Big[+\tfrac{17Mn^4}{4} - \tfrac{64Mn^3}{3} + \tfrac{481Mn^2}{12} - \tfrac{59Mn}{2} + 6M + \tfrac{5n^4}{2} - \tfrac{20n^3}{3} + \tfrac{15n^2}{2} - 3n + - \tfrac{25Mn^4}{4} + \tfrac{118Mn^3}{3} - \tfrac{883Mn^2}{12} + 49Mn - 9M - \tfrac{9n^4}{2} + \tfrac{38n^3}{3} - 13n^2 + \tfrac{9n}{2} + 2Mn^4 - 18Mn^3 + \tfrac{67Mn^2}{2} - \tfrac{39Mn}{2} + 3M + 2n^4 - 6n^3 + \tfrac{11n^2}{2} - \tfrac{3n}{2}\Big] \frac{1}{2(n-M)-1}\frac{1}{2(n-M)-3}\binom{2(n-M)}{n}\mu^3 \quad , \qquad (S4.108)$$

$$\Big[+\tfrac{5n^4}{2} - \tfrac{20n^3}{3} + \tfrac{15n^2}{2} - 3n + - \tfrac{9n^4}{2} + \tfrac{38n^3}{3} - 13n^2 + \tfrac{9n}{2} + 2n^4 - 6n^3 + \tfrac{11n^2}{2} - \tfrac{3n}{2}\Big] \frac{1}{2(n-M)-1}\frac{1}{2(n-M)-3}\binom{2(n-M)}{n}\mu^3 \quad , \qquad (S4.109)$$

$$[0]\frac{1}{2(n-M)-1}\frac{1}{2(n-M)-3}\binom{2(n-M)}{n}\mu^3 = 0 \quad . \qquad (S4.110)$$

Then, this term is zero for all possible $n$ and $M$.

Power of $\mu$ equal to 2:

$$\Big[\{(2M^2 - 2Mn + 3M - 2n + 1)\}\frac{2(n-M)-1-n}{2(n-M)-1}\frac{2(n-M)-n}{2} + \{(20M - 9n - 12 + n^2) + (M-1)\left(2M^2 - 2Mn + 9M + \tfrac{1}{2}n^2 - \tfrac{17}{2}n - 12\right)\} + \{-n + n^2 - (8n - 12M + 12)(M-1) + (M-1)(n-2M+4)(n-2M+3)\}\frac{2}{(n-2M)+2}\frac{2(n-M)+1}{(n-2M)+1}\Big]\binom{2(n-M)}{n}\mu^2 \quad , \qquad (S4.111)$$

$$\Big[\{(2M^2 - 2Mn + 3M - 2n + 1)\}\frac{((n-2M)-1)}{1}\frac{(n-2M)}{2} + \{(20M - 9n - 12 + n^2) + (M-1)\left(2M^2 - 2Mn + 9M + \tfrac{1}{2}n^2 - \tfrac{17}{2}n - 12\right)\}(2(n-M)-1) + \{-n + n^2 - (8n - 12M + 12)(M-1) + (M-1)(n-2M+4)(n-2M+3)\}\frac{2}{(n-2M)+2}\frac{2(n-M)+1}{(n-2M)+1}(2(n-M)-1)\Big]\frac{1}{2(n-M)-1}\binom{2(n-M)}{n}\mu^2 \quad , \qquad (S4.112)$$

$$\Big[\{(2M^2 - 2Mn + 3M - 2n + 1)\}\frac{((n-2M)-1)}{1}\frac{(n-2M)}{2}((n-2M)+2)((n-2M)+1) + \{2M^3 - 2M^2n + 7M^2 + \tfrac{Mn^2}{2} - \tfrac{13Mn}{2} - M + \tfrac{n^2}{2} - \tfrac{n}{2}\}(2(n-M)-1)((n-2M)+2)((n-2M)+1) + \{4M^3 - 4M^2n - 6M^2 + Mn^2 + 3Mn + 2M\}\frac{2}{1}\frac{2(n-M)+1}{1}(2(n-M)-1)\Big]\frac{1}{(n-2M)+2}\frac{1}{(n-2M)+1}\frac{1}{2(n-M)-1}\binom{2(n-M)}{n}\mu^2 \quad , \qquad (S4.113)$$

$$\Big[\cancel{16M^6} - \cancel{48M^5n} + 8M^5 + \cancel{56M^4n^2} - 24M^4n - 20M^4 - \cancel{32M^3n^3} + 32M^3n^2 + 44M^3n - 10M^3 + \cancel{9M^2n^4} - 22M^2n^3 - 35M^2n^2 + 16M^2n + 4M^2 - \cancel{Mn^5} + \tfrac{15Mn^4}{2} + 12Mn^3 - \tfrac{19Mn^2}{2} - 5Mn + 2M - \cancel{n^5} - \cancel{\tfrac{3n^4}{2}} + \cancel{2n^3} + \cancel{\tfrac{3n^2}{2}} - \cancel{n} + -\cancel{16M^6} + \cancel{48M^5n} - 40M^5 - \cancel{56M^4n^2} + 120M^4n + 68M^4 + \cancel{32M^3n^3} - 136M^3n^2 - 164M^3n + 2M^3 - \cancel{9M^2n^4} + 70M^2n^3 + 131M^2n^2 + 8M^2n - 16M^2 + \cancel{Mn^5} - \tfrac{31Mn^4}{2} - 36Mn^3 - \tfrac{9Mn^2}{2} + 11Mn + 2M + \cancel{n^5} + \cancel{\tfrac{3n^4}{2}} - \cancel{2n^3} - \cancel{\tfrac{3n^2}{2}} + \cancel{n} + 32M^5 - 96M^4n - 48M^4 + 104M^3n^2 + 120M^3n + 8M^3 - 48M^2n^3 - 96M^2n^2 -$$

$$24\,M^2n + 12\,M^2 + 8\,M\,n^4 + 24\,M\,n^3 + 14\,M\,n^2 - 6\,M\,n - 4\,M\Big]\frac{1}{(n-2M)+2}\,\frac{1}{(n-2M)+1}\,\frac{1}{2(n-M)-1}\binom{2(n-M)}{n}\mu^2 \quad , \qquad \text{(S4.114)}$$

$$\Big[-10\,M^3 - 22\,M^2n^3 - 35\,M^2n^2 + 16\,M^2n + 4\,M^2 + \frac{15\,M\,n^4}{2} + 12\,M\,n^3 - \frac{19\,M\,n^2}{2} - 5\,M\,n + 2\,M + + 2\,M^3 + 70\,M^2n^3 + 131\,M^2n^2 + 8\,M^2n - 16\,M^2 - \frac{31\,M\,n^4}{2} - 36\,M\,n^3 - \frac{9\,M\,n^2}{2} + 11\,M\,n + 2\,M + + 8\,M^3 - 48\,M^2n^3 - 96\,M^2n^2 - 24\,M^2n + 12\,M^2 + 8\,M\,n^4 + 24\,M\,n^3 + 14\,M\,n^2 - 6\,M\,n - 4\,M\Big]\frac{1}{(n-2M)+2}\,\frac{1}{(n-2M)+1}\,\frac{1}{2(n-M)-1}\binom{2(n-M)}{n}\mu^2 \quad , \qquad \text{(S4.115)}$$

$$\Big[+\frac{15\,M\,n^4}{2} + 12\,M\,n^3 - \frac{19\,M\,n^2}{2} - 5\,M\,n + 2\,M + - \frac{31\,M\,n^4}{2} - 36\,M\,n^3 - \frac{9\,M\,n^2}{2} + 11\,M\,n + 2\,M + + 8\,M\,n^4 + 24\,M\,n^3 + 14\,M\,n^2 - 6\,M\,n - 4\,M\Big]\frac{1}{(n-2M)+2}\,\frac{1}{(n-2M)+1}\,\frac{1}{2(n-M)-1}\binom{2(n-M)}{n}\mu^2 \quad , \qquad \text{(S4.116)}$$

$$[0]\,\frac{1}{(n-2M)+2}\,\frac{1}{(n-2M)+1}\,\frac{1}{2(n-M)-1}\binom{2(n-M)}{n}\mu^2 = 0 \ . \qquad \text{(S4.117)}$$

Then, this term is zero for all possible $n$ and $M$.

Power of $\mu$ equal to 1:

$$\Big[\Big\{8M - 4n - 6 - \frac{(4n-4M+6)(M-1)}{n-M+1}(n-M+1)\Big\} + \{-n + n^2 - (12n - 16M + 18)(M-1) + (M-1)(4M^2 - 4Mn - 14M + n^2 + 7n + 12) + (M-1)2(2(n-M)+3)\}\frac{2}{(n-2M)+2}\,\frac{2(n-M)+1}{(n-2M)+1}\Big]\binom{2(n-M)}{n}\mu^1 \quad , \qquad \text{(S4.118)}$$

$$\Big[\{(8M - 4n - 6) - (4n - 4M + 6)(M-1)\}((n-2M)+2)((n-2M)+1) + \{-n + n^2 - (12n - 16M + 18)(M-1) + (M-1)(4M^2 - 4Mn - 14M + n^2 + 7n + 12) + (M-1)2(2(n-M)+3)\}2\,\frac{2(n-M)+1}{1}\Big]\frac{1}{n-M+1}\,\frac{1}{((n-2M)+2)}\,\frac{1}{((n-2M)+1)}\binom{2(n-M)}{n}\mu^1 \quad , \qquad \text{(S4.119)}$$

$$[16M^4 - 32M^3n - 32M^3 + 20M^2n^2 + 44M^2n + 20M^2 - 4Mn^3 - 14Mn^2 - 14Mn - 4M - 16M^4 + 32M^3n + 32M^3 - 20M^2n^2 - 44M^2n - 20M^2 + 4Mn^3 + 14Mn^2 + 14Mn + 4M]\frac{1}{n-M+1}\,\frac{1}{((n-2M)+2)}\,\frac{1}{((n-2M)+1)}\binom{2(n-M)}{n}\mu^1 \quad , \qquad \text{(S4.120)}$$

$$[0]\,\frac{1}{n-M+1}\,\frac{1}{((n-2M)+2)}\,\frac{1}{((n-2M)+1)}\binom{2(n-M)}{n}\mu^1 = 0\ . \qquad \text{(S4.121)}$$

Also this term is zero for all possible $n$ and $M$.

Power of $\mu$ equal to $\left[\frac{n}{2}\right] - M + 2$:

$$\Bigg[\left(8\left[\frac{n}{2}\right] + 12M - 8n - 12\right) + \Bigg\{\left(4\left[\frac{n}{2}\right] - 4M - 4\right)\frac{\left[\frac{n}{2}\right]-M}{\cancel{n-\left[\frac{n}{2}\right]+1}} + (12M - 4n - 6)\frac{\left[\frac{n}{2}\right]-M}{\cancel{n-\left[\frac{n}{2}\right]+1}} + (M-1)(4M^2 - 4Mn - 14M + n^2 + 7n + 12)\frac{1}{\left[\frac{n}{2}\right]-M+1}\,\frac{1}{\cancel{n-\left[\frac{n}{2}\right]+1}} - \frac{n - n^2 + (8n - 12M + 12)(M-1)}{\cancel{n-\left[\frac{n}{2}\right]+1}}\Bigg\}\frac{2\cancel{\left(n-\left[\frac{n}{2}\right]+1\right)}}{\left(n-2\left[\frac{n}{2}\right]\right)+2}\,\frac{2\left(n-\left[\frac{n}{2}\right]\right)+1}{\left(n-2\left[\frac{n}{2}\right]\right)+1} + \Bigg\{-\frac{(1-n)n + 2(6n - 8M + 9)(M-1)}{\left[\frac{n}{2}\right]-M+1} + (M-1)\frac{(n-2M+4)}{\left[\frac{n}{2}\right]-M+2}\,\frac{(n-2M+3)}{\left[\frac{n}{2}\right]-M+1}\Bigg\}\Bigg]\binom{2\left(n-\left[\frac{n}{2}\right]\right)}{n}\binom{n-M}{\left[\frac{n}{2}\right]-M}\mu^{\left[\frac{n}{2}\right]-M+2} \quad , \qquad \text{(S4.122)}$$

$$\Bigg[\left(8\left[\frac{n}{2}\right] + 12M - 8n - 12\right)\left(\left[\frac{n}{2}\right] - M + 1\right)\left(\left[\frac{n}{2}\right] - M + 2\right) + \Bigg\{\left(4\left[\frac{n}{2}\right] - 4M - 4\right)\frac{\left[\frac{n}{2}\right]-M}{1} + (12M - 4n - 6)\frac{\left[\frac{n}{2}\right]-M}{1} + (M-1)(4M^2 - 4Mn - 14M + n^2 + 7n + 12)\frac{1}{\left[\frac{n}{2}\right]-M+1} - \frac{n - n^2 + (8Mn - 12M^2 + 12M) + (-8n + 12M - 12)}{1}\Bigg\}\left(\left[\frac{n}{2}\right] - M + 1\right)\left(\left[\frac{n}{2}\right] - M + 2\right)\frac{2}{\left(n-2\left[\frac{n}{2}\right]\right)+2}\,\frac{2\left(n-\left[\frac{n}{2}\right]\right)+1}{\left(n-2\left[\frac{n}{2}\right]\right)+1} + \Bigg\{-\frac{n - n^2 + (12n - 16M + 18)(M-1)}{1}\left(\left[\frac{n}{2}\right] - M + 2\right) + (M-1)(4M^2 - 4Mn - 14M + n^2 + 7n + 12)\Bigg\}\Bigg]\frac{1}{\left(\left[\frac{n}{2}\right]-M+2\right)}\,\frac{1}{\left(\left[\frac{n}{2}\right]-M+1\right)}\binom{2\left(n-\left[\frac{n}{2}\right]\right)}{n}\binom{n-M}{\left[\frac{n}{2}\right]-M}\mu^{\left[\frac{n}{2}\right]-M+2} \quad . \qquad \text{(S4.123)}$$

Expansions with help of WolframAlpha calculator available publicly gives:

$$\Big[\Big(12M^3 - 8M^2n - 16M^2\left[\frac{n}{2}\right] - 48M^2 + 16Mn\left[\frac{n}{2}\right] + 24Mn - 4M\left[\frac{n}{2}\right]^2 + 36M\left[\frac{n}{2}\right] + 60M - 8n\left[\frac{n}{2}\right]^2 - 24n\left[\frac{n}{2}\right] - 16n + 8\left[\frac{n}{2}\right]^3 + 12\left[\frac{n}{2}\right]^2 - 20\left[\frac{n}{2}\right] - 24\Big)\left(\left(n-2\left[\frac{n}{2}\right]\right)+2\right)\left(\left(n-2\left[\frac{n}{2}\right]\right)+1\right) + \Big\{\Big(-8M^2 + 4Mn + 4M\left[\frac{n}{2}\right] + 10M - 4n\left[\frac{n}{2}\right] + 4\left[\frac{n}{2}\right]^2 - 10\left[\frac{n}{2}\right]\Big)\left(\left[\frac{n}{2}\right] - M + 1\right) + (M-1)(4M^2 - 4Mn - 14M + n^2 + 7n + 12) - \frac{n - n^2 + (8Mn - 12M^2 + 12M) + (-8n + 12M - 12)}{1}\left(\left[\frac{n}{2}\right] - M + 1\right)\Big\}\left(\left[\frac{n}{2}\right] - M + 2\right)\frac{2}{1}\frac{\left(2\left(n-\left[\frac{n}{2}\right]\right)+1\right)}{1} + \Big\{-12M^3 + 8M^2n + 16M^2\left[\frac{n}{2}\right] + 48M^2 - 12Mn\left[\frac{n}{2}\right] - 24Mn - 34M\left[\frac{n}{2}\right] - 60M + n^2\left[\frac{n}{2}\right] + n^2 + 11n\left[\frac{n}{2}\right] + 15n + 18\left[\frac{n}{2}\right] + 24\Big\}\left(\left(n-2\left[\frac{n}{2}\right]\right)+2\right)\left(\left(n-2\left[\frac{n}{2}\right]\right)+1\right)\Big]\frac{1}{\left(\left(n-2\left[\frac{n}{2}\right]\right)+2\right)}\frac{1}{\left(\left(n-2\left[\frac{n}{2}\right]\right)+1\right)}\frac{1}{\left(\left[\frac{n}{2}\right]-M+2\right)}\frac{1}{\left(\left[\frac{n}{2}\right]-M+1\right)}\binom{2\left(n-\left[\frac{n}{2}\right]\right)}{n}\binom{n-M}{\left[\frac{n}{2}\right]-M}\mu^{\left[\frac{n}{2}\right]-M+2}. \quad \text{(S4.124)}$$

Further expansions

$$\Big[12M^3n^2 - 48M^3n\left[\frac{n}{2}\right] + 36M^3n + 48M^3\left[\frac{n}{2}\right]^2 - 72M^3\left[\frac{n}{2}\right] + 24M^3 - 8M^2n^3 + 16M^2n^2\left[\frac{n}{2}\right] - 72M^2n^2 + 32M^2n\left[\frac{n}{2}\right]^2 + 192M^2n\left[\frac{n}{2}\right] - 160M^2n - 64M^2\left[\frac{n}{2}\right]^3 - 96M^2\left[\frac{n}{2}\right]^2 + 256M^2\left[\frac{n}{2}\right] - 96M^2 + 16Mn^3\left[\frac{n}{2}\right] + 24Mn^3 - 68Mn^2\left[\frac{n}{2}\right]^2 - 12Mn^2\left[\frac{n}{2}\right] + 132Mn^2 + 80Mn\left[\frac{n}{2}\right]^3 - 156Mn\left[\frac{n}{2}\right]^2 - 244Mn\left[\frac{n}{2}\right] + 228Mn - 16M\left[\frac{n}{2}\right]^4 + 168M\left[\frac{n}{2}\right]^3 + 16M\left[\frac{n}{2}\right]^2 - 288M\left[\frac{n}{2}\right] + 120M - 8n^3\left[\frac{n}{2}\right]^2 - 24n^3\left[\frac{n}{2}\right] - 16n^3 + 40n^2\left[\frac{n}{2}\right]^3 + 84n^2\left[\frac{n}{2}\right]^2 - 28n^2\left[\frac{n}{2}\right] - 72n^2 - 64n\left[\frac{n}{2}\right]^4 - 72n\left[\frac{n}{2}\right]^3 + 180n\left[\frac{n}{2}\right]^2 + 84n\left[\frac{n}{2}\right] - 104n + 32\left[\frac{n}{2}\right]^5 - 136\left[\frac{n}{2}\right]^3 + 48\left[\frac{n}{2}\right]^2 + 104\left[\frac{n}{2}\right] - 48 + \Big\{n^2\left[\frac{n}{2}\right] - 4n\left[\frac{n}{2}\right]^2 + 3n\left[\frac{n}{2}\right] + 4\left[\frac{n}{2}\right]^3 - 6\left[\frac{n}{2}\right]^2 + 2\left[\frac{n}{2}\right]\Big\}\left(\left[\frac{n}{2}\right] - M + 2\right)2\left(2\left(n-\left[\frac{n}{2}\right]\right)+1\right) + \Big\{-12M^3n^2 + 48M^3n\left[\frac{n}{2}\right] - 36M^3n - 48M^3\left[\frac{n}{2}\right]^2 + 72M^3\left[\frac{n}{2}\right] - 24M^3 + 8M^2n^3 - 16M^2n^2\left[\frac{n}{2}\right] + 72M^2n^2 - 32M^2n\left[\frac{n}{2}\right]^2 - 192M^2n\left[\frac{n}{2}\right] + 160M^2n + 64M^2\left[\frac{n}{2}\right]^3 + 96M^2\left[\frac{n}{2}\right]^2 - 256M^2\left[\frac{n}{2}\right] + 96M^2 - 12Mn^3\left[\frac{n}{2}\right] - 24Mn^3 + 48Mn^2\left[\frac{n}{2}\right]^2 + 26Mn^2\left[\frac{n}{2}\right] - 132Mn^2 - 48Mn\left[\frac{n}{2}\right]^3 + 112Mn\left[\frac{n}{2}\right]^2 + 258Mn\left[\frac{n}{2}\right] - 228Mn - 136M\left[\frac{n}{2}\right]^3 - 36M\left[\frac{n}{2}\right]^2 + 292M\left[\frac{n}{2}\right] - 120M + n^4\left[\frac{n}{2}\right] + n^4 - 4n^3\left[\frac{n}{2}\right]^2 + 10n^3\left[\frac{n}{2}\right] + 18n^3 + 4n^2\left[\frac{n}{2}\right]^3 - 46n^2\left[\frac{n}{2}\right]^2 - 13n^2\left[\frac{n}{2}\right] + 71n^2 + 44n\left[\frac{n}{2}\right]^3 - 78n\left[\frac{n}{2}\right]^2 - 110n\left[\frac{n}{2}\right] + 102n + 72\left[\frac{n}{2}\right]^3 - 12\left[\frac{n}{2}\right]^2 - 108\left[\frac{n}{2}\right] + 48\Big\}\Big]\frac{1}{\left(\left(n-2\left[\frac{n}{2}\right]\right)+2\right)}\frac{1}{\left(\left(n-2\left[\frac{n}{2}\right]\right)+1\right)}\frac{1}{\left(\left[\frac{n}{2}\right]-M+2\right)}\frac{1}{\left(\left[\frac{n}{2}\right]-M+1\right)}\binom{2\left(n-\left[\frac{n}{2}\right]\right)}{n}\binom{n-M}{\left[\frac{n}{2}\right]-M}\mu^{\left[\frac{n}{2}\right]-M+2}. \quad \text{(S4.125)}$$

And still the remaining expansion:

$$\Big[12M^3n^2 - 48M^3n\left[\frac{n}{2}\right] + 36M^3n + 48M^3\left[\frac{n}{2}\right]^2 - 72M^3\left[\frac{n}{2}\right] + 24M^3 - 8M^2n^3 + 16M^2n^2\left[\frac{n}{2}\right] - 72M^2n^2 + 32M^2n\left[\frac{n}{2}\right]^2 + 192M^2n\left[\frac{n}{2}\right] - 160M^2n - 64M^2\left[\frac{n}{2}\right]^3 - 96M^2\left[\frac{n}{2}\right]^2 + 256M^2\left[\frac{n}{2}\right] - 96M^2 + 16Mn^3\left[\frac{n}{2}\right] + 24Mn^3 - 68Mn^2\left[\frac{n}{2}\right]^2 - 12Mn^2\left[\frac{n}{2}\right] + 132Mn^2 + 80Mn\left[\frac{n}{2}\right]^3 - 156Mn\left[\frac{n}{2}\right]^2 - 244Mn\left[\frac{n}{2}\right] + 228Mn - 16M\left[\frac{n}{2}\right]^4 + 168M\left[\frac{n}{2}\right]^3 + 16M\left[\frac{n}{2}\right]^2 - 288M\left[\frac{n}{2}\right] + 120M - 8n^3\left[\frac{n}{2}\right]^2 - 24n^3\left[\frac{n}{2}\right] - 16n^3 + 40n^2\left[\frac{n}{2}\right]^3 + 84n^2\left[\frac{n}{2}\right]^2 - 28n^2\left[\frac{n}{2}\right] - 72n^2 - 64n\left[\frac{n}{2}\right]^4 - 72n\left[\frac{n}{2}\right]^3 + 180n\left[\frac{n}{2}\right]^2 + 84n\left[\frac{n}{2}\right] - 104n + 32\left[\frac{n}{2}\right]^5 - 136\left[\frac{n}{2}\right]^3 + 48\left[\frac{n}{2}\right]^2 + 104\left[\frac{n}{2}\right] - 48 + \Big\{-4Mn^3\left[\frac{n}{2}\right] + 20Mn^2\left[\frac{n}{2}\right]^2 - 14Mn^2\left[\frac{n}{2}\right] - 32Mn\left[\frac{n}{2}\right]^3 + 44Mn\left[\frac{n}{2}\right]^2 - 14Mn\left[\frac{n}{2}\right] + 16M\left[\frac{n}{2}\right]^4 - 32M\left[\frac{n}{2}\right]^3 + 20M\left[\frac{n}{2}\right]^2 - 4M\left[\frac{n}{2}\right] + 4n^3\left[\frac{n}{2}\right]^2 + 8n^3\left[\frac{n}{2}\right] - 20n^2\left[\frac{n}{2}\right]^3 - 26n^2\left[\frac{n}{2}\right]^2 + 28n^2\left[\frac{n}{2}\right] + 32n\left[\frac{n}{2}\right]^4 + 20n\left[\frac{n}{2}\right]^3 - 74n\left[\frac{n}{2}\right]^2 + 28n\left[\frac{n}{2}\right] - 16\left[\frac{n}{2}\right]^5 + 44\left[\frac{n}{2}\right]^3 - 36\left[\frac{n}{2}\right]^2 + 8\left[\frac{n}{2}\right]\Big\} + \Big\{-12M^3n^2 + 48M^3n\left[\frac{n}{2}\right] - 36M^3n - 48M^3\left[\frac{n}{2}\right]^2 + 72M^3\left[\frac{n}{2}\right] - 24M^3 + 8M^2n^3 - 16M^2n^2\left[\frac{n}{2}\right] + 72M^2n^2 - 32M^2n\left[\frac{n}{2}\right]^2 - 192M^2n\left[\frac{n}{2}\right] + 160M^2n + 64M^2\left[\frac{n}{2}\right]^3 + 96M^2\left[\frac{n}{2}\right]^2 - 256M^2\left[\frac{n}{2}\right] + 96M^2 - 12Mn^3\left[\frac{n}{2}\right] - 24Mn^3 + 48Mn^2\left[\frac{n}{2}\right]^2 + 26Mn^2\left[\frac{n}{2}\right] - 132Mn^2 - 48Mn\left[\frac{n}{2}\right]^3 + 112Mn\left[\frac{n}{2}\right]^2 + 258Mn\left[\frac{n}{2}\right] - 228Mn - 136M\left[\frac{n}{2}\right]^3 - 36M\left[\frac{n}{2}\right]^2 + 292M\left[\frac{n}{2}\right] - 120M + n^4\left[\frac{n}{2}\right] + n^4 - 4n^3\left[\frac{n}{2}\right]^2 + 10n^3\left[\frac{n}{2}\right] + 18n^3 + 4n^2\left[\frac{n}{2}\right]^3 - 46n^2\left[\frac{n}{2}\right]^2 - 13n^2\left[\frac{n}{2}\right] + 71n^2 + 44n\left[\frac{n}{2}\right]^3 - 78n\left[\frac{n}{2}\right]^2 - 110n\left[\frac{n}{2}\right] + 102n + 72\left[\frac{n}{2}\right]^3 - 12\left[\frac{n}{2}\right]^2 - 108\left[\frac{n}{2}\right] + 48\Big\}\Big]\frac{1}{\left(\left(n-2\left[\frac{n}{2}\right]\right)+2\right)}\frac{1}{\left(\left(n-2\left[\frac{n}{2}\right]\right)+1\right)}\frac{1}{\left(\left[\frac{n}{2}\right]-M+2\right)}\frac{1}{\left(\left[\frac{n}{2}\right]-M+1\right)}\binom{2\left(n-\left[\frac{n}{2}\right]\right)}{n}\binom{n-M}{\left[\frac{n}{2}\right]-M}\mu^{\left[\frac{n}{2}\right]-M+2}, \quad \text{(S4.126)}$$

$$\left[-244Mn\left[\frac{n}{2}\right]+\cancel{228Mn}-\cancel{16M\left[\frac{n}{2}\right]^4}+168M\left[\frac{n}{2}\right]^3+16M\left[\frac{n}{2}\right]^2-288M\left[\frac{n}{2}\right]+\cancel{120M}-8n^3\left[\frac{n}{2}\right]^2-24n^3\left[\frac{n}{2}\right]-16n^3+40n^2\left[\frac{n}{2}\right]^3+84n^2\left[\frac{n}{2}\right]^2-28n^2\left[\frac{n}{2}\right]-72n^2-64n\left[\frac{n}{2}\right]^4-72n\left[\frac{n}{2}\right]^3+180n\left[\frac{n}{2}\right]^2+84n\left[\frac{n}{2}\right]-104n+32\left[\frac{n}{2}\right]^5-136\left[\frac{n}{2}\right]^3+48\left[\frac{n}{2}\right]^2+104\left[\frac{n}{2}\right]-\cancel{48}+\left\{-14Mn\left[\frac{n}{2}\right]+\cancel{16M\left[\frac{n}{2}\right]^4}-32M\left[\frac{n}{2}\right]^3+20M\left[\frac{n}{2}\right]^2-4M\left[\frac{n}{2}\right]+4n^3\left[\frac{n}{2}\right]^2+8n^3\left[\frac{n}{2}\right]-20n^2\left[\frac{n}{2}\right]^3-26n^2\left[\frac{n}{2}\right]^2+28n^2\left[\frac{n}{2}\right]+32n\left[\frac{n}{2}\right]^4+20n\left[\frac{n}{2}\right]^3-74n\left[\frac{n}{2}\right]^2+28n\left[\frac{n}{2}\right]-16\left[\frac{n}{2}\right]^5+44\left[\frac{n}{2}\right]^3-36\left[\frac{n}{2}\right]^2+8\left[\frac{n}{2}\right]\right\}+\left\{+258Mn\left[\frac{n}{2}\right]-\cancel{228Mn}-136M\left[\frac{n}{2}\right]^3-36M\left[\frac{n}{2}\right]^2+292M\left[\frac{n}{2}\right]-\cancel{120M}+n^4\left[\frac{n}{2}\right]+n^4-4n^3\left[\frac{n}{2}\right]^2+10n^3\left[\frac{n}{2}\right]+18n^3+4n^2\left[\frac{n}{2}\right]^3-46n^2\left[\frac{n}{2}\right]^2-13n^2\left[\frac{n}{2}\right]+71n^2+44n\left[\frac{n}{2}\right]^3-78n\left[\frac{n}{2}\right]^2-110n\left[\frac{n}{2}\right]+102n+72\left[\frac{n}{2}\right]^3-12\left[\frac{n}{2}\right]^2-108\left[\frac{n}{2}\right]+\cancel{48}\right\}\right]\frac{1}{\left(\left(n-2\left[\frac{n}{2}\right]\right)+2\right)}\frac{1}{\left(\left(n-2\left[\frac{n}{2}\right]\right)+1\right)}\frac{1}{\left(\left[\frac{n}{2}\right]-M+2\right)}\frac{1}{\left(\left[\frac{n}{2}\right]-M+1\right)}\binom{2\left(n-\left[\frac{n}{2}\right]\right)}{n}\binom{n-M}{\left[\frac{n}{2}\right]-M}\mu^{\left[\frac{n}{2}\right]-M+2},\qquad (S4.127)$$

$$\left[-8n^3\left[\frac{n}{2}\right]^2-24n^3\left[\frac{n}{2}\right]-16n^3+40n^2\left[\frac{n}{2}\right]^3+84n^2\left[\frac{n}{2}\right]^2-\cancel{28n^2\left[\frac{n}{2}\right]}-72n^2-64n\left[\frac{n}{2}\right]^4-72n\left[\frac{n}{2}\right]^3+180n\left[\frac{n}{2}\right]^2+84n\left[\frac{n}{2}\right]-104n+32\left[\frac{n}{2}\right]^5-136\left[\frac{n}{2}\right]^3+48\left[\frac{n}{2}\right]^2+104\left[\frac{n}{2}\right]+\left\{\cancel{+4n^3\left[\frac{n}{2}\right]^2}+8n^3\left[\frac{n}{2}\right]-20n^2\left[\frac{n}{2}\right]^3-26n^2\left[\frac{n}{2}\right]^2+\cancel{28n^2\left[\frac{n}{2}\right]}+32n\left[\frac{n}{2}\right]^4+20n\left[\frac{n}{2}\right]^3-74n\left[\frac{n}{2}\right]^2+28n\left[\frac{n}{2}\right]-16\left[\frac{n}{2}\right]^5+44\left[\frac{n}{2}\right]^3-36\left[\frac{n}{2}\right]^2+8\left[\frac{n}{2}\right]\right\}+\left\{+n^4\left[\frac{n}{2}\right]+n^4-\cancel{4n^3\left[\frac{n}{2}\right]^2}+10n^3\left[\frac{n}{2}\right]+18n^3+4n^2\left[\frac{n}{2}\right]^3-46n^2\left[\frac{n}{2}\right]^2-13n^2\left[\frac{n}{2}\right]+71n^2+44n\left[\frac{n}{2}\right]^3-78n\left[\frac{n}{2}\right]^2-110n\left[\frac{n}{2}\right]+102n+72\left[\frac{n}{2}\right]^3-12\left[\frac{n}{2}\right]^2-108\left[\frac{n}{2}\right]\right\}\right]\frac{1}{\left(\left(n-2\left[\frac{n}{2}\right]\right)+2\right)}\frac{1}{\left(\left(n-2\left[\frac{n}{2}\right]\right)+1\right)}\frac{1}{\left(\left[\frac{n}{2}\right]-M+2\right)}\frac{1}{\left(\left[\frac{n}{2}\right]-M+1\right)}\binom{2\left(n-\left[\frac{n}{2}\right]\right)}{n}\binom{n-M}{\left[\frac{n}{2}\right]-M}\mu^{\left[\frac{n}{2}\right]-M+2},\qquad (S4.128)$$

$$\left[-8n^3\left[\frac{n}{2}\right]^2-6n^3\left[\frac{n}{2}\right]+2n^3+24n^2\left[\frac{n}{2}\right]^3+12n^2\left[\frac{n}{2}\right]^2-n^2-64n\left[\frac{n}{2}\right]^4-72n\left[\frac{n}{2}\right]^3+180n\left[\frac{n}{2}\right]^2+84n\left[\frac{n}{2}\right]-104n+32\left[\frac{n}{2}\right]^5-136\left[\frac{n}{2}\right]^3+48\left[\frac{n}{2}\right]^2+104\left[\frac{n}{2}\right]+\left\{+32n\left[\frac{n}{2}\right]^4+20n\left[\frac{n}{2}\right]^3-74n\left[\frac{n}{2}\right]^2+28n\left[\frac{n}{2}\right]-16\left[\frac{n}{2}\right]^5+44\left[\frac{n}{2}\right]^3-36\left[\frac{n}{2}\right]^2+8\left[\frac{n}{2}\right]\right\}+\left\{+n^4\left[\frac{n}{2}\right]+n^4-13n^2\left[\frac{n}{2}\right]+44n\left[\frac{n}{2}\right]^3-78n\left[\frac{n}{2}\right]^2-110n\left[\frac{n}{2}\right]+102n+72\left[\frac{n}{2}\right]^3-12\left[\frac{n}{2}\right]^2-108\left[\frac{n}{2}\right]\right\}\right]\frac{1}{\left(\left(n-2\left[\frac{n}{2}\right]\right)+2\right)}\frac{1}{\left(\left(n-2\left[\frac{n}{2}\right]\right)+1\right)}\frac{1}{\left(\left[\frac{n}{2}\right]-M+2\right)}\frac{1}{\left(\left[\frac{n}{2}\right]-M+1\right)}\binom{2\left(n-\left[\frac{n}{2}\right]\right)}{n}\binom{n-M}{\left[\frac{n}{2}\right]-M}\mu^{\left[\frac{n}{2}\right]-M+2},\qquad (S4.129)$$

$$\left[-8n^3\left[\frac{n}{2}\right]^2-6n^3\left[\frac{n}{2}\right]+2n^3+24n^2\left[\frac{n}{2}\right]^3+12n^2\left[\frac{n}{2}\right]^2-n^2-32n\left[\frac{n}{2}\right]^4-8n\left[\frac{n}{2}\right]^3+28n\left[\frac{n}{2}\right]^2+2n\left[\frac{n}{2}\right]-2n+32\left[\frac{n}{2}\right]^5-136\left[\frac{n}{2}\right]^3+\cancel{48\left[\frac{n}{2}\right]^2}+104\left[\frac{n}{2}\right]+\left\{-16\left[\frac{n}{2}\right]^5+44\left[\frac{n}{2}\right]^3-\cancel{36\left[\frac{n}{2}\right]^2}+8\left[\frac{n}{2}\right]\right\}+\left\{+n^4\left[\frac{n}{2}\right]+n^4-13n^2\left[\frac{n}{2}\right]+72\left[\frac{n}{2}\right]^3-\cancel{12\left[\frac{n}{2}\right]^2}-108\left[\frac{n}{2}\right]\right\}\right]\frac{1}{\left(\left(n-2\left[\frac{n}{2}\right]\right)+2\right)}\frac{1}{\left(\left(n-2\left[\frac{n}{2}\right]\right)+1\right)}\frac{1}{\left(\left[\frac{n}{2}\right]-M+2\right)}\frac{1}{\left(\left[\frac{n}{2}\right]-M+1\right)}\binom{2\left(n-\left[\frac{n}{2}\right]\right)}{n}\binom{n-M}{\left[\frac{n}{2}\right]-M}\mu^{\left[\frac{n}{2}\right]-M+2},\qquad (S4.130)$$

$$\left[-8n^3\left[\frac{n}{2}\right]^2-6n^3\left[\frac{n}{2}\right]+2n^3+24n^2\left[\frac{n}{2}\right]^3+12n^2\left[\frac{n}{2}\right]^2-n^2-32n\left[\frac{n}{2}\right]^4-8n\left[\frac{n}{2}\right]^3+28n\left[\frac{n}{2}\right]^2+2n\left[\frac{n}{2}\right]-2n+16\left[\frac{n}{2}\right]^5-20\left[\frac{n}{2}\right]^3+4\left[\frac{n}{2}\right]+\{0\}+\left\{+n^4\left[\frac{n}{2}\right]+n^4-13n^2\left[\frac{n}{2}\right]\right\}\right]\frac{1}{\left(\left(n-2\left[\frac{n}{2}\right]\right)+2\right)}\frac{1}{\left(\left(n-2\left[\frac{n}{2}\right]\right)+1\right)}\frac{1}{\left(\left[\frac{n}{2}\right]-M+2\right)}\frac{1}{\left(\left[\frac{n}{2}\right]-M+1\right)}\binom{2\left(n-\left[\frac{n}{2}\right]\right)}{n}\binom{n-M}{\left[\frac{n}{2}\right]-M}\mu^{\left[\frac{n}{2}\right]-M+2}.\qquad (S4.131)$$

Order according to the total exponent of $n$:

$$\left[+16\left[\frac{n}{2}\right]^5-32n\left[\frac{n}{2}\right]^4+24n^2\left[\frac{n}{2}\right]^3-8n^3\left[\frac{n}{2}\right]^2+n^4\left[\frac{n}{2}\right]-8n\left[\frac{n}{2}\right]^3+12n^2\left[\frac{n}{2}\right]^2-6n^3\left[\frac{n}{2}\right]+n^4-20\left[\frac{n}{2}\right]^3+28n\left[\frac{n}{2}\right]^2-13n^2\left[\frac{n}{2}\right]+2n^3+2n\left[\frac{n}{2}\right]-n^2+4\left[\frac{n}{2}\right]-2n+\{0\}+\{0\}\right]\frac{1}{\left(\left(n-2\left[\frac{n}{2}\right]\right)+2\right)}\frac{1}{\left(\left(n-2\left[\frac{n}{2}\right]\right)+1\right)}\frac{1}{\left(\left[\frac{n}{2}\right]-M+2\right)}\frac{1}{\left(\left[\frac{n}{2}\right]-M+1\right)}\binom{2\left(n-\left[\frac{n}{2}\right]\right)}{n}\binom{n-M}{\left[\frac{n}{2}\right]-M}\mu^{\left[\frac{n}{2}\right]-M+2},\qquad (S4.132)$$

$$\left[\left(16\left[\frac{n}{2}\right]^4-32n\left[\frac{n}{2}\right]^3+24n^2\left[\frac{n}{2}\right]^2-8n^3\left[\frac{n}{2}\right]+n^4\right)\left[\frac{n}{2}\right]+\left(-8\left[\frac{n}{2}\right]^3+12n\left[\frac{n}{2}\right]^2-6n^2\left[\frac{n}{2}\right]+n^3\right)n+\left(-20\left[\frac{n}{2}\right]^3+28n\left[\frac{n}{2}\right]^2-13\,n^2\left[\frac{n}{2}\right]+2n^3\right)+\left(2\left[\frac{n}{2}\right]-n\right)n+\left(4\left[\frac{n}{2}\right]-2n\right)\right]\frac{1}{\left(\left(n-2\left[\frac{n}{2}\right]\right)+2\right)}\frac{1}{\left(\left(n-2\left[\frac{n}{2}\right]\right)+1\right)}\frac{1}{\left(\left[\frac{n}{2}\right]-M+2\right)}\frac{1}{\left(\left[\frac{n}{2}\right]-M+1\right)}\binom{2\left(n-\left[\frac{n}{2}\right]\right)}{n}\binom{n-M}{\left[\frac{n}{2}\right]-M}\mu^{\left[\frac{n}{2}\right]-M+2} \quad . \qquad \text{(S4.133)}$$

For $n$ even:

$$\left[\left(n^4-4n^4+6n^4-4n^4+n^4\right)\left[\frac{n}{2}\right]+\left(-n^3+3n^3-3n^3+n^3\right)n+\left(-\frac{20}{8}n^3+7n^3-\frac{13}{2}n^3+2n^3\right)+\left(n-n\right)n+\left(2n-2n\right)\right]\frac{1}{\left(\left(n-2\left[\frac{n}{2}\right]\right)+2\right)}\frac{1}{\left(\left(n-2\left[\frac{n}{2}\right]\right)+1\right)}\frac{1}{\left(\left[\frac{n}{2}\right]-M+2\right)}\frac{1}{\left(\left[\frac{n}{2}\right]-M+1\right)}\binom{2\left(n-\left[\frac{n}{2}\right]\right)}{n}\binom{n-M}{\left[\frac{n}{2}\right]-M}\mu^{\left[\frac{n}{2}\right]-M+2} \quad , \qquad \text{(S4.134)}$$

$$\left[(0)\left[\frac{n}{2}\right]+(0)n+\left(-\frac{5}{2}n^3+9n^3-\frac{13}{2}n^3\right)+(0)n+(0)\right]\frac{1}{\left(\left(n-2\left[\frac{n}{2}\right]\right)+2\right)}\frac{1}{\left(\left(n-2\left[\frac{n}{2}\right]\right)+1\right)}\frac{1}{\left(\left[\frac{n}{2}\right]-M+2\right)}\frac{1}{\left(\left[\frac{n}{2}\right]-M+1\right)}\binom{2\left(n-\left[\frac{n}{2}\right]\right)}{n}\binom{n-M}{\left[\frac{n}{2}\right]-M}\mu^{\left[\frac{n}{2}\right]-M+2} \quad , \qquad \text{(S4.135)}$$

$$\left[(0)\left[\frac{n}{2}\right]+(0)n+(0)+(0)n+(0)\right]\frac{1}{\left(\left(n-2\left[\frac{n}{2}\right]\right)+2\right)}\frac{1}{\left(\left(n-2\left[\frac{n}{2}\right]\right)+1\right)}\frac{1}{\left(\left[\frac{n}{2}\right]-M+2\right)}\frac{1}{\left(\left[\frac{n}{2}\right]-M+1\right)}\binom{2\left(n-\left[\frac{n}{2}\right]\right)}{n}\binom{n-M}{\left[\frac{n}{2}\right]-M}\mu^{\left[\frac{n}{2}\right]-M+2}=0 \quad . \qquad \text{(S4.136)}$$

For $n$ odd:

$$\left[\left(16\left(\frac{n-1}{2}\right)^4-32n\left(\frac{n-1}{2}\right)^3+24n^2\left(\frac{n-1}{2}\right)^2-8n^3\left(\frac{n-1}{2}\right)+n^4\right)\left(\frac{n-1}{2}\right)+\left(-8\left(\frac{n-1}{2}\right)^3+12n\left(\frac{n-1}{2}\right)^2-6n^2\left(\frac{n-1}{2}\right)+n^3\right)n+\left(-20\left(\frac{n-1}{2}\right)^3+28n\left(\frac{n-1}{2}\right)^2-13\,n^2\left(\frac{n-1}{2}\right)+2n^3\right)+\left(2\left(\frac{n-1}{2}\right)-n\right)n+\left(4\left(\frac{n-1}{2}\right)-2n\right)\right]\frac{1}{\left(\left(n-2\left[\frac{n}{2}\right]\right)+2\right)}\frac{1}{\left(\left(n-2\left[\frac{n}{2}\right]\right)+1\right)}\frac{1}{\left(\left[\frac{n}{2}\right]-M+2\right)}\frac{1}{\left(\left[\frac{n}{2}\right]-M+1\right)}\binom{2\left(n-\left[\frac{n}{2}\right]\right)}{n}\binom{n-M}{\left[\frac{n}{2}\right]-M}\mu^{\left[\frac{n}{2}\right]-M+2} \quad , \qquad \text{(S4.137)}$$

$$\left[\left((n-1)^4-4n(n-1)^3+6n^2(n-1)^2-4n^3(n-1)+n^4\right)\left(\frac{n-1}{2}\right)+\left(-(n-1)^3+3n(n-1)^2-3n^2(n-1)+n^3\right)n+\left(-\frac{20}{8}(n-1)^3+7n(n-1)^2-\frac{13}{2}n^2(n-1)+2n^3\right)+\left((n-1)-n\right)n+\left(2(n-1)-2n\right)\right]\frac{1}{\left(\left(n-2\left[\frac{n}{2}\right]\right)+2\right)}\frac{1}{\left(\left(n-2\left[\frac{n}{2}\right]\right)+1\right)}\frac{1}{\left(\left[\frac{n}{2}\right]-M+2\right)}\frac{1}{\left(\left[\frac{n}{2}\right]-M+1\right)}\binom{2\left(n-\left[\frac{n}{2}\right]\right)}{n}\binom{n-M}{\left[\frac{n}{2}\right]-M}\mu^{\left[\frac{n}{2}\right]-M+2} \quad , \qquad \text{(S4.138)}$$

$$\left[\left((n-1)-n\right)^4\left(\frac{n-1}{2}\right)+\left(n-(n-1)\right)^3n+\left(\left[-\frac{5}{2}(n-1)+7n\right](n-1)^2+\left[-\frac{13}{2}(n-1)+2n\right]n^2\right)+(-1)n+\left((2n-2)-2n\right)\right]\frac{1}{\left(\left(n-2\left[\frac{n}{2}\right]\right)+2\right)}\frac{1}{\left(\left(n-2\left[\frac{n}{2}\right]\right)+1\right)}\frac{1}{\left(\left[\frac{n}{2}\right]-M+2\right)}\frac{1}{\left(\left[\frac{n}{2}\right]-M+1\right)}\binom{2\left(n-\left[\frac{n}{2}\right]\right)}{n}\binom{n-M}{\left[\frac{n}{2}\right]-M}\mu^{\left[\frac{n}{2}\right]-M+2} \quad , \qquad \text{(S4.139)}$$

$$\left[(-1)^4\left(\frac{n-1}{2}\right)+(1)^3n+\left(\left[\left(-\frac{5}{2}n+\frac{5}{2}\right)+\frac{14}{2}n\right](n-1)^2+\left[\left(-\frac{13}{2}n+\frac{13}{2}\right)+\frac{4}{2}n\right]n^2\right)-n-2\right]\frac{1}{\left(\left(n-2\left[\frac{n}{2}\right]\right)+2\right)}\frac{1}{\left(\left(n-2\left[\frac{n}{2}\right]\right)+1\right)}\frac{1}{\left(\left[\frac{n}{2}\right]-M+2\right)}\frac{1}{\left(\left[\frac{n}{2}\right]-M+1\right)}\binom{2\left(n-\left[\frac{n}{2}\right]\right)}{n}\binom{n-M}{\left[\frac{n}{2}\right]-M}\mu^{\left[\frac{n}{2}\right]-M+2} \quad , \qquad \text{(S4.140)}$$

$$\left[\left(\frac{n-1}{2}\right)+\cancel{n}+\left(\left[\frac{5}{2}+\frac{9}{2}n\right]\left(n^2-2n+1\right)+\left[-\frac{9}{2}n+\frac{13}{2}\right]n^2\right)-\cancel{n}-2\right]\frac{1}{\left(\left(n-2\left[\frac{n}{2}\right]\right)+2\right)}\frac{1}{\left(\left(n-2\left[\frac{n}{2}\right]\right)+1\right)}\frac{1}{\left(\left[\frac{n}{2}\right]-M+2\right)}\frac{1}{\left(\left[\frac{n}{2}\right]-M+1\right)}\binom{2\left(n-\left[\frac{n}{2}\right]\right)}{n}\binom{n-M}{\left[\frac{n}{2}\right]-M}\mu^{\left[\frac{n}{2}\right]-M+2} \quad , \qquad \text{(S4.141)}$$

$$\left[\frac{n}{2}-\frac{1}{2}+\left(\left[\frac{5}{2}n^2+\cancel{\frac{9}{2}nn^2}-\frac{5}{2}2n-\frac{9}{2}n2n+\frac{5}{2}+\frac{9}{2}n\right]+\left[-\cancel{\frac{9}{2}nn^2}+\frac{13}{2}n^2\right]\right)-2\right]\frac{1}{\left(\left(n-2\left[\frac{n}{2}\right]\right)+2\right)}\frac{1}{\left(\left(n-2\left[\frac{n}{2}\right]\right)+1\right)}\frac{1}{\left(\left[\frac{n}{2}\right]-M+2\right)}\frac{1}{\left(\left[\frac{n}{2}\right]-M+1\right)}\binom{2\left(n-\left[\frac{n}{2}\right]\right)}{n}\binom{n-M}{\left[\frac{n}{2}\right]-M}\mu^{\left[\frac{n}{2}\right]-M+2} \quad , \qquad \text{(S4.142)}$$

$$\left[\frac{n}{2}-\cancel{\frac{5}{2}}+\left(\frac{5}{2}n^2-\frac{10}{2}n-\frac{18}{2}n^2+\cancel{\frac{5}{2}}+\frac{9}{2}n+\frac{13}{2}n^2\right)\right]\frac{1}{\left(\left(n-2\left[\frac{n}{2}\right]\right)+2\right)}\frac{1}{\left(\left(n-2\left[\frac{n}{2}\right]\right)+1\right)}\frac{1}{\left(\left[\frac{n}{2}\right]-M+2\right)}\frac{1}{\left(\left[\frac{n}{2}\right]-M+1\right)}\binom{2\left(n-\left[\frac{n}{2}\right]\right)}{n}\binom{n-M}{\left[\frac{n}{2}\right]-M}\mu^{\left[\frac{n}{2}\right]-M+2}, \qquad \text{(S4.143)}$$

$$[0]\frac{1}{\left(\left(n-2\left[\frac{n}{2}\right]\right)+2\right)}\frac{1}{\left(\left(n-2\left[\frac{n}{2}\right]\right)+1\right)}\frac{1}{\left(\left[\frac{n}{2}\right]-M+2\right)}\frac{1}{\left(\left[\frac{n}{2}\right]-M+1\right)}\binom{2\left(n-\left[\frac{n}{2}\right]\right)}{n}\binom{n-M}{\left[\frac{n}{2}\right]-M}\mu^{\left[\frac{n}{2}\right]-M+2}=0 \quad . \qquad \text{(S4.144)}$$

Also this term is thus zero for all possible $n$ (both even and odd) and $M$.

Power of $\mu$ equal to $\left[\frac{n}{2}\right]-M+3$:

$$\left[\left(4\left[\tfrac{n}{2}\right]+8M-4n-6\right)\left(\left[\tfrac{n}{2}\right]-M+1\right)\left(\left[\tfrac{n}{2}\right]-M+2\right)+\left\{-(n-n^2)\left(\left[\tfrac{n}{2}\right]-M+2\right)-(8n-12M+12)(M-1)\left(\left[\tfrac{n}{2}\right]-M+2\right)\right\}+(M-1)(4M^2-4Mn-14M+n^2+7n+12)\right]\frac{1}{\left[\frac{n}{2}\right]-M+2}\frac{1}{\left[\frac{n}{2}\right]-M+1}\binom{n-M}{\left[\frac{n}{2}\right]-M}\binom{2\left(n-\left[\frac{n}{2}\right]\right)}{n}\mu^{\left[\frac{n}{2}\right]-M+3}\quad, \tag{S4.145}$$

$$\left[\left(-8M^2+4Mn+4M\left[\tfrac{n}{2}\right]+14M-4n\left[\tfrac{n}{2}\right]-4n+4\left[\tfrac{n}{2}\right]^2-2\left[\tfrac{n}{2}\right]-6\right)\left(\left[\tfrac{n}{2}\right]-M+2\right)+(-n+n^2)\left(\left[\tfrac{n}{2}\right]-M+2\right)-\left(12M^2-8Mn-12M\left[\tfrac{n}{2}\right]-36M+8n\left[\tfrac{n}{2}\right]+16n+12\left[\tfrac{n}{2}\right]+24\right)(M-1)+(M-1)(4M^2-4Mn-14M+n^2+7n+12)\right]\frac{1}{\left[\frac{n}{2}\right]-M+2}\frac{1}{\left[\frac{n}{2}\right]-M+1}\binom{n-M}{\left[\frac{n}{2}\right]-M}\binom{2\left(n-\left[\frac{n}{2}\right]\right)}{n}\mu^{\left[\frac{n}{2}\right]-M+3}, \tag{S4.146}$$

$$\left[\left(-8M^2+4Mn+4M\left[\tfrac{n}{2}\right]+14M-4n\left[\tfrac{n}{2}\right]-5n+n^2+4\left[\tfrac{n}{2}\right]^2-2\left[\tfrac{n}{2}\right]-6\right)\left(\left[\tfrac{n}{2}\right]-M+2\right)+(M-1)\left(-8M^2+4Mn+22M+n^2-9n-12+12M\left[\tfrac{n}{2}\right]-8n\left[\tfrac{n}{2}\right]-12\left[\tfrac{n}{2}\right]\right)\right]\frac{1}{\left[\frac{n}{2}\right]-M+2}\frac{1}{\left[\frac{n}{2}\right]-M+1}\binom{n-M}{\left[\frac{n}{2}\right]-M}\binom{2\left(n-\left[\frac{n}{2}\right]\right)}{n}\mu^{\left[\frac{n}{2}\right]-M+3}\quad, \tag{S4.147}$$

$$\left[\left(8M^3-4M^2n-12M^2\left[\tfrac{n}{2}\right]-30M^2-Mn^2+8Mn\left[\tfrac{n}{2}\right]+13Mn+24M\left[\tfrac{n}{2}\right]+34M+n^2\left[\tfrac{n}{2}\right]+2n^2-4n\left[\tfrac{n}{2}\right]^2-13n\left[\tfrac{n}{2}\right]-10n+4\left[\tfrac{n}{2}\right]^3+6\left[\tfrac{n}{2}\right]^2-10\left[\tfrac{n}{2}\right]-12\right)+\left(-8M^3+4M^2n+12M^2\left[\tfrac{n}{2}\right]+30M^2+Mn^2-8Mn\left[\tfrac{n}{2}\right]-13Mn-24M\left[\tfrac{n}{2}\right]-34M-n^2+8n\left[\tfrac{n}{2}\right]+9n+12\left[\tfrac{n}{2}\right]+12\right)\right]\frac{1}{\left[\frac{n}{2}\right]-M+2}\frac{1}{\left[\frac{n}{2}\right]-M+1}\binom{n-M}{\left[\frac{n}{2}\right]-M}\binom{2\left(n-\left[\frac{n}{2}\right]\right)}{n}\mu^{\left[\frac{n}{2}\right]-M+3}, \tag{S4.148}$$

$$\left[\left(\cancel{8M^3}-\cancel{4M^2n}-\cancel{12M^2\left[\tfrac{n}{2}\right]}-\cancel{30M^2}-\cancel{Mn^2}+\cancel{8Mn\left[\tfrac{n}{2}\right]}+\cancel{13Mn}+\cancel{24M\left[\tfrac{n}{2}\right]}+\cancel{34M}+n^2\left[\tfrac{n}{2}\right]+2n^2-4n\left[\tfrac{n}{2}\right]^2-13n\left[\tfrac{n}{2}\right]-10n+4\left[\tfrac{n}{2}\right]^3+6\left[\tfrac{n}{2}\right]^2-10\left[\tfrac{n}{2}\right]-\cancel{12}\right)+\left(-\cancel{8M^3}+\cancel{4M^2n}+\cancel{12M^2\left[\tfrac{n}{2}\right]}+\cancel{30M^2}+\cancel{Mn^2}-\cancel{8Mn\left[\tfrac{n}{2}\right]}-\cancel{13Mn}-\cancel{24M\left[\tfrac{n}{2}\right]}-\cancel{34M}-n^2+8n\left[\tfrac{n}{2}\right]+9n+12\left[\tfrac{n}{2}\right]+\cancel{12}\right)\right]\frac{1}{\left[\frac{n}{2}\right]-M+2}\frac{1}{\left[\frac{n}{2}\right]-M+1}\binom{n-M}{\left[\frac{n}{2}\right]-M}\binom{2\left(n-\left[\frac{n}{2}\right]\right)}{n}\mu^{\left[\frac{n}{2}\right]-M+3}, \tag{S4.149}$$

$$\left[\left(+n^2\left[\tfrac{n}{2}\right]+2n^2-4n\left[\tfrac{n}{2}\right]^2-13n\left[\tfrac{n}{2}\right]-10n+4\left[\tfrac{n}{2}\right]^3+6\left[\tfrac{n}{2}\right]^2-10\left[\tfrac{n}{2}\right]\right)+\left(-n^2+8n\left[\tfrac{n}{2}\right]+9n+12\left[\tfrac{n}{2}\right]\right)\right]\frac{1}{\left[\frac{n}{2}\right]-M+2}\frac{1}{\left[\frac{n}{2}\right]-M+1}\binom{n-M}{\left[\frac{n}{2}\right]-M}\binom{2\left(n-\left[\frac{n}{2}\right]\right)}{n}\mu^{\left[\frac{n}{2}\right]-M+3}\quad, \tag{S4.150}$$

$$\left[\left(+n^2\left[\tfrac{n}{2}\right]+n^2-4n\left[\tfrac{n}{2}\right]^2-5n\left[\tfrac{n}{2}\right]-n+4\left[\tfrac{n}{2}\right]^3+6\left[\tfrac{n}{2}\right]^2+2\left[\tfrac{n}{2}\right]\right)\right]\frac{1}{\left[\frac{n}{2}\right]-M+2}\frac{1}{\left[\frac{n}{2}\right]-M+1}\binom{n-M}{\left[\frac{n}{2}\right]-M}\binom{2\left(n-\left[\frac{n}{2}\right]\right)}{n}\mu^{\left[\frac{n}{2}\right]-M+3}\quad. \tag{S4.151}$$

For $n$ even:

$$\left[\left(+\frac{n^3}{2}+n^2-n^3-n^2\frac{5}{2}-n+n^3\frac{1}{2}+n^2\frac{6}{4}+n\right)\right]\frac{1}{\left[\frac{n}{2}\right]-M+2}\frac{1}{\left[\frac{n}{2}\right]-M+1}\binom{n-M}{\left[\frac{n}{2}\right]-M}\binom{2\left(n-\left[\frac{n}{2}\right]\right)}{n}\mu^{\left[\frac{n}{2}\right]-M+3}\quad, \tag{S4.152}$$

$$\left[\left(+\frac{n^3}{2}+n^2-n^3-n^2\frac{5}{2}-\cancel{n}+n^3\frac{1}{2}+n^2\frac{3}{2}+\cancel{n}\right)\right]\frac{1}{\left[\frac{n}{2}\right]-M+2}\frac{1}{\left[\frac{n}{2}\right]-M+1}\binom{n-M}{\left[\frac{n}{2}\right]-M}\binom{2\left(n-\left[\frac{n}{2}\right]\right)}{n}\mu^{\left[\frac{n}{2}\right]-M+3}\quad, \tag{S4.153}$$

$$[0]\frac{1}{\left[\frac{n}{2}\right]-M+2}\frac{1}{\left[\frac{n}{2}\right]-M+1}\binom{n-M}{\left[\frac{n}{2}\right]-M}\binom{2\left(n-\left[\frac{n}{2}\right]\right)}{n}\mu^{\left[\frac{n}{2}\right]-M+3}=0\quad. \tag{S4.154}$$

For $n$ odd:

$$\left[\left(+n^2\frac{n-1}{2}+n^2-4n\left(\frac{n-1}{2}\right)^2-5n\frac{n-1}{2}-n+4\left(\frac{n-1}{2}\right)^3+6\left(\frac{n-1}{2}\right)^2+2\frac{n-1}{2}\right)\right]\frac{1}{\left[\frac{n}{2}\right]-M+2}\frac{1}{\left[\frac{n}{2}\right]-M+1}\binom{n-M}{\left[\frac{n}{2}\right]-M}\binom{2\left(n-\left[\frac{n}{2}\right]\right)}{n}\mu^{\left[\frac{n}{2}\right]-M+3}\quad, \tag{S4.155}$$

$$\left[\left(+n^2\frac{n-1}{2}+n^2-n(n-1)^2-5n\frac{n-1}{2}-n+\frac{1}{2}(n-1)^3+\frac{3}{2}(n-1)^2+(n-1)\right)\right]\frac{1}{\left[\frac{n}{2}\right]-M+2}\frac{1}{\left[\frac{n}{2}\right]-M+1}\binom{n-M}{\left[\frac{n}{2}\right]-M}\binom{2\left(n-\left[\frac{n}{2}\right]\right)}{n}\mu^{\left[\frac{n}{2}\right]-M+3}\quad, \tag{S4.156}$$

$$\left[\left(+n^2\frac{n-1}{2}+(n^2-n)-n(n-1)^2-5n\frac{n-1}{2}+\frac{1}{2}(n-1)^3+\frac{3}{2}(n-1)^2+(n-1)\right)\right]\frac{1}{\left[\frac{n}{2}\right]-M+2}\frac{1}{\left[\frac{n}{2}\right]-M+1}\binom{n-M}{\left[\frac{n}{2}\right]-M}\binom{2\left(n-\left[\frac{n}{2}\right]\right)}{n}\mu^{\left[\frac{n}{2}\right]-M+3}\quad, \tag{S4.157}$$

$$\left[\left(+\frac{1}{2}n^2+n-n(n-1)-\frac{5}{2}n+\frac{1}{2}(n-1)^2+\frac{3}{2}(n-1)+1\right)(n-1)\right]\frac{1}{\left[\frac{n}{2}\right]-M+2}\frac{1}{\left[\frac{n}{2}\right]-M+1}\binom{n-M}{\left[\frac{n}{2}\right]-M}\binom{2\left(n-\left[\frac{n}{2}\right]\right)}{n}\mu^{\left[\frac{n}{2}\right]-M+3}\quad, \tag{S4.158}$$

$$\left[\left(+\frac{1}{2}n^2-(n^2-n)-\frac{3}{2}n+\frac{1}{2}(n^2-2n+1)+\left(\frac{3}{2}n-\frac{3}{2}\right)+1\right)(n-1)\right]\frac{1}{\left[\frac{n}{2}\right]-M+2}\frac{1}{\left[\frac{n}{2}\right]-M+1}\binom{n-M}{\left[\frac{n}{2}\right]-M}\binom{2\left(n-\left[\frac{n}{2}\right]\right)}{n}\mu^{\left[\frac{n}{2}\right]-M+3}\quad, \tag{S4.159}$$

$$\left[\left(+\frac{1}{2}n^2-n^2+n-\frac{3}{2}n+\left(\frac{1}{2}n^2-n+\frac{1}{2}\right)+\frac{3}{2}n-\frac{3}{2}+1\right)(n-1)\right]\frac{1}{\left[\frac{n}{2}\right]-M+2}\frac{1}{\left[\frac{n}{2}\right]-M+1}\binom{n-M}{\left[\frac{n}{2}\right]-M}\binom{2\left(n-\left[\frac{n}{2}\right]\right)}{n}\mu^{\left[\frac{n}{2}\right]-M+3}\quad, \tag{S4.160}$$

$$[(0)(n-1)]\frac{1}{\left[\frac{n}{2}\right]-M+2}\frac{1}{\left[\frac{n}{2}\right]-M+1}\binom{n-M}{\left[\frac{n}{2}\right]-M}\binom{2\left(n-\left[\frac{n}{2}\right]\right)}{n}\mu^{\left[\frac{n}{2}\right]-M+3}=0\quad. \tag{S4.161}$$

This term is thus zero as well, for all possible *n* (both even and odd) and *M*.

Then, all the individual terms in (S4.71) are zero for any non-negative *n* and any *M* in the range from 1 to $[n/2]$.

Now, it remains to prove the same for the part of the overall expression containing sums, see (S4.68). The terms in the part with sums are first put to one sum (as the sums have the same range and the same exponent of $\mu$):

$$\sum_{j=3}^{\left[\frac{n}{2}\right]-M}\left[4(n-M)\binom{n-M-1}{j-1}\binom{2(n-M-j)}{n}+8(n-M)\binom{n-M-1}{j-2}\binom{2(n-M-j)+2}{n}+4(n-M)\binom{n-M-1}{j-3}\binom{2(n-M-j)+4}{n}+2(6M-2n-3)\binom{n-M}{j-2}\binom{2(n-M-j)+4}{n}+4(5M-2n-3)\binom{n-M}{j-1}\binom{2(n-M-j)+2}{n}+2(4M-2n-3)\binom{n-M}{j}\binom{2(n-M-j)}{n}-\frac{(1-n)n+4(2n-3(M-1))(M-1)}{n-M+1}\binom{n-M+1}{j-1}\binom{2(n-M-j)+4}{n}-\frac{(1-n)n+2(6n-8M+9)(M-1)}{n-M+1}\binom{n-M+1}{j}\binom{2(n-M-j)+2}{n}-\frac{2(2n-2M+3)(M-1)}{n-M+1}\binom{n-M+1}{j+1}\binom{2(n-M-j)}{n}+(M-1)\frac{(n-2M+4)}{n-M+2}\frac{(n-2M+3)}{n-M+1}\binom{n-M+2}{j+1}\binom{2(n-M-j)+2}{n}+(M-1)\frac{(n-2M+4)}{n-M+2}\frac{(n-2M+3)}{n-M+1}\binom{n-M+2}{j}\binom{2(n-M-j)+4}{n}\right]\mu^{j+1}\quad. \tag{S4.162}$$

The terms are grouped according to the second binomial coefficient:

$$\sum_{j=3}^{\left[\frac{n}{2}\right]-M}\left[4(n-M)\binom{n-M-1}{j-1}\binom{2(n-M-j)}{n}+2(4M-2n-3)\binom{n-M}{j}\binom{2(n-M-j)}{n}-\frac{2(2n-2M+3)(M-1)}{n-M+1}\binom{n-M+1}{j+1}\binom{2(n-M-j)}{n}+8(n-M)\binom{n-M-1}{j-2}\binom{2(n-M-j)+2}{n}+4(5M-2n-3)\binom{n-M}{j-1}\binom{2(n-M-j)+2}{n}-\frac{(1-n)n+2(6n-8M+9)(M-1)}{n-M+1}\binom{n-M+1}{j}\binom{2(n-M-j)+2}{n}+(M-1)\frac{(n-2M+4)}{n-M+2}\frac{(n-2M+3)}{n-M+1}\binom{n-M+2}{j+1}\binom{2(n-M-j)+2}{n}+4(n-M)\binom{n-M-1}{j-3}\binom{2(n-M-j)+4}{n}+2(6M-2n-3)\binom{n-M}{j-2}\binom{2(n-M-j)+4}{n}-\frac{(1-n)n+4(2n-3(M-1))(M-1)}{n-M+1}\binom{n-M+1}{j-1}\binom{2(n-M-j)+4}{n}+(M-1)\frac{(n-2M+4)}{n-M+2}\frac{(n-2M+3)}{n-M+1}\binom{n-M+2}{j}\binom{2(n-M-j)+4}{n}\right]\mu^{j+1}\quad. \tag{S4.163}$$

The basic binomial identities (S4.78), (S4.79), (S4.80) are applied to some of the binomial coefficients in order to obtain the same binomial coefficients (the first one of the two subsequent ones) in all the terms:

$$\sum_{j=3}^{\left[\frac{n}{2}\right]-M}\left[\left\{4(n-M)\frac{j}{n-M}\binom{n-M}{j}+2(4M-2n-3)\binom{n-M}{j}-\frac{2(2n-2M+3)(M-1)}{n-M+1}\frac{n-M+1}{j+1}\binom{n-M}{j}\right\}\binom{2(n-M-j)}{n}+\left\{8(n-M)\frac{j-1}{n-M}\frac{j}{n-M-j+1}\binom{n-M}{j}+4(5M-2n-3)\frac{j}{n-M-j+1}\binom{n-M}{j}-\frac{(1-n)n+2(6n-8M+9)(M-1)}{n-M+1}\frac{n-M+1}{n-M+1-j}\binom{n-M}{j}+(M-1)\frac{(n-2M+4)}{n-M+2}\frac{(n-2M+3)}{n-M+1}\frac{n-M+1+1}{j+1}\frac{n-M+1}{n-M+1-j}\binom{n-M}{j}\right\}\binom{2(n-M-j)+2}{n}+\left\{4(n-M)\frac{j-2}{n-M}\frac{j-1}{n-M-(j-1)+1}\frac{j}{n-M-j+1}\binom{n-M}{j}+2(6M-2n-3)\frac{j-1}{n-M-(j-1)+1}\frac{j}{n-M-j+1}\binom{n-M}{j}-\frac{(1-n)n+4(2n-3(M-1))(M-1)}{n-M+1}\frac{n-M+1}{n-M+1-(j-1)}\frac{j}{n-M-j+1}\binom{n-M}{j}+(M-1)\frac{(n-2M+4)}{n-M+2}\frac{(n-2M+3)}{n-M+1}\frac{n-M+1+1}{n-M+1+1-j}\frac{n-M+1}{n-M+1-j}\binom{n-M}{j}\right\}\binom{2(n-M-j)+4}{n}\right]\mu^{j+1}\quad, \tag{S4.164}$$

The part with the sum is then simplified thanks to cancelation of some terms in the numerators and denominators:

$$\sum_{j=3}^{\left[\frac{n}{2}\right]-M}\left[\left\{4j+2(4M-2n-3)-\frac{2(2n-2M+3)(M-1)}{j+1}\right\}\binom{n-M}{j}\binom{2(n-M-j)}{n}+\left\{8(j-1)\frac{j}{n-M-j+1}+4(5M-2n-3)\frac{j}{n-M-j+1}-\frac{(1-n)n+2(6n-8M+9)(M-1)}{n-M+1-j}+(M-1)\frac{(n-2M+4)}{j+1}\frac{(n-2M+3)}{n-M+1-j}\right\}\binom{n-M}{j}\binom{2(n-M-j)+2}{n}+\left\{4(j-2)\frac{j-1}{n-M-j+2}\frac{j}{n-M-j+1}+2(6M-2n-\right.\right.$$

$$3)\frac{j-1}{n-M-j+2}\frac{j}{n-M-j+1}-\frac{(1-n)n+4(2n-3(M-1))(M-1)}{n-M-j+2}\frac{j}{n-M-j+1}+(M-1)\frac{(n-2M+4)}{n-M+2-j}\frac{(n-2M+3)}{n-M+1-j}\Big\}\binom{n-M}{j}\binom{2(n-M-j)+4}{n}\Big]\mu^{j+1} \quad . \tag{S4.165}$$

The part with the sum is further simplified and binomial identity (S4.78) is used at some occurrences in the second binomial coefficient of the two subsequent ones:

$$\sum_{j=3}^{\left[\frac{n}{2}\right]-M}\Big[\Big\{4j+2(4M-2n-3)-\frac{2(2n-2M+3)(M-1)}{j+1}\Big\}\binom{2(n-M-j)}{n}+\frac{1}{n-M-j+1}\Big\{8(j-1)j+4(5M-2n-3)j-[(1-n)n+2(6n-8M+9)(M-1)]+(M-1)\frac{(n-2M+4)}{j+1}(n-2M+3)\Big\}\frac{2(n-M-j)+1+1}{2(n-M-j)+1+1-n}\frac{2(n-M-j)+1}{2(n-M-j)+1-n}\binom{2(n-M-j)}{n}+\Big\{4(j-2)\frac{j-1}{n-M-j+2}\frac{j}{n-M-j+1}+2(6M-2n-3)\frac{j-1}{n-M-j+2}\frac{j}{n-M-j+1}-\frac{(1-n)n+4(2n-3(M-1))(M-1)}{n-M-j+2}\frac{j}{n-M-j+1}+(M-1)\frac{(n-2M+4)}{n-M+2-j}\frac{(n-2M+3)}{n-M+1-j}\Big\}\frac{2(n-M-j)+3+1}{2(n-M-j)+3+1-n}\frac{2(n-M-j)+2+1}{2(n-M-j)+2+1-n}\frac{2(n-M-j)+1+1}{2(n-M-j)+1+1-n}\frac{2(n-M-j)+1}{2(n-M-j)+1-n}\binom{2(n-M-j)}{n}\Big]\binom{n-M}{j}\mu^{j+1} \quad . \tag{S4.166}$$

The part with sum is further simplified by factoring out the binomial coefficient and further cancellations of parts of the numerators and denominators:

$$\sum_{j=3}^{\left[\frac{n}{2}\right]-M}\Big[\Big\{4j+2(4M-2n-3)-\frac{2(2n-2M+3)(M-1)}{j+1}\Big\}+\frac{1}{\cancel{n-M-j+1}}\Big\{8(j-1)j+4(5M-2n-3)j-[(1-n)n+2(6n-8M+9)(M-1)]+(M-1)\frac{(n-2M+4)}{j+1}(n-2M+3)\Big\}\frac{2\cancel{(n-M-j+1)}}{(n-2M-2j)+2}\frac{2(n-M-j)+1}{(n-2M-2j)+1}+\Big\{4(j-2)\frac{j-1}{\cancel{n-M-j+2}}\frac{j}{\cancel{n-M-j+1}}+2(6M-2n-3)\frac{j-1}{\cancel{n-M-j+2}}\frac{j}{\cancel{n-M-j+1}}-\frac{(1-n)n+4(2n-3(M-1))(M-1)}{\cancel{n-M-j+2}}\frac{j}{\cancel{n-M-j+1}}+(M-1)\frac{(n-2M+4)}{\cancel{n-M+2-j}}\frac{(n-2M+3)}{\cancel{n-M+1-j}}\Big\}\frac{2\cancel{(n-M-j+2)}}{(n-2M-2j)+4}\frac{2(n-M-j)+3}{(n-2M-2j)+3}\frac{2\cancel{(n-M-j+1)}}{(n-2M-2j)+2}\frac{2(n-M-j)+1}{(n-2M-2j)+1}\Big]\binom{2(n-M-j)}{n}\binom{n-M}{j}\mu^{j+1} \quad . \tag{S4.167}$$

Still continue with simplification of the part with the sum. Factor out $\frac{2}{(n-2M-2j)+2}\frac{1}{(n-2M-2j)+1}$:

$$\sum_{j=3}^{\left[\frac{n}{2}\right]-M}\Big[\Big\{4j+2(4M-2n-3)-\frac{2(2n-2M+3)(M-1)}{j+1}\Big\}\frac{((n-2M-2j)+2)}{2}((n-2M-2j)+1)+\Big\{8(j-1)j+4(5M-2n-3)j-[(1-n)n+2(6n-8M+9)(M-1)]+(M-1)\frac{(n-2M+4)}{j+1}(n-2M+3)\Big\}(2(n-M-j)+1)+\Big\{4(j-2)(j-1)j+2(6M-2n-3)(j-1)j-[(1-n)n+4(2n-3(M-1))(M-1)]j+(M-1)(n-2M+4)(n-2M+3)\Big\}\frac{2}{(n-2M-2j)+4}\frac{2(n-M-j)+3}{(n-2M-2j)+3}(2(n-M-j)+1)\Big]\frac{2}{(n-2M-2j)+2}\frac{1}{(n-2M-2j)+1}\binom{2(n-M-j)}{n}\binom{n-M}{j}\mu^{j+1} \quad . \tag{S4.168}$$

Further, factor out $\frac{1}{j+1}$:

$$\sum_{j=3}^{\left[\frac{n}{2}\right]-M}\Big[\{42j(j+1)+\cancel{2}(j+1)(4M-2n-3)-\cancel{2}(2n-2M+3)(M-1)\}\frac{((n-2M-2j)+2)}{\cancel{2}}((n-2M-2j)+1)+\{8(j-1)j(j+1)+4(5M-2n-3)j(j+1)-[(1-n)n+2(6n-8M+9)(M-1)](j+1)+(M-1)(n-2M+4)(n-2M+3)\}(2(n-M-j)+1)+\{4(j-2)(j-1)j(j+1)+2(6M-2n-3)(j-1)j(j+1)-[(1-n)n+4(2n-3(M-1))(M-1)]j(j+1)+(j+1)(M-1)(n-2M+4)(n-2M+3)\}\frac{2}{(n-2M-2j)+4}\frac{2(n-M-j)+3}{(n-2M-2j)+3}(2(n-M-j)+1)\Big]\frac{1}{j+1}\frac{2}{(n-2M-2j)+2}\frac{1}{(n-2M-2j)+1}\binom{2(n-M-j)}{n}\binom{n-M}{j}\mu^{j+1} \quad . \tag{S4.169}$$

Finally, factor out also $\frac{2}{((n-2M-2j)+4)}\frac{1}{((n-2M-2j)+3)}$:

$$\sum_{j=3}^{\left[\frac{n}{2}\right]-M}\Big[\{2j(j+1)+(j+1)(4M-2n-3)-(2n-2M+3)(M-1)\}((n-2M-2j)+2)((n-2M-2j)+1)((n-2M-2j)+3)\frac{((n-2M-2j)+4)}{2}+\{8(j-1)j(j+1)+4(5M-2n-3)j(j+1)-[(1-n)n+2(6n-8M+9)(M-1)](j+1)+(M-1)(n-2M+4)(n-2M+3)\}(2(n-M-j)+1)((n-2M-2j)+3)\frac{((n-2M-2j)+4)}{2}+\{4(j-2)(j-1)j(j+1)+2(6M-2n-3)(j-1)j(j+1)-[(1-n)n+4(2n-3(M-1))(M-1)]j(j+1)+(j+1)(M-1)(n-2M+4)(n-2M+3)\}(2(n-M-j)+3)(2(n-M-j)+1)\Big]\frac{1}{j+1}\frac{2}{((n-2M-2j)+4)}\frac{1}{((n-2M-2j)+3)}\frac{2}{(n-2M-2j)+2}\frac{1}{(n-2M-2j)+1}\binom{2(n-M-j)}{n}\binom{n-M}{j}\mu^{j+1} \quad . \tag{S4.170}$$

The expression inside the sum (inside the angle brackets) is composed basically of three parts. The first part of (S4.170) expanded equals to:

$$16j^6+96j^5M-48j^5n-88j^5+240j^4M^2-240j^4Mn-440j^4M+56j^4n^2+216j^4n+180j^4+320j^3M^3-480j^3M^2n-880j^3M^2+224j^3Mn^2+864j^3Mn+720j^3M-32j^3n^3-192j^3n^2-340j^3n-170j^3+240j^2M^4-480j^2M^3n-880j^2M^3+336j^2M^2n^2+1296j^2M^2n+1080j^2M^2-96j^2Mn^3-576j^2Mn^2-1020j^2Mn-510j^2M+9j^2n^4+74j^2n^3+205j^2n^2+220j^2n+74j^2+96jM^5-240jM^4n-440jM^4+$$

$$224\,j\,M^3n^2 + 864\,j\,M^3n + 720\,j\,M^3 - 96\,j\,M^2n^3 - 576\,j\,M^2n^2 - 1020\,j\,M^2n - 510\,j\,M^2 + 18\,j\,M\,n^4 + 148\,j\,M\,n^3 + 410\,j\,M\,n^2 + 440\,j\,M\,n + 148\,j\,M - j\,n^5 - \frac{21\,j\,n^4}{2} - 40\,j\,n^3 - \frac{135\,j\,n^2}{2} - 49\,j\,n - 12\,j + 16\,M^6 - 48\,M^5n - 88\,M^5 + 56\,M^4n^2 + 216\,M^4n + 180\,M^4 - 32\,M^3n^3 - 192\,M^3n^2 - 340\,M^3n - 170\,M^3 + 9\,M^2n^4 + 74\,M^2n^3 + 205\,M^2n^2 + 220\,M^2n + 74\,M^2 - M\,n^5 - \frac{21\,M\,n^4}{2} - 40\,M\,n^3 - \frac{135\,M\,n^2}{2} - 49\,M\,n - 12\,M. \qquad \text{(S4.171)}$$

The second part expanded equals to:

$$-32\,j^6 - 176\,j^5M + 96\,j^5n + 176\,j^5 - 400\,j^4M^2 + 432\,j^4M\,n + 776\,j^4M - 108\,j^4n^2 - 420\,j^4n - 336\,j^4 - 480\,j^3M^3 + 768\,j^3M^2n + 1344\,j^3M^2 - 380\,j^3M\,n^2 - 1428\,j^3M\,n - 1084\,j^3M + 56\,j^3n^3 + 344\,j^3n^2 + 568\,j^3n + 220\,j^3 - 320\,j^2M^4 + 672\,j^2M^3n + 1136\,j^2M^3 - 492\,j^2M^2n^2 - 1764\,j^2M^2n - 1236\,j^2M^2 + 144\,j^2M\,n^3 + 828\,j^2M\,n^2 + 1224\,j^2M\,n + 350\,j^2M - 13\,j^2n^4 - 110\,j^2n^3 - 260\,j^2n^2 - 133\,j^2n + 62\,j^2 - 112\,j\,M^5 + 288\,j\,M^4n + 464\,j\,M^4 - 276\,j\,M^3n^2 - 924\,j\,M^3n - 564\,j\,M^3 + 120\,j\,M^2n^3 + 624\,j\,M^2n^2 + 744\,j\,M^2n + 40\,j\,M^2 - 22\,j\,M\,n^4 - 160\,j\,M\,n^3 - 261\,j\,M\,n^2 + 127\,j\,M\,n + 298\,j\,M + j\,n^5 + \frac{21\,j\,n^4}{2} + 16\,j\,n^3 - \frac{169\,j\,n^2}{2} - 227\,j\,n - 126\,j - 16\,M^6 + 48\,M^5n + 72\,M^5 - 56\,M^4n^2 - 168\,M^4n - 76\,M^4 + 32\,M^3n^3 + 140\,M^3n^2 + 88\,M^3n - 90\,M^3 - 9\,M^2n^4 - 50\,M^2n^3 - M^2n^2 + 260\,M^2n + 236\,M^2 + M\,n^5 + \frac{13\,M\,n^4}{2} - 20\,M\,n^3 - \frac{383\,M\,n^2}{2} - 344\,M\,n - 162\,M + 4\,n^4 + 36\,n^3 + 107\,n^2 + 117\,n + 36 \quad . \qquad \text{(S4.172)}$$

The third part expanded equals to:

$$16\,j^6 + 80\,j^5M - 48\,j^5n - 88\,j^5 + 160\,j^4M^2 - 192\,j^4M\,n - 336\,j^4M + 52\,j^4n^2 + 204\,j^4n + 156\,j^4 + 160\,j^3M^3 - 288\,j^3M^2n - 464\,j^3M^2 + 156\,j^3M\,n^2 + 564\,j^3M\,n + 364\,j^3M - 24\,j^3n^3 - 152\,j^3n^2 - 228\,j^3n - 50\,j^3 + 80\,j^2M^4 - 192\,j^2M^3n - 256\,j^2M^3 + 156\,j^2M^2n^2 + 468\,j^2M^2n + 156\,j^2M^2 - 48\,j^2M\,n^3 - 252\,j^2M\,n^2 - 204\,j^2M\,n + 160\,j^2M + 4\,j^2n^4 + 36\,j^2n^3 + 55\,j^2n^2 - 87\,j^2n - 136\,j^2 + 16\,j\,M^5 - 48\,j\,M^4n - 24\,j\,M^4 + 52\,j\,M^3n^2 + 60\,j\,M^3n - 156\,j\,M^3 - 24\,j\,M^2n^3 - 48\,j\,M^2n^2 + 276\,j\,M^2n + 470\,j\,M^2 + 4\,j\,M\,n^4 + 12\,j\,M\,n^3 - 149\,j\,M\,n^2 - 567\,j\,M\,n - 446\,j\,M + 24\,j\,n^3 + 152\,j\,n^2 + 276\,j\,n + 138\,j + 16\,M^5 - 48\,M^4n - 104\,M^4 + 52\,M^3n^2 + 252\,M^3n + 260\,M^3 - 24\,M^2n^3 - 204\,M^2n^2 - 480\,M^2n - 310\,M^2 + 4\,M\,n^4 + 60\,M\,n^3 + 259\,M\,n^2 + 393\,M\,n + 174\,M - 4\,n^4 - 36\,n^3 - 107\,n^2 - 117\,n - 36 \quad . \qquad \text{(S4.173)}$$

The parts are combined together. With the terms mutually re-combined, they give the following result:

$$16\,j^6 + 96\,j^5M - 48\,j^5n - 88\,j^5 + 240\,j^4M^2 - 240\,j^4M\,n - 440\,j^4M + 56\,j^4n^2 + 216\,j^4n + 180\,j^4 + 320\,j^3M^3 - 480\,j^3M^2n - 880\,j^3M^2 + 224\,j^3M\,n^2 + 864\,j^3M\,n + 720\,j^3M - 32\,j^3n^3 - 192\,j^3n^2 - 340\,j^3n - 170\,j^3 + 240\,j^2M^4 - 480\,j^2M^3n - 880\,j^2M^3 + 336\,j^2M^2n^2 + 1296\,j^2M^2n + 1080\,j^2M^2 - 96\,j^2M\,n^3 - 576\,j^2M\,n^2 - 1020\,j^2M\,n - 510\,j^2M + 9\,j^2n^4 + 74\,j^2n^3 + 205\,j^2n^2 + 220\,j^2n + 74\,j^2 + 96\,j\,M^5 - 240\,j\,M^4n - 440\,j\,M^4 + 224\,j\,M^3n^2 + 864\,j\,M^3n + 720\,j\,M^3 - 96\,j\,M^2n^3 - 576\,j\,M^2n^2 - 1020\,j\,M^2n - 510\,j\,M^2 + 18\,j\,M\,n^4 + 148\,j\,M\,n^3 + 410\,j\,M\,n^2 + 440\,j\,M\,n + 148\,j\,M - j\,n^5 - \frac{21\,j\,n^4}{2} - 40\,j\,n^3 - \frac{135\,j\,n^2}{2} - 49\,j\,n - 12\,j + 16\,M^6 - 48\,M^5n - 88\,M^5 + 56\,M^4n^2 + 216\,M^4n + 180\,M^4 - 32\,M^3n^3 - 192\,M^3n^2 - 340\,M^3n - 170\,M^3 + 9\,M^2n^4 + 74\,M^2n^3 + 205\,M^2n^2 + 220\,M^2n + 74\,M^2 - M\,n^5 - \frac{21\,M\,n^4}{2} - 40\,M\,n^3 - \frac{135\,M\,n^2}{2} - 49\,M\,n - 12\,M + -32\,j^6 - 176\,j^5M + 96\,j^5n + 176\,j^5 - 400\,j^4M^2 + 432\,j^4M\,n + 776\,j^4M - 108\,j^4n^2 - 420\,j^4n - 336\,j^4 - 480\,j^3M^3 + 768\,j^3M^2n + 1344\,j^3M^2 - 380\,j^3M\,n^2 - 1428\,j^3M\,n - 1084\,j^3M + 56\,j^3n^3 + 344\,j^3n^2 + 568\,j^3n + 220\,j^3 - 320\,j^2M^4 + 672\,j^2M^3n + 1136\,j^2M^3 - 492\,j^2M^2n^2 - 1764\,j^2M^2n - 1236\,j^2M^2 + 144\,j^2M\,n^3 + 828\,j^2M\,n^2 + 1224\,j^2M\,n + 350\,j^2M - 13\,j^2n^4 - 110\,j^2n^3 - 260\,j^2n^2 - 133\,j^2n + 62\,j^2 - 112\,j\,M^5 + 288\,j\,M^4n + 464\,j\,M^4 - 276\,j\,M^3n^2 - 924\,j\,M^3n - 564\,j\,M^3 + 120\,j\,M^2n^3 + 624\,j\,M^2n^2 + 744\,j\,M^2n + 40\,j\,M^2 - 22\,j\,M\,n^4 - 160\,j\,M\,n^3 - 261\,j\,M\,n^2 + 127\,j\,M\,n + 298\,j\,M + j\,n^5 + \frac{21\,j\,n^4}{2} + 16\,j\,n^3 - \frac{169\,j\,n^2}{2} - 227\,j\,n - 126\,j - 16\,M^6 + 48\,M^5n + 72\,M^5 - 56\,M^4n^2 - 168\,M^4n - 76\,M^4 + 32\,M^3n^3 + 140\,M^3n^2 + 88\,M^3n - 90\,M^3 - 9\,M^2n^4 - 50\,M^2n^3 - M^2n^2 + 260\,M^2n + 236\,M^2 + M\,n^5 + \frac{13\,M\,n^4}{2} - 20\,M\,n^3 - \frac{383\,M\,n^2}{2} - 344\,M\,n - 162\,M + 4\,n^4 + 36\,n^3 + 107\,n^2 + 117\,n + 36 + 16\,j^6 + 80\,j^5M - 48\,j^5n - 88\,j^5 + 160\,j^4M^2 - 192\,j^4M\,n - 336\,j^4M + 52\,j^4n^2 + 204\,j^4n + 156\,j^4 + 160\,j^3M^3 - 288\,j^3M^2n - 464\,j^3M^2 + 156\,j^3M\,n^2 + 564\,j^3M\,n + 364\,j^3M - 24\,j^3n^3 - 152\,j^3n^2 - 228\,j^3n - 50\,j^3 + 80\,j^2M^4 - 192\,j^2M^3n - 256\,j^2M^3 + 156\,j^2M^2n^2 + 468\,j^2M^2n + 156\,j^2M^2 - 48\,j^2M\,n^3 - 252\,j^2M\,n^2 - 204\,j^2M\,n + 160\,j^2M + 4\,j^2n^4 + 36\,j^2n^3 + 55\,j^2n^2 - 87\,j^2n - 136\,j^2 + 16\,j\,M^5 - 48\,j\,M^4n - 24\,j\,M^4 + 52\,j\,M^3n^2 + 60\,j\,M^3n - 156\,j\,M^3 - 24\,j\,M^2n^3 - 48\,j\,M^2n^2 + 276\,j\,M^2n + 470\,j\,M^2 + 4\,j\,M\,n^4 + 12\,j\,M\,n^3 - 149\,j\,M\,n^2 - 567\,j\,M\,n - 446\,j\,M + 24\,j\,n^3 + 152\,j\,n^2 + 276\,j\,n + 138\,j + 16\,M^5 - 48\,M^4n - 104\,M^4 + 52\,M^3n^2 + 252\,M^3n + 260\,M^3 - 24\,M^2n^3 - 204\,M^2n^2 - 480\,M^2n - 310\,M^2 + 4\,M\,n^4 + 60\,M\,n^3 + 259\,M\,n^2 + 393\,M\,n + 174\,M - 4\,n^4 - 36\,n^3 - 107\,n^2 - 117\,n - 36 = 0 \qquad \text{(S4.174)}$$

As can be seen, all the terms mutually cancel out. Therefore, the expression inside the sum is always zero, regardless the value of the index *j,* and of *n* and *M*. Then, the part with sums (S4.68) is zero as well.

The model function (S4.2) is thus a valid solution of the angular part of the Laplace equation in exterior in the SOS coordinates.

Then, also the assumption that the physical shape of SOS solution of Laplace equation in exterior space is the same as the solution in spherical coordinates is confirmed.